\documentclass[runningheads,onecollarge,natbib]{svjour2}
\usepackage{graphicx}
\usepackage{amssymb}
\newcommand\sun{\odot}
\journalname{Space Science Reviews}
\begin{document}

\title{The James Webb Space Telescope}

\author{
Jonathan P. Gardner$^1$ \and
John C.  Mather$^1$ \and
Mark Clampin$^2$ \and
Rene Doyon$^3$ \and
Matthew A.  Greenhouse$^1$ \and
Heidi B. Hammel$^4$ \and
John B. Hutchings$^5$ \and
Peter Jakobsen$^6$ \and
Simon J. Lilly$^7$ \and
Knox S. Long$^8$ \and
Jonathan I. Lunine$^9$ \and
Mark J. McCaughrean$^{10,11}$ \and
Matt Mountain$^8$ \and
John Nella$^{12}$ \and
George H. Rieke$^{13}$ \and
Marcia J. Rieke$^{13}$ \and
Hans-Walter Rix$^{14}$ \and
Eric P. Smith$^{15}$ \and
George Sonneborn$^1$ \and
Massimo Stiavelli$^8$ \and
H. S. Stockman$^8$ \and
Rogier A. Windhorst$^{16}$ \and
Gillian S. Wright$^{17}$
}

\institute{
\at $^1$Laboratory for Observational Cosmology, Code 665, Goddard Space Flight Center, Greenbelt MD 20771 USA
\at $^2$Laboratory for Exoplanet and Stellar Astrophysics, Code 667, Goddard Space Flight Center, Greenbelt MD 20771 USA 
\at $^3$Universit\'e de Montreal, Dept.\ de Physique, C. P. 6128 Succ. Centre-ville, Montreal QC H3C 3J7, Canada 
\at $^4$Space Science Institute, 3100 Marine Street, Suite A353, Boulder CO 80303 USA 
\at $^5$Herzberg Institute of Astrophysics, 5071 West Saanich Road, Victoria BC, V9E 2E7, Canada 
\at $^6$Astrophysics Division, RSSD, European Space Agency, ESTEC, 2200 AG Noordwijk, The Netherlands 
\at $^7$Department of Physics, Swiss Federal Institute of Technology (ETH-Zurich), ETH H\"{o}nggerberg, CH-8093 Zurich, Switzerland 
\at $^8$Space Telescope Science Institute, 3700 San Martin Dr., Baltimore MD 21218 USA 
\at $^9$Lunar and Planetary Laboratory, University of Arizona, Tucson AZ 85721 USA 
\at $^{10}$Astrophys.\ Institut Potsdam, An der Sternwarte 16, 14482 Potsdam, Germany 
\at $^{11}$University of Exeter, Physics, Stocker Road, Exeter EX4 4QL, United Kingdom 
\at $^{12}$Northrop Grumman Space Tech., 1 Space Park, Redondo Beach CA 90278 USA 
\at $^{13}$Steward Observatory, University of Arizona, 933 North Cherry Avenue, Tucson AZ 85721 USA 
\at $^{14}$Max-Planck-Institut f\"{u}r Astron., K\"{o}nigstuhl 17, Heidelberg D-69117, Germany 
\at $^{15}$NASA Headquarters, 300 E Street Southwest, Washington DC 20546 USA 
\at $^{16}$Department of Physics and Astronomy, Arizona State University, Box 871504, Tempe AZ 85287 USA 
\at $^{17}$Astronomy Technology Centre, Royal Observatory, Blackford Hill, Edinburgh EH9 3HJ, United Kingdom 
}

\date{Accepted for publication 15 May 2006.}
\authorrunning{Jonathan P. Gardner et al.}

\maketitle

\begin{abstract}
The James Webb Space Telescope (JWST) is a large (6.6m), cold
($<$50K), infrared-optimized space observatory that will be launched
early in the next decade into orbit around the second Earth-Sun
Lagrange point. The observatory will have four instruments: a
near-infrared camera, a near-infrared multi-object spectrograph,
and a tunable filter imager will cover the wavelength range, 0.6
$<$ $\lambda$ $<$ 5.0 $\mu$m, while the mid-infrared instrument
will do both imaging and spectroscopy from 5.0 $<$ $\lambda$ $<$
29 $\mu$m.

The JWST science goals are divided into four themes. The key
objective of The End of the Dark Ages: First Light and Reionization
theme is to identify the first luminous sources to form and to
determine the ionization history of the early universe. The key
objective of The Assembly of Galaxies theme is to determine how
galaxies and the dark matter, gas, stars, metals, morphological
structures, and active nuclei within them evolved from the epoch
of reionization to the present day. The key objective of The Birth
of Stars and Protoplanetary Systems theme is to unravel the birth
and early evolution of stars, from infall on to dust-enshrouded
protostars to the genesis of planetary systems. The key objective
of the Planetary Systems and the Origins of Life theme is to
determine the physical and chemical properties of planetary systems
including our own, and investigate the potential for the origins
of life in those systems. Within these themes and objectives we
have derived representative astronomical observations.

\noindent{\bf Abstract, cont.} To enable these observations, JWST consists of a telescope, an
instrument package, a spacecraft and a sunshield. The telescope
consists of 18 beryllium segments, some of which are deployed. The
segments will be brought into optical alignment on-orbit through
a process of periodic wavefront sensing and control. The instrument
package contains the four science instruments and a fine guidance
sensor. The spacecraft provides pointing, orbit maintenance and
communications. The sunshield provides passive thermal control. 
The JWST operations plan is based on that used for previous space
observatories, and the majority of JWST observing time will be
allocated to the international astronomical community through annual
peer-reviewed proposal opportunities.

\end{abstract}


\vspace{0.5in}

\noindent {\bf Table of Contents}

\noindent 1. Introduction

1.1 The End of the Dark Ages: First Light and Reionization

1.2 The Assembly of Galaxies.

1.3 The Birth of Stars and Protoplanetary Systems.

1.4 Planetary Systems and the Origins of Life.

\noindent 2. The End of the Dark Ages: First Light and Reionization

2.1 What Are the First Galaxies?

2.2 When and How Did Reionization Occur?

2.3 What Sources Caused Reionization?

2.4 Summary

\noindent 3. The Assembly of Galaxies

3.1 Previous Investigations

3.2 When and How Did the Hubble Sequence Form?

3.3 How did the Heavy Elements Form?

3.4 What Physical Processes Determine Galaxy Properties?

3.5 What Roles do Starbursts and Black Holes Play in Galaxy Evolution?

3.6 Summary

\noindent 4. The Birth of Stars and Protoplanetary Systems

4.1 How do Protostellar Clouds Collapse?

4.2 How Does Environment Affect Star Formation and Vice Versa?

4.3 What is the Initial Mass Function at sub-stellar Masses?

4.4 How do Protoplanetary Systems Form?

4.5 What are the Life Cycles of Gas and Dust?

4.6 Summary

\noindent 5. Planetary Systems and the Origins of Life

5.1 How Do Planets Form?

5.2 How Are Circumstellar Disks Like Our Solar System?

5.3 How Are Habitable Zones Established?

5.4 Summary

\noindent 6. JWST Implementation

6.1 Observatory

6.2 Observatory Performance

6.3 Observatory Design Description

6.4 Integration and Test

6.5 Instrumentation

6.6 Launch, Orbit and Commissioning

6.7 Operations

6.8 Management

\noindent 7. Summary

\clearpage

\begin{table}[t]
\caption{Table of Acronyms \label{tabacro}}
\begin{tabular}{ll}
\hline\noalign{\smallskip}
Acronym & Definition
\\[3pt]
\tableheadseprule\noalign{\smallskip}
AB & Absolute Bolometric \\
AGN & Active Galactic Nuclei \\
AI\&T & Assembly, Integration and Test \\
ALMA & Atacama Large Millimeter Array \\
AMSD & Advanced Mirror System Demonstrator \\
AU & Astronomical Unit \\
AURA & Association of Universities for Research in Astronomy \\
CDM & Cold Dark Matter \\
CMB & Cosmic Microwave Background \\
CSA & Canadian Space Agency \\
DHS & Dispersed Hartmann Sensor \\
DWS & Deep-Wide Survey \\
DSN & Deep-Space Network \\
EGG & Evaporating Gaseous Globule \\
ESA & European Space Agency \\
ESO & European Southern Observatory \\
EW & Equivalent Width \\
FGS & Fine Guidance Sensor \\
FOR & Field of Regard \\
FORS1 & Focal Reducer and Low Dispersion Spectrograph 1\\
FOV & Field of View \\
FPA & Focal-Plane Array \\
FSM & Fine Steering Mirror \\
FUSE & Far-Ultraviolet Spectroscopic Explorer \\
GOODS & Great Observatories Origins Deep Survey \\
GSC & Guide-Star Catalog \\
GSFC & Goddard Space Flight Center \\
HDF & Hubble Deep Field \\
HgCdTe & Mercury Cadmium Telluride \\
HST & Hubble Space Telescope \\
IFU & Integral Field Unit \\
IGM & Intergalactic Medium \\
IMF & Initial Mass Function \\
IR & Infrared \\
IRAS & Infrared Astronomical Satellite \\
ISIM & Integrated Science Instrument Module \\
ISM & Interstellar Medium \\
ISOCAM & Infrared Space Observatory Camera \\
ISO LWS & Infrared Space Observatory Long Wavelength Spectrograph \\
ISO SWS & Infrared Space Observatory Short Wavelength Spectrograph \\
JPL & Jet Propulsion Laboratory \\
JWST & James Webb Space Telescope \\
KBO & Kuiper Belt Object \\
L2 & Second Lagrange Point \\
LF & Luminosity Function \\
LMC & Large Magellanic Cloud \\
LRS & Low-Resolution Spectrograph \\
MIRI & Mid-Infrared Instrument \\
MSA & Micro-shutter Assembly \\
MJy/sr & Mega-Jansky per sterradian \\
NASA & National Aeronautics and Space Agency \\
NGST & Northrop Grumman Space Technology \\
NIRCam & Near-Infrared Camera \\
NIRSpec & Near-Infrared Spectrograph \\
nJy & nano-Jansky \\
\noalign{\smallskip}\hline
\end{tabular}
\end{table}

\clearpage

\begin{table}[t]
\begin{tabular}{ll}
\hline\noalign{\smallskip}
Acronym & Definition
\\[3pt]
\tableheadseprule\noalign{\smallskip}
NTT & New Technology Telescope \\
OPR & Ortho-Para Ratio \\
OS & Operating System \\
OTA & Optical Telescope Assembly \\
OTE & Optical Telescope Element \\
PAH & Polycyclic Aromatic Hydrocarbon \\
PSF & Point-Spread Function \\
PM & Primary Mirror \\
PMSA & Primary Mirror Segment Assembly \\
QA & Quality Assurance \\
QPM & Quarter Phase Mask \\
QSO & Quasi-Stellar Object \\
rms & Root Mean Squared \\
RoC & Radius of Curvature \\
SDSS & Sloan Digital Sky Survey \\
SED & Spectral Energy Distribution \\
SFR & Star-Formation Rate \\
SI & Science Instrument \\
SM & Secondary Mirror \\
SMC & Small Magellanic Cloud \\
SN & Supernova \\
SNe & Supernovae \\
SOFI & Son of ISAAC (Infrared Spectrograph and Imaging Camera) \\
SSM & Space Support Module \\
STIS & Space Telescope Imaging Spectrograph \\
STScI & Space Telescope Science Institute \\
SWG & Science Working Group \\
S\&OC & Science and Operations Center \\
TAC & Time Allocation Committee \\
TBD & To Be Determined \\
TFI & Tunable Filter Imager \\
TMA & Three Mirror Anastigmat \\
UDF & Ultra-Deep Field \\
UDS & Ultra-Deep Survey \\
ULIRG & Ultra-Luminous Infrared Galaxy \\
UV & Ultraviolet \\
VLA & Very Large Array \\
VLT & Very Large Telescope \\
WFPC2 & Wide-Field Planetary Camera 2 \\
WFE & Wavefront Error \\
WFS\&C & Wavefront Sensing and Control \\
WMAP & Wilkinson Microwave Anisotropy Probe \\
YSO & Young Stellar Object\\
\noalign{\smallskip}\hline
\end{tabular}
\end{table}

\clearpage

\section{Introduction}

The James Webb Space Telescope (JWST; table~\ref{tabacro}) will be
a large, cold, infrared-optimized space telescope designed to enable
fundamental breakthroughs in our understanding of the formation
and evolution of galaxies, stars, and planetary systems. It is a
project led by the United States National Aeronautics and Space
Administration (NASA), with major contributions from the European
and Canadian Space Agencies (ESA and CSA). It will have an
approximately 6.6 m diameter aperture, will be passively cooled to
below 50 K, and will carry 4 scientific instruments: a Near Infrared
Camera (NIRCam), a Near Infrared Spectrograph (NIRSpec), a
near-infrared Tunable Filter Imager (TFI), and a Mid Infrared
Instrument (MIRI). It is planned for launch early in the next decade
on an Ariane 5 rocket to a deep space orbit around the Sun-Earth
Lagrange point L2, about 1.5 $\times$ 10$^{6}$ km from Earth. The 
spacecraft will carry enough fuel for a 10-year mission.

In this paper, we describe the scientific capabilities and planned 
implementation of JWST. The scientific planning is based on current 
theoretical understanding, interpretation of Hubble Space Telescope 
(HST) and Spitzer Space Telescope observations (e.g., Werner et 
al. 2004 and additional papers in that volume), and results from 
studies using ground-based and other space-based facilities. 
Many classes of targets for JWST have never been observed before, 
so the scientific plans are necessarily derived from theoretical 
predictions and extrapolations from known objects. In these cases, 
we outline the basis of the predictions and describe the observatory 
capabilities needed to verify them. 

Additional results from HST and Spitzer, and other advances in
theory and observation will continue to refine the observational
plans for JWST. The scientific investigations we describe here
define the measurement capabilities of the telescope, but they do
not imply that those particular observations will be made. In this
paper we do not list all potential applications of JWST. Instead,
the scientific programs we discuss here are used to determine the
key parameters of the mission such as collecting area, spatial and
spectral resolution, wavelength coverage, etc. A mission which
provides these capabilities will support a wide variety of
astrophysical investigations. JWST is a facility-class mission, so
most of the observing time will be allocated to investigators from
the international astronomical community through competitively-selected
proposals.

The plans for JWST reflect scientific and engineering discussions, 
studies, and development over the last 16 years. At a workshop 
held in 1989, the astronomical community began discussions of 
a scientific successor to HST (B\'ely, Burrows \& Illingworth, 
1989). In the mid-1990s, the ``HST and Beyond'' committee recommended 
that NASA build an infrared-optimized telescope to extend HST 
discoveries to higher redshift and longer wavelength (Dressler 
1996). Initial studies of the mission, then called the Next Generation 
Space Telescope, were reported by Stockman et al. (1997). Its 
scientific program was given top priority by the National Academy 
of Sciences survey ``Astronomy and Astrophysics in the New Millennium'' 
(McKee \& Taylor 2001). In 2002, the Next Generation Space Telescope 
was renamed the James Webb Space Telescope in honor of the Administrator 
of NASA during the Apollo era (Lambright 1995). This paper provides 
an update to these previous studies.

The capabilities and performance specifications for JWST given here 
are preliminary. The mission is currently in its detailed 
design phase and has not yet been given official permission 
to proceed beyond that stage. NASA missions receive authority, 
and the accompanying budget, to proceed to launch only after 
they pass a nonadvocate review and transition from detailed 
design into development phases. Hence, the specifications 
given in this paper are subject to change based upon 
technical progress of individual elements of the observatory 
and available budgets. JWST is expected to receive final approval
and make the transition to the development phase in 2008.

The scientific objectives of JWST fall into four broad themes: 
The End of the Dark Ages: First Light and Reionization; The Assembly 
of Galaxies; The Birth of Stars and Protoplanetary Systems; and 
Planetary Systems and the Origins of Life.

We organize this paper as follows. The remainder of this section 
provides a scientific introduction to these themes. In sections 
2, 3, 4 and 5, we expand on the themes, and describe the scientific 
capabilities JWST will use to address them. In section 6 we describe 
the planned implementation of JWST.

\subsection{The End of the Dark Ages: First Light and Reionization}

Theory and observation have given us a simple picture of the 
early universe. The Big Bang produced (in decreasing order of 
present mass-energy density): dark energy (the cosmic acceleration 
force), dark matter, hydrogen, helium, cosmic microwave and neutrino 
background radiation, and trace quantities of lithium, beryllium, 
and boron. As the universe expanded and cooled, hydrogen molecules 
formed, enabling the formation of the first individual stars, 
at about 180 million years after the Big Bang (Barkana \& Loeb, 
2001). According to theory, the first stars were 30 to 300 times 
as massive as the Sun and millions of times as bright, burning 
for only a few million years before meeting a violent end (Bromm 
\& Larson 2004). Each one produced either a core-collapse supernova 
(type II) or a black hole. The supernovae enriched the surrounding 
gas with the chemical elements produced in their interiors, and 
future generations of stars contained these heavier elements. 
The black holes started to swallow gas and other stars to become 
mini-quasars, which grew and merged to become the huge black 
holes now found at the centers of nearly all massive galaxies 
(Magorrian et al. 1998). The supernovae and the mini-quasars 
could be individually observable by JWST.

Some time after the appearance of the first sources of light, 
hydrogen in the intergalactic medium was reionized. Results from 
the Wilkinson Microwave Anisotropy Probe (WMAP; Kogut et al. 
2003; Page et al. 2006; Spergel et al. 2006) combined with data 
on quasars at z \ensuremath{\sim} 6 from the Sloan 
Digital Sky Survey (SDSS; Fan et al. 2002) show that this reionization 
had a complex history 
(Cen 2003b). Although there are indications that galaxies produced 
the majority of the ultraviolet radiation which caused the reionization, 
the contribution of quasars could be significant.

JWST will address several key questions in this theme: What are 
the first galaxies? When and how did reionization occur? What 
sources caused reionization? JWST will conduct ultra-deep 
near-infrared surveys with spectroscopic 
and mid-infrared follow-up to find and identify the first galaxies to 
form in the early universe. It will determine the processes that 
caused reionization through spectroscopy of high-redshift quasars or 
galaxies, studies of the properties of galaxies during that epoch.

\subsection{The Assembly of Galaxies.}

Theory predicts that galaxies are assembled through a process 
of the hierarchical merging of dark matter concentrations (e.g., 
White \& Frenk 1991; Cole et al. 1994). Small objects formed 
first, and were drawn together to form larger ones. This dynamical 
build-up of massive systems is accompanied by chemical evolution, 
as the gas (and dust) within the galaxies are processed through 
successive generations of stars. The interaction of these luminous 
components with the invisible dark matter produces the diverse 
properties of present-day galaxies, organized into the Hubble 
Sequence of galaxies. This galaxy assembly process is still occurring 
today, as the Magellanic Clouds fall into the Milky Way, and 
as the Andromeda Nebula heads toward the Milky Way for a possible 
future collision. To date, galaxies have been observed back to 
times about one billion years after the Big Bang. While most 
of these early galaxies are smaller and more irregular than present-day 
galaxies, some early galaxies are very similar to those seen 
nearby today.

Despite all the work done to date, many questions are still open. 
We do not really know how galaxies are formed, what controls 
their shapes, and what makes them form stars. We do not know 
how the chemical elements are generated and redistributed through 
the galaxies, and whether the central black holes exert great 
influence over the galaxies. We do not know the global effects 
of violent events as small and large parts join together in collisions.

JWST will address several key questions in this theme: When and
how did the Hubble Sequence form? How did the heavy elements form?
What physical processes determine galaxy properties? What are the
roles of starbursts and black holes in galaxy evolution? To answer
these questions, JWST will observe galaxies back to their earliest
precursors, so that we can understand their growth and evolution.
JWST will conduct deep-wide imaging and spectroscopic surveys 
of thousands of galaxies to study morphology, composition and 
the effects of environment. It will conduct detailed studies 
of individual ultra-luminous
infrared galaxies (ULIRGs) and active galactic nuclei (AGN) 
to investigate what powers these energetic sources.

\subsection{The Birth of Stars and Protoplanetary Systems.}

While stars have been the main topic of astronomy for thousands 
of years, only in recent times have we begun to understand them 
with detailed observations and computer simulations. A hundred 
years ago, we did not know that stars are powered by nuclear 
fusion, and 50 years ago we did not know that stars are continually 
being formed. We still do not know the details of how stars are 
formed from clouds of gas and dust, why most stars form in groups, 
or how planetary systems form. Young stars within a star-forming 
region interact with each other chemically, dynamically and radiatively 
in complex ways. The details of how they evolve and liberate 
the ``metals'' back into space for recycling into new generations 
of stars and planets remains to be determined through a combination 
of observation and theory.

We also know that a substantial fraction of stars, solar-type 
and later, have gas-giant planets, although the discovery of 
large numbers of these planets in very close orbits around their 
stars was a surprise. The development of a full theory of planet 
formation requires substantially more observational input, including 
observations of young circumstellar disks and older debris disks 
in which the presence of planets can be traced.

JWST will address several key questions in this theme: How do 
protostellar clouds collapse? How does environment affect star 
formation and vice versa? What is the initial mass function of 
stars at sub-stellar masses? How do proto-planetary systems form? 
How do gas and dust coalesce to form planetary systems?
JWST will observe stars at all phases of their evolution, 
from infall onto dust-enshrouded 
protostars, through the formation of planetary systems,  
penetrating the dust to determine the physical processes that 
produce stars, planets and debris disks.

\subsection{Planetary Systems and the Origins of Life.}

Understanding the origin of the Earth and its ability to support 
life is an important objective for astronomy. Key parts of the 
story include understanding the formation of planetesimals, and 
how they combine to form larger objects. We do not know how planets 
reach their present orbits, and how the large planets affect 
the smaller ones in solar systems like our own. We want to learn 
about the chemical and physical history of the small and large 
objects that formed the Earth and delivered the necessary chemical 
precursors for life. The cool objects and dust in the outer Solar 
System are evidence of conditions in the early Solar System, 
and are directly comparable to cool objects and dust observed 
around other stars.

JWST will address several key questions in this theme: How do 
planets form? How are circumstellar disks like our Solar System? 
How are habitable zones established? JWST will determine the 
physical and chemical properties of planetary systems including 
our own, and investigate the potential for the origins of life 
in those systems. JWST will use coronagraphy to investigate exosolar planets 
and debris disks, and will compare these observations to 
objects within our own Solar System.

\section{The End of the Dark Ages: First Light and Reionization}

The key objective of The End of the Dark Ages: First Light and 
Reionization theme is to identify the first luminous sources 
to form and to determine the ionization history of the early 
universe.

The emergence of the first individual sources of light in the universe marks 
the end of the ``Dark Ages'' in cosmic history, a 
period characterized by the absence of discrete sources of light 
(Rees 1997). Understanding these first sources is critical, since 
they greatly influenced subsequent structures. The current leading 
models for structure formation predict a hierarchical assembly 
of galaxies and clusters. The first sources of light act as seeds 
for the successive formation of larger objects, and by studying 
these objects we will learn the processes that formed the nuclei 
of present day giant galaxies.

This epoch is currently under intense theoretical investigation. 
The formation of structure in the Dark Ages is easier to study 
theoretically than similar processes occurring at other epochs 
because: i) the formation of the first structures is directly 
linked to the growth of linear perturbations, and ii) these objects 
have known elemental abundances set by the end-product of the 
primordial nucleosynthesis. By studying this epoch, it is possible 
to probe the power spectrum of density fluctuations emerging 
from recombination at scales smaller than those accessible by 
current cosmic microwave background experiments.

Some time after the appearance of the first sources of light, 
hydrogen in the universe is reionized. We do not know the time 
lag between these two events, nor whether reionization is brought 
about by the first light sources themselves or by subsequent 
generations of objects. Reionization is by itself a period in 
cosmic history that is as interesting as the emergence of first 
light. The epoch of reionization is the most recent global phase 
transition undergone by the universe after the Big Bang.

Before the epoch of first light, ionizing photons and metals 
are essentially absent. Thus, hydrogen molecules can form and 
survive to become the primary cooling agent of the first perturbation 
to collapse. We expect the stars that formed from this process 
to be very massive and very hot. Historically, this primordial 
stellar population has been given the name of population III 
stars. The most overdense peaks in the perturbations emerging 
from recombination will collapse first. Hierarchical-clustering 
models predict that these overdense peaks sit in larger overdense 
regions that have larger mass but lower contrast. We may expect 
the first stars to be markers of, and possibly reside in, star 
clusters or even small galaxies.

At the end of their short lives, some very massive first stars 
will leave black holes as remnants. These black holes will begin 
accreting gas and form mini-AGN. In more exotic models, black 
holes can form directly from the collapse of perturbations instead
of from stellar remnants. Thus, in these models the first sources
of light would be powered by gravitational accretion, rather than
by nuclear fusion.

Soon after the first light sources appear, both high-mass stars 
and accretion onto black holes become viable sources of ionizing
radiation. We do not know which are primarily responsible for
reionizing hydrogen in the surrounding intergalactic medium (IGM).
AGN produce a highly energetic synchrotron power spectrum, and
would reionize helium as well as hydrogen. Because observational
evidence reveals that helium is reionized at a much later time,
hydrogen was probably reionized by starlight at earlier epochs.
However, it is possible that helium recombines after being reionized
for the first time together with hydrogen. A second reionization
of helium would occur during the epoch when quasar activity peaks.
More recently, a combination of observations by the WMAP of the
cosmic microwave background polarization (Kogut et al. 2003; 
Page et al. 2006; Spergel et al. 2006) with
spectra of z $>$ 6 quasars found by the Sloan Digital Sky Survey
(SDSS) (Fan et al. 2001; 2002) has revealed the possibility that
there were two reionization epochs for hydrogen (Cen 2003a; 2003b).
In these models, the completion of the reionization epoch that is
seen at z $\sim$ 6 would be that of the second reionization, with
the first reionization taking place during or after the peak of
the first light epoch at higher redshifts. Although the observations
allow for other possibilities, in general, there is evidence that
the reionization history of the universe was complex (e.g., Gnedin
2004).

\subsection{What Are the First Galaxies?}

When did the first luminous sources arise and what was their 
nature? What were their clustering properties?

\begin{figure*}
\centering
\includegraphics[width=1.00\textwidth]{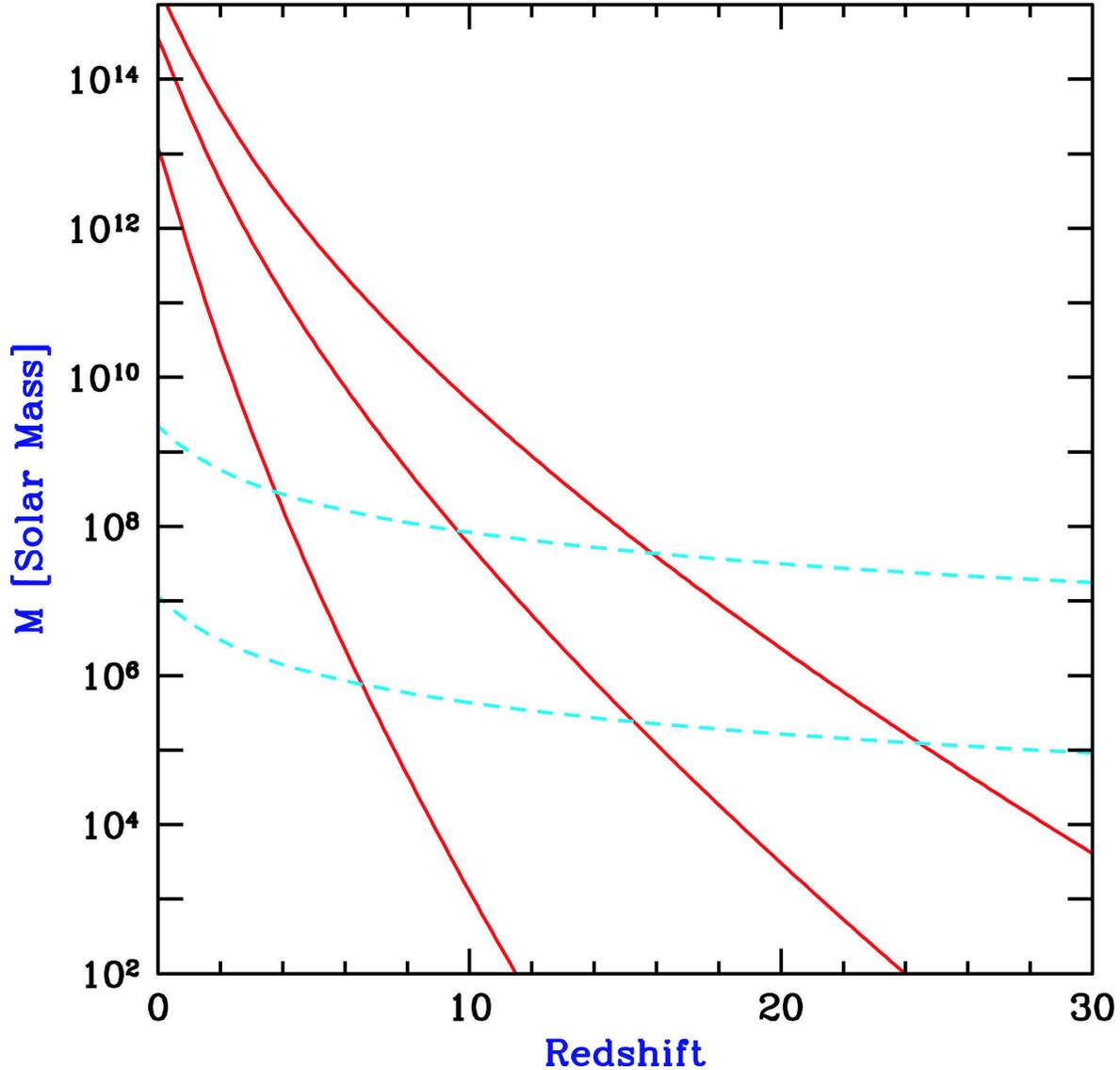}
\caption{
The mass of collapsing dark matter halos
in the early universe. The red solid curves show the mass of collapsing
halos corresponding to 1, 2 and 3$\sigma $ fluctuations (in order from
bottom to top.) The blue dashed curves show the mass corresponding to the
minimum temperature required for efficient cooling with primordial atomic
species only (upper curve) or with the addition of molecular hydrogen (lower
curve). The intersection of these lines indicate that the epoch of formation
for the first galaxies is likely to be 10 $<$ z $<$ 20 (From Barkana \& Loeb 2001).}
\label{fig001}
\end{figure*}

In standard Cold Dark Matter (CDM) cosmology, galaxies are assembled 
through hierarchical merging of building blocks with smaller 
mass. The first such building blocks, with M \ensuremath{\geq} 10$^{4}$ M$_{\ensuremath{\sun}}$ 
form in these models at z \ensuremath{\gtrsim} 15 (see Fig. \ref{fig001}; Barkana 
\& Loeb 2001; Couchman \& Rees 1986; Haiman \& Loeb 1997; Ostriker 
\& Gnedin 1996; Haiman, Thoul \& Loeb 1996; Abel et al. 1998; 
Abel et al. 2000).

While we do not know whether the first sources of light are powered 
by nuclear energy from fusion reactions in stars, or by gravitational 
accretion (Haiman \& Loeb 1999), it is possible that population 
III stars are responsible for the reionization of hydrogen at 
z \texttt{>} 6 (Madau \& Shull 1996; see also Haiman \& Loeb 1999; 
Gnedin \& Ostriker 1997; Chiu \& Ostriker 2000). Efficient cooling 
by H$_{2}$ molecules and an early, vigorous formation of massive 
objects could result in reionization at redshifts as early as 
z \ensuremath{\sim} 20 (Cen 2003b; Haiman \& Holder 2003).

The very first stars (population III) have zero metallicity. 
In the absence of any metals, cooling is dominated by the less 
effective H$_{2}$ cooling process, which leads to the formation 
of very massive objects, with masses exceeding 100 M$_{\ensuremath{\sun}}$ 
(Bromm et al. 1999; Bromm et al. 2002) and possibly going as 
high as 500 M$_{\ensuremath{\sun}}$. The spectral energy distribution (SED) 
of these massive stars resembles a black body with an effective 
temperature around 10$^{5}$ K (Bromm et al. 2001). Due to their 
high temperatures, these stars are very effective at ionizing 
both hydrogen and helium. It should be noted that, even at lower 
mass, zero-metallicity stars are expected to be much hotter than 
their solar metallicity analogs (Tumlinson \& Shull 2000).

Two consequences of the high effective temperature of 
zero-metallicity stars are their effectiveness in ionizing hydrogen 
(and helium) and their low optical-to-UV fluxes. Both tend to 
make the direct detection of the stellar continuum much harder 
than the detection of the associated HII region. In the surrounding 
HII region, electron temperatures exceed 20,000 K and 45\% of 
the total luminosity is emitted through the Lyman \ensuremath{\alpha} 
line, resulting in a Lyman \ensuremath{\alpha} equivalent width (EW) of 
3000 {\AA} (Bromm et al. 2001). The helium lines are also 
strong, with the intensity of HeII \ensuremath{\lambda}1640 comparable 
to that of H\ensuremath{\beta} (Panagia et al. 2003; Tumlinson et al. 
2001).

An interesting feature of these models is that the HII region
emission longward of Lyman \ensuremath{\alpha} is dominated by a
strong two-photon nebular continuum. The
H\ensuremath{\alpha}/H\ensuremath{\beta} ratio for these models is
3.2. Both the red continuum and the high
H\ensuremath{\alpha}/H\ensuremath{\beta} ratio could be incorrectly
interpreted as a consequence of dust extinction, even though no
dust is present in these systems.

By estimating the brightness of the sources that enriched the IGM
to 10$^{-2}$ Z$_{\sun}$ (Miralda-Escud\'{e} \& Rees 1998), one
finds that a combined surface brightness of about AB = 32 mag
arcsec$^{-2}$ is needed. (The AB magnitude system is defined to be
AB = 31.4 - 2.5log(f$_{\ensuremath{\nu}}$), where f$_{\ensuremath{\nu}}$
is in nJy, Oke 1974.) This surface brightness is about 2 orders of
magnitude brighter than the surface brightness derived below for
reionization (see Section 2.2). For reasonable luminosity
functions, these sources would be either detected directly,
or by exploiting amplification by gravitational lensing from an
intervening cluster of galaxies. Their large number offers the
promising prospect of identifying first light by observing a decrease
in the number of sources seen at increasing redshifts (after properly
accounting for the effects of sample completeness.)

The deepest images of the universe include the Hubble Ultra-Deep
Field in the optical (Beckwith et al. 2003), which reaches AB =
29.0 mag in the I band, HST near-infrared images of the UDF, which
reach AB = 28.5 in the J and H bands (Bouwens et al. 2005a), and
the Spitzer Great Observatories Origins Deep Survey (Dickinson
2004), which reaches AB = 26.6 mag at 3.6 $\mu$m. Galaxies are
detected in these observations at 6 $<$ z $<$ 7 (e.g., Yan et al.
2005) with potential candidates at even higher redshift. The
rest-frame ultra-violet (UV) luminosity function of z $\sim$ 6
galaxies is intrinsically fainter than that at z $\sim$ 3 (Dickinson
et al. 2004; Bouwens et al. 2005b), showing that the global
star-formation rate is climbing. However, the detection of galaxies
with stellar populations as old as 400 to 500 Myr at z $\sim$ 6.5
(Egami et al. 2005; Mobasher et al. 2005; Eyles et al. 2005) indicate
that the first galaxies formed much earlier, perhaps in the range
7.5 $<$ z $<$ 13.5. The Spitzer detections point to the importance
of using mid-infrared observations for galaxy age determinations
through stellar population model fitting.

The number of supernovae expected before reionization also strongly
depends on the assumptions made about the nature of the ionizing
sources. Based on relatively normal stellar populations and a
metallicity of 5 \ensuremath{\times} 10$^{-3}$ Z$_{\sun}$ at the
end of reionization, one arrives at an estimate of about one SN
arcmin$^{-2}$ at any give time (Miralda-Escude \& Rees 1997).
However, if the ionizers are very massive population III stars and
the metallicity at the end of reionization is lower than 5
\ensuremath{\times} 10$^{-3}$ Z$_{\sun}$, the SN rate would be one
hundred, or even one thousand times smaller. SNe with very massive
population III progenitors could be much brighter than regular type
II SNe.

\subsubsection{Observations}

The direct detection of an individual population III star is not
feasible even for JWST, as a 1000 M$_{\sun}$ star at z=30 would
have an AB magnitude of \ensuremath{\sim} 36. However, JWST can
detect super star clusters or dwarf galaxies made of population
III stars, as well as supernovae with population III progenitors.
In order to directly detect these first luminous objects and to
identify the redshift when they appear, we need to study the
evolution of the number density of objects N(z), and the evolution
of the star formation rate (SFR(z)) as a function of redshift. A
complementary method is to study the evolution of the mean metallicity
of galaxies \ensuremath{\langle}Z\ensuremath{\rangle}(z). Once
candidate first light objects are identified, JWST will study them
in detail to confirm their nature.

Number evolution: There will be no objects more distant than 
the first objects that formed and so N(z) will reach zero beyond 
the redshift of formation of the first sources. A strong upper 
limit on the number density of objects at redshifts greater than 
that of the most distant object observed is a likely indication 
that first light objects were detected.

Evolution of the star-formation rate: In addition to using 
ultra-violet emission as an indicator of star formation in galaxies,
one can determine the star formation rate as a function of redshift 
by measuring the SN rate.

Metallicity evolution: The metallicity of first light objects 
should be zero, while non-zero metallicity would indicate that 
the object formed from gas that had already been enriched. For 
the brightest objects, JWST will be able to obtain spectra. At 
low metallicity, the ratios of oxygen lines to Balmer lines, 
such as [OIII]/H\ensuremath{\beta}, are a linear measure of metallicity.

Confirmation: A small sample of candidate first light objects 
will be studied in detail, in order to place strong upper limits 
on their metal content and to prove the absence of an older stellar 
population by measuring their optical rest-frame spectral energy 
distribution (SED). Alternatively, identifying the age of an 
older stellar population sets a lower limit to the redshift of 
the first star formation.

JWST will need two observing programs to make these measurements: 
an ultra-deep imaging survey, and in-depth follow-up of candidate 
high-redshift sources with low-resolution spectroscopy and mid-infrared 
photometry. The depth of the ultra-deep survey will be built 
up in several epochs, so that supernovae can be identified for 
subsequent observations. A wider survey, described in the next 
section, could also use multiple epochs to search for supernovae.

\paragraph{Ultra-Deep Imaging Survey}

\begin{figure*}
\centering
\includegraphics[width=1.00\textwidth]{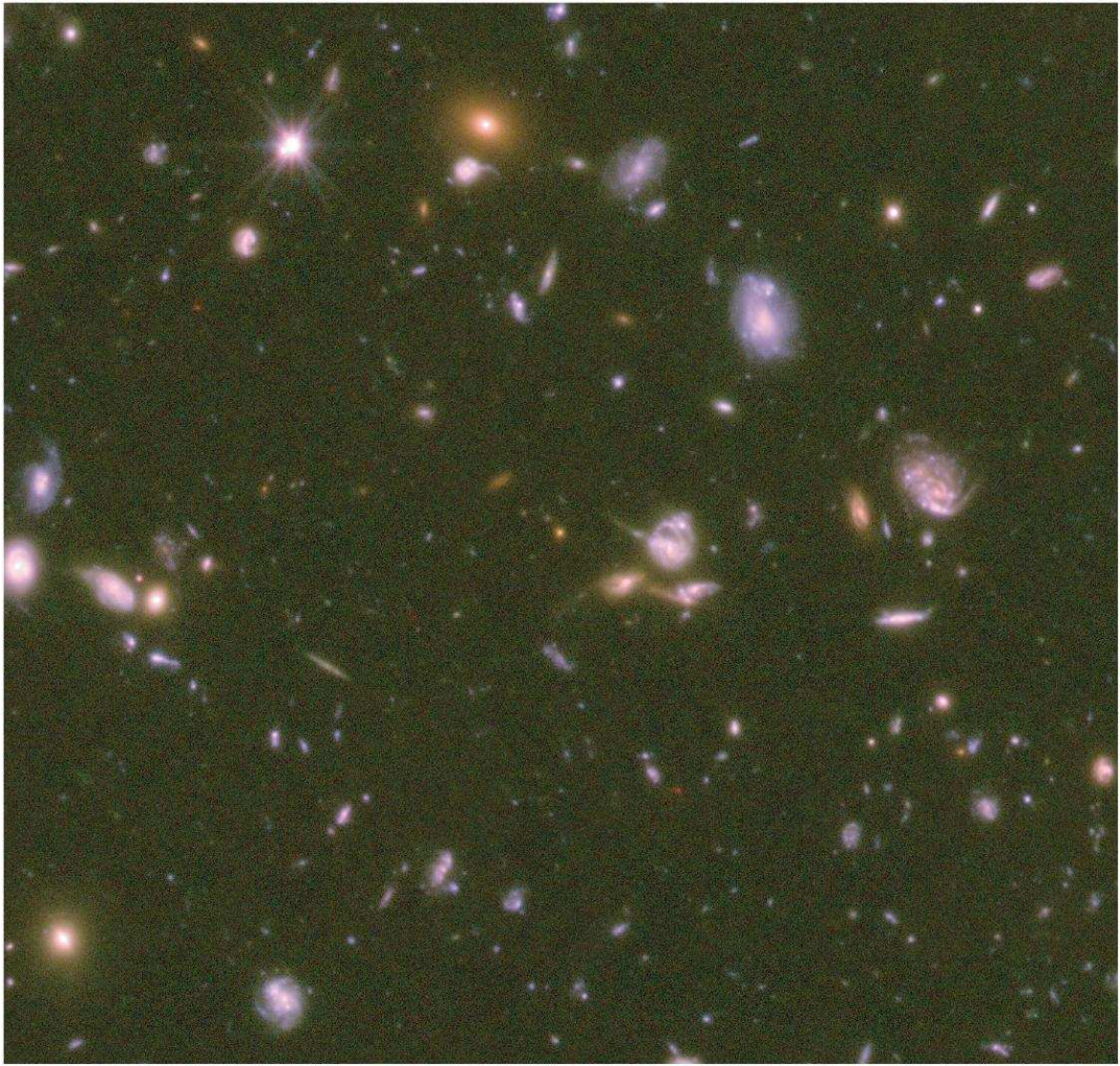}
\caption{
A simulated JWST galaxy field. The three colors correspond to 0.7,
0.9 and 2.0 microns. The Hubble Ultra-Deep Field (UDF) was imaged with the
Advanced Camera for Surveys onboard HST. The UDF will probably be the
deepest survey before JWST. For this simulation, we have taken the HST UDF
and convolved it with the JWST point-spread function and scaled it to a 20
hour exposure. (From Cohen et al., in preparation).}
\label{fig002}
\end{figure*}

To identify a sample of high redshift galaxies, JWST will make 
an ultra-deep imaging survey using several broad-band filters 
(Fig. \ref{fig002}). The Lyman break technique will identify objects at 
increasing redshifts up to z = 20 or higher. For dwarf galaxies 
with 10$^{6}$ M$_{\ensuremath{\sun}}$ of zero-metallicity massive stars at 
z $\sim$ 20, the expected AB magnitude at emitted wavelengths 
just longward of Lyman \ensuremath{\alpha} is \ensuremath{\sim}31 mag. A similar 
value (AB = 31 mag) is obtained by redshifting the brightest 
local super star clusters to z = 20. To enable this survey, JWST 
will have the sensitivity to reach AB = 31 mag in a feasible 
(although long) exposure time, about 100 to 200 hours exposure per filter, 
depending on the signal-to-noise ratio needed. The expected 
number densities are about 1 object arcmin$^{-2}$, thus a significant 
sample requires deep observations over several 10 arcmin$^{2}$. 
Another driver for a large area is that it is necessary to cover 
a volume of at least 50 Mpc on the side in order to average over 
cosmic structures. Such a volume is obtained at z = 15 over a \ensuremath{\Delta}z 
= 3 for an area of 35 arcmin$^{2}$. By focusing on volumes with \ensuremath{\Delta}z/z 
= 0.2, one obtains roughly the same comoving volume per unit 
redshift at all redshifts z \texttt{>} 5.

The sample will allow the derivation of N(z). The expected density
of first light sources is much lower than the density of sources
needed to enrich the metals in the IGM so that one should be able
to see a drop in number counts. The intensity of the non-ionizing
continuum can be calibrated to star formation rate (SFR) to yield
SFR(z). The required observations are deep broad-band imaging in
the near infrared, with mid-infrared follow-up observations. One
practical way to carry out this survey is to reach the depth of AB
= 31 mag for one field and integrate only to the depth of AB = 30
mag for an additional three fields. This survey could be done with
a total of 2 to 4 \ensuremath{\times} 10$^{6}$s exposure time,
depending on the number of filters required.

\paragraph{In-Depth Study of First Light Sources}

The brightest first light source candidates (or those amplified 
by intervening gravitational lenses) will be suitable for more 
detailed follow-ups. Near-infrared spectroscopy at R = 100 will 
be needed to verify the photometric redshifts. This will only 
be possible at a limit much brighter than that of the deep imaging. 
Observations at rest-frame wavelengths longer than 0.4 \ensuremath{\mu}m 
(i.e., at observer's wavelengths up to 8.4 \ensuremath{\mu}m for z \texttt{<} 
20) will establish the absence of an older generation of stars, 
confirming the nature of the sources as first generation objects. 
Spectroscopic follow up at R = 1000 aimed at measuring the Balmer 
line intensities will provide star formation rates and estimates 
of the dust content. Metal lines can be used to derive metallicities 
and the mean metallicity as a function of redshift.

This program combines deep near-infrared spectroscopy and deep 
mid-infrared imaging, using total integration times of up to 
\ensuremath{\sim} 10$^{6}$ s. It is possible that in order to 
achieve the required signal-to-noise ratio it will 
be necessary to exploit the gravitational lensing effect of a 
cluster of galaxies.

\paragraph{Supernova Search in Galaxy Surveys}

\begin{figure*}
\centering
\includegraphics[width=1.00\textwidth]{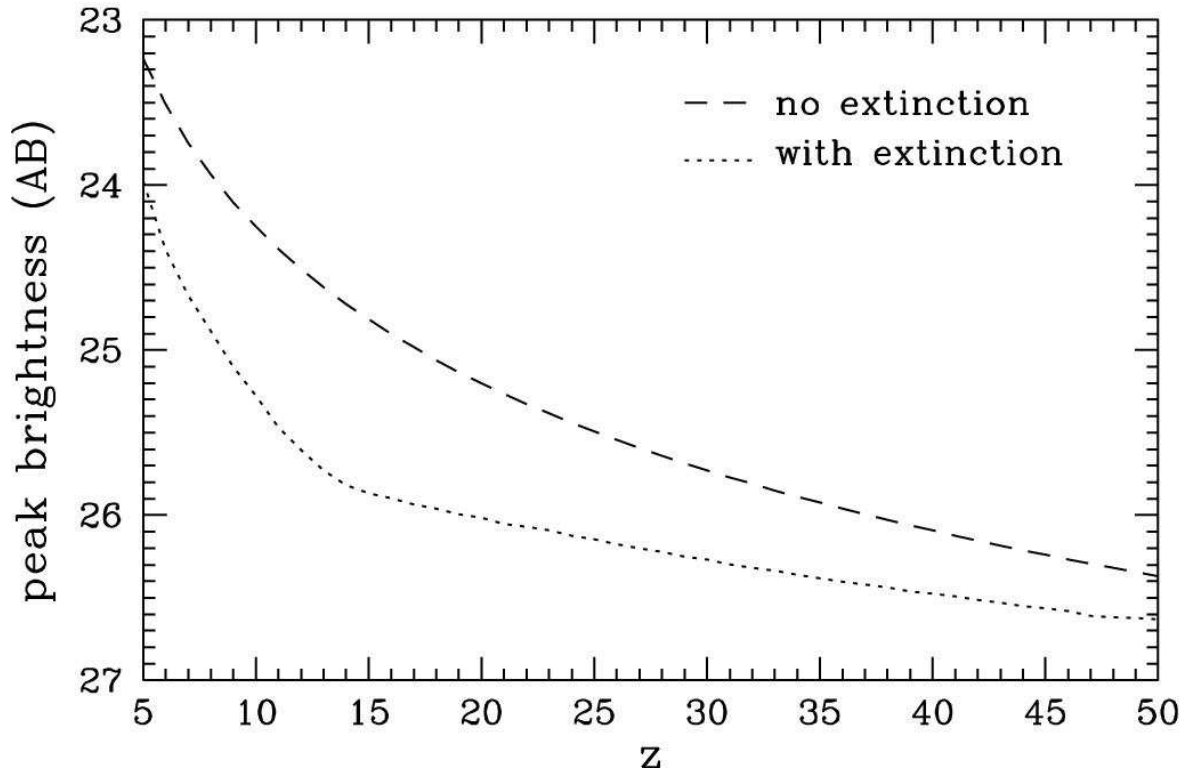}
\caption{
Predicted peak brightness of population III SNe as a function of
redshift. The observed peak brightness of a 250 M$_{\sun }$ SN in the
spectral region just longward of Lyman $\alpha $ is plotted assuming no
extinction and with a worst case extinction. Although the number of SN
expected in JWST surveys could be very low, they are bright enough that they
can be easily seen at any redshift. A 200 M$_{\sun }$ SN would be 1.7 mag
fainter and a 175 M$_{\sun }$ SN would be 3.5 mag fainter (From Weinmann
\& Lilly 2005).}
\label{fig003}
\end{figure*}

Individual population III stars are too faint to be detected, 
but supernovae can be identified up to very high redshift, since 
they could peak at levels brighter than AB = 27 mag (Fig. \ref{fig003}; 
Weinmann \& Lilly 2005). Although predictions are model dependent, 
the brightest known SNe would be visible to z \texttt{>} 30, and 
it is possible that population III stars will produce bright 
type II supernovae. Detection of a number of SNe at high redshift will 
require multiple visits and will yield a SN-based star formation 
rate. The redshift of each supernova will in general be determined 
photometrically, although spectroscopy may be possible on the 
brightest sources. The expected number of population III SNe 
is very uncertain; predictions range between 2500 (Wise \& Abel 
2003) and 50 (Mackey et al. 2003) yr$^{-1}$ deg$^{-2}$, while Weinmann 
\& Lilly (2005) argue that the rate is 4 yr$^{-1}$ deg$^{-2}$. Large 
areas need to be surveyed in order to obtain a significant sample. 
This program requires broad-band near-infrared imaging. Because 
of time dilation, the time between visits of the search field 
will need to be up to 6 months or more. While the first two visits 
produce only one search epoch, each successive visit produces 
another search epoch. Thus it is convenient to search the same 
field for an extended period of time. This could be accomplished 
by combining the SN search program with the ultra-deep observation, 
or with the wider surveys described in the next section. JWST 
will have a continuous viewing zone around each of the ecliptic 
poles which will enable repeated observations throughout the 
year.

\subsection{When and How Did Reionization Occur?}

Was reionization a single event? What is the ionization history 
of the universe prior to the final reionization?

The most direct observational evidence of re-ionization is the 
detection of a Gunn-Peterson trough (Gunn \& Peterson 1965) in 
the spectrum of high redshift quasars. Neutral hydrogen clouds 
along the line of sight (the Lyman \ensuremath{\alpha} forest) produce 
increasing absorption as the redshift increases. At z \ensuremath{\sim} 
5, some signal is detected shortward of the Lyman \ensuremath{\alpha} 
line, suggesting that the universe is fully ionized at z = 5 and 
that re-ionization was completed at still higher redshifts.

\begin{figure*}
\centering
\includegraphics[width=1.00\textwidth]{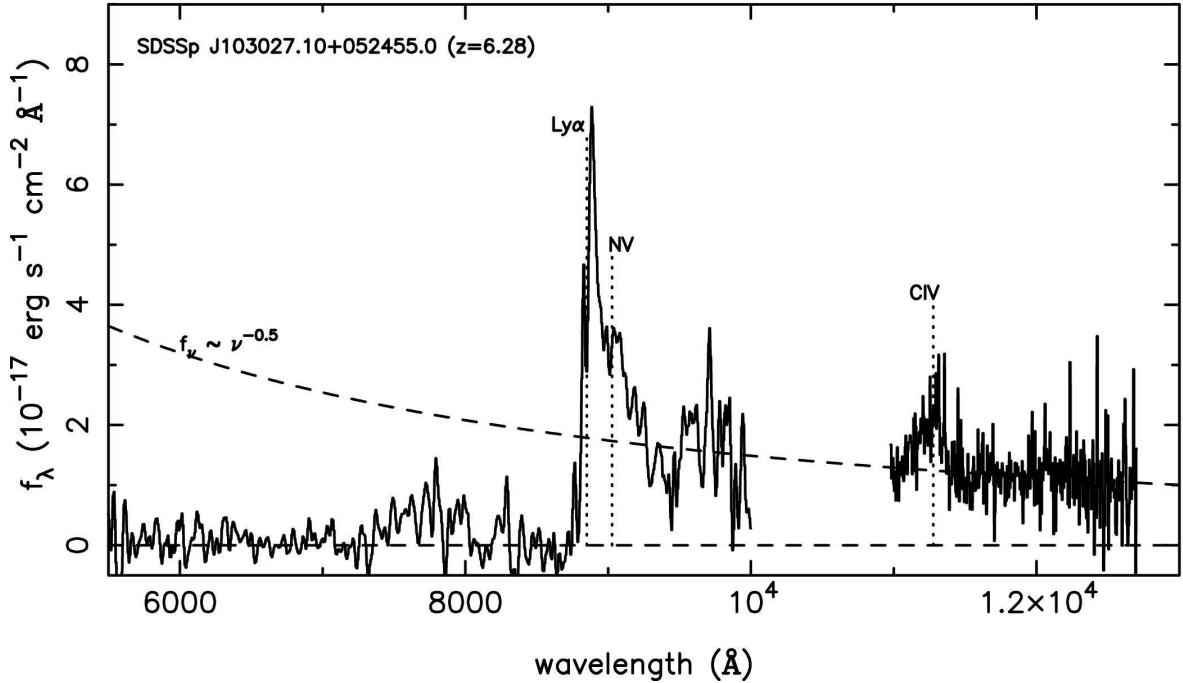}
\caption{
Spectrum of quasar SDSSp J103027.10 0552455.0 at z = 6.28. The
absence of flux over the 300 {\AA } region shortward of the Lyman $\alpha $
line is a possible indication of a Gunn-Peterson trough, indicating that the
fraction of neutral hydrogen has increased substantially between z = 5.7 and
z = 6, and that the universe is approaching the epoch of complete reionization
at z = 5.7 (From Fan et al. 2001).}
\label{fig004}
\end{figure*}

Fan et al. (2001; 2003; 2004) detected high redshift quasars 
using the Sloan Digital Sky Survey, including some at z \texttt{>} 6. 
One QSO at z = 6.28 shows a drop in continuum flux just shortward 
of Lyman \ensuremath{\alpha} by a factor 150 (see Fig. \ref{fig004}). This is evidence 
that a Gunn-Peterson trough has been detected in this object 
(Becker et al. 2001; Fan et al. 2002). Other QSOs at slightly 
lower redshift show a much smaller continuum drop. Variation 
in the QSO properties indicates that the reionization did not 
occur abruptly at the same time throughout the universe. Haiman 
\& Holder (2003) argue for an extended ``percolation'' period 
of reionization. 

\begin{figure*}
\centering
\includegraphics[width=1.00\textwidth]{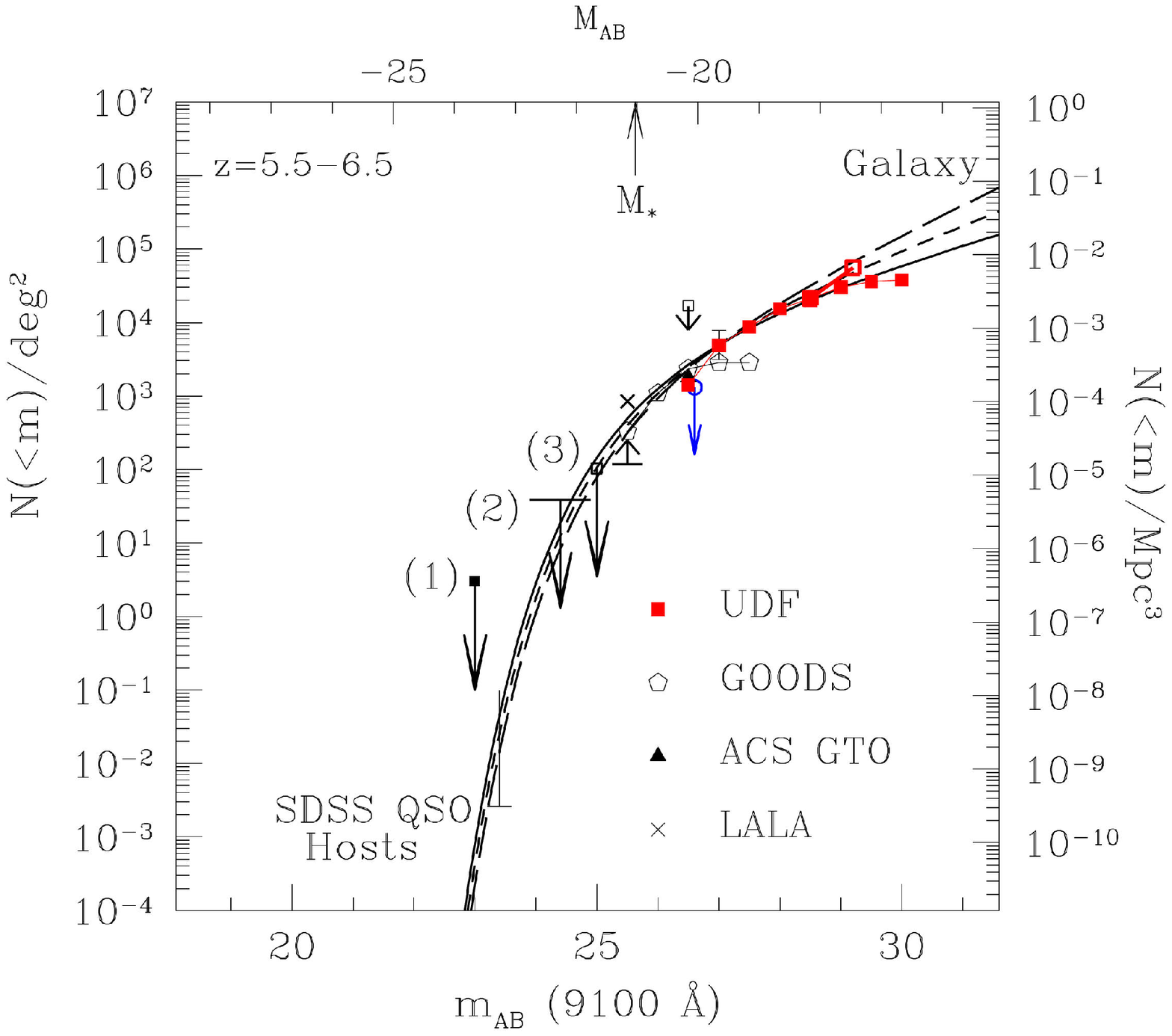}
\caption{
Cumulative galaxy counts for z $\sim $ 6. Galaxy counts for z $%
\sim $ 6 are predicted on the basis of lower-redshift measurements. In this
figure, the AB magnitudes refer to a band from 9100 to 9800 {\AA } in the
observed frame. Different curves refer to different cosmologies and
different normalizations of the luminosity function. At AB = 28, one expects a
few z $\sim $ 6 galaxies per square arcmin but this number is uncertain by
about an order of magnitude. JWST will go fainter by three magnitudes and
reach completely uncharted territory (From Yan \& Windhorst 2004b).}
\label{fig005}
\end{figure*}

We cannot conclude that the reionization epoch has been determined 
on the basis of these few objects, particularly since even a 
very modest local neutral hydrogen column density could produce 
the observed Gunn-Peterson troughs. However, these detections 
open up the possibility that re-ionization was completed at the 
relatively low redshift of z \ensuremath{\sim} 6. There are few constraints 
on the number density of galaxies at redshift greater than 6. 
By extrapolating the luminosity function of Lyman break galaxies 
at lower redshift (Steidel et al. 1999) one can obtain predictions 
for the number of galaxies at z \ensuremath{\sim} 6 (Fig. \ref{fig005}; Yan \& 
Windhorst 2004b), which are at the level of a few AB = 28 galaxies 
per arcmin square.

In the years before the launch of JWST, progress with HST and 
large ground based telescopes will allow us to study the bright 
end of the luminosity function of galaxies at z \texttt{>} 6. However, 
these facilities are unlikely to push to z \texttt{>} 8, measure the 
internal properties of these objects, or characterize the population 
of galaxies.

\begin{figure*}
\centering
\includegraphics[width=1.00\textwidth]{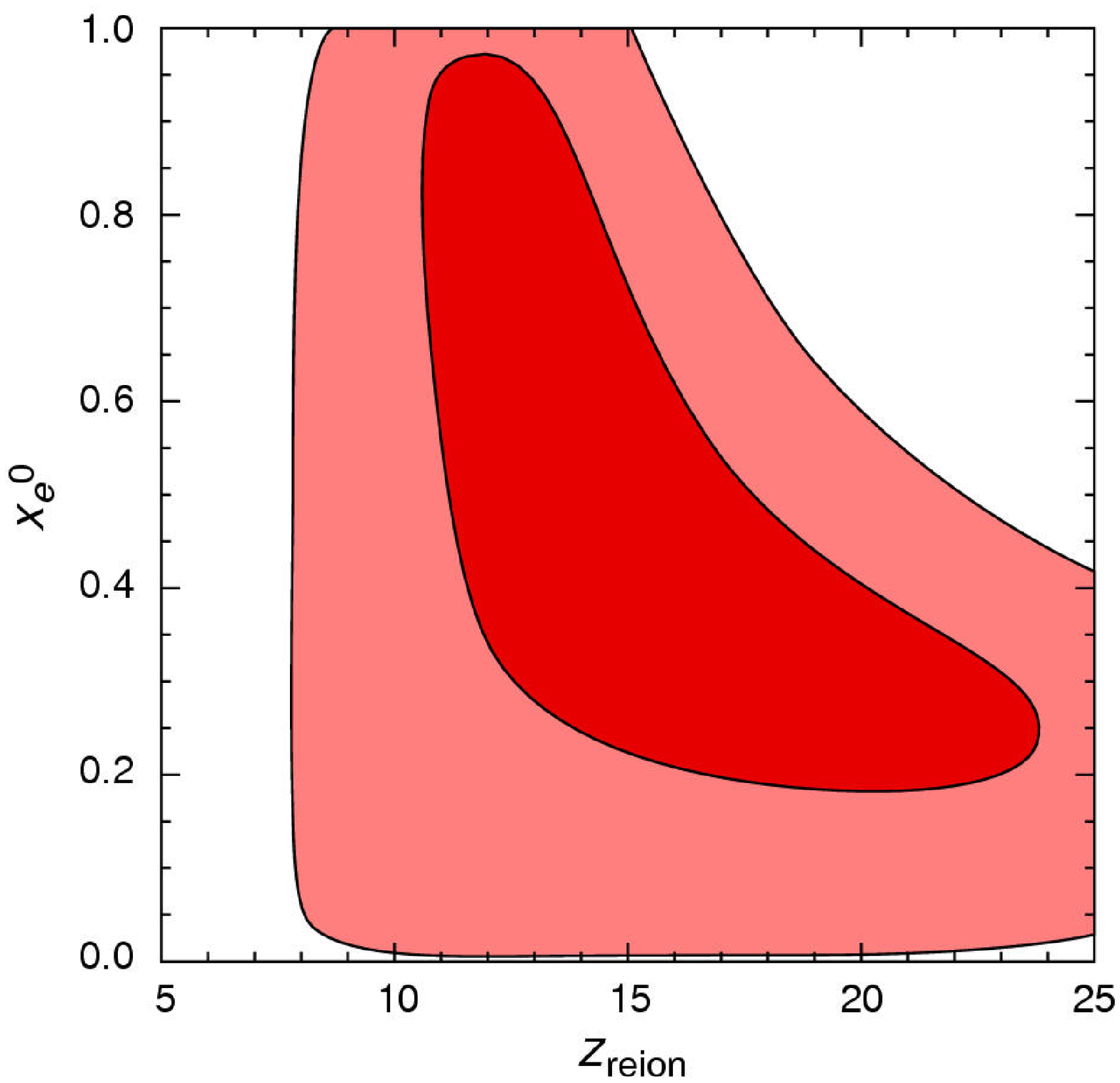}
\caption{
WMAP constraints on the reionization history. The plot shows the 68\%
and 95\% joint 2-d marginalized
confidence level contours for a model in which the Universe is partially 
reionized with an ionization fraction $x_e^0$ at $z_{reion}$, and then 
fully reionized at z=7. The WMAP data are inconsistent with a single
epoch of reionization at z $\sim$ 6, and argue for a complex reionization
history. (From Spergel et al. 2006).
}
\label{fig006}
\end{figure*}

The correlations between the cosmic microwave background temperature 
and polarization, as measured by WMAP, support an earlier 
reionization of hydrogen, giving $z_{reion} = 10.9^{+2.7}_{-2.3}$ under
the assumption of a single epoch of full reionization (Fig. \ref{fig006}; 
Spergel et al. 2006). This may be an indication that hydrogen at 
least partially recombined after the first epoch of reionization, 
only to be reionized again a lower redshift. In contrast to the 
reionization of hydrogen, the epoch of helium reionization 
has been firmly identified at z \ensuremath{\sim} 3 through the detection 
of a Gunn-Peterson trough in quasar spectra (Jakobsen et al. 
1994; Davidsen et al. 1996; Heap et al. 2000). 

Even though one often refers to the epoch of reionization as 
if it were a sudden transition, the time elapsed between the 
epochs when 10\% and 90\% of hydrogen was reionized can last a 
significant fraction of the age of the universe. The WMAP detection 
of a significant Compton opacity is evidence of either an extended 
reionization process, or of two distinct reionization epochs 
(Cen 2003a, 2003b; Haiman \& Holder 2003; Holder et al. 2003, 
Stiavelli et al. 2004; Page et al. 2006, Spergel et al. 2006). 
Regardless of the specifics of the reionization 
process, inhomogeneities along the line of sight may create dispersion 
in optical depth shortwards of Lyman \ensuremath{\alpha}. Moreover, only 
a very low residual fraction of neutral hydrogen is needed to 
produce a Gunn-Peterson trough in the spectra of high redshift 
quasars. In addition, the opacity near Lyman \ensuremath{\alpha} would 
be modified in the neighborhood of ionizing sources (Miralda-Escud\'{e} 
\& Rees 1994), in analogy to the proximity effect in QSOs (M{\o}ller 
\& Kjaergaard 1992).

It is possible to compute the minimum surface brightness required 
to reionize the universe, under the assumptions that the universe
was reionized by hot population III stars, and that all UV photons
can escape the system. This minimum surface brightness of ionizing
sources at z \texttt{>} 6 is AB \ensuremath{\cong} 29 mag arcmin$^{-2}$
in a redshifted \ensuremath{\lambda} = 1400 {\AA} band (Stiavelli
et al. 2004), when counted as the typical ionizing flux seen per
unit area. For a luminosity function similar in shape to that of
z = 3 Lyman break galaxies and with M$^{*}$ not fainter than
\ensuremath{-}15 mag, this implies a few sources per square arcmin
with AB = 31 or brighter.

While models differ significantly in the details of how the reionization 
was started by these various possible first light populations 
at 15 \texttt{<} z \texttt{<} 25, they all converge to produce roughly 
the same cosmic star-formation history of population II stars 
in the mini halos of dwarf galaxies at 6 \texttt{<} z \texttt{<} 10. This 
is simply the consequence of the need to fit the nearly complete 
Gunn-Peterson troughs now seen in the spectra of at least four 
SDSS quasars in the range 6.05 \texttt{<} z \texttt{<} 6.43 (Fan et al. 
2003). While these indicate non-zero flux shortward of 0.8 \ensuremath{\mu}m, 
there is essentially zero flux longwards of 0.810 \ensuremath{\mu}m. 
Hence, there was significant HI in front of these quasars at 
z \texttt{>} 5.7, although the HI-fraction at z = 6 was still very 
small (of order 10$^{-4}$ to 10$^{-5}$).

In most models, the conclusion of this reionization epoch is 
modeled by dwarf galaxies producing an increasing number of population 
II stars at 6 \texttt{<} z \texttt{<} 11. Most models are similar 
in their predictions of the cosmic star formation history at 
6 \texttt{<} z \texttt{<} 8, in order to match the SDSS Gunn-Peterson 
troughs seen at z = 6. For example, the Cen (2003a) models predict 
an increase in the cosmic star-formation history of a full factor 
of 10 over 16 \texttt{>} z \texttt{>} 11 and another factor of 10 increase 
over 11 \texttt{>} z \texttt{>} 6. In other words, most of the population 
II stars that we see today were born in dwarf galaxies, but most 
were not born until about z = 8 (consistent with the oldest ages 
of population II measured today of 12.8 Gyr), and it was likely 
the high-mass end of those population II stars that completed 
the epoch of reionization by z = 6. In WMAP cosmology (Spergel 
et al. 2003), there was only 300 Myr between at 6 \texttt{<} z \texttt{<} 
8 and another 170 Myr at 8 \texttt{<} z \texttt{<} 10, so the stellar 
population that was formed in those galaxies, and whose O, B 
and A stars helped complete the reionization of the universe 
by z = 6, is still visible as the low-mass population II stars 
seen today.

\subsubsection{Observations}

The epoch of reionization is revealed through signatures in the 
Lyman \ensuremath{\alpha} forest: a black Lyman \ensuremath{\alpha} Gunn-Peterson 
trough, islands in the Lyman \ensuremath{\alpha} forest, and appearance 
of a Lyman \ensuremath{\alpha} damping wing. In addition to these techniques, 
the epoch of reionization can be identified as the redshift at 
which there is fast evolution of the properties of Lyman \ensuremath{\alpha} emitters. 
However, a sharp transition in the Lyman \ensuremath{\alpha} luminosity 
function can be suppressed if, for instance, a relatively long 
reionization onset is coupled to a smooth increase in metal content. 
Sources at higher redshifts will have increasingly more absorbed 
Lyman \ensuremath{\alpha} but also increasingly stronger intrinsic equivalent 
widths because of the lower metallicity. It is easy to build 
models where the two effects cancel out. Alternative methods, 
not sensitive to this limitation, use the evolution 
of the ratio between Lyman \ensuremath{\alpha} and Balmer lines.

Three observing programs are needed to firmly establish the epoch 
of reionization and to probe the possibility that a first reionization 
took place at very high redshifts. A starting sample of Lyman-break 
selected galaxies will be obtained from the ultra-deep observations 
required to identify the first light sources.

\paragraph{Lyman \ensuremath{\alpha} Forest Diagnostics}

JWST will make deep spectroscopic observations of QSOs or bright
galaxies to study the Lyman $\alpha$ forest. High signal-to-noise,
R $\sim$ 1000, near-infrared spectra of the brightest high-redshift
QSOs or galaxies will reveal the presence of a Gunn-Peterson trough
or of a Lyman \ensuremath{\alpha} damping wing. The targets will
be the brightest known high redshift objects, perhaps from JWST
surveys, or perhaps found by other means. High signal to noise is needed to
discriminate between optical depths \ensuremath{\tau} of a few and
\ensuremath{\tau} \texttt{>}\texttt{>} 10. A damping wing should
be present for a few million years, before the ionizing radiation
is sufficient to create a large Str\"{o}mgren sphere around each
ionizing source. Failure to detect a damping wing does not necessarily
imply that the universe is ionized. R = 100 spectra will be able
to determine the presence of a Lyman \ensuremath{\beta} ``island''.
This is relevant if reionization occurs relatively abruptly. In
this case, objects at redshifts between the redshift of reionization,
z$_{reion}$, and z =
(\ensuremath{\lambda}$_{\ensuremath{\alpha}}$/\ensuremath{\lambda}$_{\ensuremath{\beta}}$)
(1+z$_{reion}$) \ensuremath{-} 1, will show an island of normal,
finite, forest absorption between the Lyman \ensuremath{\alpha}
and the Lyman \ensuremath{\beta} forests. Here,
\ensuremath{\lambda}$_{\ensuremath{\alpha}}$ and
\ensuremath{\lambda}$_{\ensuremath{\beta}}$ are the rest frame
wavelengths of Lyman \ensuremath{\alpha} and Lyman \ensuremath{\beta},
respectively.

If there are indeed two distinct reionization epochs, the (possibly 
partial) recombination following the first reionization may be 
detectable in continuum spectra of high redshift objects as an 
absorption signature in the region shortward of Lyman \ensuremath{\alpha}.

\paragraph{Survey for Lyman \ensuremath{\alpha} Sources}

When the universe was still neutral, Lyman \ensuremath{\alpha} was efficiently 
scattered over a large volume or absorbed by dust. The faintest 
Lyman \ensuremath{\alpha} sources and those with narrow Lyman \ensuremath{\alpha} 
emission will therefore not be visible before reionization. Thus, 
at reionization, one expects a fast evolution of the faint end 
of the Lyman \ensuremath{\alpha} luminosity function of star-forming objects 
(Malhotra \& Rhoads 2004). To detect a transition in the properties 
of Lyman \ensuremath{\alpha} sources at the epoch of reionization, JWST 
will select Lyman \ensuremath{\alpha} emitters at a variety of increasing 
redshifts by using the narrow-band excess technique in near-infrared 
images. Given the high probability of interlopers, the sources 
would need to be confirmed either by detecting a second emission 
line with images at another wavelength, spectroscopically, or 
by using the Lyman-break technique. The aim is to detect rapid 
evolution of the Lyman \ensuremath{\alpha} luminosity function at one 
or two specific redshifts. Such evolution would be indicative 
of reionization. Line intensities will be fainter than 6 x 10$^{-18}$ 
erg cm$^{-2}$ s$^{-1}$. The need to verify the identification of a line 
as Lyman \ensuremath{\alpha} requires one to attempt the detection of 
a second line, e.g., H\ensuremath{\beta}. This will in general be 30 
times fainter than Lyman \ensuremath{\alpha}. An alternative method for 
finding Lyman \ensuremath{\alpha} emitters would be to search in the spectral 
domain with spectroscopy of blank areas.

By following the properties of Lyman \ensuremath{\alpha} emitters to the 
highest redshifts, we will be able to identify a period of partial 
recombination that would appear as a statistical brightening 
followed by dimming of Lyman \ensuremath{\alpha} sources in the intermediate 
non fully-ionized period. This might be more sensitive than the 
equivalent test based on the absorption of the ionizing continuum 
photons, since for the latter a very small neutral fraction is 
already sufficient to produce very high opacity. 

\paragraph{The Ratio between Lyman \ensuremath{\alpha} and Balmer Lines}

If neither the metallicity nor the dust content of the universe 
changes abruptly at reionization, then detection of a rapid change 
in the Lyman \ensuremath{\alpha} to H\ensuremath{\alpha} (or H\ensuremath{\beta}) ratio 
can be used to identify the reionization epoch. By measuring 
the hydrogen Balmer lines in addition to Lyman \ensuremath{\alpha}, it 
is possible to determine the amount by which Lyman \ensuremath{\alpha} 
is suppressed due to either scattering or absorption. Any rapid 
evolution in this ratio as a function of redshift might indicate 
a change in the mean ionization state of the universe. This measurement 
requires R = 1000 spectroscopy of Lyman \ensuremath{\alpha}, and H\ensuremath{\alpha} 
or H\ensuremath{\beta}, and requires measurements of line intensities 
down to 2 x 10$^{\ensuremath{-}}$$^{19}$ erg cm$^{\ensuremath{-}}$$^{2}$ s$^{\ensuremath{-}}$$^{1}$ at \ensuremath{\lambda} 
\texttt{>} 3 \ensuremath{\mu}m. This flux limit corresponds to a Lyman \ensuremath{\alpha} 
intensity of 6 x 10$^{\ensuremath{-}}$$^{18}$ erg cm$^{\ensuremath{-}}$$^{2}$ s$^{\ensuremath{-}}$$^{1}$ 
(detected at 1 \ensuremath{\mu}m) and a flux ratio of 30 between Lyman \ensuremath{\alpha} 
and H\ensuremath{\beta}.

\subsection{What Sources Caused Reionization?}

What were the sources responsible for reionization? Were they 
powered by nuclear fusion or gravitational accretion? How is 
the evolution of galaxies and black holes affected by the possibly 
extended period of reionization?

It is often assumed that the population III stars were responsible 
for the reionization of hydrogen, mainly because it is not clear 
how seed black holes could form in the absence of stars. This 
is supported by the measured luminosity function of z \ensuremath{\sim} 
6 quasars, which does not produce a sufficient number of ionizing 
photons to keep the universe ionized (Fan et al. 2004; Yan \& 
Windhorst 2004a), and by observations of the soft X-ray background, 
which set limits on accretion by black holes at high redshift 
(Dijkstra, Haiman \& Loeb 2004). However, the observed local 
black-hole mass -- host galaxy bulge velocity dispersion relation 
(Ferrarese \& Merritt 2000; Gebhardt et al. 2000) clearly indicates 
that the evolution of AGN and their galaxy hosts are closely 
related. Although Walter et al. (2004) conclude that this M$_{BH}$ 
-- \ensuremath{\sigma}$_{bulge}$ relation is unlikely to hold at high redshift, 
this result is controversial (Shields et al. 2003), so determining 
the relative contributions of fusion and accretion to reionization 
and investigating the relationship between galaxies and black 
holes during this epoch will connect the first light sources 
to the processes that assembled galaxies after reionization.

Nuclear processing of only $\sim$ 10$^{-6}$ of the baryons 
would be sufficient to reionize the universe (Loeb \& Barkana 
2001), leading to a minimum average metallicity of the universe 
at reionization of \ensuremath{\sim}10$^{-3}$ Z$_{\sun}$. It is not clear what 
the mean metallicity of objects observed at these redshifts would 
be. Indeed, the metallicity of the first objects and that of 
the Inter-Galactic Medium (IGM) could be very different. If population 
III stars are formed in halos of sufficiently low mass, they 
can enrich the IGM by SN-driven winds (Madau et al. 2001; Mori 
et al. 2002). When a halo undergoes a SN-driven outflow, the 
ejection of metals can be very effective. However, it is not 
clear how effective this process is when averaged over all halos. 
It is possible that the most massive halos retain most of their 
metals and have much higher metallicities at the epoch of reionization, 
as seen in the nearly-Solar metallicities in z \ensuremath{\sim} 6 QSOs 
(Freudling et al. 2003).

If the power source for reionization is not nuclear fusion but 
rather gravitational accretion onto massive black holes, the 
higher efficiency of gravitational accretion requires a smaller 
fraction of material to be processed. This scenario does not 
place any constraint on the metallicity of the universe at reionization. 
Even if reionization is caused by stellar UV radiation, it is 
natural to expect that some fraction of these stars will leave 
black holes as remnants (Heger \& Woosley 2002; Madau \& Rees 
2001). Both scenarios would lead to the presence of seed black 
holes at the end of re-ionization, with implications for the 
formation of AGN and galaxies (Silk \& Rees 1998; Stiavelli 1998).

Barkana \& Loeb (2000) predict a distinct drop in the cosmic 
star formation rate around the reionization redshift. As the 
intergalactic medium is photoionized, the temperature increases, 
which suppresses the formation of low-mass galaxies. The luminosity 
function of galaxies should show a much steeper faint-end slope 
before reionization than after. This may have already been seen in the 
Hubble Ultra-Deep Field (Yan \& Windhorst 2004b, Bouwens et al. 
2004).

\subsubsection{Observations}

When the reionization epoch is identified, one needs to find 
a population of objects that have sufficient ionizing continuum 
to ionize all of the hydrogen. Once these sources are identified, 
one can derive their properties and determine their nature and 
energy source. A combination of spectroscopic diagnostics (line 
shapes, line widths, and line intensity ratios) and photometry 
can be used to distinguish between stellar and non-stellar photo-ionization. The 
ionizing continuum can be derived indirectly by estimating its 
slope and intensity. This slope can be derived from the ratio 
between hydrogen and helium lines. The hydrogen Balmer lines 
can provide the intensity.

\paragraph{Determine the Source Nature}

Identification of the nature of the ionizing sources requires 
a combination of diagnostics: line shapes, line widths, line 
ratios, shape of the continuum. We expect the intrinsic line 
widths of AGN-powered sources to be broader than those of sources 
ionized by stellar radiation. The line shapes may also help in 
distinguishing primordial HII regions from mini-AGN. Mid-infrared 
photometry can help distinguish the flat ultraviolet-optical 
continuum of a star-bursting galaxy from the redder quasars. 
This program requires a combination of deep near-infrared R = 
1000 spectroscopy and mid-infrared imaging.

\paragraph{Measuring the Ionizing Continuum}

In order to measure the ionizing continuum of a class of sources, 
we need to measure their hydrogen and helium Balmer lines. Comparison 
between these lines provides an estimate for the steepness, or 
hardness of the ionizing continuum. The hydrogen Balmer line 
intensity provides the normalization. Taken together, the normalization 
and slope provide a measurement of the rate of production of 
ionizing photons for any given class of sources under the assumption 
that the escape fraction is known. The escape fraction can be 
measured from deep imaging observations, or estimated from the 
line equivalent widths. This program requires near-infrared spectroscopy 
of very faint objects. The expected observed surface brightness 
of the sources responsible for reionization ranges between AB = 27 
and 29 mag arcmin$^{-2}$, counted as the typical ionizing flux per 
unit area over which they are detected. The former applies to 
the case of metal-enriched reionization sources with dust and 
low escape fraction of ionizing UV, the latter applies to zero-metallicity 
ionizing sources with 100\% escape fraction for a more extended 
reionization period. This program requires near-infrared R = 
1000 spectroscopy of high redshift galaxies. It is likely to 
be satisfied by the same data set that was used to determine 
the nature of the reionizing sources.

\begin{table}[t]
\caption{JWST\ Measurements for the End of the Dark Ages Theme\label{tab001}}
\begin{tabular}{p{1.0in}p{0.8in}p{1in}p{1.0in}p{0.0in}}
\hline\noalign{\smallskip}
{Observation} &
{Instrument} &
{Depth, Mode} &
{Target} &
\\[3pt]
\tableheadseprule\noalign{\smallskip}
\raggedright Ultra-deep survey (UDS) &
NIRCam &
1.4 nJy at 2$\mu$m &
10 arcmin$^{2}$ &
\\
&&&& \\
\raggedright In-depth study &
NIRSpec &
\raggedright 23 nJy, R$\sim$100 &
\raggedright Galaxies in UDS area &
\\
&&&& \\
&
MIRI &
\raggedright 23 nJy at 5.6$\mu$m &
\raggedright Galaxies in UDS area &
\\
&&&& \\
\raggedright Lyman $\alpha $ forest diagnostics&
NIRSpec&
\raggedright $2\times 10^{-19}$ erg cm$ ^{-2}$ s$^{-1}$, R$\sim $1000&
\raggedright Bright z$>$7 quasar or galaxy &
\\
&&&& \\
\raggedright Survey for Lyman $\alpha $ sources&
TFI&
\raggedright 2$\times $10$^{-19}$ erg
cm$^{-2}$ s$^{-1}$, R$\sim $100&
\raggedright 4 arcmin$^{2}$ containing known high-z object &
\\
&&&& \\
\raggedright Transition in Lyman $\alpha $/Balmer&
NIRSpec&
\raggedright 2$\times $10$^{-19}$ erg cm$^{-2}$ s$^{-1}$, R$\sim $1000&
\raggedright UDS or wider survey area &
\\
&&&& \\
\raggedright Measure ionizing continuum&
NIRSpec&
\raggedright 2$\times $10$^{-19}$ erg cm$^{-2}$ s$^{-1}$, R$\sim $1000&
\raggedright Same data as above &
\\
&&&& \\
\raggedright Ionization source nature&
NIRSpec&
\raggedright 2$\times $10$^{-19}$ erg cm$^{-2}$ s$^{-1}$, R$\sim $1000&
\raggedright Same data as above &
\\
&&&& \\
&MIRI&23 nJy at 5.6$\mu $m&
&
\\
&&&& \\
\raggedright LF of dwarf galaxies&
NIRCam&
1.4 nJy at 2$\mu $m&
UDS data& \\
\noalign{\smallskip}\hline
\end{tabular}
\end{table}

\paragraph{Luminosity Function of Dwarf Galaxies}

The Luminosity Function (LF) of dwarf galaxies over the redshift 
range 6 \texttt{<} z \texttt{<} 10 will reveal the completion of reionization 
and the birth of population II stars. High-mass population II 
stars likely completed the reionization at 8 \texttt{>} z \texttt{>} 6, 
and low-mass population II stars are still visible today (Yan 
\& Windhorst 2004b).

Dwarf galaxies at 6 \texttt{<} z \texttt{<} 10 are best found with the 
Lyman break or dropout technique. Finding objects in this redshift 
range requires high sensitivity at wavelengths 0.8 \texttt{<} \ensuremath{\lambda} 
\texttt{<} 1.3 \ensuremath{\mu}m. JWST will measure any structural properties 
of these objects at wavelengths \texttt{>} 2.0 \ensuremath{\mu}m,
where it will be diffraction limited. While JWST will not be
diffraction limited at shorter wavelengths, it will be critical
for the study of the conclusion of the epoch of reionization at 6
\texttt{<} z \texttt{<} 10, that objects can be detected in the
0.8 \texttt{<} \ensuremath{\lambda} \texttt{<} 1.3 \ensuremath{\mu}m
range, and that basic properties such as colors and total fluxes
can be measured with sufficient signal to noise. This program requires
near-infrared ultra-deep imaging as for the ultra-deep survey.

\subsection{Summary}

Table~\ref{tab001} summarizes the measurements needed for the
End of the Dark Ages theme. They include:

\textbullet{}
{\it Ultra-deep survey (UDS).} The UDS will be the deepest NIRCam
survey, probably done in Treasury or Legacy mode. The survey will
use a full set of broad-band NIRCam filters, with exposure times
optimized to find high-redshift objects using the drop-out technique.
If done in the continuous viewing zone, the observations could be
scheduled in several epochs to find high redshift supernovae.

\textbullet{}
{\it In-depth study.} Follow-up observations of very high redshift
objects found in the ultra-deep survey will be used to investigate
their nature. NIRSpec low resolution spectroscopy will be used to
search for continuum breaks and emission lines. MIRI photometry of
high-redshift objects will give age estimates, relying on upper
limits for very young populations.

\textbullet{}
{\it Lyman $\alpha$ forest diagnostics.} Spectra of the brightest
high-redshift objects will be used to look for Gunn-Peterson troughs
and determine the epoch and nature of reionization.

\textbullet{}
{\it Survey for Lyman-$\alpha$ sources.} A narrow-band imaging program
will search for Lyman $\alpha$- emitting companions to known high-z
objects. The properties of these objects are expected to be different
before and after reionization.

\textbullet{}
{\it Transition in Lyman $\alpha$/Balmer.} This program will
determine the epoch of reionization through the effect on the galaxy
population by measuring spectral lines of galaxies before and after
reionization. It needs to see both Lyman $\alpha $ and H$\alpha $, in
galaxies at a range of redshifts.

\textbullet{}
{\it Measure ionizing continuum.} Ratios of the hydrogen and helium Balmer
lines will reveal the hardness of the ionizing continuum.

\textbullet{}
{\it Ionization source nature.} Near-infrared line widths and mid-IR
photometry will separate star-formation from AGN as source of
ionizing continuum.

\textbullet{}
{\it LF of dwarf galaxies.} The number of dwarf galaxies as a
function of redshift changes as the universe is re-ionized.

\section{The Assembly of Galaxies}

The key objective of The Assembly of Galaxies theme is to determine 
how galaxies and the dark matter, gas, stars, metals, morphological 
structures, and active nuclei within them evolved from the epoch 
of reionization to the present day.

Galaxies are basic building blocks of the universe. Material 
within galaxies undergo the vast cycle of stellar birth, life 
and death that results in the production of the heavy elements, 
the formation of planets and the emergence of life. Most of the 
astrophysical complexity of the universe is manifest within galaxies, 
and the formation of galaxies represents a key link between this 
complexity and the relative simplicity of the early universe. 
On the one hand, the most basic properties of galaxies reflect 
the distribution of dark matter in the universe, which are believed 
to result from very simple quantum processes operating during 
the earliest moments of the Big Bang. On the other hand, the 
subsequent complex astrophysical behavior of the baryonic material 
within these dark matter halos produces the morphological symmetries 
and diverse properties of present-day galaxies. Therefore, understanding 
the processes that formed the present-day population of galaxies 
is central to cosmology, to astrophysics and to our understanding 
of the emergence of life in the universe.

The cold dark matter (CDM) cosmological model provides a conceptual framework 
for understanding the formation of galaxies through the hierarchical 
assembly of progressively more massive objects. However, many 
of the most basic questions about this process remain unanswered 
due to the difficulty of observing faint objects at high redshifts. 
The origins of the most fundamental scaling relations for galaxies 
are not well understood, and the CDM paradigm has not yet been
tested on galactic scales. On the theoretical side, the ``semi-analytic''
models for galaxy formation and evolution include many free
parameters, while numerical gravito-hydrodynamic simulations do
not yet have the resolution and dynamic range needed to simultaneously
model individual star-formation events and the growth of a galaxy
in its cosmological environment.

It is clear that the formation and early evolution of galaxies 
is a complex and multifaceted problem that no single observation 
or theory will solve. Essential elements of an understanding 
of galaxy assembly will almost certainly include the following:

{\textbullet} 
The fundamental physics of the very early universe, including 
the origin of density fluctuations and the nature of the dark 
matter and dark energy;

{\textbullet} 
The hierarchical assembly of matter through gravitational instability;

{\textbullet} 
The formation of stars under a wide range of conditions, including 
some quite different from those encountered today;

{\textbullet} 
The origin and growth of black holes at the centers of galaxies;

{\textbullet} 
The feedback of energy and radiation produced by the first galaxies 
or pre-galactic objects on the surrounding material, including 
the re-ionization of the intergalactic medium;

{\textbullet} 
The exchange of material between galaxies and the surrounding 
reservoir of baryons.

Coupled with these physical processes, a host of observational 
issues must be understood, including the effects of dust obscuration 
and the inevitable observational selection effects in wavelength, 
and point-source and surface-brightness sensitivity.

Progress requires new observational data, both to characterize 
the galaxy population at different epochs, and to understand 
the astrophysics of key processes that are occurring in the universe 
at early times. JWST will address the most pressing of these 
observational questions.

To gain an understanding of the extremely distant universe requires 
a systematic and comprehensive approach. Objects must be detected 
and identified (i.e., recognized as being at high redshift), and 
then characterized in terms of their physical properties, and 
of the physical processes occurring in and around them. They 
must be placed in the context of a global understanding of the 
other objects and other phenomena going on at the same epochs. 
It is also essential to understand which objects at one epoch 
evolve into which objects at a subsequent epoch, and to understand 
the relationship at all times between the visible baryonic material 
and the underlying dark matter.

\subsection{Previous Investigations}

During the mid-1990s, the simultaneous use of efficient multi-object
spectrographs on large telescopes, the first 8 to 10 m telescopes
and HST observations led to a dramatic advance in our direct
observational knowledge of the galaxy population at earlier epochs.

At z \ensuremath{\sim} 1, the universe appears roughly similar at optical 
and near-infrared wavelengths to that seen today. There is a 
full range of Hubble types including spirals and ellipticals 
(e.g., Driver et al. 1995; Schade et al. 1995; Abraham et al. 
1996; Brinchmann et al. 1998), a well-developed luminosity function 
of quiescent red galaxies (Lilly et al. 1995), approximately 
the same number density of large spiral disks, ``normal'' Tully-Fisher 
rotation curves in these disks (Vogt et al. 1996; 1997), and 
so on. The metallicities of the star-forming gas are close to 
solar. Some clear evolutionary effects are apparent, as luminous 
galaxies at z \ensuremath{\sim} 1 have signatures of vigorous star-formation 
activity, such as blue colors, strong emission lines, irregular 
morphologies. These indications are usually seen locally only 
in smaller galaxies, the so-called ``down-sizing'' 
effect (Cowie et al. 1995). In addition, the overall luminosity 
density in the ultraviolet, and in emission lines, is about a 
factor of five higher at z \ensuremath{\sim} 1 than it is locally.

Extending beyond z \ensuremath{\sim} 1, the known galaxies at z \ensuremath{\sim} 
3 are generally blue with compact or irregular morphologies. 
Most of these galaxies have been selected in the ultraviolet, 
and it is not yet clear whether there is a real absence of well-developed 
spiral or quiescent elliptical galaxies at this redshift; nor 
is it clear when such galaxies first appear (see e.g., Giavalisco 
et al. 1996, Abraham et al. 1999, Zepf 1997, Dickinson 2000, 
Franx et al. 2003) Recent Spitzer results have begun to address 
this question by examining the population at z \ensuremath{\sim} 2 (Yan et al. 
2004; Labb\'{e} et al. 2005). There are indications in these data that 
some galaxies have substantial old stellar populations by z $\sim$ 3, but that
there is not a large, previously hidden population of old galaxies (Barmby et al. 2004). 
Spitzer 24 $\mu$m detections of extremely red galaxies at z $\sim$ 2 show two 
populations, merger-induced dusty starbursts and galaxies with old
stellar populations (Yan et al. 2004; Chary et al. 2004).

The first samples selected through deep K-band imaging appear to
show large numbers of red galaxies at redshifts approaching z
\ensuremath{\sim} 2 (McCarthy et al. 2004, Abraham et al. 2004),
although their stellar masses are sufficiently uncertain that it
is not yet clear what fraction of the z \ensuremath{\sim} 1 population
these represent. The handful of ultraviolet-selected galaxies
studied in detail at z \ensuremath{\sim} 3 show evidence for
significantly sub-solar metallicities (Z \ensuremath{\sim} 0.3
Z$_{\ensuremath{\sun}}$) and for galactic winds of several hundred
km s$^{-1}$, indicating substantial ejection of enriched material
into the intergalactic medium.

Beyond z \ensuremath{\sim} 3, our picture of the galaxy population becomes 
very fragmentary as we approach the epoch at which reionization 
appears to have been completed (z \ensuremath{\sim} 6 to 7). Small samples 
of galaxies are known, generally found through their strong Lyman \ensuremath{\alpha} emission 
(Hu et al. 2002, 2004; Rhoads et al. 2003) or by extensions of 
the Lyman break ``drop-out'' technique to longer 
wavelengths (Dickinson et al. 2004, Bouwens et al. 2003, Yan 
\& Windhorst 2004b), but the systematic and
detailed study of these exceedingly faint objects is difficult, and relies
on the brightest end of the luminosity function (e.g., Bouwens et al. 2005).

Results from COBE showed that the extragalactic background light 
has equal energy in the far-infrared as in the optical and near-infrared, 
and that the absorption and re-radiation of light by dust has 
played a major role in shaping the appearance of the universe 
(Puget et al. 1996; Fixsen et al. 1998). Much less is currently 
known about the sources responsible for the far-IR/sub-mm background 
than the optical sources described above. At 850 \ensuremath{\mu}m, 
about 50\% of the background has been resolved (e.g., Hughes et 
al. 1998; Barger et al. 1998; Eales et al. 1999, 2000), and these 
sources are extremely luminous heavily dust-enshrouded galaxies 
with luminosities greater than several 10$^{12}$ L$_{\ensuremath{\sun}}$, comparable 
to or greater than the local ultra-luminous infrared galaxies 
(ULIRGs) discovered by IRAS. Although little is known reliably 
about their redshifts, it is clear that these are about 100 times 
more common at high redshift (z \texttt{>} 1) than they are locally. 
At 15 \ensuremath{\mu}m, deep counts are available from ISOCAM surveys, 
and Chary \& Elbaz (2001) show that the rapid evolution required 
to account for these must flatten at z = 1, so as not to overproduce 
the background seen beyond 100 \ensuremath{\mu}m (see also Lagache, 
Dole \& Puget 2003). This is broadly similar to the behavior 
seen in the ultraviolet, with a possibly steeper rise at low 
redshifts. It is not yet known definitively whether the energy 
source in these obscured objects is a massive burst of star-formation 
or accretion onto a black hole in an active galactic nucleus. 
It is tempting to associate these objects with major mergers 
of young galaxies, since the low redshift ULIRGs appear to be 
triggered by such events.

From the above it is clear that the redshift range 1 \texttt{<} z 
\texttt{<} 7 is the time when the galaxy population acquired most 
of its present-day characteristics, when a large fraction of 
the stars we see today were formed, and when a large fraction 
of the metals were produced. Accordingly, this is the period 
when the most important astrophysical processes in galaxy formation 
and evolution occurred.

\subsection{When and How Did the Hubble Sequence Form?}

Where were stars in the Hubble Sequence galaxies formed? When 
did luminous quiescent galaxies appear? How does this process 
depend on the environment?

To answer these questions, we need observations of the morphologies, 
stellar populations, and star-formation rates in a very large 
sample of galaxies observed in deep imaging and spectroscopic 
surveys. This investigation has substantial overlap with the 
chemical enrichment of galaxies, the measurement of masses, and 
the nature of the highly obscured luminous galaxies.

JWST will characterize the star-formation rates in individual 
galaxies, ideally as a function of their mass, environment, and 
cosmic epoch. JWST will also determine when the long-lived stars 
in a typical galaxy were formed, whether {\em in situ} or in smaller 
galaxies that subsequently merged together to form a large galaxy. 
Direct characterization of the merging rate of galaxies will 
provide another angle on this question.

The emergence of quiescent red galaxies, which have completed 
their major episodes of star formation, at least for the time 
being, will tell us why star formation ceases in some galaxies. 
The importance of chaotic star formation in starbursts, as compared 
with the steady-state star formation in stable galactic disks, 
will reveal the modes of star formation that dominate different 
phases of galactic evolution, and that develop the morphological 
components in the galaxies.

Quantities such as the disk-size function, as well as color gradients 
within galactic disks at different redshifts will show directly 
how galactic disks grew, while the merger rate of disk galaxies 
will reveal the rate at which stars, originally formed in disks, 
are redistributed into the spheroids.

\subsubsection{Observations}

Except in objects with very high levels of dust extinction, the
star-formation rate of massive stars in a galaxy can best be
estimated from measurements of the H\ensuremath{\alpha} emission
line, complemented by those of other emission lines, the ultraviolet
continuum and the bolometric luminosity at longer wavelengths. JWST
should have the capability to spectroscopically measure the
H\ensuremath{\alpha} emission (5 \ensuremath{\times} 10$^{-19}$
erg s$^{-1}$ cm$^{-2}$) that would be produced by a star-formation
rate of only 1 M$_{\ensuremath{\sun}}$/yr at z \ensuremath{\sim}
5 (Kennicutt 1999).

The existence of older stellar populations is best revealed by
continuum imaging at rest wavelengths \ensuremath{\lambda} \texttt{>}
0.5 \ensuremath{\mu}m, or even at \ensuremath{\lambda} \texttt{>}
1 \ensuremath{\mu}m, if possible. Based on the Local Group and
Milky Way Galaxy, the deepest near-infrared imaging should be able
to detect the Small Magellanic Cloud (with M$_{V}$ = --16.2 mag)
if placed at z \ensuremath{\sim} 5, where it would be unresolved
and have AB \ensuremath{\sim} 30.3 mag at 3.5 \ensuremath{\mu}m.

With imaging data that span the rest-frame UV and optical with 
at least 5 filters, redshifts for essentially all galaxies above 
a faint flux threshold (typically \ensuremath{\geq} 10$\sigma$) can be estimated 
using photometric redshift techniques (e.g., Hogg et al. 1998). 
These techniques have a typical accuracy \ensuremath{\delta}$_{z}$/(1+z) \texttt{<} 
0.1, and with only a few percent of the estimates falling far 
from the actual redshift. Confirmation of these will be possible 
using either R \ensuremath{\sim} 100 or R \ensuremath{\sim} 1000 near-infrared 
spectra, as required.

\begin{figure*}
\centering
\includegraphics[width=1.00\textwidth]{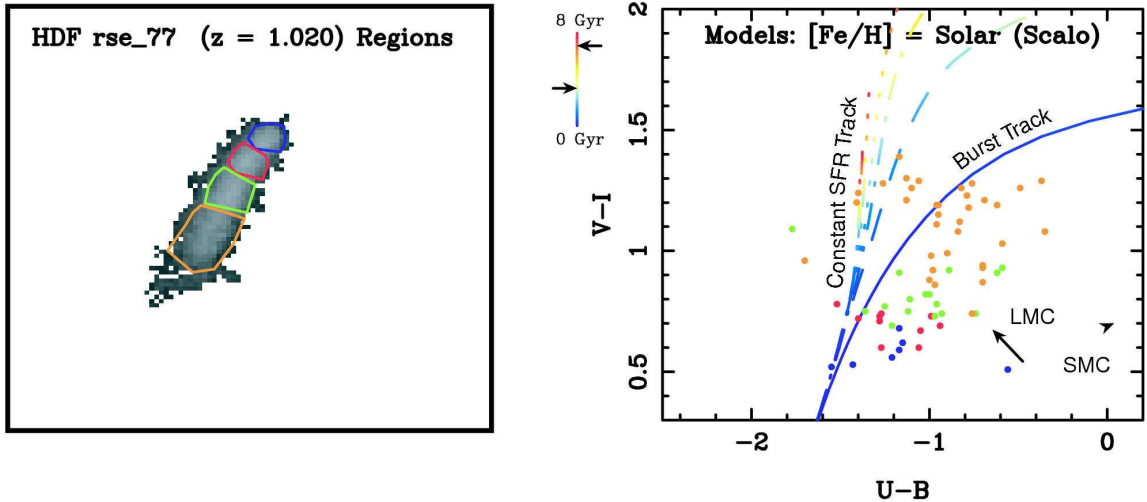}
\caption{
Galaxies in deep HST images are separated on the basis of color
into regions with different star-formation histories. On the left, we show
four different color regions in the galaxy image. On the right, these
regions are placed, pixel by pixel on a color-color diagram and compared to
model predictions to determine the ages of the regions. The arrows
labeled LMC and SMC are the reddening curve from the Large and Small Magellanic
Clouds, respectively. (From Abraham et al. 1997).}
\label{fig007}
\end{figure*}

Broad-band colors on their own can reveal information on the 
ages and reddening of individual components within a galaxy (Fig.~\ref{fig007}; 
Abraham et al. 1997, 1999), possibly revealing the physical causes 
of episodes of star-formation, such as sequential star-bursts.

If the JWST galaxy surveys are conducted in the same regions 
as existing HST observations such as the HDF, GOODS and the UDF, 
the data will allow a full representation of the spectral energy 
distribution of each galaxy, and of the distinct morphological 
components within it. Full SEDs will be obtained from the Lyman 
limit at 912 {\AA} out to a solid anchor in any older population 
in the rest-frame 0.6 \ensuremath{\mu}m region, even for galaxies at 
redshifts as high as z \ensuremath{\sim} 7. Only with JWST can the relationship 
between old and young stellar populations be understood fully, 
and only with JWST can a full characterization of the star-formation 
process at high redshift be made.

The properties of galaxies today depend on their environments 
and there is strong observational evidence for a morphology-density 
relation, showing a clear difference between stellar populations 
in the field and in rich clusters (e.g., Dressler 1984). It is 
not completely understood how these differences came about, and 
if they were established early in the evolutionary history of 
galaxies, perhaps in groups prior to the establishment of the 
full-blown clusters. Carrying out the above studies in a range 
of environments would show when and why these differences arose.

\subsection{How did the Heavy Elements Form?}

Where and when are the heavy elements produced? To what extent 
do galaxies exchange material with the intergalactic medium?

The average metallicity of the universe and of the objects in 
it as a function of epoch provides a fundamental metric reflecting 
the development of structure and complexity on galactic scales. 
Metallicity is observable and ``long-lived'' in the sense that 
heavy atomic nuclei, once produced, are not readily destroyed. 
The production of heavy elements is also one of only two cosmologically 
significant producers of luminosity in the universe, along with 
gravitational accretion energy.

\begin{figure*}
\centering
\includegraphics[width=1.00\textwidth]{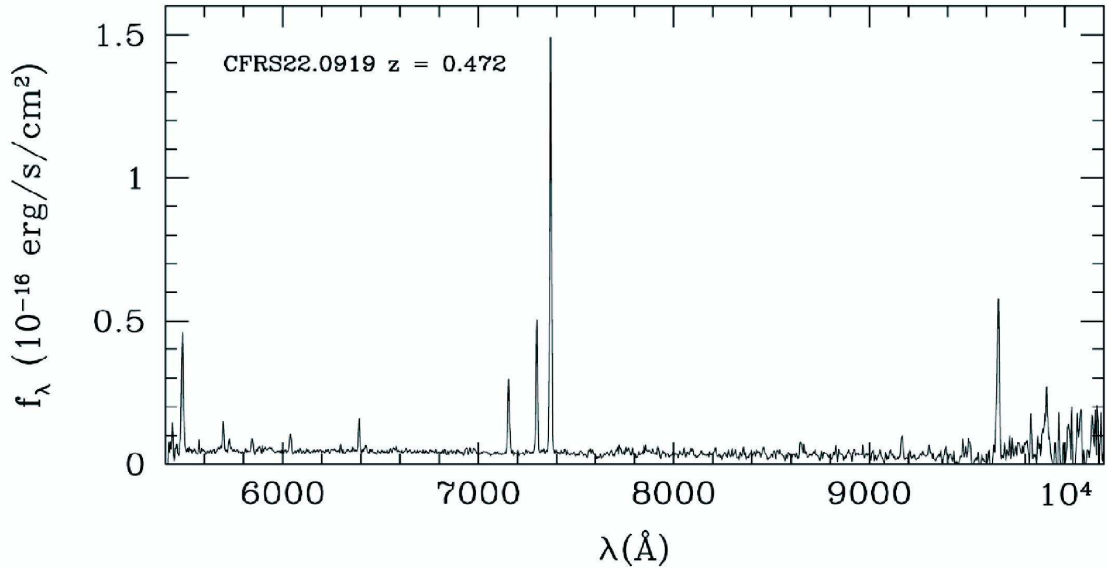}
\caption{
Spectrum of a galaxy at z $\sim $ 0.5, taken as part of the
Canada-France Redshift Survey, shows O[II] to H $\alpha $, the lines that
make up the R$_{23}$ index (From Lilly, Carollo \& Stockton 2003).}
\label{fig008}
\end{figure*}

For many years, the metallicities of gas at high redshifts have 
been studied through the analysis of absorption line systems 
seen in quasar spectra (e.g., Hamann \& Ferland 1999). The lines 
of sight to quasars probe random regions of the universe. The 
study of the metallicities of material in galaxies at high redshift 
is at a much earlier stage of development. This is more relevant 
for models of the chemical evolution of galaxies and for the 
use of metallicity estimates to constrain the present-day descendents 
of high redshift galaxies. The emission-line gas in star-forming 
regions is relevant for planetary and astro-biological studies, 
since it is likely to be representative of the material out of 
which the stars and planets are made. The metallicities of star-forming 
gas, especially of the [O/H] abundance, can be measured using 
diagnostics such as the R$_{23}$ index (Pagel et al. 1979), which 
is based on strong emission lines such as [OII] \ensuremath{\lambda}3727, 
H\ensuremath{\beta}, [OIII] \ensuremath{\lambda}\ensuremath{\lambda} 4959, 5007, H\ensuremath{\alpha}, 
[NII] \ensuremath{\lambda}6583, [SII] \ensuremath{\lambda}\ensuremath{\lambda} 6717, 6731 (Fig.~\ref{fig008}). 
Such measurements require R \ensuremath{\sim} 1000 to separate H\ensuremath{\alpha} 
and [NII].

\subsubsection{Observations}

JWST will be able to measure the H\ensuremath{\alpha}, H\ensuremath{\beta} and 
[OII]3727 and [OIII]5007 emission lines from compact low-extinction 
galaxies at z \ensuremath{\sim} 5 that are forming stars at the rate 
of 3 M$_{\ensuremath{\sun}}$/yr. This star-formation rate is comparable 
to that of the Milky Way today and requires a line sensitivity 
of 5\ensuremath{\times}10$^{-19}$ erg s$^{-1}$ cm$^{-2}$ in the 2 to 4 \ensuremath{\mu}m 
range at 10\ensuremath{\sigma}. In order to assemble sufficient samples 
for statistical determinations, JWST will have a multi-object spectrograph, 
and will be able to make these 
measurements with high multiplexing gain.

Extensive studies of the local universe have recently been extended 
to z \texttt{>} 1 using ground-based facilities. However, beyond 
z \ensuremath{\sim} 0.5 these observations become progressively more 
difficult from the ground as the emission lines are redshifted 
into the infrared. Emission-line measurements have only been 
made for a fraction of galaxies at redshifts much greater than 
z \ensuremath{\sim} 1. With JWST, all of these lines will be observable 
in the near infrared over the redshift range 1.7 \texttt{<} z \texttt{<} 
6, enabling metallicities of individual star-forming galaxies 
to be measured to a precision of about 0.2 dex.

Metallicities will be determined over this range of redshifts 
to tie in with the observations of ``first light'' and the very 
first enrichment, and to trace the development of metallicity 
through the epoch when most of the stars and metals were made.

Measurement of the gas metallicity in a very large number of 
faint galaxies (i.e., the metallicity distribution function) and 
comparison with the metallicities in neutral absorption line 
gas will allow JWST to address the origin of the enriched intergalactic 
medium, the enrichment histories of different types of galaxies, 
and the degree to which merging or accretion of galaxies alters 
the metallicity of growing galaxies.

\subsection{What Physical Processes Determine Galaxy Properties?}

When and how are the global scaling relations for galaxies established? 
Do luminous galaxies form through the hierarchical assembly of 
dark matter halos?

\paragraph{Global scaling relations:}

Despite the variety of galaxy properties observed today, galaxies 
obey a number of remarkably tight scaling relations between basic 
properties of luminosity, size, kinematics and metal enrichment. 
These include the Tully-Fisher relation for disk galaxies (Tully 
\& Fisher 1977) and the ``fundamental plane'', and projections 
thereof, for spheroids (Faber \& Jackson 1976; Kormendy 1977; 
Bender et al. 1992). More recently, a surprising relationship 
between the mass of the central black hole and the properties 
of the surrounding spheroid (e.g., the velocity dispersion) has 
been established (Ferrarese \& Merritt 2000; Gebhardt et al. 
2000; Tremaine et al. 2002). It is not known how or when these 
were established and whether they represent an asymptotic (late-epoch) 
state or whether they are obeyed at essentially all epochs (once 
allowance is made for the evolution of the stellar population).

Simulations of galaxy formation have managed to reproduce the 
slopes, but not the normalizations of these dynamical relations. 
The compatibility of scaling relations based on color or metallicity 
with models in which most stars are formed outside of their eventual 
parent galaxies is not completely clear. Determination of the 
nature of these scaling laws at 1 \texttt{<} z \texttt{<} 7, and of the 
scatter about them would likely reveal what physical process 
was responsible.

\paragraph{Hierarchical Assembly:}

In the standard CDM paradigm, the mass function of bound structures
develops with time, as smaller objects are assembled hierarchically
into larger ones, leading to an increase in the characteristic mass
M$^{*}$ in the Press-Schechter mass function (Press \& Schechter
1974; Schechter 1976; Percival 2001). JWST images and spectra will
study the evolution and organization of baryons in galaxies at high
redshift, but will not reveal the underlying structures of non-luminous
matter, which make up the gravitationally bound dark matter halos.
It is the development of these halos which is the fundamental test
of the CDM theory of galaxy formation.

\subsubsection{Observations}

\paragraph{Global scaling relations:}

Construction of the global scaling relations requires deep imaging 
for structural parameters plus high-resolution spectroscopy for 
kinematical data. Disk rotation curves at high spatial resolution 
can be measured in H\ensuremath{\alpha} in the observed near-infrared. 
This is much superior to the lines that are accessible in the 
observed optical band, such as Lyman \ensuremath{\alpha}, which are strongly 
affected by dust and radiative transfer effects in the interstellar 
medium, and in outflow regions, making them essentially useless 
for dynamical studies.

An important stellar kinematical diagnostic will come from the 
CO bandheads at 2.2 \ensuremath{\mu}m. These are particularly useful 
as they appear very quickly in young stellar populations, are 
largely independent of metallicity, and are little affected by 
dust extinction. For these reasons, they may well be the kinematic 
diagnostic of choice at high redshift. JWST will have sufficient 
mid-infrared sensitivity to measure the stellar velocity dispersions 
using the CO bandheads at z \ensuremath{\sim} 3 in at least the brighter 
of the Lyman break galaxies. Scaling from an optical magnitude of
AB = 24.5 mag in the R band, and the spectral energy distribution
of a present-day Scd galaxy, this observation needs a continuum sensitivity at
9 \ensuremath{\mu}m of AB = 21.3 mag at 10\ensuremath{\sigma} per
resolution element.

\paragraph{Hierarchical Assembly:}

The dark matter mass function of bound objects at very high redshifts 
can be uniquely measured in two ways with JWST. First, the dynamics 
of groups of galaxies or sub-galactic fragments can be used to 
determine the typical masses of halos (Zaritsky \& White 1994). 

These measurements require observations of emission lines in the
rest-frame optical, such as [OII] 3727, [OIII] 5007 and
H\ensuremath{\alpha}. These are very difficult to measure from the
ground when redshifted into the near infrared.

Second, JWST will measure halo masses through the gravitational
bending of light. Using this weak-lensing method, ground-based
programs have measured the mass within 200-500 kpc of galaxies at
redshifts of z \ensuremath{\sim} 0.1 (McKay et al. 2002) and z
\ensuremath{\sim} 1 (Wilson et al. 2001). Using the superior
resolution of HST, these measurements are likely to be extended
into 30 to 50 kpc for galaxies at z \ensuremath{\sim} 1 (e.g., Rhodes
et al. 2004; Rhodes 2004). While there are some hints of variable
halo structures for galaxies of different luminosity and total halo
mass, the radial penetration of these surveys, and the ability to
compare galaxies of different morphologies are limited by statistics.
We expect that HST will establish the statistical mass functions
for spiral and elliptical galaxies at z \ensuremath{\sim} 1, but
not much beyond that, because of its limited sensitivity and sampling
at \ensuremath{\lambda} \texttt{>} 1.6 \ensuremath{\mu}m.

JWST will extend the equivalent measurements of galaxies to z 
\ensuremath{\sim} 2.5 and thus determine the development of the dark 
matter halos during the peak growth of galaxies and star formation. 
JWST will require near-infrared imaging with high spatial resolution 
and sensitivity to achieve this greater depth. Background galaxies 
with a size comparable to the resolution of JWST will be measured 
at \ensuremath{\sim} 20$\sigma$.

The same near-infrared sensitivity and resolution will also make 
JWST superior to those of ground-based facilities and HST for 
the study of dark matter structures on larger scales, e.g., 1 
to 10 arcminutes or 2 to 20 Mpc (co-moving) at z \ensuremath{\sim} 3. 
These volumes measure the clustering of dark matter on cluster 
or even supercluster scales, and would extend the study of the 
mass function into the linear regime. The goal of these observations 
would be to verify the growth of structure between z \ensuremath{\sim} 
1000 (the CMB large-scale structure) and z \ensuremath{\sim} 2.5, i.e., 
during the period that dark matter dominated the cosmological 
expansion of the universe prior to the beginning of dark energy 
dominance at z \ensuremath{\sim} 1.

\subsection{What Roles do Starbursts and Black Holes Play in Galaxy Evolution?}

What are the redshifts and power sources of the high-redshift 
ultra-luminous infrared galaxies (ULIRGs)? What is the relation 
between the evolution of galaxies and the growth and development 
of black holes in their nuclei?

\paragraph{ULIRGs:}

The optical identification of high-redshift ULIRGs, found at 
sub-mm wavelengths, is extremely difficult with ground-based 
8-10 m telescopes. The objects are very faint, and the detected 
images are at the confusion limits of the sub-mm telescopes. 
At present, none of the deepest field samples are securely identified 
at a level greater than 50\%. Intensive efforts with ground-based 
telescopes will improve this before JWST's launch, but it is 
almost certain that many currently-known sub-mm sources will 
still be unidentified by the time JWST is launched. Spitzer observations
have revealed the power of the mid-infrared in ULIRG and AGN identification (Ivison
et al. 2004; Egami et al. 2004; Frayer et al. 2004). Analogs of 
known z \ensuremath{\sim} 2 ULIRGs, if they exist at z \texttt{>} 5, will 
have remained unidentified from the ground until JWST, even though 
they may well already be present in today's sub-mm samples. The 
Atacama Large Millimeter Array (ALMA) will resolve the confusion 
in the sub-mm, but deep imaging with JWST at \ensuremath{\lambda} \texttt{>} 
2 \ensuremath{\mu}m is needed to identify these sub-mm sources.

\paragraph{AGN:}

One of the most surprising discoveries in the study of galaxies 
in the last ten years has been that the masses of central black 
holes are tightly correlated with the bulge stellar population 
in present-day galaxies (e.g., Tremaine et al. 2002). These estimates 
have been extended using proxy indicators to redshifts z \ensuremath{\sim} 
2 in QSOs, and there are indications that this correlation still 
holds at high redshift (Shields et al. 2003; but see also Walter 
et al. 2004). Furthermore, the host galaxies of QSOs at redshifts 
z \texttt{>} 2 appear to be in very high states of star formation, 
while the peak in the quasar number density at z \ensuremath{\sim} 2 
suggests that the formation of the central black hole is contemporaneous 
with the production of the bulk of the stellar population. However, 
the existence of some bright QSOs at redshifts above 6, with 
spectra that differ little from those with the lowest redshift, 
suggests that some massive black holes and their associated stellar 
populations have formed early in the history of the universe 
(Fan et al. 2001; Freudling et al. 2003). The close connection 
between central black holes and spheroid populations must be 
intimately connected with galaxy formation and evolution, and 
with the events that trigger and fuel active galactic nuclei 
(AGN) over cosmic time.

Black hole masses have been measured by echo or reverberation 
mapping, maser kinematics, nuclear gas dynamics, nuclear star 
dynamics, and emission-line widths in AGN broad-line regions. 
They show a good level of agreement and are probably correct 
to within a factor of two or three. Many of these methods will 
be applicable at high redshifts with JWST. Bulge stellar populations 
are characterized by the bulge luminosity profiles, velocity 
dispersion, and overall flux, with appropriate mass-to-light 
ratios according to the stellar populations.

There are many questions that remain about the formation and 
evolution of super-massive black holes. We do not know if the 
seed black holes are primordial, if they form through the high-mass 
end of the population III mass function, and if they form over 
a wide range of redshifts. We do not know if their evolution 
traces the hierarchical growth of structure, or through merging 
within an initial stellar population. We do not know the role 
of angular momentum, and the role of central engine accretion 
mechanisms in their growth. Finally, we do not know the redshift 
dependence of black hole mass growth.

\begin{figure*}
\centering
\includegraphics[width=1.00\textwidth]{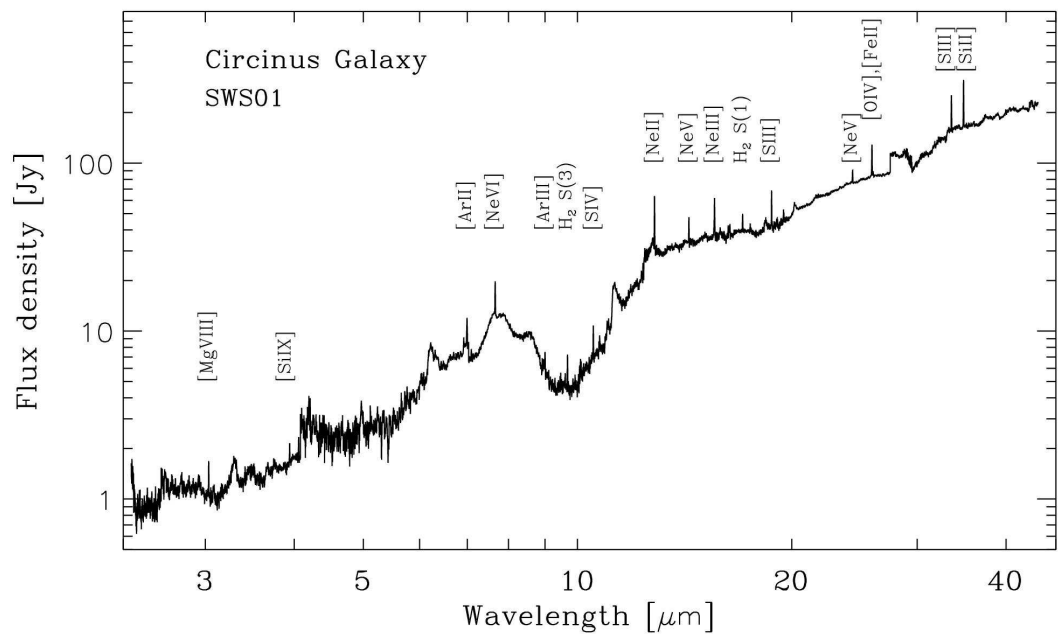}
\caption{
Mid-infrared spectrum of the Circinus galaxy taken with ISO shows
an abundance of emission lines useful for diagnosing the energy sources
which power ULIRGs (From Moorwood et al. 1996).}
\label{fig009}
\end{figure*}

\subsubsection{Observations}

Mid-infrared imaging will test whether mergers are the cause 
of the energy injection in high redshift ULIRGs. This will penetrate 
the dust obscuration that is known to be present in these obscured 
galaxies and will sample the oldest stellar populations in these 
objects, rather than just knots of recent star-formation. At 
the median redshifts of z \ensuremath{\sim} 2 to 3 expected for many 
of the sub-mm selected ULIRGs, JWST images at 4 \ensuremath{\mu}m will 
sample the stellar populations in these galaxies at wavelengths 
longwards of 1 \ensuremath{\mu}m in the rest-frame, allowing the best 
possible identification of mergers.

While many of these questions will be addressed using the same 
types of observations outlined above for non-active galaxies, 
JWST will also observe a range of active galaxy types and luminosities.

Near-infrared spectroscopy with JWST will have the capability 
to measure redshifts for identifications that cannot be secured 
from ground-based spectroscopy. Most ULIRGs at z \ensuremath{\sim} 4 
are too faint to be observed from the ground at \ensuremath{\lambda} \texttt{<} 
2 \ensuremath{\mu}m. The H\ensuremath{\alpha} line, which would be expected to 
be the strongest line in these highly obscured but vigorously 
star-forming galaxies, redshifts out of the ground-based K-band 
window at z \texttt{>} 2.6, but will be readily observable with JWST 
near-infrared spectroscopy to z \ensuremath{\sim} 6.5. Beyond z \texttt{>} 
2.6, it may be possible to observe lines shortward of H\ensuremath{\alpha} 
from the ground (e.g., [OIII] 4939,5007, H\ensuremath{\beta} and [OII] 
3727), but these will be extremely faint in these highly reddened 
objects, and even these will have left the K-band by z \ensuremath{\sim} 
5. Use of R = 1000 spectroscopy will yield kinematic information on the merging 
system and will separate H\ensuremath{\alpha} and [NII] allowing some 
estimate of metallicity to be made.

\begin{table}[t]
\caption{JWST\ Measurements for the Assembly of Galaxies Theme\label{tab002}}
\begin{tabular}{p{1.0in}p{0.8in}p{1in}p{1.0in}p{0.0in}}
\hline\noalign{\smallskip}
{Observation} &
{Instrument} &
{Depth, Mode} &
{Target} &
\\[3pt]
\tableheadseprule\noalign{\smallskip}
\raggedright Deep-wide survey (DWS)&
NIRCam&
3 nJy at 3.5$\mu$m&
100 arcmin$^2$&
\\
&&&& \\
\raggedright Metallicity determination&
NIRSpec&
\raggedright 5\ensuremath{\times}10$^{-19}$ erg s$^{-1}$ 
cm$^{-2}$, R\ensuremath{\sim}1000&
\raggedright Galaxies in DWS&
\\
&&&& \\
\raggedright Scaling relations&
MIRI &
\raggedright 11 \ensuremath{\mu}Jy at 9\ensuremath{\mu}m, R\ensuremath{\sim}3000&
\raggedright Lyman 
Break galaxies at z\ensuremath{\sim}3&
\\
&&&& \\
&
NIRCam&
3 nJy at 3.5 \ensuremath{\mu}m&
DWS data&
\\
&&&& \\
\raggedright Obscured galaxies&
MIRI&
23 nJy at 5.6\ensuremath{\mu}m&
ULIRGs&
\\
&&&& \\
&
NIRSpec&
\raggedright 5\ensuremath{\times}10$^{-19}$ erg s$^{-1}$ cm$^{-2}$, R\ensuremath{\sim}1000&
\raggedright ULIRGs and AGN&
\\
&&&& \\
&
MIRI &
\raggedright 1.4\ensuremath{\times}10$^{-16}$ erg s$^{-1}$ cm$^{-2}$ at 24 \ensuremath{\mu}m, 
R\ensuremath{\sim}2000&
\raggedright ULIRGs and AGN& \\
\noalign{\smallskip}\hline
\end{tabular}
\end{table}

Finally, at the long-wavelength end of the mid-infrared, high 
resolution spectra will allow the detection of narrow emission 
lines such as [NeVI] 7.66 \ensuremath{\mu}m to z \ensuremath{\sim} 2.5, while 
spectra at lower resolution will allow measurement of the equivalent 
width of the 7.7 \ensuremath{\mu}m polycyclic aromatic hydrocarbon (PAH) 
feature at redshifts as high as z \ensuremath{\sim} 2.5 and of the 3.3 
\ensuremath{\mu}m feature to redshifts of z \ensuremath{\sim} 6. These emission 
line and PAH features are good diagnostics of the energy sources 
in the center of these systems (Soifer et al. 2004; Armus et al. 2004). Star-bursts have strong PAH features, 
while AGN have much weaker features, because the PAHs are themselves 
destroyed and the hot dust continuum is stronger. [NeVI] is also 
much stronger in AGN-powered systems. JWST will allow application 
of these same diagnostics which have proven most useful in the low-redshift 
ULIRG systems.

JWST will be able to measure the [NeVI] line in an ultra-luminous 
obscured galaxy with the bolometric luminosity of Arp 220, 1.3\ensuremath{\times}10$^{12}$ 
L$_{\ensuremath{\sun}}$, at z \ensuremath{\sim} 2, assuming a line/bolometric luminosity 
ratio as in the Circinus galaxy (Fig.~\ref{fig009}).

\subsection{Summary}

Table~\ref{tab002} summarizes the measurements needed for the
Assembly of Galaxies theme. They include:

\textbullet{}
{\it Deep-Wide Survey (DWS).} A deep-wide multi-filter NIRCam survey
will be used for faint galaxy identification and morphology. Galaxies
would be assigned to approximate redshift bins using photometric
redshifts over the range 1 \texttt{<} z \texttt{<} 6. The stellar
populations that make up the morphological features in the galaxies
would be identified on the basis of their broad-band colors. This
program is designed to detect all galaxies brighter than the SMC
at z=5.

\textbullet{}
{\it Metallicity Determination.} Follow-up multi-object 
spectroscopy of hundreds or thousands of galaxies in the DWS 
will reveal the buildup of heavy elements as galaxies are assembled. 
The depth is sufficient to determine R$_{23}$ from emission line 
ratios for a galaxy with SFR = 3 M$_{\sun}$/yr at z = 5.

\textbullet{}
{\it Scaling relations.} MIRI spectroscopy of the CO bandhead at
rest wavelength 2.2 microns will measure the velocity dispersion and
put the galaxy on the fundamental plane or Tully-Fisher relation.
The depth is sufficient to measure the stellar velocity dispersion
for an R=24.5 mag Lyman Break galaxy at z = 3. In addition, a weak
lensing analysis of the DWS data will reveal the relationship
between the masses of galactic halos and their star light out to
z\ensuremath{\sim}2.5.

\textbullet{}
{\it Obscured galaxies.} Imaging of ULIRGs will penetrate the
obscuring dust to reveal the presence of merger-induced starbursts.
Redshift identification of highly obscured systems can be done with
H\ensuremath{\alpha} out to z\ensuremath{\sim}6.5. R\ensuremath{\sim}1000
spectroscopy will also reveal the kinematics of merging systems.
MIRI spectroscopy will determine the energy sources that power
these objects. The depth is sufficient to measure [NeVI] in a
z\ensuremath{\sim}2 ULIRG with Arp220 bolometric luminosity, assuming a
Circinus spectrum.

\section{The Birth of Stars and Protoplanetary Systems}

The key objective of The Birth of Stars and Protoplanetary Systems 
theme is to unravel the birth and early evolution of stars, from 
infall on to dust-enshrouded protostars, to the genesis of planetary 
systems.

\begin{figure*}
\centering
\includegraphics[width=1.00\textwidth]{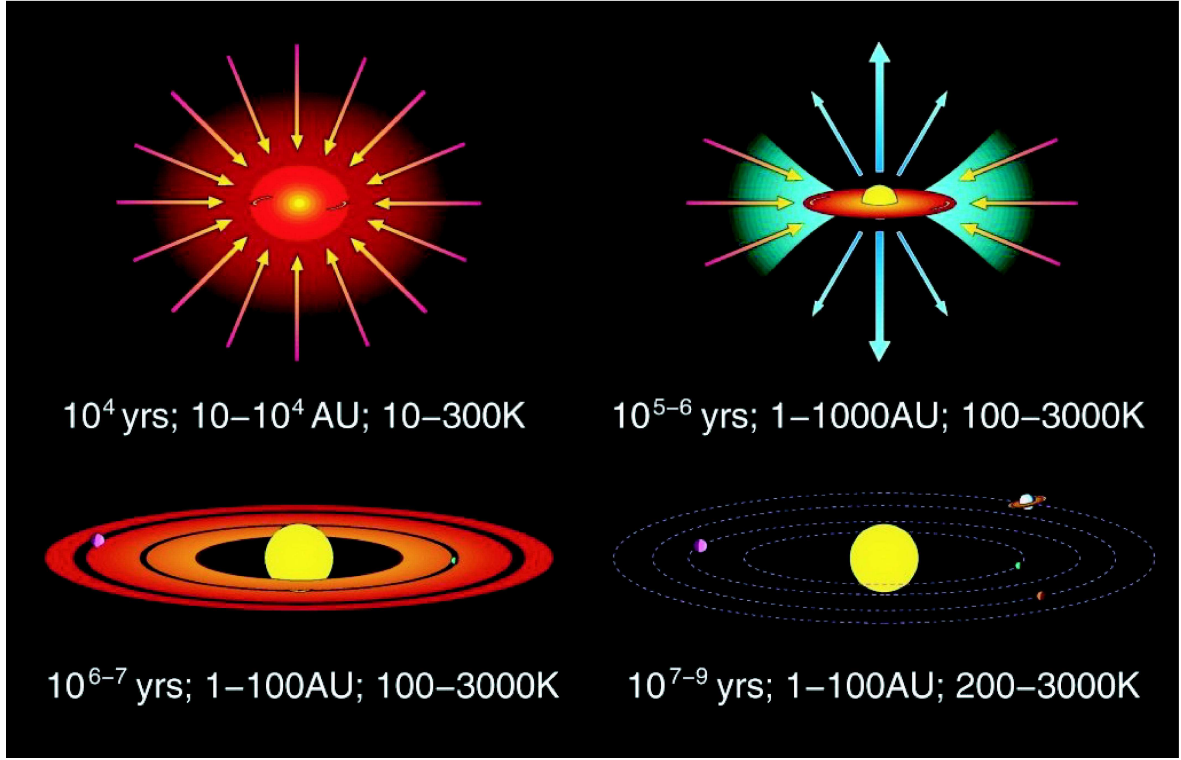}
\caption{
The formation of a single, isolated low-mass star and its
planetary system. Following a deeply embedded protostellar collapse phase
(Class 0 YSO; top-left), a circumstellar disk and collimated outflow are
established, which renders the central star visible at most orientations
(Class I/II; top-right). After accretion from the envelope is terminated,
perhaps by environmental influences, planetesimals and protoplanets form in
the passive disk via sedimentation and agglomeration (Class III;
bottom-left), later leaving a mature planetary system in orbit around the
star (bottom-right). The range of temperatures (10 K to 3000 K) involved and
the associated circumstellar dust extinction implies that the bulk of the
radiation seen in the early phases comes out at near-infrared through
millimeter wavelengths. Typical size scales (1 to 1000 AU, or 0.002 to 2
arcsec at 500 pc) imply that high spatial resolution is required for such
studies (After Shu et al. 1987).}
\label{fig010}
\end{figure*}

The formation of stars and planets is a complex process, even 
in the well-developed paradigm for a single, isolated low-mass 
star (see, e.g., Shu et al. 1987; Fig.~\ref{fig010}). We now know, however, 
that things are even more complicated, as stars very rarely form 
in isolation. The current picture of star formation starts on 
large scales, as molecular cloud cores cool and fragment to form 
highly dynamic clusters of protostars, spanning the mass spectrum 
from O stars to planetary-mass brown dwarfs. Within those clusters, 
individual young sources are often encircled by disks of warm 
gas and dust, where material aggregates to form proto-planetary 
systems. These disks are the source of highly-collimated jets 
and outflows, which transfer energy and angular momentum from 
the infalling material into the surrounding medium, and clear 
away the remainder of the birth core. On larger scales, the intense 
ultraviolet flux and strong winds of the most massive stars can 
disperse an entire molecular cloud, while simultaneously ionizing 
and evaporating the circumstellar disks of the surrounding lower-mass 
stars.

Young stars, brown dwarfs, and circumstellar disks emit the bulk 
of their radiation at near- and mid-infrared wavelengths, and 
at the earliest stages, the shorter wavelength emission is absorbed 
by the dust surrounding them. To probe these obscured regions 
and detect emission from gas and dust at temperatures ranging 
from 3000 K to 100 K, imaging and spectroscopic observations 
from roughly 1 to 30 \ensuremath{\mu}m are required. High sensitivity, 
high spatial resolution, and a large dynamic range are needed 
to study the physical properties, composition, and structure 
of faint stellar companions, disks, and protoplanets immediately 
adjacent to their much brighter neighbors. Finally, a large field of view 
is needed to ensure that the diverse range of sources, phenomena, 
and their interactions within a given star-forming complex can 
be captured and disentangled.

\subsection{How do Protostellar Clouds Collapse?}

How do clouds of gas and dust collapse down to the dense cores 
that form stars? What is the early evolution of protostars?

\paragraph{Clouds and Cores:}

Stars form in small (\ensuremath{\sim}  0.1 pc) regions undergoing gravitational 
collapse within larger molecular clouds. These dense cores have 
densities n$_{H_2}$ \texttt{>} 10$^{4}$ cm$^{-3}$, roughly a hundred times greater 
than ambient cloud material. Standard theory predicts that these 
cores collapse from the inside out (e.g., Shu 1977; Terebey et 
al. 1984), in which the center forms first and the outer envelope 
rains down upon it. The collapse propagates at an effective sound 
speed of about 0.3 km s$^{-1}$, accounting for gas pressure and 
support due to magnetic fields and turbulence. The slowly collapsing 
and slowly rotating core approximates a singular isothermal sphere, 
breaking down in the center where a protostar and a more rapidly 
rotating disk are found. However, there are alternatives to the 
standard picture. Ambipolar diffusion, due to incomplete coupling 
of magnetic fields to the gas, can result in rigid, rather than 
differential, rotation of the cloud core (Mouschovias \& Palelogou 
1981; Crutcher et al. 1994). Furthermore, cores may be externally 
pressure-confined (Alves et al. 2001), or may be altogether more 
chaotic and dynamic structures formed in the intersections of 
fractal clouds (Bate, Bonnell, \& Bromm 2003; MacLow \& Klessen 
2004). These different models predict different density distributions 
for star-forming cores. By measuring those density distributions 
for cores in a wide range of environments and evolutionary states, 
we can hope to understand the relative roles that magnetic fields, 
turbulence, and rotation play while the clouds collapse to form 
stars.

Observations of optically-thin dust emission at millimeter continuum 
wavelengths have been used to trace the structure of dense cores, 
but the inversion of a measured intensity profile into a density 
profile is difficult, as it relies on an assumed underlying temperature 
profile and three-dimensional structure. For example, it has 
not yet been possible to distinguish unambiguously between flattened 
and peaked central cores (Ward-Thompson et al. 1994, 1999; Evans 
et al. 2001; Zucconi et al. 2001). The low spatial resolution 
of current sub-millimeter telescopes (about 10 arcsec) is also 
a problem that will be partly alleviated by new sub-millimeter 
interferometers such as the Sub-Millimeter Array and ALMA.

\begin{figure*}
\centering
\includegraphics[width=0.48\textwidth]{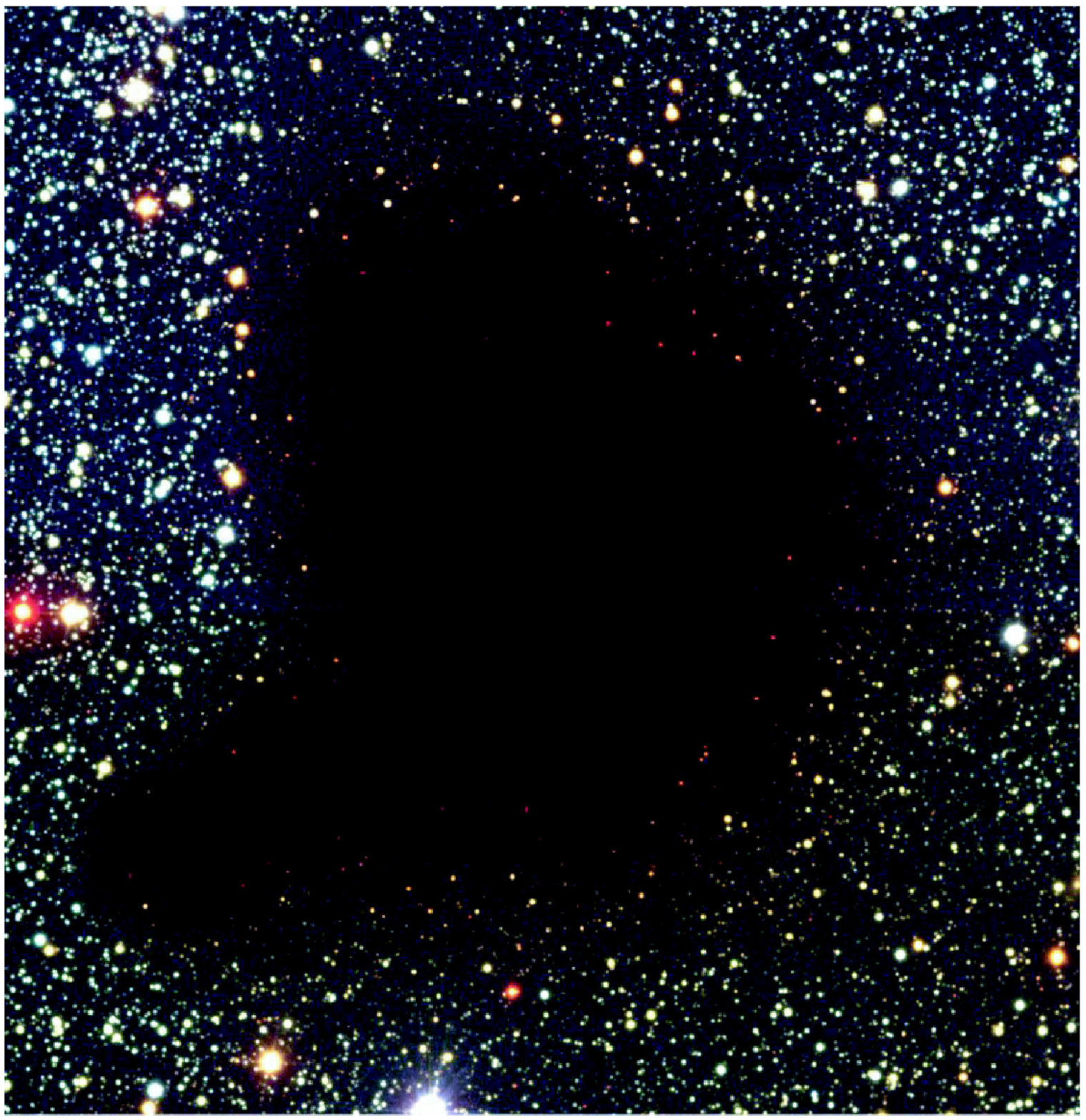}
\hfill
\includegraphics[width=0.48\textwidth]{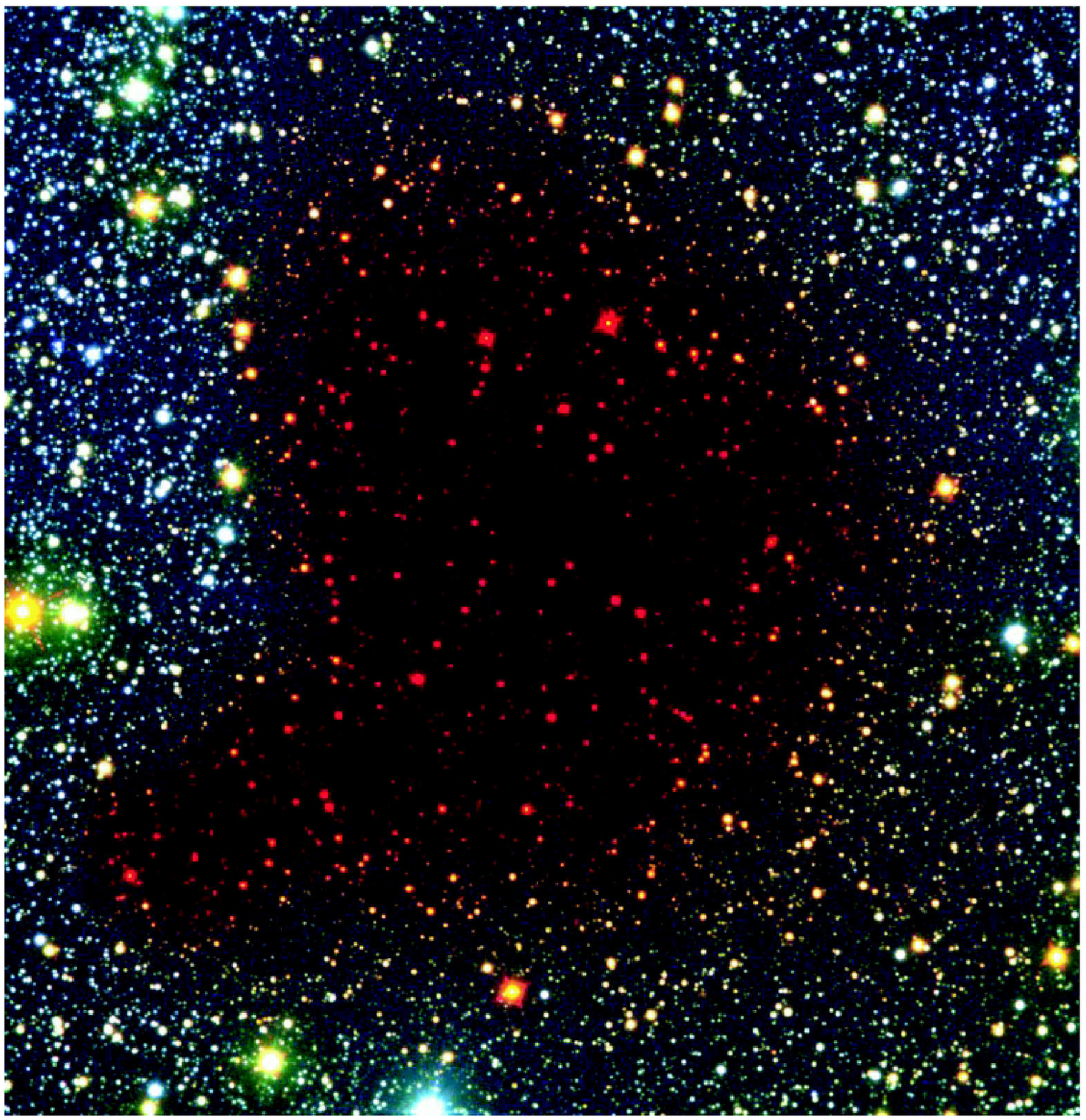}
\caption{
The low-mass dark cloud Barnard 68 imaged at optical and
near-infrared wavelengths using FORS1 on the ESO VLT and SOFI on the ESO
NTT. The left panel shows a color composite of optical B, V, and I images,
while the right panel shows a composite of the B and I images with a
near-infrared K$_{S}$ image. The images cover 4.9 x 4.9 arcmin, or 0.18 x
0.18 pc. At optical wavelengths, the small cloud is completely opaque
because of the obscuring effect of dust particles in its interior. Since the
light from stars behind the cloud is only visible at the infrared
wavelengths, they appear red. Using other infrared images and measuring the
extinction on a star-by-star basis, the dust column density profile of Barnard 68
and similar dark clouds could be measured (From Alves et al. 2001).}
\label{fig011}
\end{figure*}

\begin{figure*}
\centering
\includegraphics[width=1.00\textwidth]{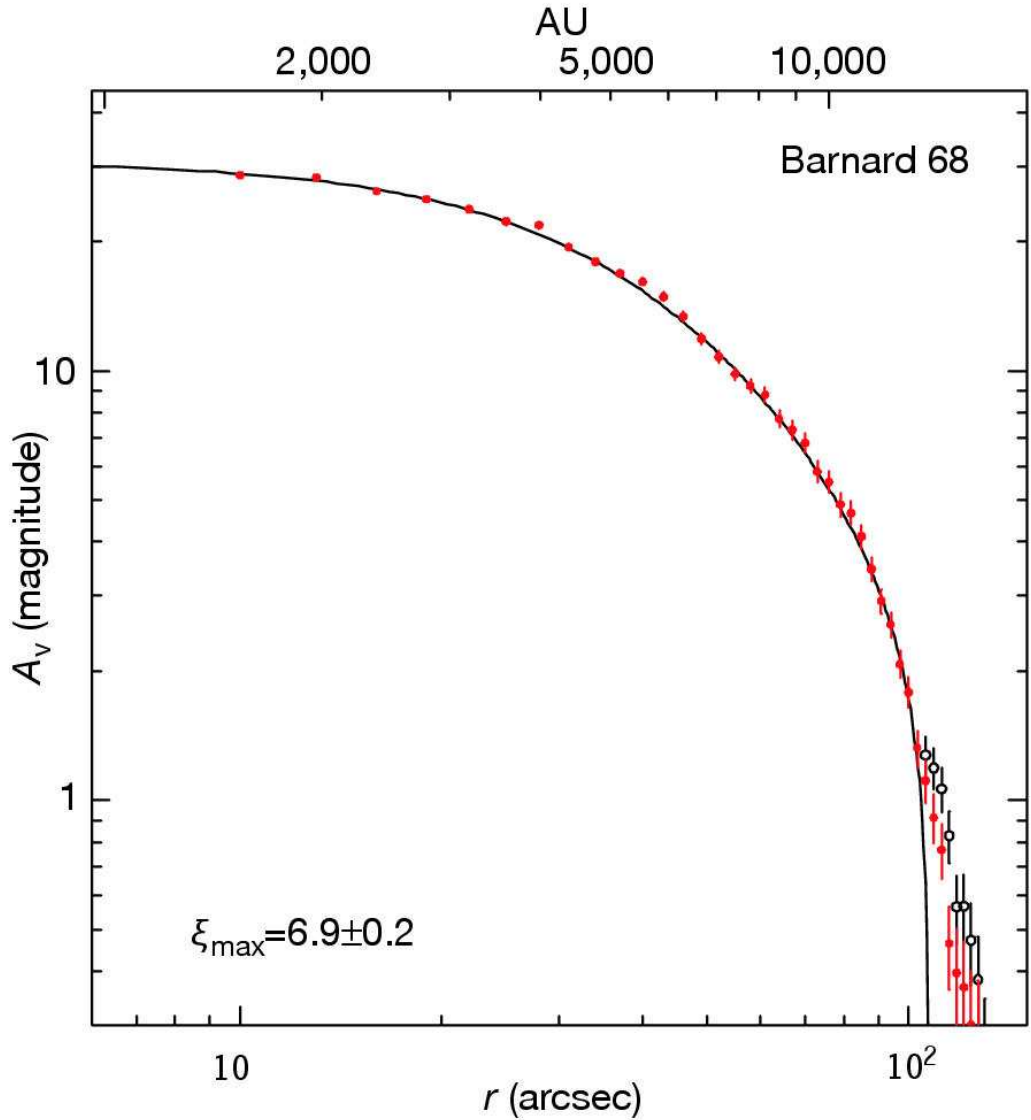}
\caption{
Azimuthally averaged radial dust column density profile for Barnard 68.
The red circles show the averaged profile of a subsample of data that does
not include the southeast prominence of the cloud (see Fig.~\ref{fig011}), while the
open circles include the prominence. The solid line represents the best fit
of a theoretical Bonnor-Ebert sphere to the data. The close match suggests
that the internal structure of the cloud is well characterized by a
self-gravitating, pressure-confined, isothermal sphere, and the cloud
appears to be near hydrostatic equilibrium (From Alves et al. 2001).}
\label{fig012}
\end{figure*}

An alternative technique involves mapping the extinction seen 
along various lines of sight through a cloud core, by measuring 
the near-infrared colors of discrete background field stars shining 
through it. The extinction can be directly related to the dust 
column density, so a 2D projection of the core density profile 
can be deduced, assuming a fixed gas-to-dust ratio. Used on ground-based 
telescopes, a typical maximum depth of AB = 22 mag in the K-band 
can be reached with seeing-limited resolution, providing a resolution 
in the resulting extinction map of 10 to 15 arcsec through extinctions 
of up to A$_{V}$ \ensuremath{\sim} 60 mag in dark clouds (e.g., Lada et al. 
1994; Alves et al. 1998; Alves et al. 2001; see Fig.~\ref{fig011} and Fig.~\ref{fig012}). 
The much greater sensitivity and substantially improved spatial 
resolution of the JWST will yield more much detailed profiles 
through greater column density.

Another approach maps the attenuation of the diffuse mid-infrared 
background produced by the interstellar radiation field or by 
hot sources in the same star-forming complex as the core. This 
background is particularly bright in the 6.2 and 7.7 \ensuremath{\mu}m 
PAH emission features, where dust extinction is also near a minimum. 
In this manner, Bacmann et al. (2000) used ISOCAM to measure 
extinction profiles in pre-stellar cores with a spatial resolution 
of 10 arcsec through extinction values of up to A$_{V}$ \ensuremath{\sim} 
50 mag. Again, JWST's mid-infrared spatial resolution and high 
sensitivity will enable mapping through much higher extinctions 
and with greater fidelity.

\paragraph{Protostars:}

\begin{figure*}
\centering
\includegraphics[width=1.00\textwidth]{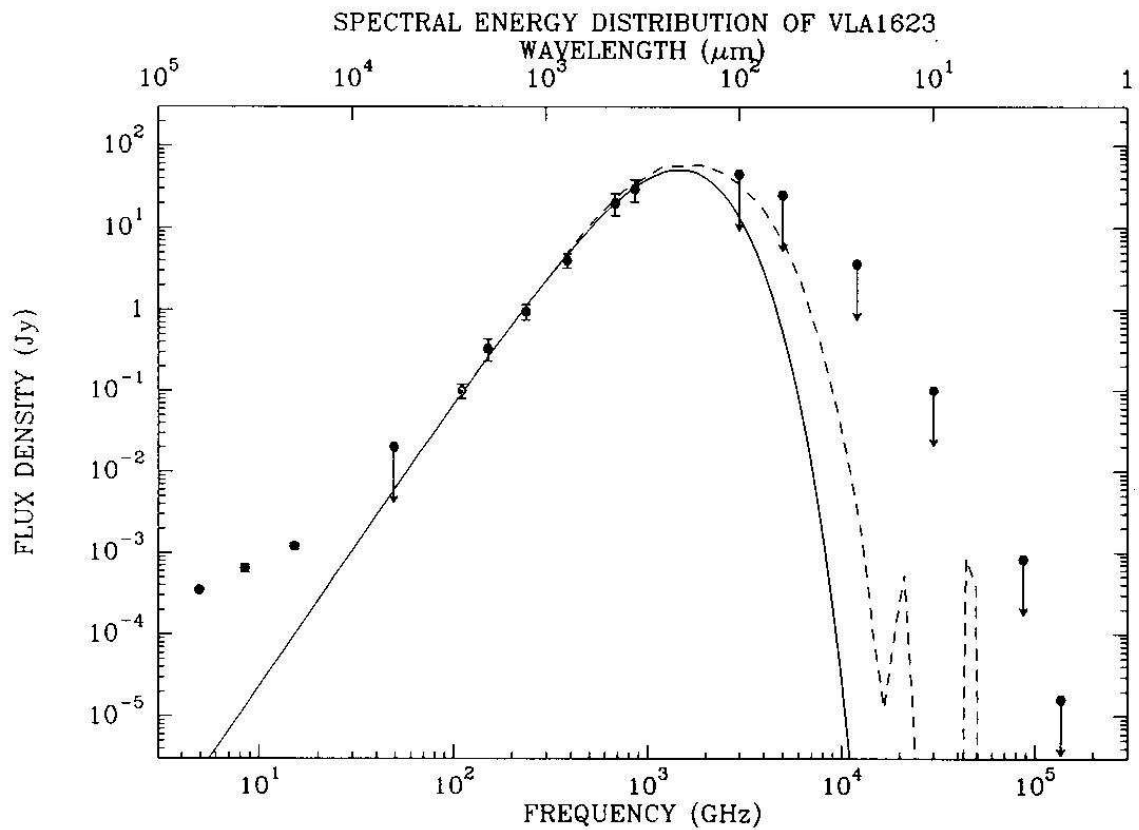}
\caption{
The spectral energy distribution of the Class 0 protostar. VLA
1623, in Ophiuchus. Despite a nominal blackbody temperature of only $\sim $
20 K, radiative transfer models predict significant mid-infrared flux as
emission from the warm core is scattered off dust grains in the inner
envelope (From Andr\'{e} et al. 1993).}
\label{fig013}
\end{figure*}

Once self-gravitating molecular cloud cores have formed, they 
can collapse to form protostellar seeds, which gain material 
via continuing accretion. The earliest category of protostar, 
the Class 0 object (Andr\'{e}, Ward-Thompson, \& Barsony 1993), 
is deeply embedded in, and obscured by, the massive envelope 
from which it is accreting, and its spectral energy distribution 
is dominated by this cold (\ensuremath{\sim} 20 K) material. As a result, 
these young (\ensuremath{\sim}  10$^{4}$ yrs) sources emit the bulk of their 
flux at millimeter and sub-millimeter wavelengths, and are generally 
undetected at shorter wavelengths to date (Fig.~\ref{fig013}).

Detecting and studying the 10 \ensuremath{\mu}m to 20 \ensuremath{\mu}m emission 
from protostars is important. Radiative transfer models (Wolfire 
\& Cassinelli 1986, 1987; Andr\'{e} et al. 1993; Boss \& Yorke 1995) 
predict that there should be a warm `shoulder' in the mid-infrared 
in the otherwise single \ensuremath{\sim} 20 K blackbody SED, and that 
protostars should be roughly 1000 times brighter than the blackbody 
at some wavelengths, since radiation from the warm central source 
is scattered off dust grains in the inner envelope into the line-of-sight. 
The degree of scattering is a strong function of the density 
distribution in the envelope, so the departures from the single 
blackbody SED at mid-infrared wavelengths would be an important 
diagnostic of envelope structure, most critically the power law 
of the density distribution. 

Cernicharo et al. (2000) confirmed these predictions with ISOCAM 
detections of a few luminous Class 0 protostars. Imaging in selected 
narrow bands (5.3, 6.6, 7.5 \ensuremath{\mu}m) between ice and silicate 
absorption features, warm material (\ensuremath{\sim} 700 K) was observed 
through effective extinctions of A$_{V}$ \ensuremath{\sim} 80 to 100 mag, 
and with flux coming from within 4 AU of the accreting protostars. 
More detailed observations of this kind are required for a much 
wider range of protostellar luminosities, in order to constrain 
the central protostellar parameters in envelope models, so that 
density distributions can be extracted more accurately from the 
Class 0 envelope observations.

The dynamics of the protostellar collapse can be diagnosed through 
imaging and spectroscopy of shocks, which form as material accretes 
onto the inner envelope and disk, and as the vertical velocity 
component is dissipated. The models of Yorke \& Bodenheimer (1999) 
predict at least two shock fronts at 500 to 1000 AU from the 
protostar, with positions changing as a function of evolution 
in the system, i.e., moving further from the source as the disk 
grows, but disappearing once the accretion terminates.

Finally, it is now clear that the majority of stars form in 
binaries or high-order multiples, but their origin is not well 
understood. While some theoretical predictions of fragmentation 
models are supported indirectly by statistical studies of evolved 
binary systems at optical and near-infrared wavelengths, direct 
observations of the binary formation phase itself became possible 
only recently with the advent of large, sensitive millimeter 
interferometers. However, as noted above, millimeter wavelength 
observations can only probe extended envelopes, not the protostellar 
cores themselves. Deep high-resolution imaging at 10 \ensuremath{\mu}m 
is therefore needed to observe the central hydrostatic cores 
in simultaneously turbulent, rotating, fragmenting, and collapsing 
protostellar clouds. In combination with detailed kinematic data 
supplied by future millimeter interferometers such as ALMA, such 
data will provide crucial tests of binary fragmentation models, 
allowing us to determine true initial binary fractions and separations, 
and how these properties change as stars evolve through the pre-main-sequence 
phases.

\subsubsection{Observations}

\paragraph{Clouds and Cores:}

Measuring the larger-scale structure in clouds can be done from 
the ground, but to probe the centers of pre-stellar cores and 
Class 0 envelopes, substantially higher sensitivity and spatial 
resolution are required to detect the much fainter and redder 
stars through the compact core. To carry out these observations, 
at least a factor of two increase in extinction penetration is 
required relative to present ground-based observations, i.e., 
up to A$_{V}$ \ensuremath{\sim} 120 mag, which requires an additional 7 
mag of sensitivity at 2 \ensuremath{\mu}m.

To achieve this, JWST will reach a 10\ensuremath{\sigma} point-source 
limiting sensitivity of at least K$_{AB}$ = 29 mag, or 9 nJy at 
2 \ensuremath{\mu}m. Diffraction-limited spatial resolution is required 
to ensure high-fidelity mapping of the extinction profile with \ensuremath{\sim} 1-2 
arcsec resolution, i.e., 200-500 AU for clouds at a few hundred 
parsecs distance. Finally, in order to map the full extinction 
profile of a typical 0.1 pc radius cloud core at the same distance, 
a field-of-view of 2 to 4 arcmin is required.

The centers of cores and Class 0 objects have even more extinction, 
and thus mid-infrared observations using the extended background 
emission technique are required. A typical mid-infrared background 
flux of 10 MJy sr$^{-1}$ yields an unattenuated surface brightness 
of 240 \ensuremath{\mu}Jy arcsec$^{-2}$.

To observe this surface brightness through A$_{V}$ \ensuremath{\sim} 300 
mag of extinction (or A$_{7}$$_{\ensuremath{\mu}}$$_{m}$ \ensuremath{\sim} 7 mag, assuming 
a standard extinction law) requires a 10\ensuremath{\sigma} surface-brightness 
sensitivity of 1 \ensuremath{\mu}Jy arcsec$^{-2}$ over the 6.7 to 7.7 \ensuremath{\mu}m 
region where the background is bright and the dust extinction 
low; binning of the pixels to 1 arcsec can be used to increase 
the surface brightness sensitivity. In order to map the central 
regions of cores and connect the results with those obtained 
for the lower-density outskirts in the near-infrared, JWST will 
have a mid-infrared field-of-view larger than 1 arcmin (i.e., 
0.15 pc at 500 pc).

\paragraph{Protostars:}

In order to characterize the density structure in the envelopes 
and cores of Class 0 sources, broad-band fluxes from 10 to 20 \ensuremath{\mu}m 
and narrow-band imaging in the 5 to 7 \ensuremath{\mu}m extinction windows 
are required. To detect such young protostars and protostellar 
cores at a distance of \ensuremath{\sim} 150 pc (the distance of the nearest 
star-forming regions like Taurus and Chamaeleon), JWST will reach 
sensitivities of 1 \ensuremath{\mu}Jy at 6 \ensuremath{\mu}m and 10 \ensuremath{\mu}Jy 
at 15 \ensuremath{\mu}m (Fig.~\ref{fig013}). Going out to a distance of 500 pc, 
and thus encompassing a much wider range of star-forming environments, 
sensitivities of 0.1 \ensuremath{\mu}Jy and 1 \ensuremath{\mu}Jy will be required 
at 6 and 15 \ensuremath{\mu}m, respectively.

High spatial resolution (\texttt{<}1 arcsec FWHM) is required at 
mid-infrared wavelengths, as Class 0 protostars should be slightly 
extended due to dust scattering, with diameters of order 1000 
AU, i.e., 6 to 2 arcsec in star-forming regions at 150 to 500 
pc. In addition, the sources need to be resolved out against 
the larger (2000 to 5000 AU) envelope in which they are embedded 
(e.g., Boss \& Yorke 1995).

Resolving binary protostars at the peak of the separation distribution 
for pre-main sequence stars (\ensuremath{\sim}  30 AU) requires an angular 
resolution less than 0.25 arcsec at 6 \ensuremath{\mu}m, where hot dust 
emission from the inner parts of accretion disks around slightly 
more evolved (Class I) protostars dominates. The preferred wavelength 
range to detect embedded protostars directly is 20 to 25 \ensuremath{\mu}m; 
diffraction-limited imaging (\ensuremath{\sim} 0.8 arcsec FWHM) would 
resolve binary protostars at separations that are slightly beyond 
the peak of the period distribution. ISO and Spitzer do not have 
sufficient angular resolution to distinguish the protostars from 
extended emission by externally heated small grains in the envelopes, 
or to resolve typical binary protostellar separations. 

For brighter protostars and cores, JWST will use mid-infrared 
integral-field spectroscopy of a range of narrow-band features 
(extinction windows, shock tracers, and PAHs) and the intervening 
continuum simultaneously, thus yielding a large range of diagnostics, 
while ensuring accurate spatial registration when calculating 
temperatures and sizes of the sources. Assuming the necessary 
spectral resolution of R = 2000 at 15 \ensuremath{\mu}m, a continuum 
sensitivity of 7 x 10$^{-19}$ erg cm$^{-2}$ s$^{-1}$ per resolution element is needed 
to detect a 10 \ensuremath{\mu}Jy source, such as a typical Class 0 protostar 
in Taurus-Auriga, at a distance of about 140 pc.

\subsection{How Does Environment Affect Star Formation and Vice Versa?}

How do very massive stars form? How do stellar winds and ionizing
radiation from massive stars affect nearby star formation?

The formation of massive stars produces intense winds and ionizing 
radiation which impacts the surrounding molecular cloud material 
and the nascent circumstellar disks of adjacent low-mass stars. The 
mechanism by which massive stars form is not yet known. The standard 
disk-accretion scenario appropriate for low-mass stars cannot 
be simply scaled up, since predictions are that the radiation 
pressure from the growing central source would build up so quickly 
that no more material could accrete, limiting the mass of the 
source (Yorke \& Sonnhalter 2002).

Various alternatives have been proposed to solve this problem. 
McKee \& Tan (2002) have suggested that very high accretion rates 
can overcome the radiation pressure, allowing a massive star 
to form ``normally'' via infall and a disk. However, to date, 
no definitive evidence for circumstellar disks around massive 
stars exists, leaving this hypothesis unconfirmed. Alternatively, 
Bonnell et al. (1998) proposed that massive stars might form 
by the agglomeration of colliding low-mass stars in the very 
deep potential well at the center of a dense young cluster. In 
this model, stellar densities of 10$^{6}$ to 10$^{8}$ stars per cubic 
parsec are required, on the order of 100 to 1000 times the density 
seen in the 0.1 pc cores of present-day young embedded clusters.

The required high densities would be achieved by shrinking such 
a cluster by a factor of 10 in linear size, i.e., to a core radius 
of 0.01 pc, or \ensuremath{\sim}2 arcsec at 1 kpc: such densities are 
predicted in some models of early cluster evolution (see, e.g., 
Kroupa, Petr, \& McCaughrean 1999; Stahler, Palla, \& Ho 2000). 
There are a number of observational consequences of massive star 
formation via this collisional route (Bally \& Zinnecker 2005), 
including high-luminosity infrared flaring and impulsive wide-angle 
outflows, in addition to the simple fact of a very high density 
of low-mass stars crammed into a small region, all of which should 
be traceable with the JWST.

Once born, massive stars can be disturbing neighbors. On small 
scales, ionizing photons and strong winds from O and B stars 
can destroy disks around young low-mass stars (Johnstone et al. 
1998; Bally et al. 2000). On larger scales, they can simultaneously 
evaporate and compress surrounding molecular material, either 
halting or triggering further star formation in surrounding molecular 
material (Larosa 1983; Bertoldi 1989; Lefloch \& Lazareff 1994; 
Hester et al. 1996).

\begin{figure*}
\centering
\includegraphics[width=1.00\textwidth]{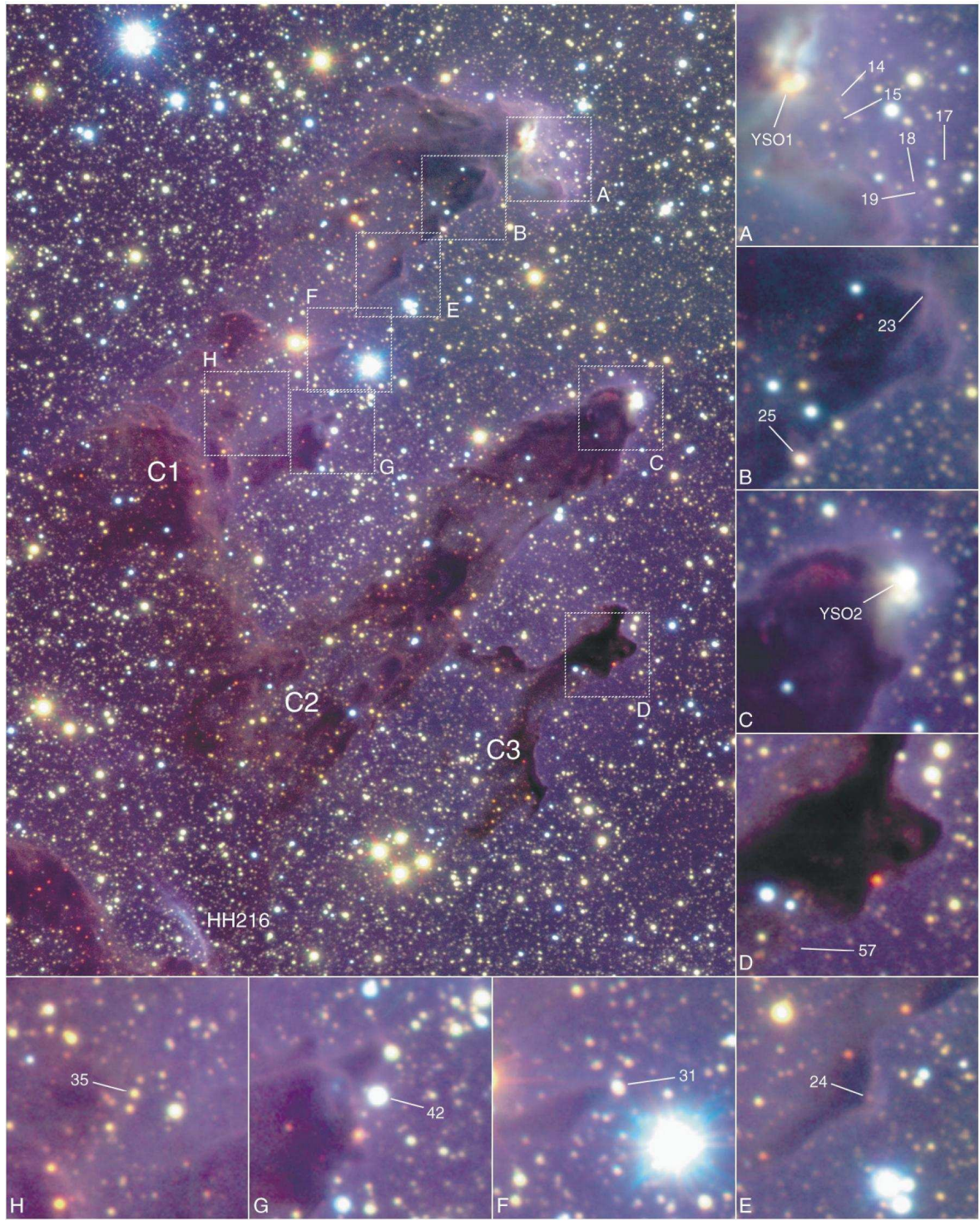}
\caption{
Near-infrared (1 to 2.5 $\mu $m) image of the M16 Elephant Trunks
made using the ESO VLT. The J$_{S}$-band data are shown as blue, H-band as
green, and K$_{S}$-band as red. The image covers 158 $\times$ 214 arcsec$^2$ 
(1.5 $\times$ 2.0 pc$^2$ at 1.9 kpc); north is up, east left. Subimages show more
detail, including evaporating gaseous globules from Hester et al. (1996) found to be associated with
point sources; E23, a globule with no near-infrared point source, but thought
to contain an embedded protostar driving a collimated jet; YSO1 and YSO2,
massive sources in the tips of C1 and C2, respectively; and HH 216, an
optically-visible Herbig-Haro object (From McCaughrean \& 
Andersen 2002).}
\label{fig014}
\end{figure*}

An example of these environmental impacts is seen in M 16, the 
Eagle Nebula where parsec-scale molecular trunks are evaporated 
and disrupted by O and B stars of the adjacent NGC 6611 cluster 
(Hillenbrand et al. 1993). A population of small 1000 AU-scale 
(0.5 arcsec at M 16) dense globules is seen on the surfaces of 
the trunks (Hester et al. 1996). These evaporating gaseous globules 
may contain young stars about to be revealed as the O 
and B stars evaporate their birth clouds. Near-infrared observations 
show that about 10 to 20\% contained embedded low-mass stellar 
and brown dwarf candidates (McCaughrean \& Andersen 2002; see Fig.~\ref{fig014}), 
in addition to the higher-mass young stellar objects (YSOs) in 
the tips of the ablating columns (see also Thompson, Smith, \& 
Hester 2002; Sugitani et al. 2002).

Although columns like these may be associated with recent star 
formation, it remains unknown whether it is triggered by the 
passage of an ionization front from the surrounding O and B stars, 
or whether pre-existing YSOs are revealed as the parent cores 
are eroded away (see also Smith, Stassun \& Bally 2005). Do O 
and B stars have an active role in creating new stars via radiative 
implosion, or are they simply destructive: exposing stars prematurely, 
perhaps reducing their final masses and destroying their disks? 
To what extent is the mass of a star (and by extension, the whole 
stellar IMF) determined by when and how its envelope of accreting 
material is stripped away by nearby massive stars rather than 
processes local to the star itself?

\subsubsection{Observations}

In order to solve the massive star formation paradox, deep, high 
spatial resolution imaging is required in the mid-infrared to 
see through the high extinction into the very dense, very young 
cluster cores. These stars are obscured by the considerable density 
of gas and dust that accompany them: a prototypical hyper-compact 
H II region might contain a volume density of 10$^{7}$ cm$^{-3}$ in a 
region 0.01 pc (2000 AU) in radius, yielding a column density 
to the center of 3 $\times$ 10$^{23}$ cm$^{-2}$ or A$_{V}$ \ensuremath{\sim} 150 mag.

On the basis of extinction alone, observations at 8 and 13 \ensuremath{\mu}m 
would be preferred, but some trade-off with spatial resolution 
will be sought, given the potentially very high density of point 
sources that would be expected in the agglomeration hypothesis where, 
for example, as many as 50 to 100 low-mass stars might occupy 
a region only 4 arcsec across.

The best compromise is found at somewhat shorter wavelengths. 
At 3.8 and 4.8 \ensuremath{\mu}m, A$_{V}$ \ensuremath{\sim} 150 mag would be reduced 
to 6 and 4 mag, respectively. The aim would be to measure the 
full stellar initial mass function and thus in order to detect 
a fiducial 0.1 M$_{\sun}$, 1 Myr old star at 1 kpc, seen through this 
dust column, JWST will reach point source limits of \ensuremath{\sim}  
2 and 10 \ensuremath{\mu}Jy, respectively (Baraffe et al. 1998). Given 
the impact of crowding, JWST must combine good sensitivity with 
large dynamic range and excellent, stable imaging quality. With 
diffraction-limited imaging at these wavelengths, \ensuremath{\sim} 0.15 
arcsec resolution should yield about 30 fully-sampled resolution 
elements across 4 arcsec, equivalent to 4000 AU at 1 kpc distance.

To determine the impact of massive stars on their environments, 
comprehensive surveys will be made of dark clouds and elephant 
trunks in regions with recent massive star formation, to reveal 
populations of young stars and study their properties. By examining 
the masses and ages of the sources as a function of their distance 
from the ionization front, it will be possible to determine whether 
O and B stars simply reveal pre-existing star formation or trigger 
it directly (see, e.g., Smith et al. 2005). 

Sensitive mid-infrared observations are needed to penetrate the 
dust in such trunks, and reveal pre-main sequence stars and brown 
dwarfs, as well as very young protostellar sources. Broad-band 
imaging at 3 to 20 \ensuremath{\mu}m is needed to provide a census of 
the photometric properties of deeply embedded and/or very young 
sources, while follow-up spectroscopy is necessary to compare 
the mass, ages, and evolutionary status of the sources in a given 
trunk with models. In addition, narrow-band line imaging in a 
variety of ionized, atomic, and molecular tracers is needed to 
delineate the rate at which photoionization, photoevaporation, 
and shocks are propagating into the molecular cloud core, and 
to determine the balance between implosion and destruction. All 
of these observations will be conducted at high spatial resolution 
to permit unambiguous identification of a given source with a 
given globule: typical sizes are 200 to 2000 AU or 0.1 to 1 arcsec 
at 2 kpc distance. 

Young sources embedded in such regions will vary considerably 
in flux as a function of mass, wavelength, extinction, and circumstellar 
excess emission. For example, the sensitivity required to detect 
a 1 Myr old, 0.02 M$_{\sun}$, low-mass brown dwarf (Baraffe et al. 
1998) with typical infrared excess emission due to a circumstellar 
disk (Kenyon \& Hartmann 1995) would be \ensuremath{\sim}  55 nJy, 2.3 \ensuremath{\mu}Jy, 
and 34 \ensuremath{\mu}Jy at 3.5, 4.8, and 20 \ensuremath{\mu}m, respectively, 
assuming a distance of 2 kpc and an extinction of A$_{V}$ = 100 
mag. Under similar conditions, a 0.075 M$_{\sun}$ source at the star 
to brown dwarf boundary would require sensitivities of 0.65, 
24, and 370 \ensuremath{\mu}Jy, respectively, at the same wavelengths. 
Broad-band imaging photometry is needed in the near- and mid-infrared 
down to the 0.02 M$_{\sun}$ limit, with multiobject spectroscopy at 
R $\sim$ 100 for the shorter near-infrared wavelengths, and R $\sim$ 1000 
at the longer. In the mid-infrared, integral field spectroscopy 
at R $\sim$ 2000 of selected individual sources down to the star-brown 
dwarf boundary is required.

\subsection{What is the Initial Mass Function at sub-stellar Masses?}

Does cloud fragmentation explain low-mass star formation, and is
there a lower limit to the mass? How does the sub-stellar initial
mass function depend on metallicity or environment?

The initial mass function (IMF) is a key product of star formation 
(Salpeter 1955; Miller \& Scalo 1979). Remarkably, the IMF is 
almost entirely feature-free, all the way from the most massive 
stars down to \ensuremath{\sim}  0.3 M$_{\sun}$. Below 0.3 M$_{\sun}$, studies show 
that the IMF flattens somewhat, but continues to increase. Below 
0.1 M$_{\sun}$ (100 M$_{\rm JUP}$), the mass function starts to decline, and 
microlensing observations (Alcock et al. 1998) clearly indicate 
that our Galaxy is not full of sub-Jupiter mass brown dwarfs. 
Thus, somewhere below 0.1 M$_{\sun}$, the physics of star formation 
produces a turnover and decline in the IMF, perhaps with some 
lower-limit boundary condition.

The classical theory of opacity-limited fragmentation has long 
predicted a significant boundary around 0.003 to 0.01 M$_{\sun}$, equivalent 
to 3 to 10 M$_{\rm JUP}$, below which it is believed that cores become 
opaque to their own radiation, and therefore cannot cool and 
fragment any further (Hoyle 1953; Low \& Lynden-Bell 1976; Rees 
1976; Silk 1977; Boss 1988). More recent work by Boss (2001) 
has shown that magnetic fields may lower this limit to about 
1 M$_{\rm JUP}$. However, the whole fragmentation scenario at low masses 
may have to be replaced by a more complex model involving a wide 
range of physical processes, including supersonic turbulence 
(Padoan \& Nordlund 2002), dynamical interactions between protostars 
(Bate, Bonnell, \& Bromm 2002), and feedback due to strong bipolar 
outflows (Adams \& Fatuzzo 1996), or ionizing radiation from 
massive stars (Palla \& Stahler 2000).

Thus the form of the substellar IMF can yield important clues 
in our understanding of the star formation process. Studies of 
embedded and young open clusters in the 10 to 100 M$_{\rm JUP}$ regime 
yield surprisingly different results. On one hand, a rising substellar 
IMF has been found in the 100 Myr Pleiades (Bouvier et al. 1998; 
Zapatero-Osorio et al. 1999) and other young open clusters, while 
in the 1 Myr old Trapezium Cluster there is a strong turnover 
of the IMF across the stellar to substellar boundary, with a 
smaller proportion of brown dwarfs down to 5 to 10 M$_{\rm JUP}$ (Hillenbrand 
\& Carpenter 2000; Luhman et al. 2000; Muench et al. 2002; Lucas 
\& Roche 2000; McCaughrean et al. 2002).

\begin{figure*}
\centering
\includegraphics[width=1.00\textwidth]{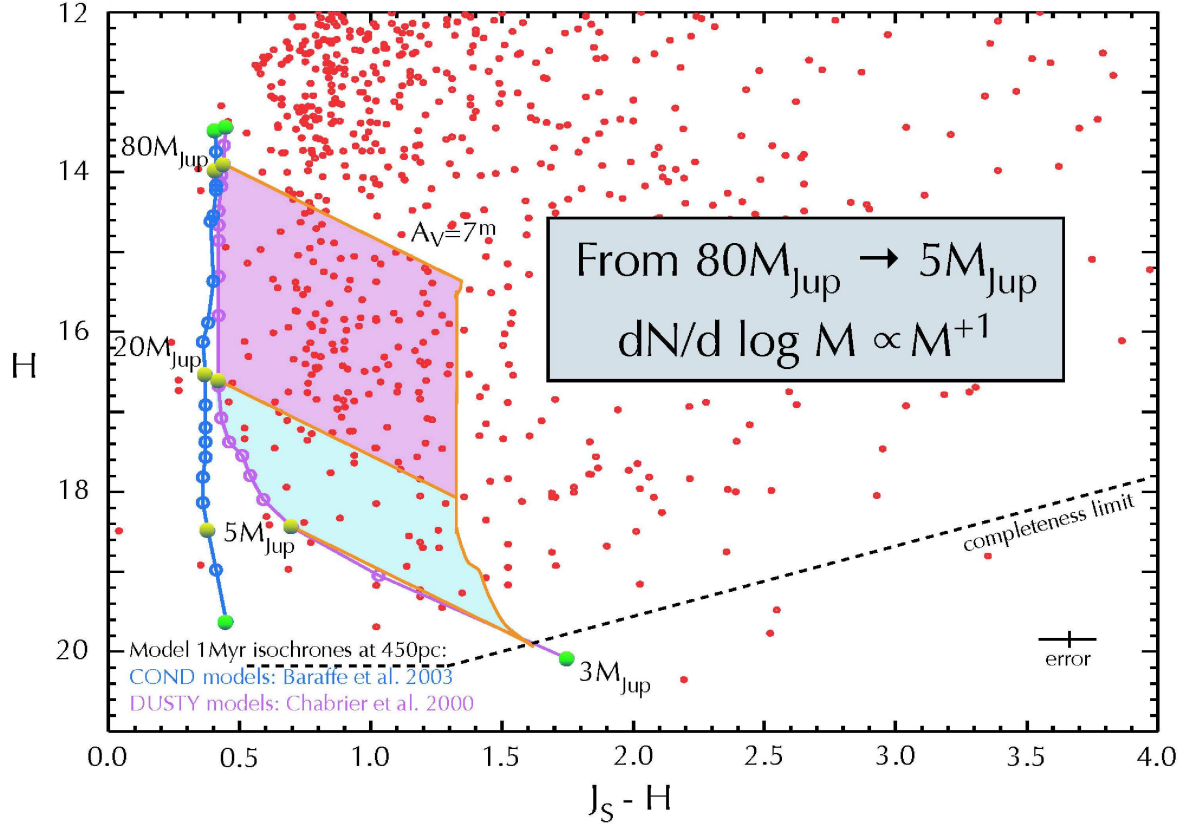}
\caption{
H vs. (J$_{S}$-H) color-magnitude diagram for the Orion Trapezium
Cluster. An extinction-limited sample of sources in the range 0.005 to 0.08 
M$_{\sun}$ (5 to 80 M$_{\rm JUP}$), assumed to be 1 Myr old, shows that the brown
dwarf end of the IMF is falling, as roughly characterized by the form
dN/dlogM $\sim $ M$^{+1}$ (From McCaughrean et al. 2002).}
\label{fig015}
\end{figure*}

Searching for and characterizing sources at 1 M$_{\rm JUP}$ and below 
in nearby star-forming regions would allow us to constrain the 
physics of any lower-limit boundary (Fig.~\ref{fig015}). In addition, 
free-floating objects less massive than 10 M$_{\rm JUP}$ can serve as 
important proxies for true planets, yielding crucial insights 
into their early evolution and complementing JWST studies that 
will be made of mature giant planets in orbit around nearby stars.

Furthermore, the near-infrared sensitivity and high spatial resolution 
of JWST will enable us to search for brown dwarfs in distant 
clusters with greatly differing metallicity, including globular 
clusters, clusters in the inner and outer galaxy, and in the 
Magellanic Clouds. In this way, we will be able to determine 
whether the substellar IMF is universal, or if it depends on 
the local Jeans mass.

\subsubsection{Observations}

In order to clearly identify the bottom of the IMF, it is important 
that we probe to well below 1 M$_{\rm JUP}$. However, sources with this 
mass at 1 Myr have effective temperatures just below 1000 K, 
and thus the bulk of the flux from objects at this mass and lower 
is emitted in the thermal infrared longward of 3 \ensuremath{\mu}m,
making ground-based observations prohibitive. JWST will be able to
detect such sources readily, permitting surveys at 3 to 10
\ensuremath{\mu}m in nearby young clusters to 1 M$_{\rm JUP}$ and
below.

Broadband imaging observations will identify very low luminosity 
candidates and will compare their properties with pre-main sequence 
evolutionary models. Imaging surveys at 2 to 5 \ensuremath{\mu}m will 
be supplemented with 10 \ensuremath{\mu}m photometry to separate the 
circumstellar emission due to disks from the intrinsic source 
luminosity. Wide-field imaging is required for nearby clusters: 
a typical cluster is about 1 pc in diameter or 3.5 arcmin at 
1 kpc. At larger distances, high spatial resolution will eliminate 
crowding and resolve the lowest-mass sources from adjacent brighter, 
more massive neighbors.

Multiobject spectroscopy at 2 to 5 \ensuremath{\mu}m will distinguish 
between true cluster members and field interlopers. These observations 
will also break the age-reddening-mass degeneracy found in young 
clusters, by determining spectral types, making it possible to 
accurately deredden the sources and locate them in theoretical 
pre-main sequence Hertzsprung-Russell diagrams. Typically, within a given cluster, 
there will be 100 to 1000 sources with masses in the range 1 
to 100 M$_{\rm JUP}$ (see Fig.~\ref{fig015}).

R $\sim$ 1000 multi-object spectra across the 2 to 5 \ensuremath{\mu}m regime 
will allow spectral typing with the H$_{2}$O and methane absorption 
bands, atomic Na, K, and Ca lines, and H$_{2}$ collisionally-induced 
absorption. Surface gravity indicators will be used to eliminate 
older, high-gravity non-cluster members and spectral-type based 
effective temperatures will be assigned to the true cluster members.

In extremely crowded regions with bright nebulosity, contrast 
and light leakage in the spectrograph becomes issues, reducing 
sensitivity to the lowest-mass sources. Narrow-band imaging provides 
an excellent alternative, using photometry in a set of about 
10 select narrow-band features to construct spectral indices 
with respect to the atmospheric models. These data could be compared 
to an optimal set of spectral templates across the 2 to 5 \ensuremath{\mu}m 
regime as a function of mass and age (e.g., Burrows et al. 2001; 
Baraffe et al. 2003).

The mass limit that can reasonably be reached in a given cluster 
is a function of its age, distance, and the foreground and intracluster 
reddening, all of which can vary considerably. Table~\ref{tab003} shows the 
sensitivities required to reach mass limits of 1, 10, and 100 
M$_{\rm JUP}$, assuming a 1 Myr old cluster at 500 pc (e.g., Orion), 
5 kpc (inner galaxy), and 50 kpc (Magellanic Clouds), with 10 
magnitudes of visual extinction in each case.

\begin{table}[t]
\caption{Predicted Fluxes for Sub-Stellar Objects\label{tab003}}
\begin{tabular}{lcccc}
\hline\noalign{\smallskip}
{Mass}&
{T$_{\rm eff}$}&
{2.2 \ensuremath{\mu}m}&
{3.8 \ensuremath{\mu}m}&
{4.8 \ensuremath{\mu}m} \\

{(M$_{\rm JUP}$)}&
{(K)}&&& \\[3pt]

\tableheadseprule\noalign{\smallskip}

&&\multicolumn{3}{c}{0.5 kpc distance} \\
&&&&\\

1&941&290&270&1300 \\
10&2251&34000&33000&28000 \\
100&2856&950000&820000&760000 \\

&&&&\\
\hline
&&&&\\
&&\multicolumn{3}{c}{5 kpc distance} \\
&&&&\\

1&941&2.9&2.7&13 \\
10&2251&340&330&280 \\
100&2856&9500&8200&7600 \\

&&&&\\
\hline
&&&&\\
&&\multicolumn{3}{c}{50 kpc distance} \\
&&&&\\

1&941&0.03&0.03&0.13\\
10&2251&3.4&3.3&2.8\\
100&2856&95&82&76\\

&&&&\\
\noalign{\smallskip}\hline
\end{tabular}

{NOTE --- Predicted Fluxes in nJy at near-infrared wavelengths for 
1 M$_{\rm JUP}$, 10 M$_{\rm JUP}$, and 100 M$_{\rm JUP}$ sources are modeled at 1 Myr 
age, and at distances of 0.5, 5, and 50 Kpc. A typical extinction 
of 10 magnitudes at V-band is assumed throughout. (Burrows et 
al. 1997; Burrows et al. 2001; Baraffe et al. 1998; Chabrier 
et al. 2000; Marley et al. 2002; Baraffe et al. 2003).}

\end{table}

For the nearby clusters at \ensuremath{\sim} 500 pc, JWST will carry 
out accurate broad-band near- and mid-infrared photometry, as 
well as classification spectroscopy at 2 to 4 \ensuremath{\mu}m down 
to 1 M$_{\rm JUP}$, reaching a limiting continuum sensitivity of 300 
nJy. This spectroscopy will be done with either multiobject spectroscopy 
or narrow-band R \ensuremath{\sim} 100 imaging in (approximately) 10 
bands.

Towards the inner galaxy, imaging surveys down to 1 M$_{\rm JUP}$ are 
needed, with classification spectroscopy down to 10 M$_{\rm JUP}$. Finally, 
at the Magellanic Clouds, imaging down to 10 M$_{\rm JUP}$ is required, 
with spectroscopy to 100 M$_{\rm JUP}$ or 0.1 M$_{\sun}$, just above the star-brown 
dwarf boundary. The massive 30 Doradus cluster in the Large Magellanic 
Cloud is the closest object we have to a starburst template; 
a detailed study of its low-mass content is crucial.

In all young clusters, 3 to 10 \ensuremath{\mu}m imaging is required 
to measure excess thermal emission indicative of disks around 
brown dwarfs, and thus assess their mode of formation and whether 
or not they may build planetary systems (cf. McCaughrean et al. 
1996; Muench et al. 2001; Natta \& Testi 2001; Liu et al. 2003).

High spatial resolution images of clusters are needed to measure 
the binary frequency function to separations of 30 AU (0.06 arcsec 
at 500 pc), the peak of the main sequence binary separation distribution. 
We will determine whether the frequency and properties of field 
binaries and multiples can be reproduced by mixing the populations 
of clusters and low-mass star-forming regions in different proportions 
(Kroupa et al. 1999; Scally et al. 1999).

Similarly, high spatial resolution is required to conduct high 
accuracy proper motion measurements to analyze the internal dynamical 
state of a cluster, to determine what fraction of it will remain 
bound, and what effect dynamical mass segregation has had on 
the measured mass function (Kroupa 1998). Assuming a centroid 
accuracy for point sources equal to the Gaussian sigma of the 
point spread function divided by the signal-to-noise ratio, a 
50$\sigma$ measurement will yield 0.8 milliarcsec precision at 2 \ensuremath{\mu}m 
or 0.4 AU at 500 pc. Measurements at 3 epochs over 3 years will 
permit individual proper motions to be measured for sources down 
to 1 M$_{\rm JUP}$ to an accuracy of about 1 km s$^{-1}$ in nearby clusters, 
resolving the typical cluster velocities of 3 to 4 km s$^{-1}$ (Jones 
\& Walker 1988). Uncertainties in the geometrical distortion 
correction may limit this technique.

\subsection{How do Protoplanetary Systems Form?}

How do circumstellar disks form and evolve? What determines their
physical sizes? How do dust grains within the disks form planets?

The existence of disks around young low-mass stars was firmly 
established in the 1990s, initially based on indirect measurements, 
such as spectral energy distributions, asymmetric wind profiles, 
polarization mapping (Beckwith \& Sargent 1993; Strom et al. 
1993), and more recently through direct imaging (McCaughrean 
et al. 2000; Wilner \& Lay 2000). Well-resolved direct images 
of circumstellar disks reveal their internal density and temperature 
structure, and show how disks are affected by their ambient environment. 
Since circumstellar disks are both a product and a mediator of 
the star formation process, as well as the progenitors of planetary 
systems, it is clear that a fuller understanding of the evolution 
of circumstellar disks will play a key role in our understanding 
of these central topics.

\begin{figure*}
\centering
\includegraphics[width=1.00\textwidth]{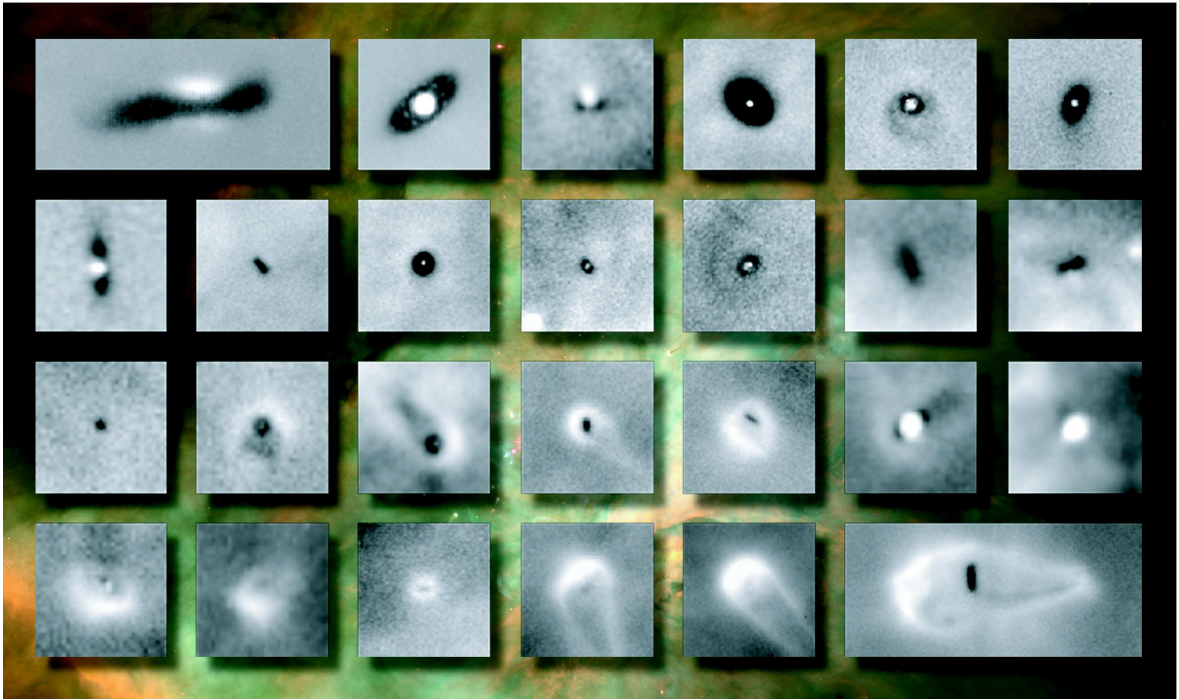}
\caption{
A collection of young circumstellar disks seen as silhouettes
against the bright background emission of the Orion Nebula H II region,
imaged using HST. All of these sources are associated with young (0.5 to 2
Myr) low-mass (0.3 to 1.5 M$_{\sun}$) members of the Trapezium Cluster. In
several cases, where the disk is oriented close to edge-on and the central
star cannot be seen directly, its presence is betrayed by polar reflection
nebulae. A number of the disks are also seen to be embedded in ionized
tadpole-shaped nebulae (`proplyds'), which are created as the disk is heated
and ablated by the central massive stars in the cluster. The square panels
are 1.67 $\times$ 1.67 arcsec$^2$ or 750 $\times$ 750 AU$^2$ in size; the larger panels are
proportional. The disks range from 50 to 500 AU in radius. The images were
either made through the H$\alpha $ filter of WFPC2 or the [O III]
filter of STIS. Data from McCaughrean \& O'Dell (1996); Bally, O'Dell, \& McCaughrean
(2000).}
\label{fig016}
\end{figure*}

Young circumstellar disks have been directly imaged at optical through 
mid-infrared wavelengths in a number of nearby star-forming 
regions, including Taurus-Auriga (Burrows et al. 1996; Koresko 
1998; Stapelfeldt et al. 1998; Krist et al. 1999; Padgett et 
al. 1999), as well as closer to home (Koerner et al. 1998; Jayawardhana 
et al. 1998; Schneider et al. 1999). The largest sample of young 
circumstellar disks imaged to date is in the Orion Nebula. HST 
optical emission-line surveys of the region (O'Dell et al. 1993; 
O'Dell \& Wen 1994, Bally et al. 1995; O'Dell \& Wong 1996; Bally 
et al. 2000) have shown that many of the approximately 2000 young 
(\ensuremath{\sim}  1 Myr old) stars of the Trapezium Cluster are either 
surrounded by compact ionized nebulae, called `proplyds', which 
are interpreted as circumstellar disks externally ionized by 
the central O and B stars, or are surrounded by dark silhouettes, 
which are disks seen in projection against the bright H II region 
or within the proplyds (O'Dell \& Wen 1994; McCaughrean \& O'Dell 
1996; Fig.~\ref{fig016}).

The approximately 50 silhouette disks that have been observed 
have diameters ranging from the HST resolution limit of 0.1 arcsec 
up to 2 arcsec (50 to 1000 AU). The disks are truncated at the 
outer edge, due either to internal evolution or external effects 
such as photoionization or star-disk interactions in the dense 
cluster environment (McCaughrean \& O'Dell 1996). The distribution 
of disk sizes shows that disks inside ionized proplyds tend to 
be a little smaller than those of the pure (non-ionized) silhouettes 
(Rodmann 2002), suggesting that they are being rapidly eroded 
by the O and B stars (Johnstone et al. 1998).

Can these disks form planets before being destroyed? One way 
of answering this question is to look for evidence of growth 
in the dust grain population, which would indicate that planetesimal 
formation is already underway (Beckwith et al. 2000). By comparing 
the diameter of a silhouette disk at UV, optical, and near-infrared 
wavelengths, an estimate can be made of the dominant particle 
size in its outer reaches. HST and ground-based adaptive-optics 
studies to date have proven ambiguous, with contradicting suggestions 
of ISM-like grains and much larger 5 \ensuremath{\mu}m grains (cf. McCaughrean 
et al. 1998; Throop et al. 2001; Shuping et al. 2003). However, 
due to the relatively limited spatial resolution of HST at the 
critical near-infrared wavelengths, it has not been possible 
to attempt these studies for any but the single largest disk. 

\subsubsection{Observations}

JWST will provide high spatial resolution imaging of a substantially 
larger sample of silhouette disks in selected near-infrared wavelengths 
where the background H II region is particularly bright. These 
observations will cover a broad wavelength range to yield maximum 
leverage with respect to the dust extinction. Most critical are 
the 1.87 \ensuremath{\mu}m Pa\ensuremath{\alpha} line, inaccessible from the ground, 
and the 4.05 \ensuremath{\mu}m Br\ensuremath{\alpha} line, where adequate sensitivity 
is hard to achieve from the ground.

Narrow-band (R $\sim$ 100) diffraction-limited JWST imaging will yield 
angular resolutions of 0.07 and 0.15 arcsec in the P\ensuremath{\alpha} 
and Br\ensuremath{\alpha} lines, respectively, corresponding to 30 and 
70 AU at the Orion distance of 500 pc. These resolutions imply 
that for the more than 10 silhouette disks with diameters of 200 
AU (0.4 arcsec) or greater, it will be possible to measure the 
outer-disk radial profiles and in combination with similar resolution 
HST images in the H\ensuremath{\alpha}, [O III], and [O II] lines, a statistical 
assessment of the grain sizes in their outer reaches can be made. 
An important goal is to understand whether grain growth is inhibited 
or promoted if a disk is embedded in an ionized proplyd.

Extrapolating from measured H\ensuremath{\alpha} fluxes around the Orion 
silhouette disks and the known foreground dust extinction, predicted 
Pa\ensuremath{\alpha} line fluxes in the outer parts of the nebula are 
\ensuremath{\sim}10$^{-13}$ erg s$^{-1}$ cm$^{-2}$ arcsec$^{-2}$ or 10$^{-16}$ erg s$^{-1}$ 
cm$^{-2}$ pixel$^{-1}$, assuming 0.03 arcsec pixels. In the disk centers, 
the fluxes will be roughly 10 times fainter, about 10$^{-17}$ erg 
s$^{-1}$ cm$^{-2}$ pixel$^{-1}$, and accurate mapping of the disk profiles 
requires imaging at 100$\sigma$ to this flux level through a 1\% 
narrow-band filter in the near infrared.

A clean, stable, and well-characterized PSF is also needed
to reduce the `PSF-blurring' that ultimately limits such studies 
(McCaughrean \& O'Dell 1996). In the more general case where the 
disk is not seen edge-on, the observations will be taken using 
coronagraphy to occult the central star, which is relatively 
bright in the near-infrared.

\subsection{What are the Life Cycles of Gas and Dust?}

How do gas-phase molecules interact with dust grains in quiescent
cloud cores? How does the formation of a star and planetary system
affect the astrochemical evolution of the gas and dust? What is
the origin of water and organic materials in a planetary system?

\begin{figure*}
\centering
\includegraphics[width=1.00\textwidth]{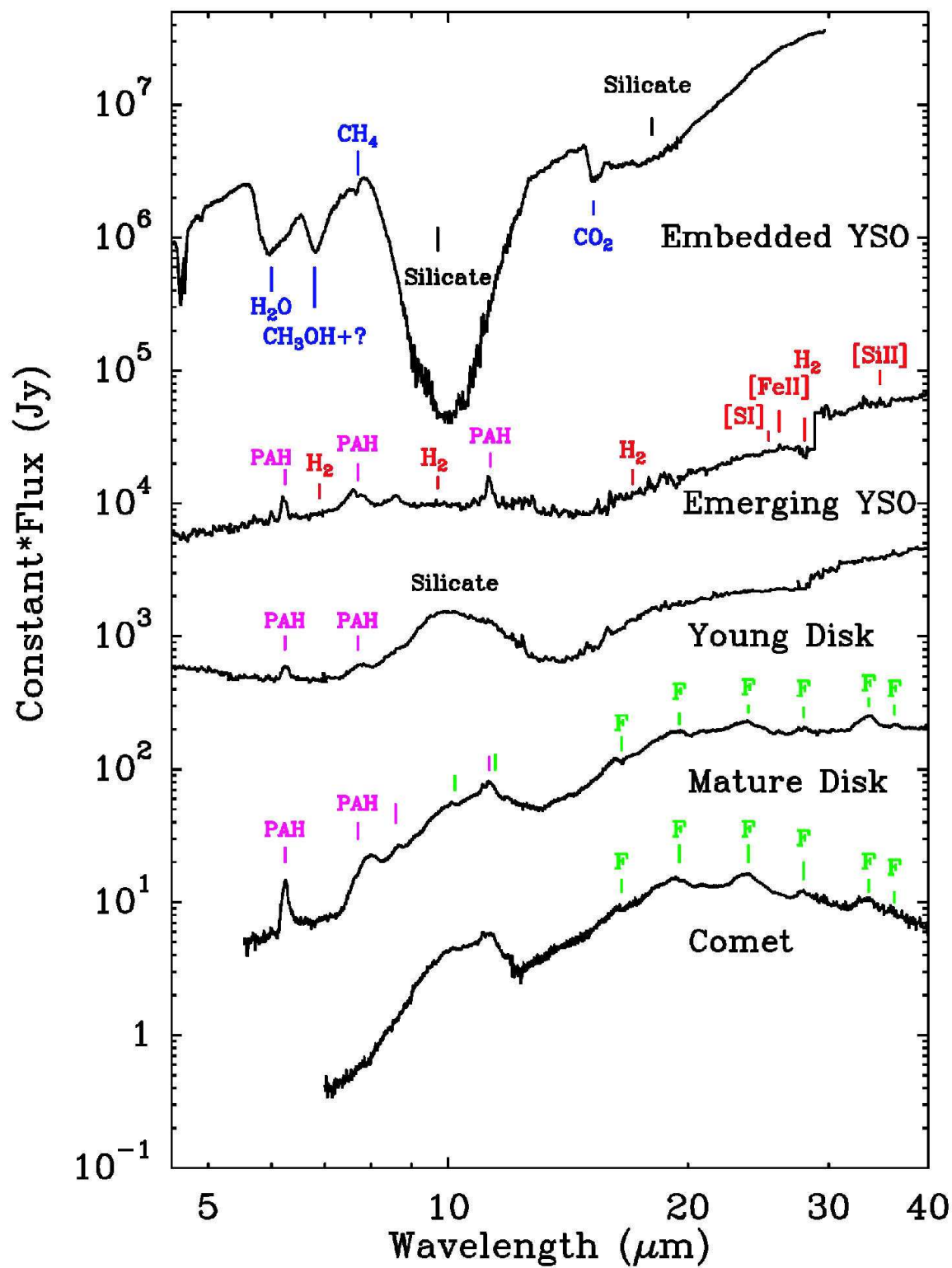}
\caption{
A series of ISO SWS mid-infrared spectra of young stars and
circumstellar disks at different stages in their evolution (Gibb et 
al.\ 2000; van den Ancker et al.\ 2000a,b; Malfait et al.\ 1998). From top to
bottom, in a rough evolutionary sequence, the spectra change from being
dominated by solid state absorption features and shocked gas emission lines,
to PAH features and photo-dissociation region lines, to amorphous and
crystalline silicates with H I recombination lines. An ISO SWS spectrum of
Comet Hale-Bopp is shown for comparison (Crovisier et al.\ 1997).}
\label{fig017}
\end{figure*}

Generations of both low- and high-mass stars have converted primordial 
hydrogen and helium into successively heavier elements, including 
carbon, oxygen, and nitrogen, elements that make up life on Earth. 
Returned to the interstellar medium via winds and supernovae, 
these elements may eventually be incorporated into molecular 
clouds and later into new stars and their circumstellar disks. 
A major goal of astrochemistry and astrobiology is to trace the 
life cycles of gas and dust from pre-stellar cores to planetary 
systems (van Dishoeck \& Blake 1998; Waters 2000; Ehrenfreund 
\& Charnley 2000; see Fig.~\ref{fig017}).

In cold quiescent cloud cores, gas-phase molecules stick to dust 
grains, forming icy mantles on silicate cores. This freeze-out 
is predicted to be very efficient, and it is a paradox that gas-phase 
molecules are seen in clouds at all. Recent infrared and millimeter 
observations yield indirect evidence that there is substantial 
depletion, with up to 90\% of the heavy elements frozen out onto 
grains (Lada et al. 1999; Kramer et al. 1999).

When the heavy elements are frozen out, the chemical composition 
is initially modified via grain-surface reactions. Then, as a 
central protostar develops and heats up its envelope, the ices 
evaporate back to the gas phase and a rich mixture of organic 
compounds develops. Later, disk accretion events lead to FU 
Ori-type outbursts, producing heating and UV flux. As the envelope 
material is processed in this way, its thermal and irradiation 
history is imprinted through irreversible changes in band profiles 
in spectra, via features including the solid CO$_{2}$ bending mode 
at 15 \ensuremath{\mu}m, the OCN$^{-}$ band at 4.62 \ensuremath{\mu}m, and the unidentified 
6.85 \ensuremath{\mu}m feature (Ehrenfreund et al. 1998; Gerakines et 
al. 1999; Schutte \& Greenberg 1997; Schutte \& Khanna 2003).

Later, gas and dust from the envelope is incorporated into a 
circumstellar disk surrounding the young star, where it is further 
processed by UV radiation, X-rays, and thermal processes. In 
the cold disk midplane, molecules will freeze out again on the 
dust, while gas-phase molecules such as H$_{2}$, CO, CH$_{4}$, and C$_{2}$H$_{2}$ 
are expected in the warmer regions. Disk spectra should be dominated 
by features from PAHs, ices, and silicates, both in crystalline 
and amorphous form. These particles are the building blocks of 
planets and, as they include water ices and organic materials, 
potentially of life. By studying the astrochemical evolution 
in detail, it is possible to trace the formation history of planetesimals 
and other solid bodies, their composition, their processing, 
and their possible future evolution.

Gas plays an important role in the formation of giant planets. 
Warm (\ensuremath{\sim} 100 K) molecular gas located 1 to 50 AU from 
the central star in nearby debris disks can be traced using the 
pure rotational lines of H$_{2}$: J = 5-3 S(3) at 9.662 \ensuremath{\mu}m, 
J = 4-2 S(2) at 12.278 \ensuremath{\mu}m, and J = 3-1 S(1) at 17.035 \ensuremath{\mu}m. 
Simultaneous measurements of these lines would provide a profile 
of the temperature and mass of the gas as a function of radius. 
Dust-to-gas mass ratios can be derived by comparing the H$_{2}$ 
results with dust measurements. These H$_{2}$ measurements in different-age 
disks are sensitive to small masses of molecular gas and will 
constrain the age at which gas giants can form. The measurements 
can also show how gas is cleared from the disks and on what time 
scale.

To determine the evolution of these various tracers as they cycle 
back and forth between the gas and solid phases, we require high 
dynamic range, medium resolution near- and mid-infrared spectroscopy 
of individual sources in star-forming regions spanning a range 
of ages and environments. Spectra from the ISO SWS have shown 
signs of material evolution in the disks around more massive 
stars (Malfait et al. 1998; Meeus et al. 2001; Bouwman et al. 
2001) and the H$_{2}$ S(1) lines at 17 and 28 \ensuremath{\mu}m have been 
tentatively detected by ISO in two nearby debris disks (Thi et 
al. 2001). JWST will allow this work to be extended to a much 
more extensive sample, including proto-solar-type stars.

\subsubsection{Observations}

Spectroscopy of individual sources covering 3 to 28.3 \ensuremath{\mu}m 
is needed at spectral 
resolutions of R = 1000 to 3000, depending on the source and 
the tracers under study, with smoothing to R \ensuremath{\sim} 100 (with 
an attendant gain in signal-to-noise) acceptable for detecting 
very broad features at the faintest possible levels. 

Tracers may be seen in absorption against bright continuum sources. 
For example, in order to map the freeze-out processes in a dark 
cloud core, solid-state absorption bands are seen in the spectra 
of background field sources. To detect ices frozen on grains 
in the midplanes of circumstellar disks, the central young star 
itself can be used as the continuum source in edge-on geometries. 
Conversely, the tracers may appear in emission, as in the case 
of gas-phase molecules, PAHs, ices, and silicates in both amorphous 
and crystalline forms from warm material in the inner regions 
of disks.

\begin{table}[t]
\caption{JWST\ Measurements for the Birth of Stars and Protoplanetary Systems Theme\label{tab004}}
\begin{tabular}{p{1.0in}p{0.8in}p{1in}p{1.0in}p{0.0in}}
\hline\noalign{\smallskip}
{Observation} &
{Instrument} &
{Depth, Mode} &
{Target} &
\\[3pt]
\tableheadseprule\noalign{\smallskip}
\raggedright Cloud Collapse&
NIRCam&
9 nJy at 2 \ensuremath{\mu}m&
\raggedright e.g., Barnard 68&
\\
&&&& \\
&\raggedright MIRI (imaging)&
1 \ensuremath{\mu}Jy arcsec at 7 \ensuremath{\mu}m&
\\
&&&& \\
\raggedright Evolution of Protostars&
\raggedright MIRI (imaging)&
\raggedright 0.1 \ensuremath{\mu}Jy at 6 \ensuremath{\mu}m, 
1 \ensuremath{\mu}Jy at 15 \ensuremath{\mu}m&
\raggedright e.g., Taurus-Auriga&
\\
&&&& \\
&
\raggedright MIRI (spectra)&
\raggedright 7x10$^{-19}$ erg cm$^{-2}$ s$^{-1}$ at 15 \ensuremath{\mu}m&
\raggedright Class 0 protostars 
in Taurus-Auriga&
\\
&&&& \\
\raggedright Massive Stars&
NIRCam&
\raggedright 2 \ensuremath{\mu}Jy at 3.8 \ensuremath{\mu}m, 10 \ensuremath{\mu}Jy 
at 4.8 \ensuremath{\mu}m&
\raggedright e.g., Eagle Nebula&
\\
&&&& \\
&
NIRCam&
\raggedright 55 nJy at 3.5 \ensuremath{\mu}m, 2.3 \ensuremath{\mu}Jy at 4.8 \ensuremath{\mu}m&
\\
&&&& \\
&NIRSpec& 
fixed slits; IFU&
\raggedright e.g., Sources in the Eagle Nebula&
\\
&&&& \\
&
\raggedright MIRI (imaging)&
\raggedright 34 \ensuremath{\mu}Jy at 20 \ensuremath{\mu}m&
&
\\
&&&& \\
IMF&
NIRCam&
\raggedright 2.9 nJy at 2.2 \ensuremath{\mu}m&
\raggedright e.g., Orion Nebula, 30 Doradus&
\\
&&&& \\
&
\raggedright NIRSpec or TFI&
\raggedright 290 nJy at 2.2 \ensuremath{\mu}m&
&
\\
&&&& \\
&
MIRI&
&& \\
&&&& \\
\raggedright Protoplanetary Systems&
TFI&
\raggedright 10$^{-13}$ erg s$^{-1}$ cm$^{-2}$ arcsec$^{-2}$ 
at 1.87 \ensuremath{\mu}m; coronagraph&
\raggedright e.g., Orion Nebula&
\\
&&&& \\
Astrochemistry&
\raggedright MIRI (spectra)&
\raggedright 2.6x10$^{-}$$^{17}$ erg s$^{-1}$ cm$^{-2}$ arcsec$^{-2}$ 
at 17 \ensuremath{\mu}m&
\raggedright Proto-stars and Debris Disks& \\
&&&& \\

\noalign{\smallskip}\hline
\end{tabular}
\end{table}

To study the mineralogy of the dust in nearby debris disks and 
of their precursors around young stars, a 5\ensuremath{\sigma} limiting 
sensitivity of 10 \ensuremath{\mu}Jy is needed at 25 \ensuremath{\mu}m with 
a smoothed spectral resolution of R \ensuremath{\sim} 100 and with azimuthal 
averaging around the disk. This observation will detect a Vega-like 
system at a distance of 40 pc. 

To study the evolution of gas in disks, a line flux sensitivity 
of 2.6 x 10$^{-17}$ erg s$^{-1}$ cm$^{-2}$ arcsecond$^{-2}$ at 17 \ensuremath{\mu}m 
would enable the detection of about half an Earth mass of 100 
K molecular gas in a debris disk at 30 pc. It is also desirable 
that JWST spectroscopy be sensitive to the lowest H$_{2}$ transition 
line at 28.22 \ensuremath{\mu}m, which will be the brightest line and 
sensitive to the coldest gas, although the sensitivity at this 
wavelength will be severely limited by the detector technology. 
High spectral resolution is essential at this wavelength to obtain 
adequate contrast of the emission line relative to the continuum.

Finally, integral field spectroscopic capability in the mid-infrared 
with JWST will allow accurate mapping of the emission and absorption 
components in a given system. A field of at least 3 arcsec will 
match the scales in typical nearby debris disks and more massive 
disks around young stars in nearby star-forming regions. A high-contrast 
long slit for near-infrared spectroscopy will allow the faint 
disk material to be well separated from the potentially bright 
central source, and to map out spatial structures in the disks.

\subsection{Summary}

Table~\ref{tab004} summarizes the measurements needed for the Birth of Stars 
and Protoplanetary Systems theme. They include:

\textbullet{}
{\it Cloud collapse.} NIRCam imaging of background stars through
dark dust clouds will map the extinction and thus the density of
the clouds. MIRI imaging will reveal the densest inner cores, which
are not reachable even with NIRCam.

\textbullet{}
{\it Evolution of protostars.}
MIRI imaging 
of ``Class 0'' protostars will show their density and structure. 
These stars are still embedded in the dense dust clouds, and 
are not visible at shorter wavelengths.
MIRI spectroscopy of protostars will provide 
diagnostics of extinction windows, shock tracers and PAHs.

\textbullet{}
{\it Massive stars.} High spatial resolution imaging of crowded
star-forming regions will test the agglomeration hypothesis for
massive star formation Deep imaging of ``elephant trunks'' will
reveal the low-mass stars forming within and test models of the
effects of massive stars on their environment. NIR spectroscopy
will reveal masses, age and evolutionary status of sources forming
in the ``elephant trunks''. MIRI imaging will see through the
densest dust columns to reveal the pre-main sequence stars and
brown dwarfs forming within.

\textbullet{}
{\it IMF.} This program will search for the low-mass end of the
IMF in nearby young star-forming clusters, by detecting 1 M$_{\rm
JUP}$ objects in the Orion Nebula. Scattered light from the bright
stars within the nebula are probably the limiting factor. It will
also photometrically detect 1 M$_{\rm JUP}$ objects in clusters in
the inner galaxy, and 10 M$_{\rm JUP}$ object in 30 Doradus in the
LMC. Spectroscopic follow-up of the more massive objects will reach
the sub-brown dwarf stage in the inner galaxy, and the faintest M
dwarfs in the LMC. MIRI observations of the low-mass end of the
IMF will be used to determine the frequency of disks around brown
dwarfs, by searching for excess thermal emission. Can planets form
around brown dwarfs?

\textbullet{}
{\it Protoplanetary systems.} TFI imaging will be used to resolve
dust disks in silhouette against a brighter background. HST has
done this for some disks in Orion. JWST measurements of these same
disks will constrain the grain sizes, compositions and profiles.

\textbullet{}
{\it Astrochemistry.} MIRI spectroscopy of organic molecules (plus
H$_{2}$, H$_{2}$O) in proto-stars will be used to study the origins
of the ``stuff of life''.

\section{Planetary Systems and the Origins of Life}

The key objective of the Planetary Systems and the Origins of 
Life theme is to determine the physical and chemical properties 
of planetary systems including our own, and to investigate the potential 
for the origins of life in those systems.

To trace the origins of the Earth and life in the universe, we 
need to study planet formation and evolution, including the structure 
and evolution of circumstellar material. The search for evidence 
of life in our Solar System and beyond is fundamental to the 
understanding of our place in the cosmos. JWST observations of 
objects in our own Solar System and planetary systems around 
other stars will provide data crucial for understanding the origin 
of planetary systems, and the potential for stable habitable regions around 
other stars. 

\subsection{How Do Planets Form?}

What are the physical processes that lead to planets? How common
are giant planets and what is the distribution of their orbits?
How do giant planets affect the formation of terrestrial planets?

\paragraph{Formation:}

The formation of multiple objects is a common outcome of star 
formation, including binary or higher-order star systems, a central 
star orbited by brown dwarfs and/or planets, or a star with a 
remnant disk of particulates. Brown dwarfs and giant planets 
might arise from two different formation mechanisms. Brown dwarfs 
may represent direct collapse of gas, from a molecular cloud 
clump or from a disk of material (Chabrier \& Baraffe 2000), 
while giant planets could form from a two-stage process in which 
growth of a rock-ice core triggers the rapid accretion of gas 
(Lunine et al. 2004). In this model, brown dwarf companions are 
the low-mass tail of binary star formation, while giant planets 
are the high-mass end of a process that also makes Earths and 
Neptunes, and the two processes may produce distinct initial 
mass functions. Even if there is an overlap in the masses generated 
by the two processes, they would be distinguished by the metallicities 
of the objects generated relative to the parent star. Giant planet 
formation by two-stage accretion increases the metallicity of 
the planets by a factor of several relative to the parent star, 
but direct collapse does not need to do so.

The formation of giant planets is a signpost, detectable with JWST, 
of a process that may also generate terrestrial planets. In contrast, 
direct collapse formation of brown dwarfs may signal systems 
in which terrestrial planet formation is rare or impossible, 
because of the required disk mass, angular momentum and subsequent 
disk evolution.

\paragraph{Frequency and Orbits:}

Planets of Uranian mass or larger exist in orbits that are detectable 
by the current radial velocity surveys around 8\% of F, G, and 
K type stars in the solar neighborhood. Extrapolating to larger 
semi-major axes, and guided by our own dynamically crowded outer 
solar system (the region from 5 to 30 AU), one can infer that 
about 15\% of mature F, G, and K type stars should possess at 
least one giant planet (Marcy et al. 2005). Theoretical studies 
of the inward migration of giant planets through interactions with the gaseous 
disk, or with massive remnant particulate disks, suggest that 
many more stars generate giant planets during their pre-main 
sequence phase (Trilling et al. 2002). Many or most of these 
giant planets are lost through inward migration and merging with 
the central star, or by ejection. We do not know how many nearby 
stars possess giant planets in orbits too large for detection 
by the radial velocity approach, or with periods too long for 
planned space-borne astrometric surveys. Giant planet formation 
may be a process favored in the colder outer regions of protoplanetary 
disks, where water ice exists. Alternatively, giant planets could 
be formed over a wide range of semi-major axes, even in the warm 
inner parts of disks.

Ground-based surveys have uncovered eight planets that pass 
in front of their star as seen from Earth (transiting planets)
(e.g., Henry et al. 2000; Charbonneau et al. 2000; Konacki 
et al. 2005; Alonso et al. 2004; Sato et al. 2005; for a 
review of surveys see Gillon et al. 2005). 
These transiting planets are the only solar-system-aged 
extrasolar planets that can be physically characterized 
until direct imaging is possible. 
HST and Spitzer have detected the atmospheres of two of 
these planets, including sodium, an extended atmosphere 
of hydrogen, and thermal emission (Charbonneau et al. 2002, 2005; 
Vidal-Madjar et al. 2003; 
Deming et al. 2005). These atmosphere 
detections are made possible by a differential measurement 
of the planet and starlight combined (during primary transit 
or outside of secondary eclipse) compared with only the starlight 
(outside of primary transit or during secondary eclipse). 
In addition to these atmosphere measurements, radii of all 
transiting planets can be determined, and the wide range of 
planet densities has uncovered two of the biggest puzzles 
in extrasolar planet science so far: an anomalously low 
density for one of the planets that cannot be explained 
by the standard planet evolution theory
and an 
anomalously high density for another planet, 
implying a core several times more massive than 
those of the giant planets in our own solar system (e.g., Gaudi 2005). 

\paragraph{Effect of Giant Planets on Terrestrial Planets:}

Given a particular distribution of orbital semi-major axes of 
extrasolar giant planets, one can assess the dynamical consequences 
of these bodies on both the stability of the orbits of putative 
terrestrial planets in the habitable zones of low-mass stars, 
and on the delivery of water to the habitable zone from colder 
regions. Jupiter and Saturn, forming within a few million years 
of the birth of our protoplanetary disk, both ejected remnant 
planetesimals from their own orbits, and increased the inclinations 
and eccentricities of planetesimals, so that they reached the 
region where the Earth and other terrestrial planets formed later 
(roughly 50 - 100 million years based on radioisotopic dating). 
If this is a general characteristic of giant planet formation, 
then in many disks both processes may lead to delivery of colder, 
water- and organic-rich planetesimals to the inner planet-forming 
region, and accelerate the growth of terrestrial planets (Morbidelli et 
al. 2000). On the other hand, radial migration of giant planets 
inward through the zone of formation of the terrestrial planets 
would sweep material there into the parent star, along with the 
migrating planet (Mardling \& Lin 2004). It is even possible that 
giant planets do both in a given system: an early generation 
of rapidly forming Jovian mass objects sweeping inward through 
the disk in the first million years or so, followed by formation 
of a second generation of giant planets that triggers the formation 
of terrestrial planets and delivers volatiles.

\subsubsection{Observations}

JWST will provide broad and narrow-band photometry of giant planets 
and brown dwarfs in bound systems as well as spectra of isolated 
giant planets and brown dwarfs. JWST will detect giant planets 
in large orbits that are unreachable by other techniques, 
provide spectroscopic and photometric information on their thermal 
properties, and constrain their mass and age. The statistics 
of the metallicities of isolated brown dwarfs and giant planets 
will be compared with those of similar-age low-mass stars (F, 
G, K) to determine whether the isolated population is biased 
toward higher metallicities, and hence formed in a process wherein 
large amounts of metals are incorporated during formation.

JWST will make coronagraphic images at 2.7 \ensuremath{\mu}m and
4.44 \ensuremath{\mu}m of nearby stars to find mature Jovian
companions and of more distant stars to find young Jovian companions.
The 2.7-\ensuremath{\mu}m to 4.44-\ensuremath{\mu}m flux ratio is
extremely sensitive to the temperature of planetary companions and
provides a rapid diagnosis of an object's age and mass, along with
a rough estimate of the orbital parameters (Sudarsky et al. 2003).
One modeling uncertainty that could dramatically affect this ratio
is the existence of clouds of various condensable species, ranging
from silicates to water ice, according to the background
temperature-pressure profile. Extensive additional modeling work
will be required to understand the impact of clouds on the
observations, which in general will be to render the flux ratio
less sensitive to temperature.

Following-up the discoveries, JWST will make additional 
coronagraphic images at R \ensuremath{\sim} 100 spectral 
resolution to determine effective temperature and radius. The 
spectrum of an extrasolar planet reveals not only temperature 
and composition, but also gravity and the presence of clouds 
as well. The relative differences between absorption features 
will provide diagnostic information on all of these parameters 
(Burrows et al. 2003). With the mass obtained from the temperature of the planet
and the age of the companion star, we can then infer the planet's 
radius. Again, this determination will be affected by the presence 
of clouds, and the clouds in turn will be revealed by the relative 
changes in absorption features compared with cloud-free giant 
planets. JWST coronagraphy will also make it possible to study 
bound giant planets in Jovian-type orbits around nearby stars 
as well.

\begin{figure*}
\centering
\includegraphics[width=1.00\textwidth]{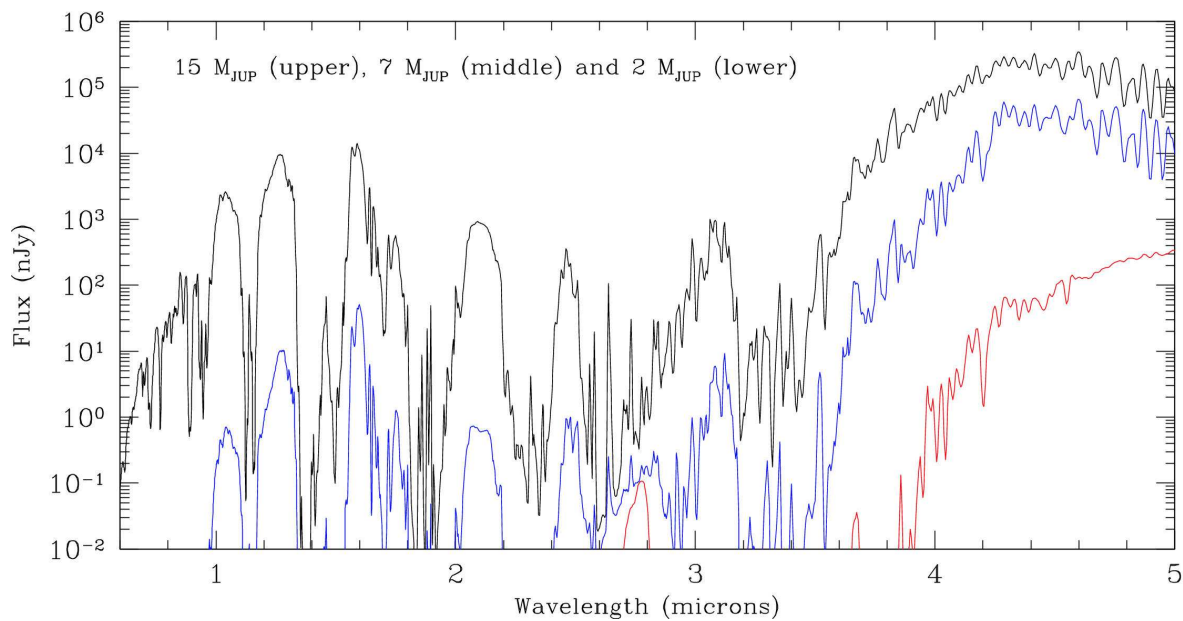}
\caption{
Detectability of extra-solar giant planets and brown dwarfs.
We plot model spectra (Sudarsky et al.\ 2003) of a 15, 7 and 2 M$_{\rm JUP}$
extrasolar giant planet, free-floating at 10 parsecs from Earth, with an age
of 5 billion years. The 5-$\mu$m window makes the
smallest mass planet detectable with JWST, while the 15 M$_{\rm JUP}$ object
(transitional to the T dwarfs) is bright enough to be studied
spectroscopically with narrow-band imaging or spectroscopy.}
\label{fig018}
\end{figure*}

\begin{figure*}
\centering
\includegraphics[width=1.00\textwidth]{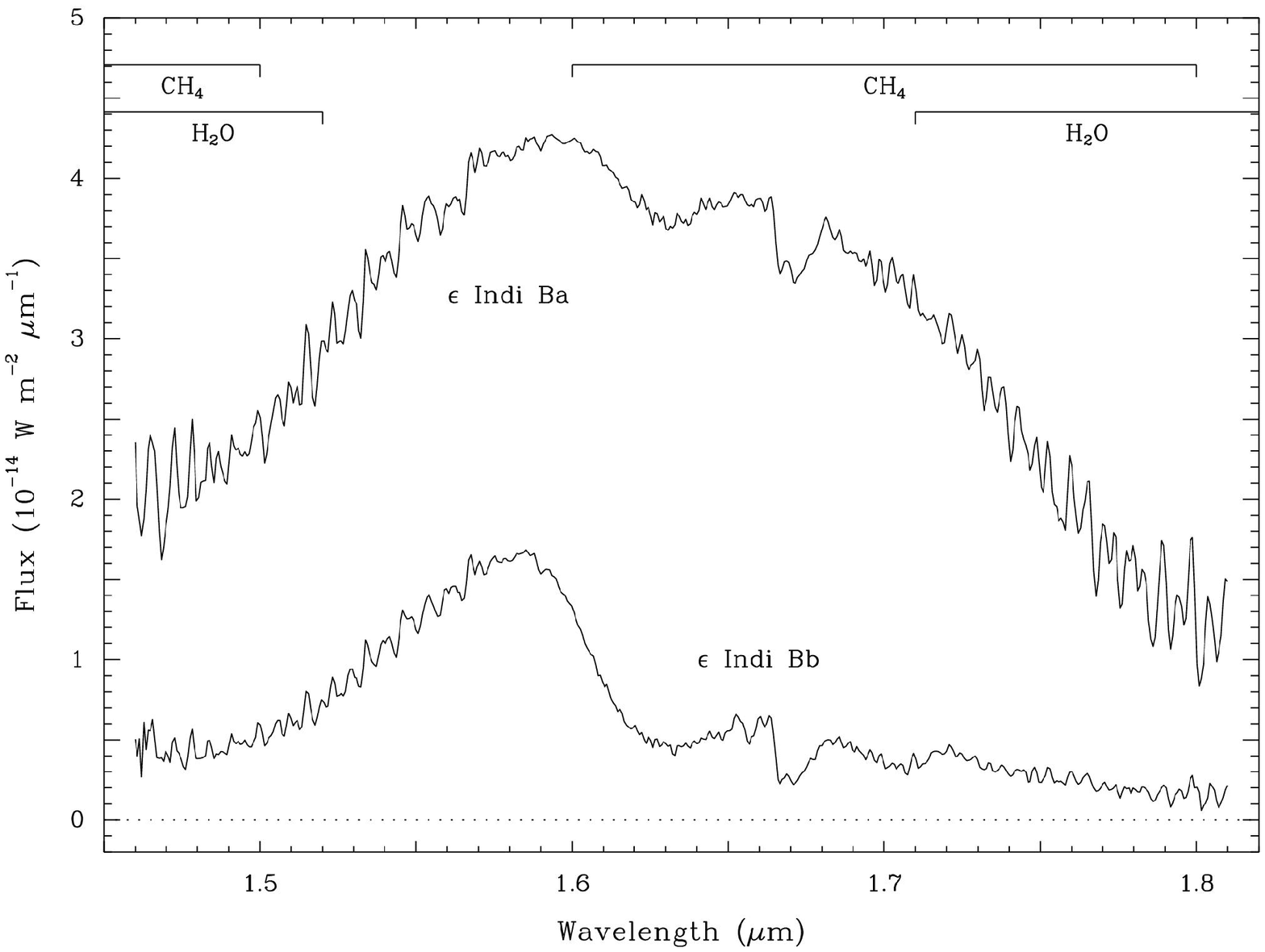}
\caption{
Spectra of the closest T-dwarfs. The closest brown dwarfs to
Earth, $\epsilon $ Indi B a and b (originally thought to be one object),
show distinctly different spectra (McCaughrean et al.\ 2004). The differences
between the two low-resolution spectra relate primarily to effective
temperature (1250 K vs. 850 K) through a range of potential effects
including molecular composition, presence of grains, and the effect of
physical temperature on the band shapes themselves.}
\label{fig019}
\end{figure*}

Isolated or widely-separated giant planets and sub-brown dwarfs
can be observed without coronagraphy, making their detection and
study much easier, and JWST will do R \ensuremath{\sim} 1000
spectroscopy at 1 to 5 \ensuremath{\mu}m of these objects
(Fig.~\ref{fig018}). With this spectral resolution, it is possible
to unambiguously determine basic physical parameters of a brown
dwarf atmosphere such as gravity, composition, the temperature
profile (gradient, inversions), and the effect of clouds. Much of
the information that may be contained in key features such as the
methane absorption feature is poorly understood, since telluric
absorption renders these studies difficult with ground-based data.
JWST will provide much higher signal-to-noise spectra of T-dwarfs
down through warm Jupiters with full coverage through the regions
obscured by telluric absorption (Fig.~\ref{fig019}).

\begin{figure*}
\centering
\includegraphics[width=1.00\textwidth]{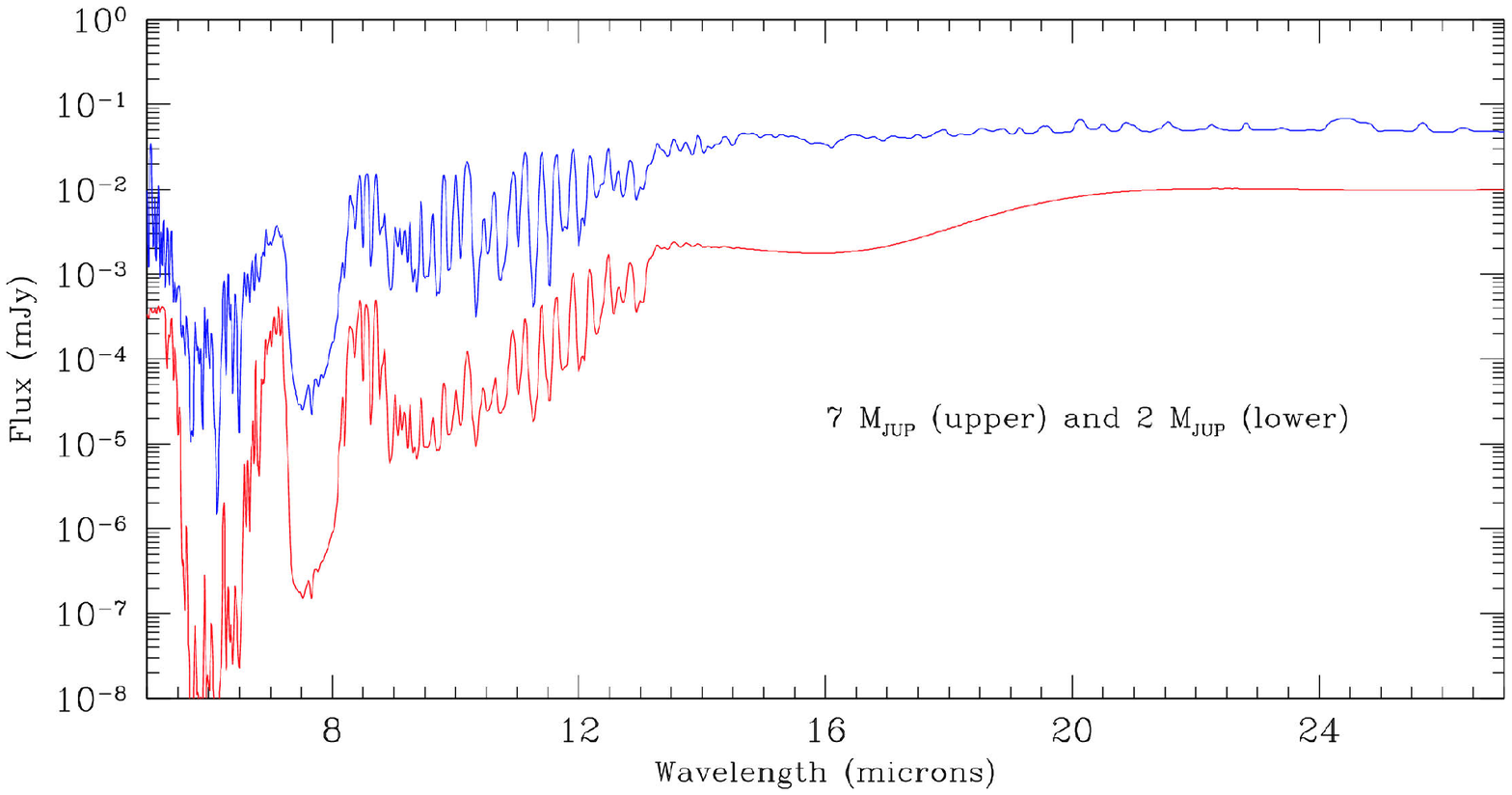}
\caption{
Detectability of Jovian-sized exo-planets in the mid infrared. We plot model spectra of
isolated giant planets of 7 and 2 M$_{\rm JUP}$, 10 parsecs from Earth, around a
five billion years old star by Sudarsky et al. (2003). The 2 M$_{\rm JUP}$ object is detectable in
JWST broad-band coronagraphic imaging; the 7 M$_{\rm JUP}$ object is bright enough that spectra at
R $\sim $ 3000 can be collected, enabling atmospheric structure and
composition to be inferred.}
\label{fig020}
\end{figure*}

JWST will have the coronagraphic sensitivity to detect a Jupiter 
analog around a Solar-type star out to \ensuremath{\sim} 30 pc. As shown 
in Fig.~\ref{fig020}, this would be a broadband detection, taking advantage 
of both the long-wavelength end of the 5 \ensuremath{\mu}m excess emission, 
as well as the more typically blackbody emission at longer wavelengths.

\begin{figure*}
\centering
\includegraphics[width=1.00\textwidth]{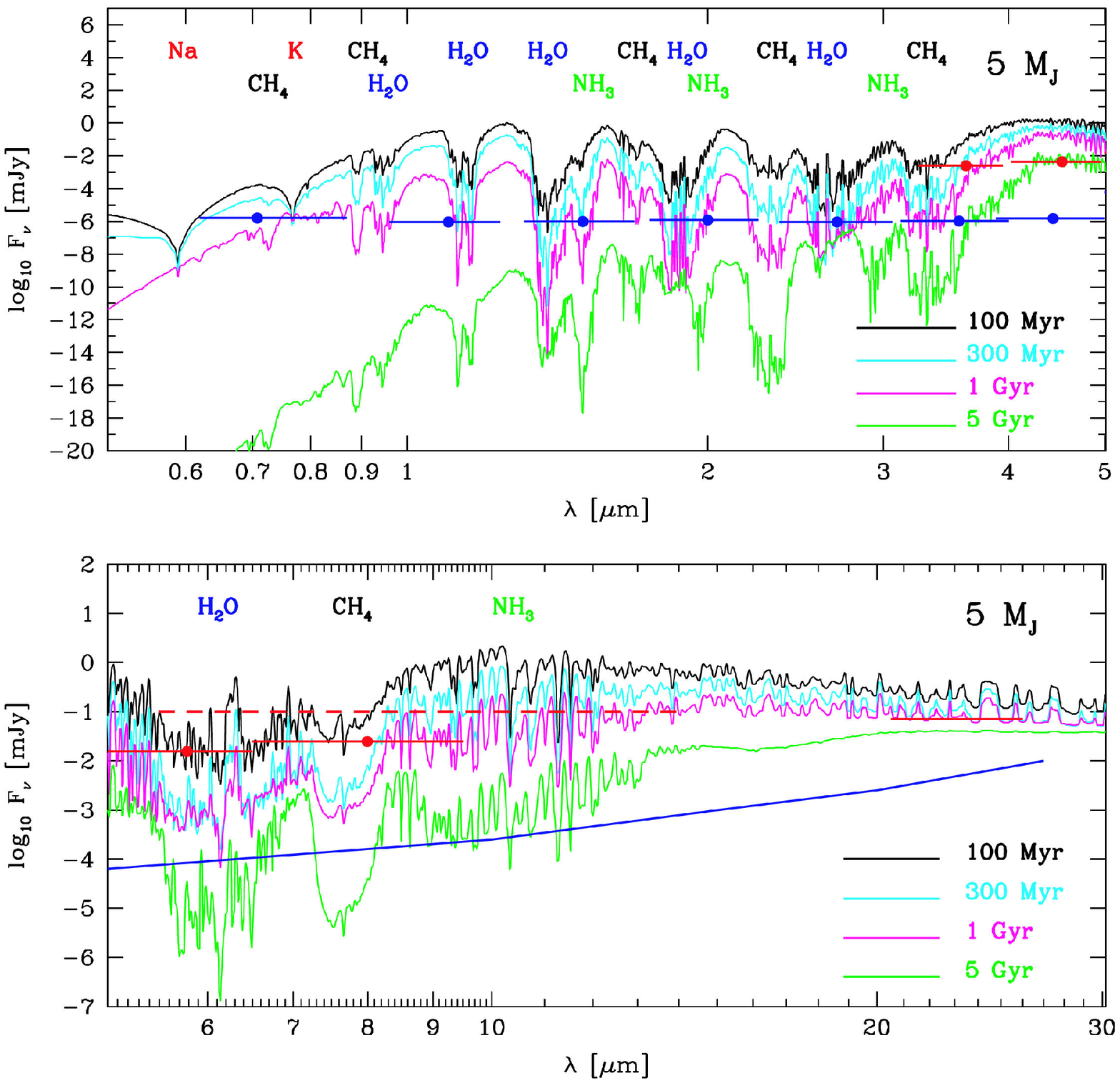}
\caption{
Guide to spectral features in extrasolar giant planets. A 5 M$_{\rm JUP}$ 
planet of varying ages is shown, with major species responsible
for the spectral features labeled. Approximate Spitzer (red) and
JWST (blue) sensitivities are shown. Detection of these features
in the near- and mid-infrared samples different regions of the
atmosphere, allowing abundance to be separated from temperature
profile and the effects of clouds (From Burrows et al. 2003).}
\label{fig021}
\end{figure*}

JWST will make spectra at 5 to 29 \ensuremath{\mu}m of 1 M$_{\rm \rm JUP}$ and larger 
objects within 10 pc. This broad wavelength range is near 
to, but generally longward, of the Planck-function peak (Fig.~\ref{fig020}). 
JWST spectra of the nearest systems can provide detailed insights 
into the nature of giant planet atmospheres, including abundances 
of ammonia and methane that are key indicators of atmospheric 
mixing and temperature profiles. Fig.~\ref{fig021} shows that many spectral 
features are available for filter and spectroscopic study in 
the near and mid infrared of objects with masses between that 
of Jupiter and brown dwarfs.

JWST will be able to measure density and atmospheric properties of
transiting extrasolar planets. These measurements are only possible
with the large aperture and stable space environment, together with
the excellent calibration due to the precisely predictable on/off
nature of both the primary and secondary eclipses. Short period
giant transiting planets will come from the numerous ground-based
surveys, while the Kepler space mission will detect Earth-sized to
giant planets with semi-major axes out to 1 AU. JWST will use
near-infrared R $\sim$ 1000 spectroscopy to disperse the light from
bright stars at a high cadence. JWST will detect atomic and molecular
absorption features O$_2$, CO$_2$, CO, CH$_4$, H$_2$O, Na, and K.
Extremely precise photometry will be obtained by binning the spectral
data together, as was done with HST (Charbonneau et al. 2002). The
measurement of a planet's radius and atmosphere will tell us about
its composition and evolution. They will help us understand the
exotic planets not found in our own solar system: including hot
giant planets close to the star, super-Earth-mass rocky planets,
and giant planets within 1 AU of the parent star. JWST will also
be capable of detecting planetary moons and rings in transit.

\subsection{How Are Circumstellar Disks Like Our Solar System?}

What comparisons, direct or indirect, can be made between our 
Solar System, circumstellar disks (forming solar systems), and 
remnant disks?

Detailed observations of disks around other stars, both during 
and after planet formation and disk clearing, provides a global 
view of the distribution of major condensables and the effect 
of planets on the distribution of the dusty, small body material. 
Disks are both the product of star formation and the initial 
step in the formation of planets.

The remnant of the circumstellar disk that formed our Solar System
is observable today as the smaller planets, moons, asteroids and
comets, along with the zodiacal light and interplanetary gas and
dust. We have samples of this material in meteorites and interplanetary
dust particles. Studies of samples provide accurate chemical and
(radioactive isotope) age determinations for events, but are
difficult to relate to specific locations in our own protoplanetary
disk because of dynamical stirring of material. Further, the
primitive material in our meteorite collection does not seem to
correspond in detail to the abundances in the Earth's mantle and
crust, suggesting that much of the ``primitive inventory'' is
missing or sequestered in the core (Drake \& Righter 2002).

Spectroscopy and photometry of small Solar System bodies, particularly 
comets, at wavelengths and sensitivities unavailable from the 
ground can identify isotopic ratios and molecular and elemental 
abundances. These can be compared with remnant and planet-forming 
disks, providing direct measurements of the smaller components 
of circumstellar disk formation. For example, IR and radio spectroscopic 
observations of cometary parent molecules, the species that sublimate 
directly from the nucleus, suggest a strong similarity in chemical 
composition between cometary nuclei and the icy dust in protostellar 
environments (Irvine et al. 2000). In both cases, composition 
is dominated by H$_{2}$O ice, while CO and CO$_{2}$ are usually the 
next most abundant. The spectra also show CH$_{3}$OH, H$_{2}$CO, and 
CH$_{4.}$ In addition, C$_{2}$H$_{6}$ and C$_{2}$H$_{2}$ have also been observed 
in comets, but they have not yet been detected in icy ISM grains. 
High concentrations of deuterated species are another indication 
that comets may retain pristine interstellar matter (Meier \& 
Owen 1999).

\begin{figure*}
\centering
\includegraphics[width=1.00\textwidth]{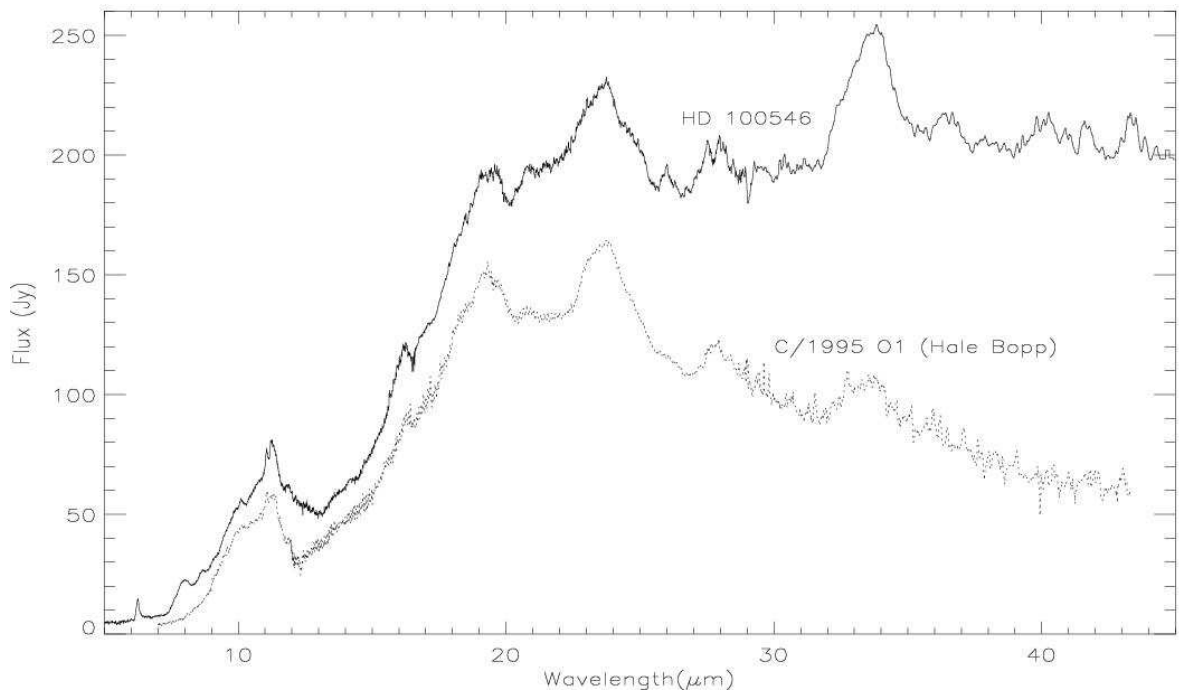}
\caption{
Comets and circumstellar disks. An ISO spectrum of the Herbig
Ae/Be star HD 100546 (upper curve) shows emission features in the star's
circumstellar disk as predicted if the disk is heated by radiation from the
central source. The particle composition in the circumstellar disk appears
to be remarkably similar to that of comet Hale-Bopp (lower curve). JWST will
only reach the shorter wavelengths depicted here, but the similarities are
most prominent at those wavelengths (From Malfait et al. 1998).}
\label{fig022}
\end{figure*}

Since the mid 1980s apparition of Comet Halley, it has been 
known that the silicate mineralogy in cometary dust is similar 
to that in circumstellar disks (Spinrad 1986). This was illustrated 
with ISO observations of Hale-Bopp, which showed that the mid-infrared 
reflectivity of the comet's dust is strikingly similar to that 
in a protoplanetary disk surrounding the young star HD 100546 
(Fig.~\ref{fig022}; Malfait et al. 1998). In particular, strong emission 
features of carbon- and oxygen-rich dust are seen in both spectra, 
with the most prominent being attributed to crystalline silicates. 
This observation confirms the key role of comets in understanding 
the chemical nature of dust in debris disks.

\subsubsection{Observations}

\paragraph{Circumstellar Disks:}

JWST will resolve the details of nearby debris disk structures to determine 
the dynamical effects of planets and through spectra, the radial 
and even azimuthal distribution of major elements and their molecular 
or mineralogical carriers. JWST will also be able to constrain 
the radial temperature distribution on the surface and in the 
interiors of the disks.

\begin{figure*}
\centering
\includegraphics[width=1.00\textwidth]{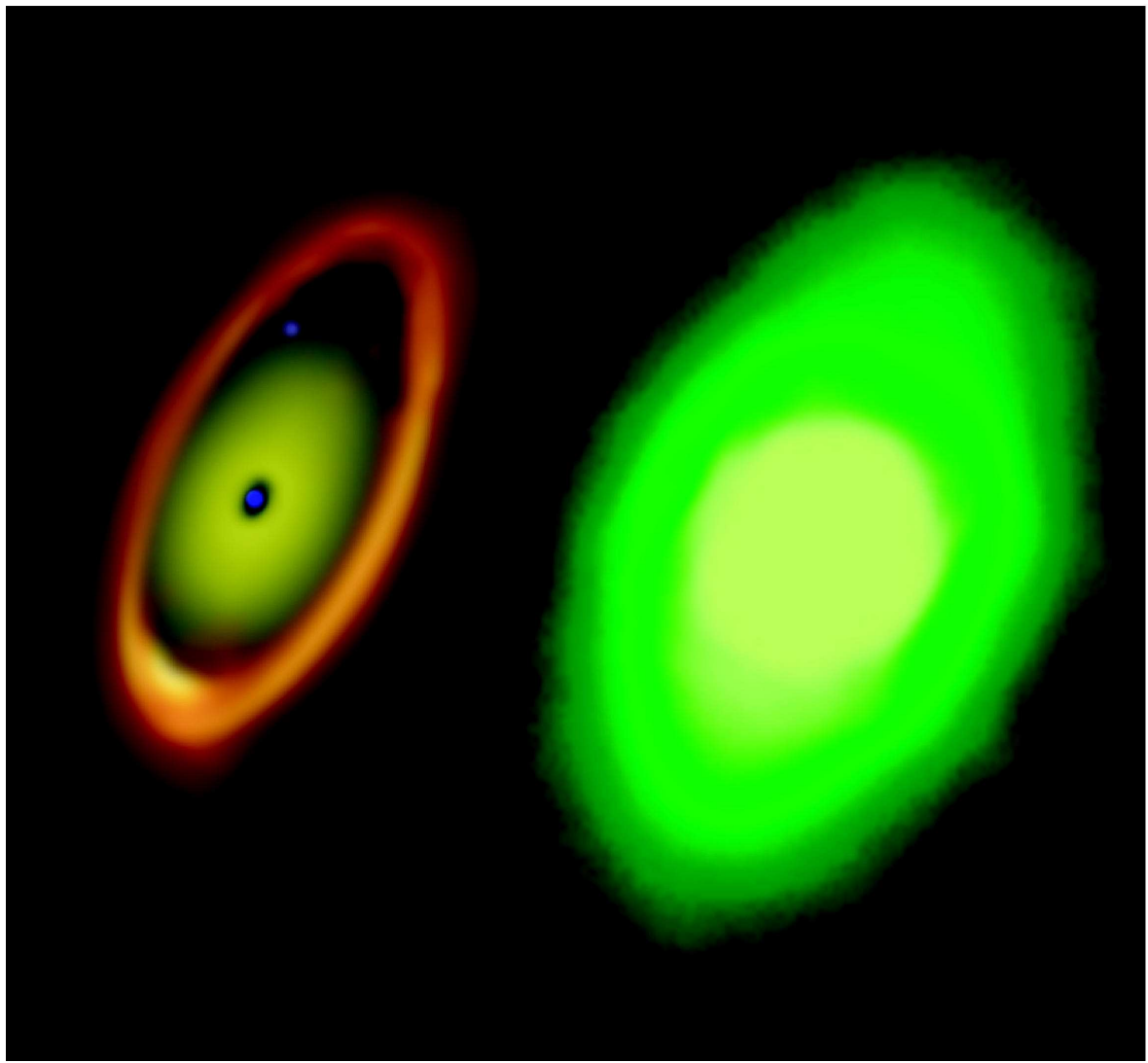}
\caption{
Dusty debris disks. We show the Fomalhaut debris disk as it
appears to Spitzer at 24 microns (right; Stapelfeldt et al. 2004), and in a
simulated JWST image. Structure within the disk is clearly resolvable by
JWST.}
\label{fig023}
\end{figure*}

JWST will test models of debris-disk evolution in the presence of planets 
out to distances of 40 to 50 pc. At these distances, it will provide 
the same spatial resolution available with Spitzer on 
the Vega disk (8 parsecs away), which has been sufficient to determine the 
first-order effects of a giant planet on disk morphology (Su et al. 2005). 
Around Vega, Fomalhaut, and other nearby disk systems, details of 
the perturbations in disk structure on 10 AU scales are likely 
to be discernable (Fig.~\ref{fig023}), so that the nature of debris
disks can be constrained.

\begin{figure*}
\centering
\includegraphics[width=1.00\textwidth]{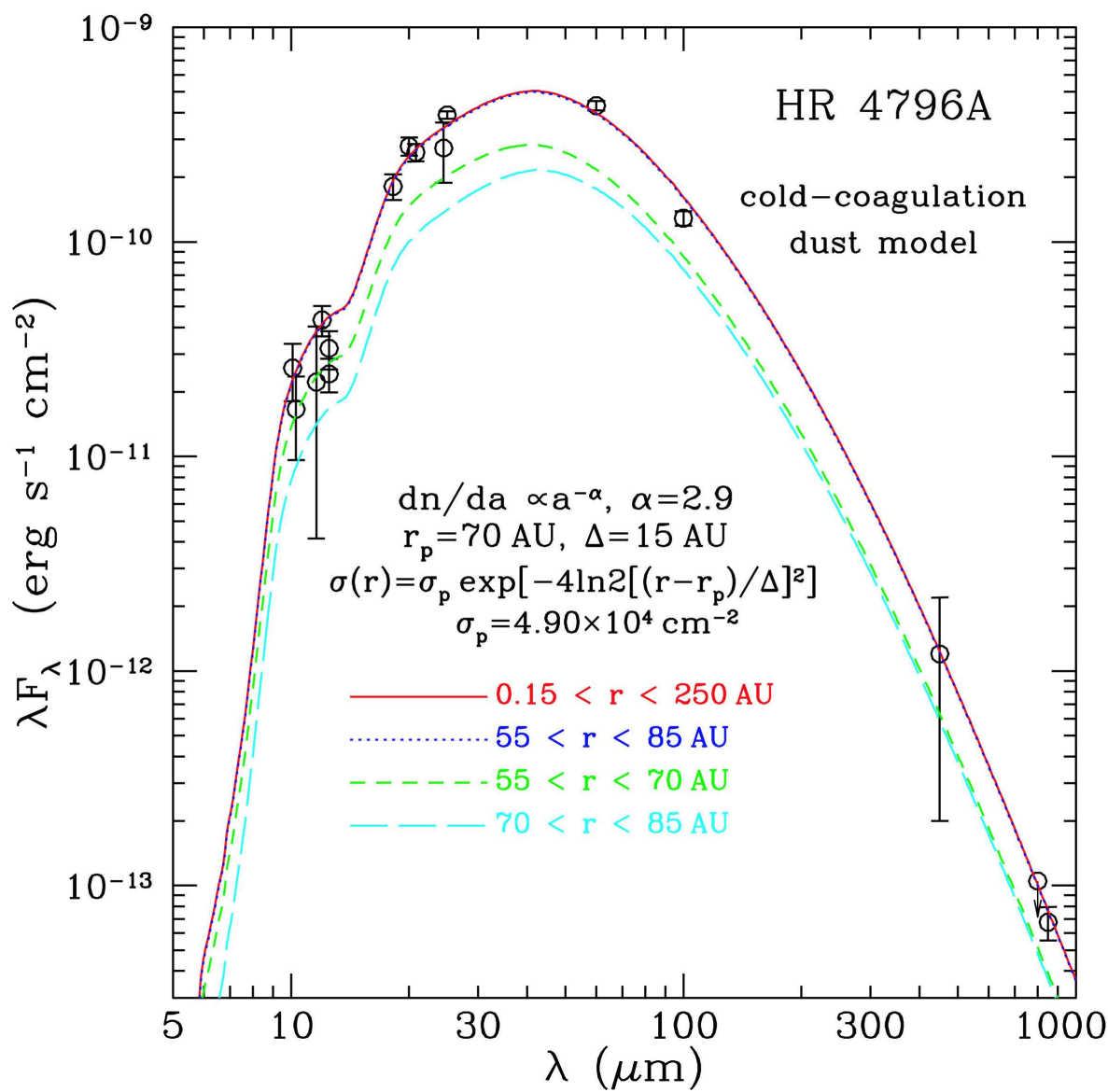}
\caption{
Models and observations of a dusty disk. Models of the spectral
energy distribution the dusty disk around HR4796 are compared with
observations. Emissions from different ranges of semi-major axis are
plotted. Most of the action in terms of peak emission and interesting
spectral features lies beyond 5 $\mu$m, as is the case for most dust disks
(Li \& Lunine 2003).}
\label{fig024}
\end{figure*}

Spectroscopic resolution of 1000 in the near and mid-infrared 
will resolve the spectral signatures of key ices and silicates 
in the disks, and coronagraphic capability is necessary to block 
the light of the central star and observe planets perturbing 
the disk structures. The grain temperatures in the disks are 
generally low enough that the peak emission will be in the mid-infrared 
(Fig.~\ref{fig024}), and there is little dust at much warmer temperatures.

\paragraph{Comets:}

Comets are remnants of Solar System formation, and
their current composition and physical properties provide a constraint 
on the conditions in the solar nebula 4.6 billion years ago. 
Comets were the building blocks of the giant planets' cores. 
Low resolution infrared spectroscopy of cometary dust will uncover 
mineralogical signatures, which can be compared with those seen 
in protostellar and planetary debris disks around nearby young 
stars and solar analogs, and potentially reveal the isotopic 
ratios of some major elements.

Observations of comets with JWST will enable investigations of 
the chemical composition of cometary ice and dust with unprecedented 
sensitivity. Near- and mid-IR spectroscopy of cometary comae 
can be used to measure abundances of H$_{2}$O, CO, CO$_{2}$, and CH$_{3}$OH 
in even relatively faint comets. Near-IR spectrometry with R $\sim$ 1000 resolution 
will be used to measure the ratio of ortho and para 
H$_{2}$O separately (Crovisier et al. 1997), possibly providing 
an indication of the comet's formation temperature (Mumma et 
al. 1987). Likewise, mid-IR spectroscopy can determine the mineralogy 
of cometary dust grains in virtually any comet passing through 
the inner Solar System. Finally, JWST's ability to image cometary 
nuclei at both mid and near-infrared wavelengths with high spatial 
resolution and sensitivity will allow high accuracy measurements 
of sizes and albedos for essentially every observable comet, 
including those of both the short- and long-period dynamical 
classes. The results from cometary programs can be combined with 
those from programs investigating circumstellar disks and star 
formation regions to build a complete picture of planetary system 
formation and evolution.

JWST will measure the CO$_{2}$ abundance in comets that come 
within \ensuremath{\sim}3 AU of the Earth and Sun. Depending on 
the circumstances, JWST can measure CO$_{2}$ emission in either 
the \ensuremath{\nu}$_{3}$ band near 4.3 \ensuremath{\mu}m or the \ensuremath{\nu}$_{2}$ band 
near 15 \ensuremath{\mu}m, both of which are exceptionally strong (Goody 
\& Yung 1989). Only JWST can detect CO$_{2}$; strong absorption 
in the terrestrial atmosphere prevents ground-based infrared 
observations. The molecule does not have a permanent electric 
dipole moment so it does not emit in the radio.

Although we now have several means of measuring H$_{2}$O and CO 
directly in comets from ground-based observatories (Mumma et 
al. 2003), JWST will do so with much higher sensitivity. From 
the ground, H$_{2}$O can only be observed in intrinsically faint 
intercombination bands; atmospheric absorption prevents observations 
in the much stronger fundamental bands. By observing H$_{2}$O in 
the fundamental bands with a resolving power of \ensuremath{\sim}100, 
JWST will not only detect much lower activity levels, but will 
also measure the ortho-to-para ratio (OPR) in many comets. Since 
the OPR is not modified in cometary comae, nor is its value expected 
to change from the warming of a nucleus during its short-duration 
trek through the inner Solar System, the OPR probably reflects 
the comet's temperature at the time of its formation (Mumma et 
al. 1987). However, cosmic rays may alter the OPR in the outer layers 
of comets in the Oort cloud (Mumma et al. 1993). In that case 
the OPR may provide an unambiguous method for identifying the 
comet as dynamically new.

Measuring the HDO/H$_{2}$O ratio in long- and short-period comets 
(or the latter's larger cousins in the Kuiper Belt) will supplement 
the D/H determinations made to date (Meier et al. 1998). This 
is key to understanding the relationship between water in comets 
and that on the Earth and Mars (Lunine et al. 2003). JWST's high 
sensitivity spectral observations in the near- and mid-IR will 
enable this measurement in certain comets and the largest KBOs.

Similarly, the CO abundance can be measured directly from ground-based 
facilities, but JWST would again provide at least an order of 
magnitude improvement in sensitivity. Thermal emission from a 
warm telescope and the Earth's atmosphere limits the capabilities 
of even large ground-based infrared facilities. Radio observations 
of CO are much less sensitive than IR observations, because of 
CO's small electric dipole moment.

Cometary molecules have spectral lines across the entire JWST 
spectral range. Primary vibration-rotation lines are concentrated 
in the 1-5 \ensuremath{\mu}m band. We also need observations at
15 \ensuremath{\mu}m to study CO$_{2}$.

Spitzer has neither the spectral resolution nor the wavelength 
range required for gas-phase cometary studies. In addition, JWST's 
spatial resolution will permit great advances over Spitzer for 
the study of solid phase materials in both comets and circumstellar 
material.

\paragraph{Moving Targets:}

To observe an appropriate sample of comets, JWST must be able 
to track an object that moves with respect to the background 
stars at a minimum rate of 0.030 arcsec/sec. Even with the Solar 
and anti-Solar exclusion angles, this capability allows tracking 
of long-period or hyperbolic comets such as Halley and Hale-Bopp 
to within about 2 AU of the sun, at which point they are very 
active. It also permits studies of short-period comets such as 
Wirtanen, Encke, and Hartley 2.

As most comets tend to be relatively bright, the exposure times 
will be short. A slight discrepancy between the true rate and 
the JWST tracking rate is not an issue. However, for distant 
cometary nuclei (and KBOs), the exposure times are longer. In 
that case, differences between the true and implemented rates 
can cause the object to drift across pixels, smearing the image 
and degrading the signal. The track rate of the object must be 
followed to an accuracy of a small fraction of a pixel in a typical 
imaging frame time, e.g., 0.005 arcsec in 1000 sec (for very faint 
and typically slow objects).

Moving target tracking is currently under study by the JWST Project 
to determine if this capability is affordable.

\subsection{How Are Habitable Zones Established?}

What are the sources of water and organics for planets in habitable zones? How 
are systems cleared of small bodies? What are the planetary evolutionary 
pathways by which habitability is established or lost? Does our 
Solar System harbor evidence for steps on these pathways?

\paragraph{Water and Organics:}

Some geochemical evidence suggests that Earth's water did not 
come from locally formed planetesimals at 1 AU (Morbidelli et 
al. 2000). However, the source of water is uncertain. Asteroids 
are a dynamically plausible source and could be isotopically 
consistent, if chondrites are a typical sample of the section 
of the primordial asteroid belt that supplied water to the Earth 
(Robert 2001). However, other geochemical evidence severely 
limits the amount of chondritic material the Earth might have 
acquired (Drake \& Righter 2002). Cometary HDO/H$_{2}$O values measured 
in three long-period comets are twice that of Earth's ocean water 
(Meier et al. 1998). Also, the D/H ratio of short period comets 
and their presumed Kuiper Belt progenitors remains unknown. By 
measuring isotopic ratios in comets and larger Kuiper Belt bodies, 
JWST can solve this part of the puzzle, removing a major uncertainty 
in the source of water for our own planet.

Similarly, the source of the early abundance of Martian water 
is uncertain; as on Earth, it could be local, asteroidal, or 
cometary (Lunine et al. 2003). The continued search for extant 
Martian water inventories is of relevance to this problem as 
well.

Comets remain a highly plausible source of Earth's organics (Pierazzo 
\& Chyba 1999) and the inventory of organics derived from high 
sensitivity infrared spectra will be of value in constraining 
the theories. JWST measurements of the composition and structure 
in protoplanetary disks around other stars will extend the quantification 
of the source of water and organics to putative habitable worlds 
around stars other than the Sun.

\paragraph{Small Bodies:}

Studies of active gas-dust disks and remnant disks from JWST, 
with Spitzer disk studies as a foundation (Meyer et al. 2004), 
will better quantify how solid debris and gas is cleared from 
such disks. JWST will address the timing, the role of the planets, 
and the amount of remnant gas and dust during the early history 
of planetary systems. Measurement of the isotopic, elemental 
and molecular abundances in icy bodies in the outer Solar System, 
as well as in large bodies such as Titan, will provide a body 
of chemical data that will allow us to determine the relationship 
of these various bodies to a set of putative primitive reservoirs, 
constrained as well by the extrasolar disk observations. 

Also, understanding the physical characteristics, including chemistry, 
of KBOs as a function of subclass (Plutino, Scattered Disk, Classical) 
provides important clues to the dynamical evolution of the Kuiper 
Belt. JWST will be particularly important for mid-infrared measurements 
of these objects.

\paragraph{Pathways to habitability:}

How do planets come to be habitable? The role of the giant planets 
in dynamical stability, timing of terrestrial planet formation, 
and supply of volatiles has been discussed above. Do the properties 
of the precursor gas-dust disks, and their co-evolution with 
forming planets, also determine habitability through planetary 
system architecture and planetary masses? Are there young systems 
that seem, in terms of disk architecture or presence of giant 
planets, to be on a trajectory to nurture the development of 
habitable planets?

\paragraph{Evidence in our Solar System:}

Mars is a world on which water once flowed, lakes stood, oceans 
might have come and gone, and perhaps with them life. Why did 
Mars become cold and dry? Was loss of water a consequence of 
the atmospheric erosion of carbon dioxide and other greenhouse 
gases, or the irreversible production of carbonates on the surface 
from water and atmospheric carbon dioxide? Where are these carbonates 
today? Are there patches where clues to the evolutionary drying 
of Mars can be found? What was the original inventory of water 
on Mars?

Saturn's moon Titan is a Mercury-sized world rich in organic 
molecules, has a dense atmosphere, and possesses a water-ice 
and rock interior that could, under other circumstances, supply 
abundant water for life. Could Titan have been habitable earlier 
in its history? How does the surface-atmosphere exchange of mass 
and energy work on an organic-rich but abiotic world, which is 
subject to weak solar forcing but strong seasonal variations? 
Are there surface and lower atmospheric changes on Titan that 
might be missed by Cassini in its four-year tour?

Prior to the origin of life, chemical processes may have led 
to a substantial level of complexity (Lunine 2005). Depending 
on the nature of the prebiotic environment, available building 
blocks may have included amino and hydroxy acids, purines, pyrimidines, 
assorted sugars and fatty acids. These could have combined to 
form polymers of largely random sequence and mixed stereochemistry 
(handedness) (Botta 2004). Remote detection of 
stereochemical orientation is almost impossible, although amino 
acid detection might be possible if sufficient concentrations 
exist on optically exposed surfaces.

\subsubsection{Observations}

\paragraph{Comets:}

Comets, through collisions with the terrestrial planets, might 
represent a significant source of volatile material in the inner 
Solar System. A substantial portion of the Earth's oceans and 
organic material were probably provided by cometary bombardment, 
tying comets directly to the origin of life. JWST's spatial and 
spectral resolution will allow studies of gas phase processes 
in the inner comae of active comets, revealing their composition.

\paragraph{KBOs:}

JWST will obtain near-IR and mid-IR measurements of the brightness 
of KBOs to separate effects due to size and albedo and constrain 
their physical size. Spectra will be obtained for the larger 
and closer KBOs to identify major molecules and isotopic ratios 
for the largest objects, including Triton, Pluto, Charon, Quaoar, 
and Varuna. In this context, Neptune's satellite Triton is a 
captured Kuiper Belt Object, as indicated by Triton's retrograde 
orbit and Neptune's dilapidated remnant regular satellite system 
(McKinnon et al. 1995).

\begin{figure*}
\centering
\includegraphics[width=1.00\textwidth]{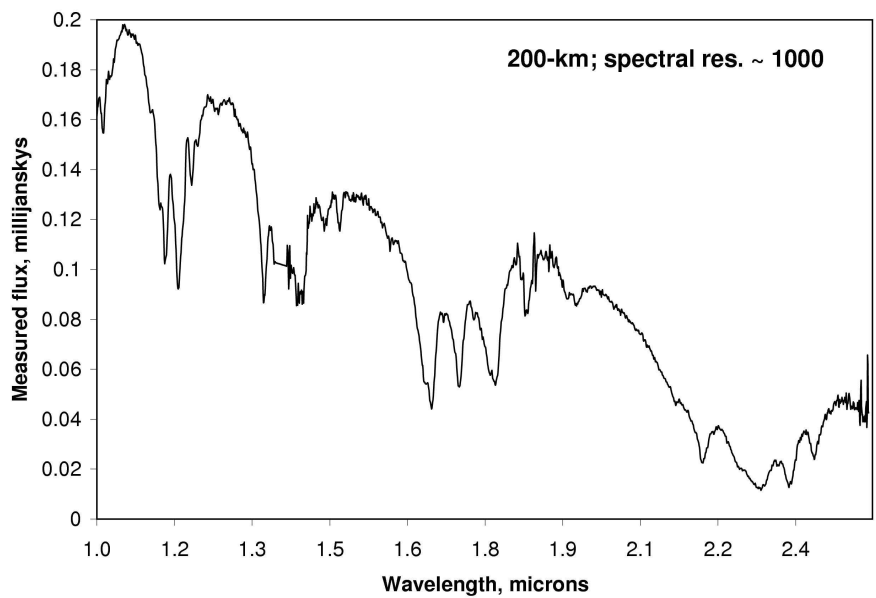}
\caption{
Simulated KBO spectrum. The simulated spectrum of a Kuiper
Belt Object with a 200-km radius using a spectrum borrowed from that of
Pluto (Cruikshank et al. 1997). This spectrum is well above JWST's
sensitivity limit. Many or all of the major ices can be identified with
absorption features out to 2 $\mu$m (Schmitt et al. 1998).}
\label{fig025}
\end{figure*}

JWST will obtain high resolution, near-infrared spectra of bodies
hundreds of kilometers in size, typical of those detected in surveys
of the Kuiper Belt, to determine the presence of various ices,
including deuterated species and other isotopic bands, as shown in
the simulation in Fig.~\ref{fig025}. Ratios of isotopes, including
$^{13}$CO/$^{12}$CO,
C$^{18}$O/C$^{16}$O,
$^{13}$CH$_{4}$/$^{12}$CH$_{4}$,
$^{15}$NH$_{3}$/$^{14}$NH$_{3}$,
CH$_{3}$D/CH$_{4}$,
and HDO/H$_{2}$O,
may be detectable on Pluto depending on their abundances. This
information will reveal the relationship of comets and the bound
water in carbonaceous chondrites to the ices on the surfaces of
the largest KBOs.

\begin{figure*}
\centering
\includegraphics[width=1.00\textwidth]{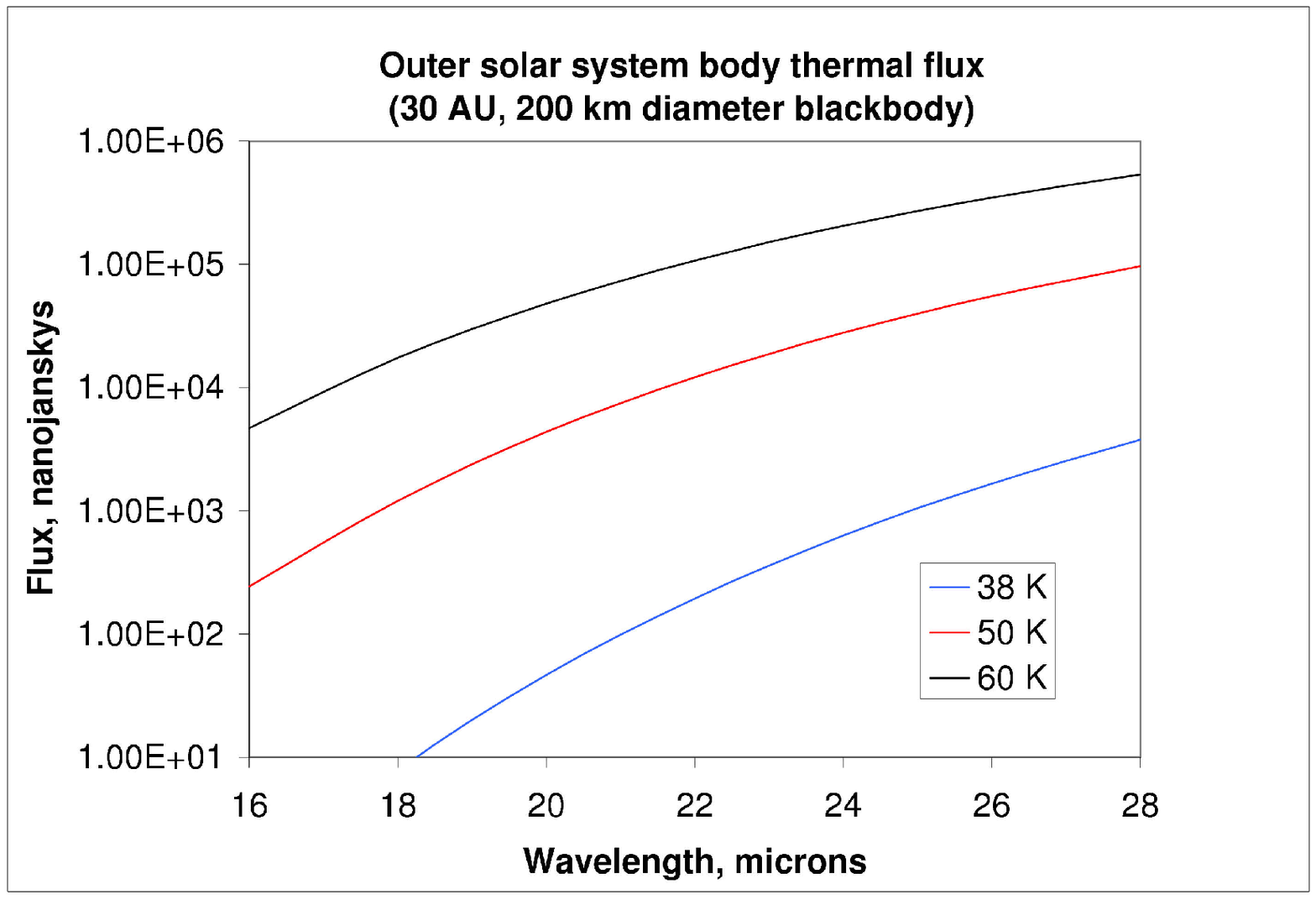}
\caption{
Blackbody flux versus wavelength in
the mid-infrared for a 200-km radius Kuiper Belt object at three different
possible temperatures, located 30 AU from JWST.}
\label{fig026}
\end{figure*}

JWST will be able to detect KBOs in the mid-infrared down to typical
discovery sizes of hundreds of kilometers, in order to provide a
separate determination of size and brightness, when combined with
the optical brightness (Fig.~\ref{fig026}). These observations will
provide information on the dynamical state of the Kuiper Belt, in
particular whether the size distribution suggests a population
built by accretion or by collisional grinding. They will provide
an indication of the color variation and possible existence of
multiple classes of Kuiper Belt objects (e.g., volatile-rich versus
quiescent).

Target tracking requirements for Kuiper Belt objects are typically
less than those of comets (discussed above), although more accurate
tracking rates are needed because of the long exposure times.
Current ultra-deep surveys with 8-meter ground-based telescopes
achieve R-magnitudes above 26 (Sheppard et al. 2005). JWST will
enable deeper searches for new objects and higher sensitivity
characterization of known ones. Photometry of near-infrared reflected
light and of thermal emission at 20 \ensuremath{\mu}m will constrain
simultaneously the albedo and the radius. Spectroscopic studies at
resolutions of \ensuremath{\sim} 1000, comparable to the best
ground-based studies of Pluto and Triton, will be extended to
smaller and more distant bodies. R = 3000 spectroscopy of Triton
and Pluto will constrain isotopic ratios in the water ice and other
components, as well as providing surface temperature monitoring
through the nitrogen overtone band. While the near infrared is the
most familiar territory in this regard, because of the ground-based
spectra of Triton and Pluto, rotational features will appear in
the mid-infrared and will provide compositional and isotopic data
not attainable in the near-IR overtone region. Mid-infrared detection
leading to the separation of the size and albedo of such objects
is of fundamental importance, and requires the high mid-infrared
sensitivity that JWST provides.

\paragraph{Titan:}

JWST will observe Titan to establish a long-time baseline of 
atmospheric and surface changes connected with the 2004-2008 
Cassini mission survey, creating a 10 year or longer baseline 
of space-borne near-infrared observations of Titan's surface 
and atmosphere. JWST will use near-infrared spectrometry with 
spectral resolution a factor of six higher than on Cassini to 
determine the types of organic species present on the surface. 
Thus, while Cassini gets better spatial resolution, JWST will 
achieve higher spectral resolution over the mid-latitude regions 
of Titan. These data will reveal whether surface changes or secular 
atmospheric changes are in evidence over a decadal timescale.

\subsection{Summary}

\begin{table}[t]
\caption{JWST\ Measurements for the Planetary Systems and Origins of Life Theme\label{tab005}}
\begin{tabular}{p{1.0in}p{0.8in}p{1in}p{1.0in}p{0.0in}}
\hline\noalign{\smallskip}
{Observation} &
{Instrument} &
{Depth, Mode} &
{Target} &
\\[3pt]
\tableheadseprule\noalign{\smallskip}
\raggedright
Isolated Extra-Solar Giant Planets&
NIRCam&
\raggedright 4 nJy at 2 \ensuremath{\mu}m
&
\raggedright Star-forming regions &
\\
&&&& \\
Bound planets&
\raggedright TFI or NIRCam&
\raggedright 200 \ensuremath{\mu}Jy at 4 \ensuremath{\mu}m&
\raggedright Nearby 
Stars&
\\
&&&& \\
\raggedright In-depth study&
\raggedright TFI&
\raggedright 60 nJy, R = 100&
\raggedright Nearby Stars&\\
&NIRSpec&
1 \ensuremath{\mu}Jy, R = 3000&
&\\
&MIRI&
1 \ensuremath{\mu}Jy at 15 \ensuremath{\mu}m&&
\\
&&&& \\
\raggedright Transiting planets&
NIRSpec&
R$\sim$1000&
Kepler discoveries&
\\
&&&& \\
\raggedright Circumstellar disks&
MIRI&
\raggedright 1 $\mu$Jy at 15 \ensuremath{\mu}m&
\raggedright
Debris 
Disks&
\\
&&&& \\
Comets&
NIRSpec&
\raggedright 1 \ensuremath{\mu}Jy, R = 3000 at 2 \ensuremath{\mu}m&
10/yr&\\
&MIRI&
\raggedright 1 \ensuremath{\mu}Jy, 
R=3000 at 15 \ensuremath{\mu}m.&&
\\
&&&& \\
\raggedright Kuiper Belt Objects&
NIRCam&
\raggedright 4 nJy at 2 \ensuremath{\mu}m&
10 arcmin$^{2}$&\\
&NIRSpec&
\raggedright 1 \ensuremath{\mu}Jy, R = 3000 at 2 \ensuremath{\mu}m&&\\
&MIRI&
\raggedright 1 \ensuremath{\mu}Jy at 25 \ensuremath{\mu}m&&\\
\\
&&&& \\
\raggedright Satellites&
NIRSpec&
\raggedright 1 \ensuremath{\mu}Jy, R = 3000 at 2 \ensuremath{\mu}m&
Titan& \\
&MIRI&
1 \ensuremath{\mu}Jy at 15 \ensuremath{\mu}m&&\\
&&&& \\

\noalign{\smallskip}\hline
\end{tabular}
\end{table}

Table~\ref{tab005} summarizes the measurements needed for the
Planetary Systems and Origins of Life theme. They include:

\textbullet{}
{\it Isolated extra-solar giant planets.} Deep NIRCam imaging (done
as part of IMF studies) will detect isolated giant planets and
brown dwarfs, or planets in large orbits.

\textbullet{}
{\it Bound planets.}
Using TFI and NIRCam coronagraphy, JWST can observe bound 
giant planets, and characterize their ages and masses through 
spectral features.

\textbullet{}
{\it In-depth study.} Subject to the discovery of planets appropriate
for follow-up (that is, at the right distance from their star, not
too faint, etc.), JWST will get spectra with NIRSpec, narrow-band
imaging with TFI, and mid-IR (coronagraphic) photometry with MIRI.

\textbullet{}
{\it Transiting planets.} Using rapid readout of the NIRSpec detectors,
JWST will measure differential light curves of transiting planets
discovered by Kepler. These spectra will contain atmospheric
signatures of O$_2$, CO$_2$, CO, CH$_4$, H$_2$O, Na, and K.

\textbullet{}
{\it Circumstellar disks.} Resolved mid-IR spectra of circumstellar
disks will reveal the signature of key ices and silicates, constraining
the constituents of terrestrial planets. Resolved, coronagraphic
imaging of disks may reveal structures due to shepherding planets.

\textbullet{}
{\it Comets} NIRSpec spectroscopy of comets will be used to measure
the abundances of H$_2$O, CO, CO$_2$, and more complex molecules
like CH$_3$OH. Ortho and para H$_2$O can be separated, indicating
the comets formation temperature. HDO/H$_2$O will constrain models
of the source of water on the Earth and Mars. MIRI spectroscopy
will reveal the mineralogy of the comets.

\textbullet{}
{\it Kuiper belt objects.} NIRCam imaging will be used to find and
identify KBOs, characterizing the population's dynamics and
sub-classes. NIRSpec spectroscopy and MIRI photometry and spectroscopy
will be used to study individual objects, revealing molecules and
isotopic ratios, size and albedo, and temperature.

\textbullet{}
{\it Satellites.} Using NIRSpec and MIRI, JWST will observe weather
on Titan to establish a 10-year baseline follow-on to the Cassini
mission, monitoring seasonal changes as the Saturnian system
transitions from summer to autumn. NIRSpec's spectral resolution
is a factor of 6X higher than Cassini's, and will reveal the organic
species present on the surface.

\section{JWST Implementation}

To make the scientific measurements described in sections 2 through
5, JWST will be a large cold telescope, with a wide field of view,
exceptional angular resolution and sensitivity, and wide wavelength
coverage in both imaging and spectroscopy. It will be launched
early in the next decade to an orbit around the Earth-Sun second
Lagrange point (L2). The scientific objectives impose observational
requirements that can only be met with JWST; no other existing or
planned ground-based telescope or space telescope mission can meet
these requirements. In this section we describe the design of the
JWST observatory and mission.

\begin{figure*}
\centering
\includegraphics[width=1.00\textwidth]{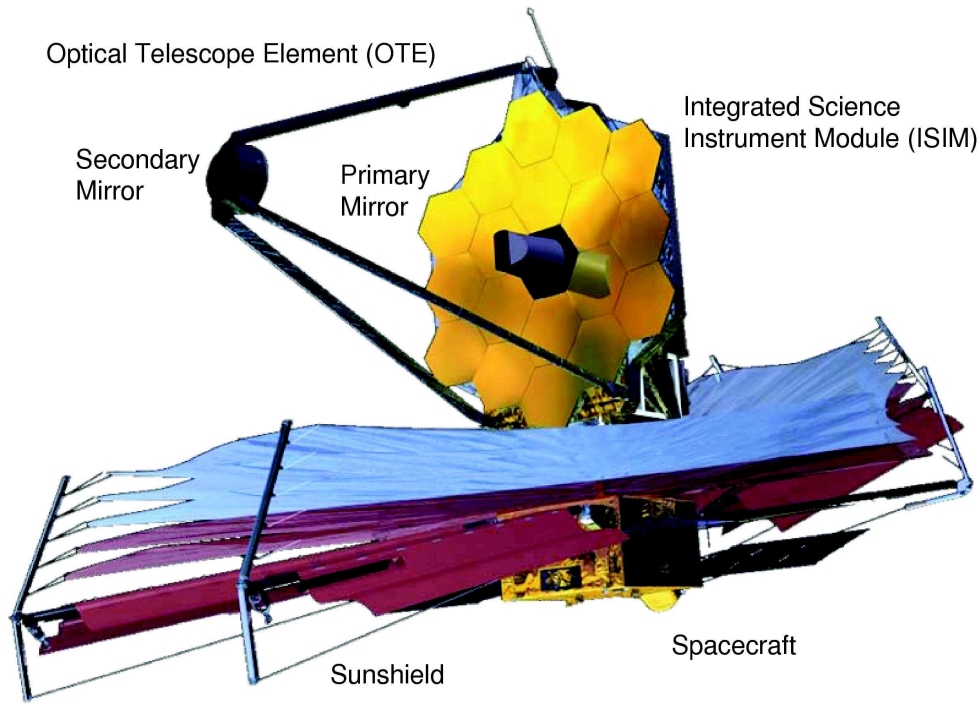}
\caption{
The JWST Observatory. The optical telescope element contains the
primary and secondary mirrors, the integrated science instrument
module (ISIM) element contains the instrumentation, and the spacecraft
element consists of the spacecraft and the sunshield.} 
\label{fig027} \end{figure*}

The JWST Project is organized into three segments: observatory,
ground and launch. The observatory is composed of an optical
telescope element, an integrated science instrument module
(ISIM) containing the scientific instruments, a spacecraft, and a
sunshield (Fig.~\ref{fig027}). The telescope (Feinberg 2004) is a deployable
optical system that provides diffraction-limited performance at 2
\ensuremath{\mu}m using active wavefront sensing and control
(WFS\&C). The ISIM (Davila et al. 2004; Greenhouse et al. 2004)
contains four science instruments (SIs): a near-infrared camera
(NIRCam; Horner \& Rieke 2004), a near-infrared multi-object
spectrograph (NIRSpec; Zamkotzian \& Dohlen 2004), a mid-infrared
camera (MIRI; Wright et al. 2004), and a near-infrared tunable
filter imager (TFI; Rowlands et al. 2004b). The ISIM also contains
a fine guidance sensor (FGS; Rowlands et al. 2004a) to provide
active control of pointing. The spacecraft provides the pointing
platform and housekeeping functions for the observatory. The
wavelength range of JWST and the SIs spans 0.6 to 29 microns,
limited at the short end by the gold coatings on the primary mirror
and at the long end by the detector technology. The sunshield shades
the telescope and ISIM from solar illumination to allow zodiacal-light-limited
performance at $\lambda < 10 \mu$m and high sensitivity out to 29
\ensuremath{\mu}m, and provides a stable thermal environment for
the telescope and ISIM.

\subsection{Observatory}

Table~\ref{tab006} summarizes the key features of the observatory
architecture (Nella et al. 2004). The primary mirror uses semi-rigid
primary mirror segments mounted on a stable and rigid backplane
composite structure. The architecture is termed ``semi-rigid'',
because it has a modest amount of flexibility that allows for
on-orbit compensation of segment-to-segment radius of curvature
variations. Semi-rigid segments have a high degree of inherent
optical quality and stability, allowing verification of optical
performance by 1-g end-to-end ground testing before launch.

\begin{table}[t]
\caption{Observatory Key Features\label{tab006}}
\begin{tabular}{ll}
\hline\noalign{\smallskip}
{Element} & 
{Feature}
\\[3pt]
\tableheadseprule\noalign{\smallskip}
Optical Telescope&Three mirror anastigmat f/20 design\\
Element&Fine steering mirror (FSM) to provide line-of-sight\\
&\hspace{0.5in}stabilization \texttt{<}7.3 milliarcsec\\
&Four separate deployments\\
&Semi-rigid, adjustable hexagonal mirror segments and \\
&\hspace{0.5in}graphite composite backplane structure\\
&\\
Primary Mirror&25 m$^{2}$ collecting area\\
&Primary mirror deploys in two steps (2-chord fold)\\
&18 semi-rigid hexagonal segments with set-and-monitor\\
&\hspace{0.5in}wavefront control actuators\\
&Mirror segment material is Beryllium \\
&\\
Secondary Mirror&Tripod configuration for support structure\\
&Deployment using a single redundant actuator\\
&Semi-rigid optic with 6 degrees of freedom alignment\\
&\\
Aft Optics&Fixed baffle contains tertiary mirror and FSM\\
&\\
ISIM&Simple semi-kinematic mount; 8 m$^{2}$ of thermal radiator\\
&\hspace{0.5in}area, and 19.9 m$^{3}$ volume.\\
&Contains all science instruments and fine guidance sensor\\
&\\
Tower&Integral 1 Hz passive vibration isolators\\
&Isolates the telescope from spacecraft both thermally and\\
&\hspace{0.5in}vibrationally\\
&\\
Sunshield &Five layer ``V'' groove radiator design reduces solar\\
&\hspace{0.5in}energy to 10's of mW\\
&Folded about telescope during launch\\
&Sized (\ensuremath{\sim}19.5m $\times$ 11.4m) and shaped to limit solar\\
&\hspace{0.5in}radiation induced momentum buildup.\\
&\\
Spacecraft Bus &Chandra-based attitude control subsystem\\
&Two-axis gimbaled high-gain earth-pointing antenna with\\
&\hspace{0.5in}omnis, Ka and S band\\
&471 Gbit solid state recorder to store 2 days of science\\
&\hspace{0.5in}and engineering data\\
&Propellant for \texttt{>}10 years\\
\noalign{\smallskip}\hline
\end{tabular}
\end{table}

The telescope optics are made of beryllium. Spitzer and IRAS had beryllium
mirrors and its material properties are known at temperatures as
low as 10 K. Beryllium has an extremely small variation in its
coefficient of thermal expansion over temperatures of 30 to 80 K,
making the telescope optics intrinsically stable to small temperature
variations. Beryllium fabrication and figuring procedures were
designed using the results from the Advanced Mirror System Demonstrator
(AMSD) program (Feinberg 2004; Stahl, Feinberg \& Texter 2004). An
engineering demonstration unit will be used as a pathfinder for
the flight mirror processing.

The sunshield enables passive cooling and provides a stable cryogenic
environment by minimizing the amount of solar energy incident onto
the telescope and ISIM. The observatory will not use active wavefront
control during observations.

\subsection{Observatory Performance}

Table~\ref{tab007} summarizes the predicted performance for the
JWST Observatory.

\begin{table}[t]
\caption{Performance of the JWST Observatory\label{tab007}}
\begin{tabular}{p{1.0in}l}
\hline\noalign{\smallskip}
{Parameter} & 
{Capability}
\\[3pt]
\tableheadseprule\noalign{\smallskip}
Wavelength &0.6 to 29 $\mu$m\\
&\\
\raggedright Image quality&Strehl ratio of 0.8 at 2 $\mu$m.\\
&\\
Telescope FOV&Instruments share \ensuremath{\sim} 166 square arcminutes FOV.\\
&\\
Orbit&Lissajous orbit about L2\\
&\\
\raggedright Celestial Sphere Coverage&100\% annually \\
&39.7\% at any given time\\
&100\% of sphere has at least 51 contiguous days visibility\\
&30\% for \texttt{>} 197 days\\
&Continuous viewing zone \texttt{<}5 degrees from\\
&\hspace{0.5in}each ecliptic pole\\
&\\
\raggedright Observing Efficiency&Observatory \ensuremath{\sim} 80\%. Overall efficiency \texttt{>} 70\%.\\
&\\
Mission Life&Commissioning in less than 6 months\\
&5 year minimum lifetime after commissioning\\
&10 years fuel carried for station keeping\\
\noalign{\smallskip}\hline
\end{tabular}
\end{table}

\subsubsection{Image quality}

The imaging performance of the telescope will be diffraction limited
at 2 $\mu$m, defined as having a Strehl ratio \texttt{>} 0.80 (e.g.,
B\'ely 2003). JWST will achieve this image quality using image-based
wavefront sensing and control (WFS\&C) of the primary mirror. There
will also be a fine guidance sensor in the focal plane and a fine
steering mirror to maintain pointing during observations.

The Strehl ratio specification is used to determine the allowed
optical wavefront error (WFE) and its allocation to low (0 to 5
cycles/aperture), mid (5 to 30 cycles/aperture) and high frequencies
($>$ 30 cycles/aperture). Point spread function (PSF) stability is
needed to reliably separate the optical PSFs from different targets
or for the same target at different observations, and to ensure radiometric
stability. A Monte Carlo
analysis was performed for the optical design by varying the spatial
characteristics of the errors, and worst-case analyses were also performed
showing the design and build tolerances are suitable for the science. 
The allocated top-level WFE is 150
nm root-mean-squared (rms) through to the NIRCam focal plane, and
includes both the effect of 7.0 milliarcsec image motion, 
most of which is line-of-sight
jitter and 51 nm of drift instability.

\subsubsection{Sky coverage and continuous visibility}

Field of regard (FOR) refers to the fraction of the celestial sphere 
that the telescope may point towards at any given time. A large 
FOR increases the number of days per year of target visibility, 
provides the ability to visit targets repeatedly for time variability 
studies, flexibility to schedule observations, to revisit failed 
observations, and to respond to targets of opportunity. JWST's 
FOR is limited by the size of the sunshield.

\begin{figure*}
\centering
\includegraphics[width=1.00\textwidth]{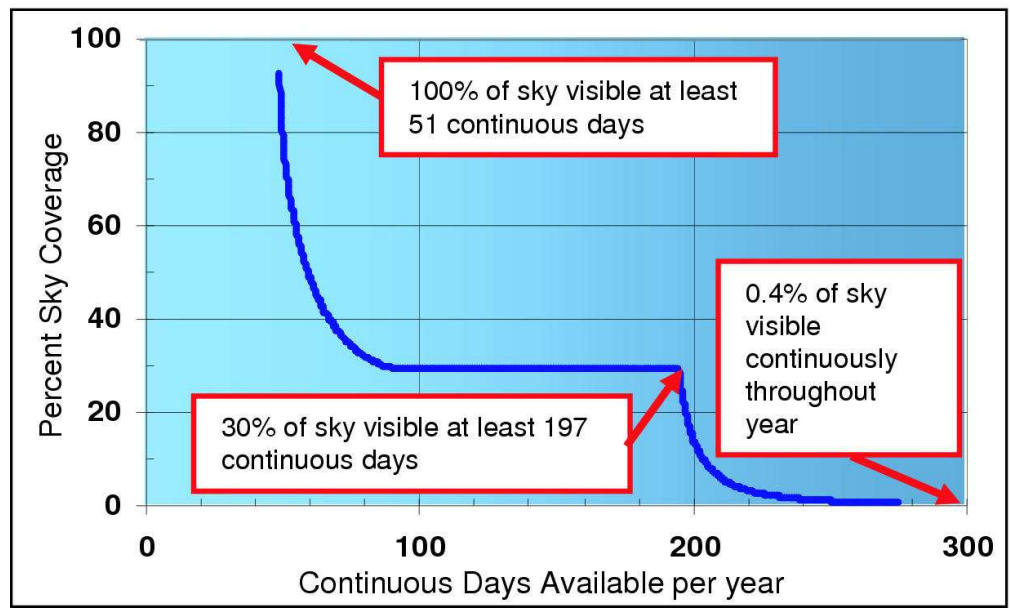}
\caption{
Sky coverage and continuous visibility. There is a continuous viewing 
zone within 5$^{\circ}$ of each ecliptic pole. Thirty percent of the sky
is viewable for at least 197 days per year, and all of the sky will have 
at least 51 days of continuous visibility each year.
}
\label{fig028}
\end{figure*}

\begin{figure*}
\centering
\includegraphics[width=1.00\textwidth]{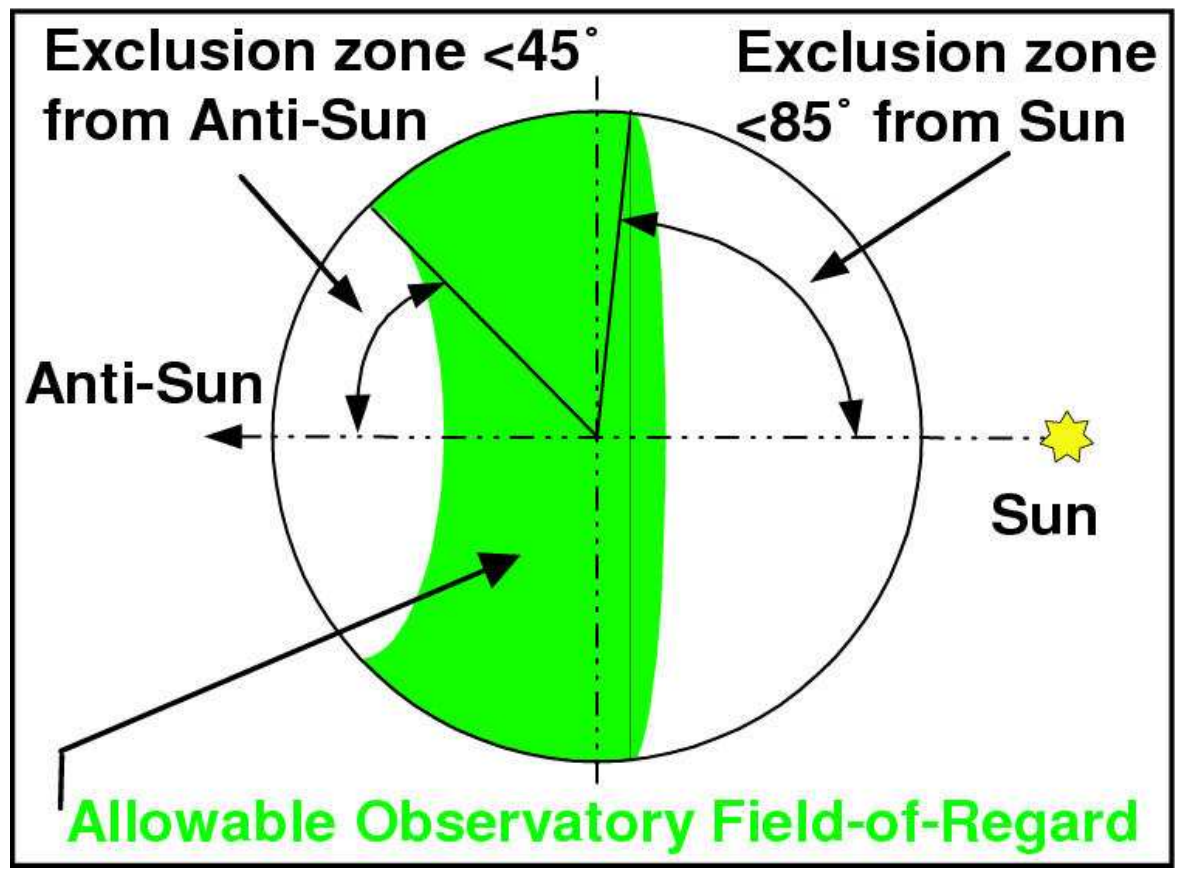}
\caption{
Observatory field of regard (FOR). JWST will be able to see 
about 40\% of the sky at any given time, in an annulus around 
the anti-Sun direction.
}
\label{fig029}
\end{figure*}

Sky coverage performance is shown in Fig.~\ref{fig028}. A continuous
viewing zone within 5\ensuremath{^\circ} of both the north and
south ecliptic poles is available throughout the year. Thirty
percent of the sky can be viewed continuously for at least 197
continuous days. All regions of the sky have at least 51 days of
continuous visibility per year. The architecture provides an
instantaneous FOR at any epoch of approximately 40\% of the sky
(Fig.~\ref{fig029}) This FOR extends 5\ensuremath{^\circ} past the
ecliptic pole, and provides 100\% accessibility of the sky during
a one-year period. In addition, a nominal 5 degree angular safety
margin will be maintained when determining the allowable Observatory
pointing relative to the Sun.

In order to take full advantage of the FOR, JWST 
will be able to observe any point within it at any allowable 
roll angle, with a probability of acquiring a guide star of at 
least 95\% under nominal conditions. This will ensure that most 
of the required targets will be observable without special 
scheduling or other procedures.

\subsubsection{Sensitivity and stray light limitations}

\begin{figure*}
\centering
\includegraphics[width=1.00\textwidth]{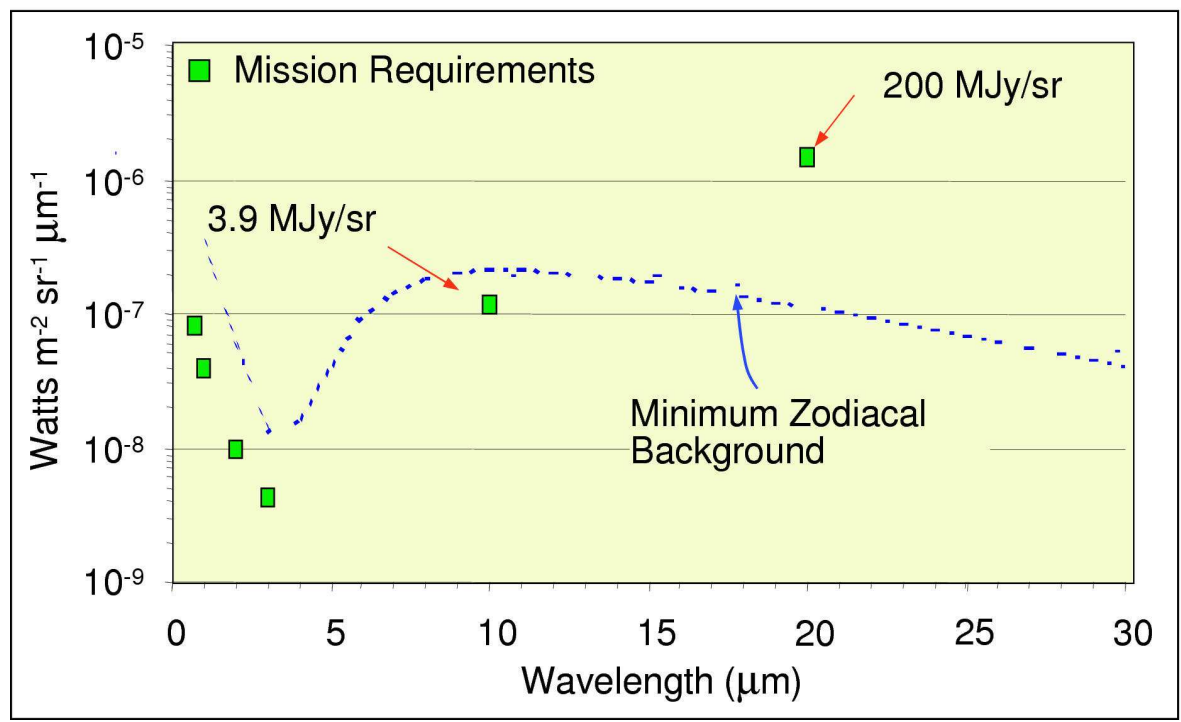}
\caption{
Scattered and self-emission background. JWST will be zodiacal
background limited at $\lambda$ \texttt{<} 10 $\mu$m, and will be limited by
thermal emission from the telescope at longer wavelengths.}
\label{fig030}
\end{figure*}

Many JWST observations will be background limited. The background 
is a combination of in-field zodiacal light, scattered thermal 
emission from the sunshield and telescope, scattered starlight, 
and scattered zodiacal light. Over most of the sky, the zodiacal 
light dominates at wavelengths $\lambda < 10 \mu m$. 
The flux sensitivities of the JWST instruments are given in 
Section 6.5. Fig.~\ref{fig030} gives 
the wavelength dependence of the maximum effective background 
radiance for JWST architecture (the sum of all individual thermal 
emissions and scattered sources). At $\lambda < 10 \mu$m, broad-band
imaging sensitivity will be zodiacal-light limited.

\subsubsection{Observational efficiency}

The observational efficiency determines the scientific productivity
of the mission. Efficiency is the fraction of the total mission
time spent actually ``counting photons,'' rather than occupied by
overhead activities such as slewing, stabilizing, calibrating, and
adjusting parts of the observatory. The observatory is designed to
be stable and simple to operate, and 19\% overhead is allocated to
the spacecraft and observatory, including wavefront sensing and
control operations. An additional 6\% overhead is allocated to the
instrumentation, including standard-star calibrations, and 5\% is
allocated to scheduling, including failed guide-star acquisitions.
The overall efficiency is expected to be 70\%, although this depends
on the specific observing programs that are implemented. In comparison
to missions in low-Earth orbit, there are significant efficiency
advantages of the L2 orbit. JWST will use parallel instrument
calibration (i.e., darks) to achieve this overall efficiency.

The observatory will be fully commissioned 180 days after launch.
The semi-rigid mirror segment design is expected to be stable, so
that wavefront control adjustments will not be needed more than
once per week, minimizing the overhead for this activity. Wavefront
quality is monitored by using the science target imagery in addition
to dedicated observations of bright stars, reducing the overhead.
Time intervals used for momentum control are minimized by the
sunshield design, which approximately balances the observatory
torque from solar radiation pressure over most of the field of
regard.

The thermal stability of the observatory is predicted to be good
enough that optical adjustments will not be necessary even when
large changes in the Sun angle are commanded. Included in the
efficiency overhead is an estimate for safe mode based on the
design and on-orbit performance for the Chandra Observatory. The
largest contributors to the observatory inefficiency are slewing
to new targets and the dither of the locations of a single target
on the focal plane. This indicates that the JWST architecture is
optimized in the sense that the overheads associated with pointing
to targets are the dominant contributors to inefficiency.

\subsection{Observatory Design Description}

\subsubsection{Optical Telescope Element (OTE)}

\begin{figure*}
\centering
\includegraphics[width=1.00\textwidth]{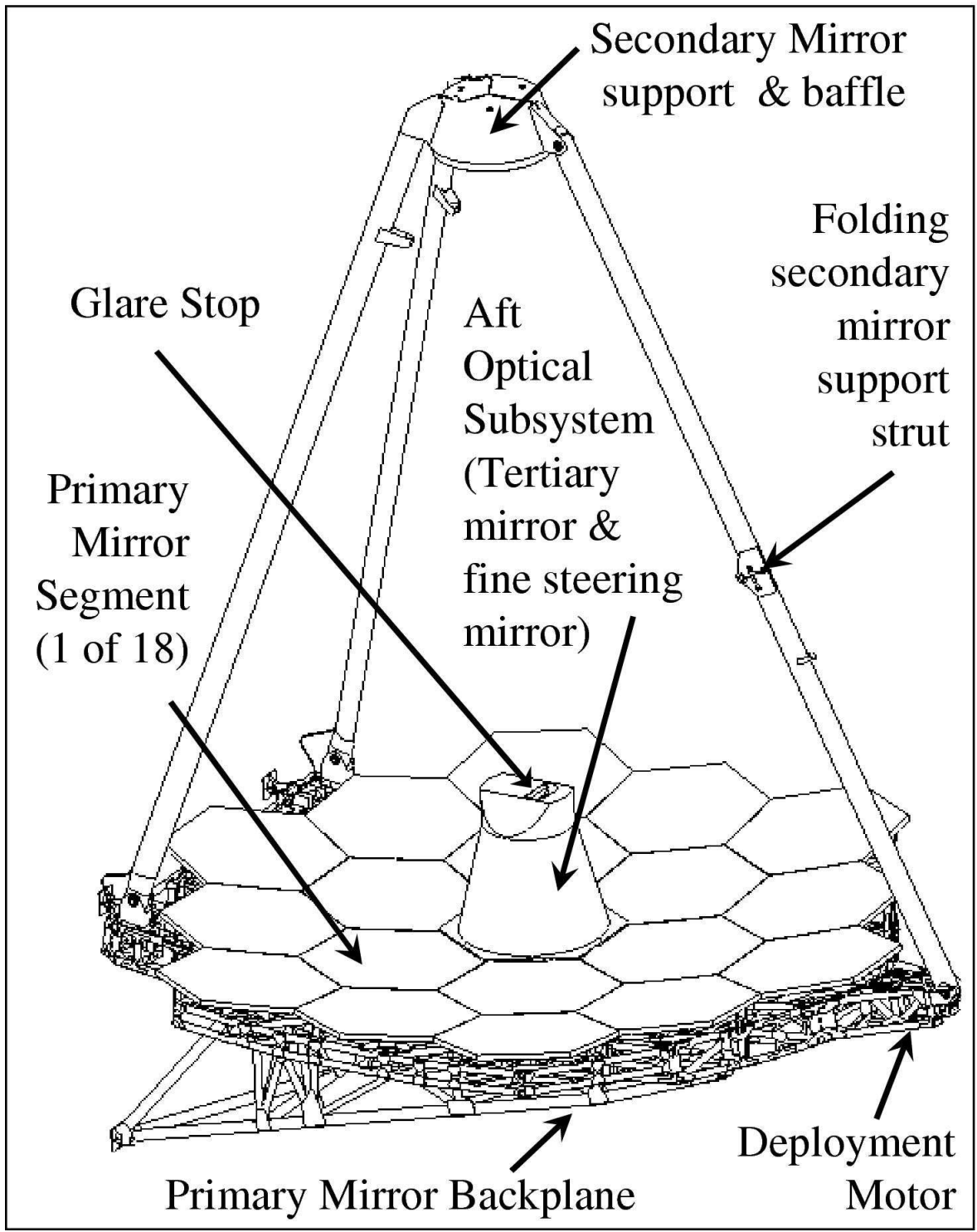}
\caption{
Isometric view of the JWST telescope. The JWST telescope is a three-mirror
anastigmat made up of 18 hexagonal segments.}
\label{fig031}
\end{figure*}

The telescope was designed to provide a well-corrected image to
the ISIM. It has 132 degrees of freedom of adjustment mechanisms
and a composite structure so that passive figure control and passive
disturbance attenuation is possible. Fig.~\ref{fig031} shows an
isometric view of the telescope's primary and secondary mirrors.

\begin{figure*}
\centering
\includegraphics[width=1.00\textwidth]{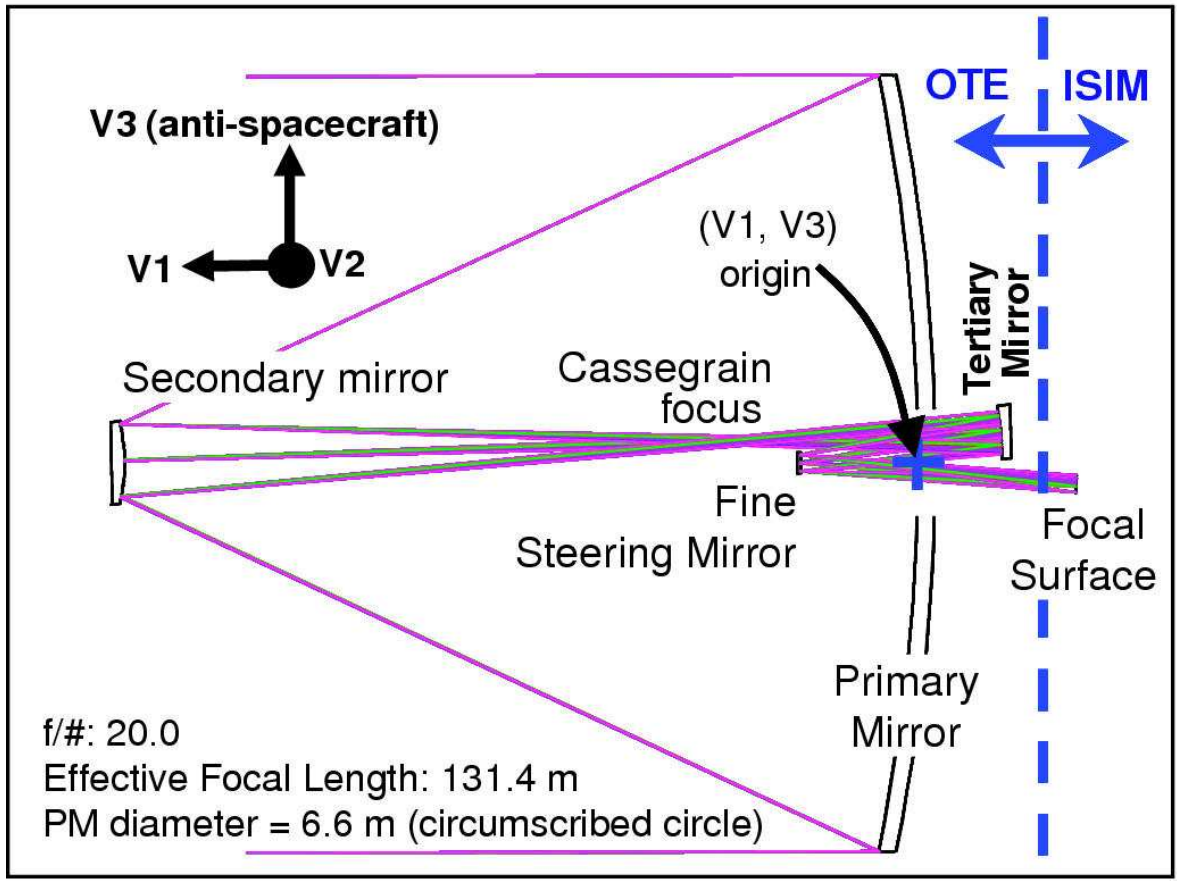}
\caption{
The telescope optical layout. The JWST optical telescope has an effective f/number of 20 and
an effective focal length of 131.4 m.}
\label{fig032}
\end{figure*}

The optical telescope has a 25 m$^{2}$ collecting area, three-mirror
anastigmat (Fig.~\ref{fig032}), made up of 18 hexagonal segments.
It has a Strehl ratio of approximately 0.8 at \ensuremath{\lambda}
= 2 \ensuremath{\mu}m. This configuration provides excellent image
quality over a large field of view (FOV), and accommodates the
science instruments and the guider. The telescope has an effective
f/number of 20, and an effective focal length of 131.4 meters. Each
ISIM instrument reimages the telescope focal plane onto its detectors,
allowing for independent selection of detector plate scale for
sampling of the optical PSF. A fine steering mirror (FSM) is used
for image stabilization. The FSM is located at the image of the
pupil, after the tertiary mirror but forward of the focal plane
interface to the ISIM. The FSM, coupled with the low structural
noise spacecraft, suppresses line-of-sight jitter to allow
diffraction-limited performance at 2 \ensuremath{\mu}m. The secondary
mirror is supported by a deployable tripod support structure, which
latches into position following deployment. This structure provides
the necessary jitter stability with minimum beam blockage.

\begin{figure*}
\centering
\includegraphics[width=1.00\textwidth]{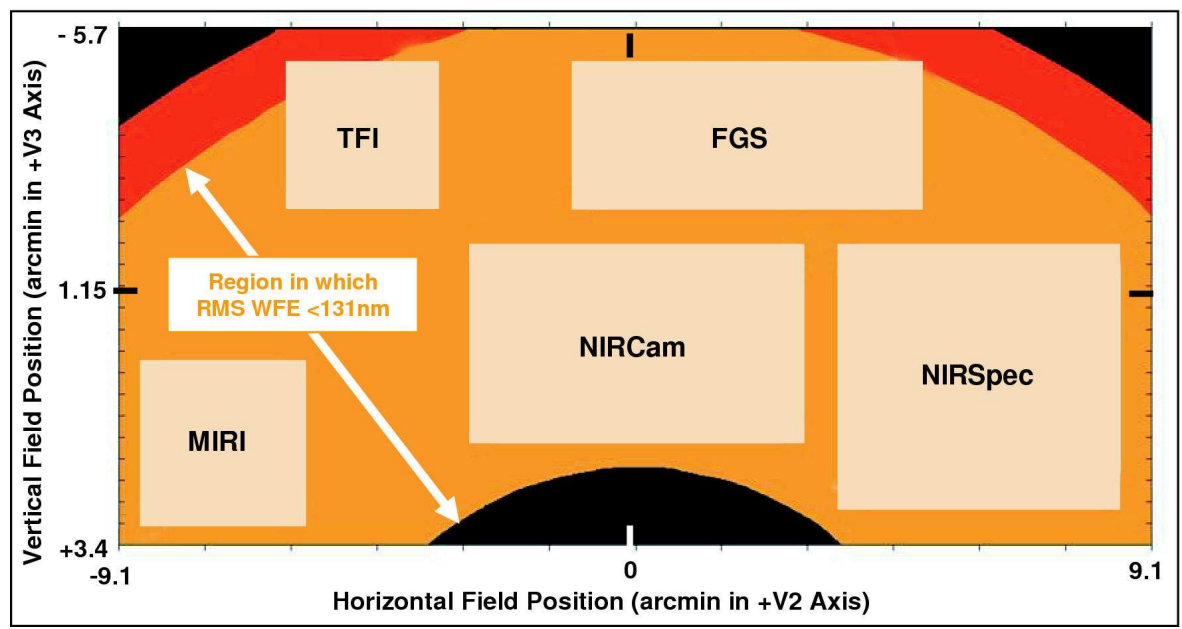}
\caption{
Placement of the ISIM instruments in the telescope field of view. 
}
\label{fig033}
\end{figure*}

Fig.~\ref{fig033} shows the placement of the individual ISIM
instrument detectors on the telescope's field of view. The telescope
design and control is optimized to equalize performance over the total FOV.
The wavefront error (WFE) that results from residuals in the optical
design is a small portion of the total telescope WFE. This allows
significant WFE allocation to the manufacturing processes and 
on-orbit environmental degradation, while still delivering 131 nm
rms WFE over the NIRCam field, as required for diffraction-limited
performance at 2 $\mu$m. The outer black area in the figure
represents portions of the field of view that have some amount of
vignetting.

The NIRCam detectors are placed in a spatial region 
with the lowest residual WFE to take full advantage of imaging 
performance. MIRI and NIRSpec are positioned in an area with 
slightly larger WFE, but still well within the requirements for 
instrument performance. Similar considerations exist for placement 
of the FGS and TFI focal planes within the telescope FOV. The telescope 
mirrors are gold coated, providing a broad spectral bandpass, from 
0.6 to 29 \ensuremath{\mu}m.

\subsubsection{Mirror segments}

\begin{figure*}
\centering
\includegraphics[width=1.00\textwidth]{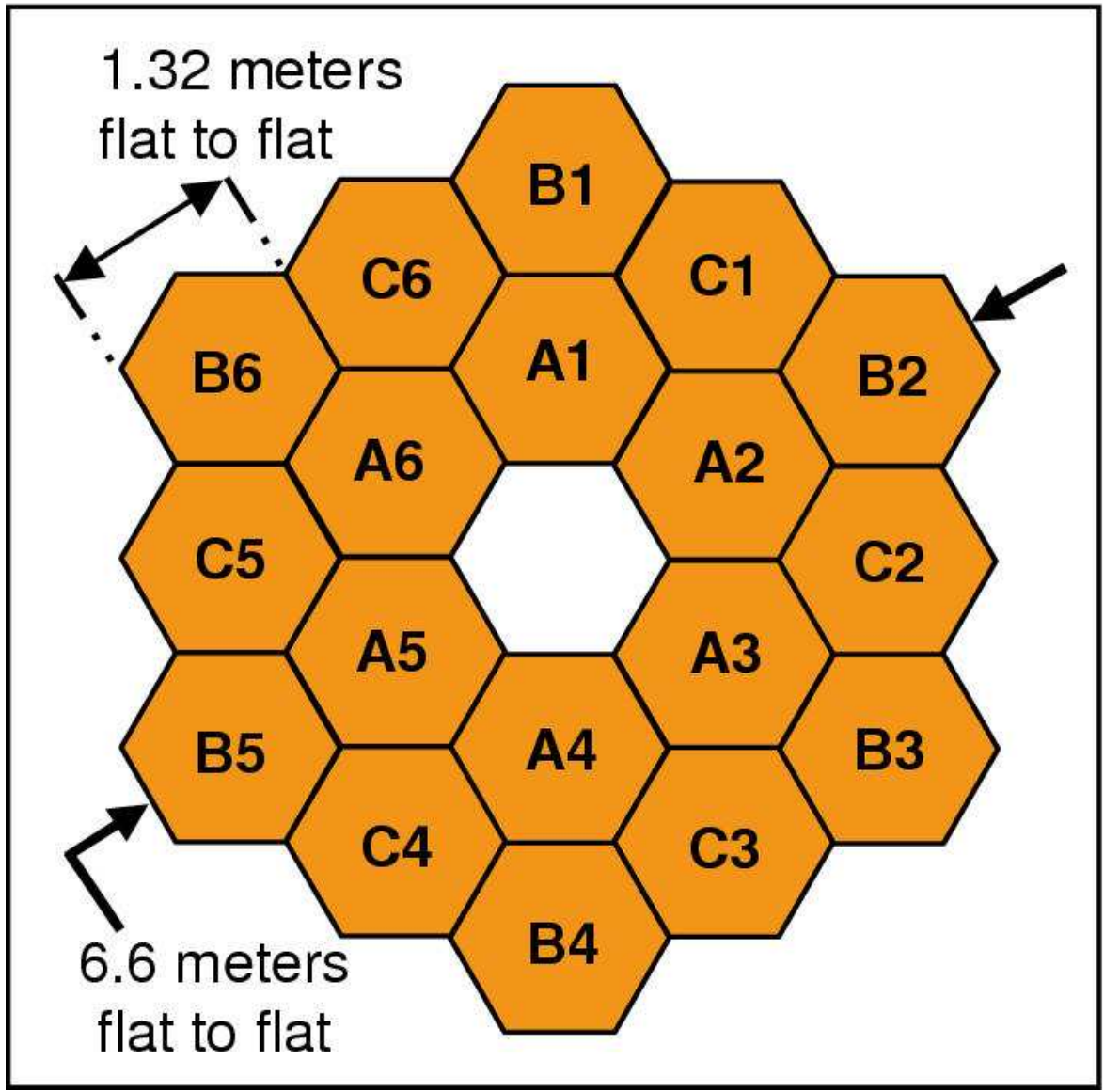}
\caption{
Primary mirror dimension and prescription. The primary mirror is made up of 
18 hexagonal segments, each 1.32 m flat to flat. There are three different
segment aspheric prescriptions, designated A, B and C, with six segments of each prescription.
The segments with the same prescription are interchangeable.}
\label{fig034}
\end{figure*}

The primary mirror is composed of 18 individual beryllium mirror segments 
(Fig.~\ref{fig034}). When properly phased relative to each other, these 
segments act as a single mirror. Primary mirror phasing is achieved 
via six degree of freedom rigid body motion of the individual
segments, and an additional control for the mirror segment radius
of curvature. The six degrees of freedom are decenter (x, y), tip,
tilt, piston (z), and clocking. Each mirror segment is hexagonal
with a 1.32 m flat-to-flat dimension. There are three separate
segment types with slightly different aspheric prescriptions
depending on placement. The three prescriptions are identified in
Fig.~\ref{fig034} by letters A, B, and C. Numbers 1 through 6
represent the six-fold symmetry of the hexagonal packing of the
primary mirror. All segments within a type (letter) are completely
interchangeable. The mid- and high-spatial-frequency figure errors
are caused by manufacturing errors within the segments. The low
frequency errors are corrected by the primary mirror adjustments
on orbit.

\begin{figure*}
\centering
\includegraphics[width=1.00\textwidth]{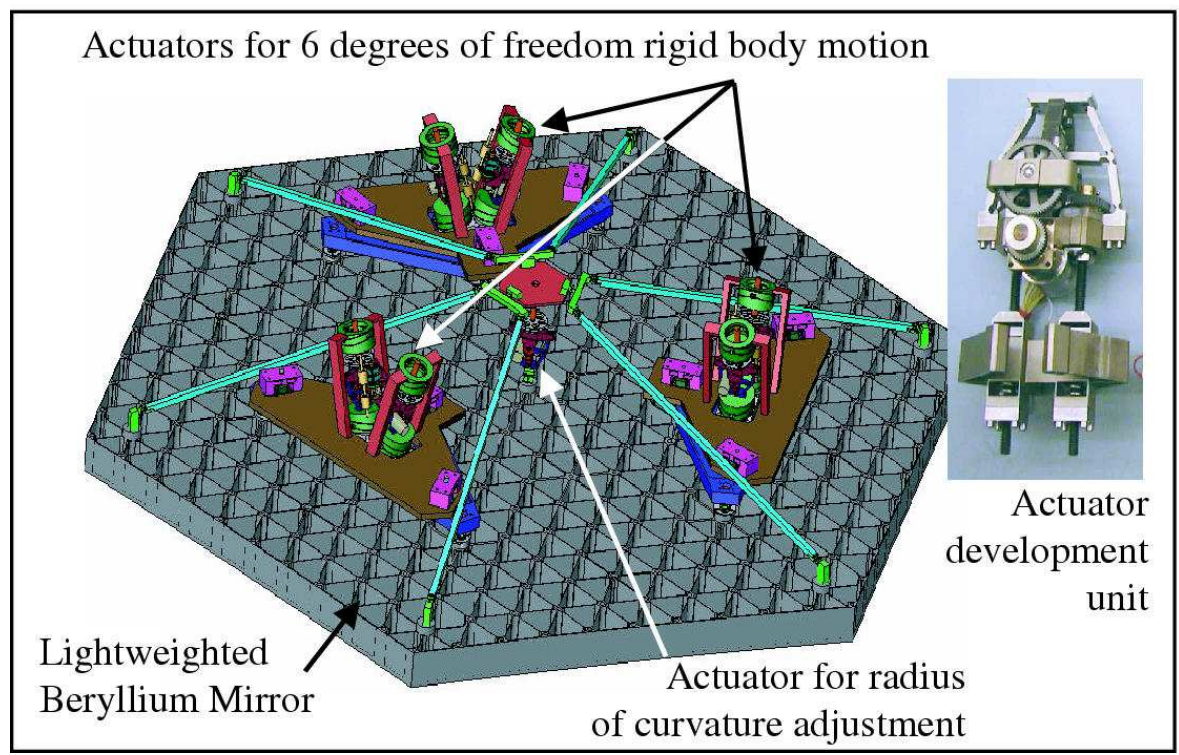}
\caption{
Rear view of a primary mirror segment. Each segment is supported by six actuators, providing
rigid body motion with six degrees of freedom. In addition, radius-of-curvature adjustment is
provided by a seventh actuator in the center of the segment.}
\label{fig035}
\end{figure*}

Fig.~\ref{fig035} shows the rear portion of the mirror segments
and the seven actuators. The six actuators providing rigid body
motion are arranged in three bipods (a hexapod) to form a kinematic
attachment to the backplane. Each bipod attaches to a triangular
structure shown in Fig.~\ref{fig035}, which is attached to the
isogrid structure of the mirror segment. This structure spreads
the loads over the surface of the mirror. The other end of each
actuator attaches through a secondary structure and flexure to the
backplane. The seventh actuator controls the segment radius of
curvature and is independent of the rigid body actuators and the
backplane structure. The actuators operate at cryogenic and ambient
temperatures, and have both coarse- and fine-positioning capability.
The semi-rigid segments have very low mirror figure changes due to
1-g effects, thus supporting high-fidelity ground testing and a
minimum number of actuators for each segment. The secondary mirror
has 6 actuators which provide 6 degree of freedom rigid body control,
although only 5 degree of freedom control is actually needed.

\subsubsection{Integrated Science Instrument Module}

The ISIM contains the science instruments for the observatory, 
and the support electronics for the science instruments: NIRCam,
NIRSpec, TFI, and MIRI (described in Section 6.5). A cryo-cooler
will be used for cooling MIRI and its Si:As detectors. The
near-infrared detector arrays in the other instruments are passively
cooled HgCdTe. In addition to the science instruments, the ISIM
contains the fine guidance sensor (FGS) and the computer that
directs the daily science observations based on plans received from
the ground. The science instruments and FGS have non-overlapping
FOVs as shown in Fig.~\ref{fig033}. Simultaneous operation of all
science instruments is possible; this capability will be used for
parallel calibration, including darks and possibly sky flats. FGS
is used for guide star acquisition and fine pointing. Its FOV and
sensitivity are sufficient to provide a greater than 95\% probability
of acquiring a guide star for any valid pointing direction and roll
angle.

\paragraph{Optical interface:}

The optical interface to the ISIM is a curved focal surface
located approximately 196 cm behind the primary mirror vertex. 
The ISIM is supported off the rear of the telescope backplane by four 
kinematic mounts (two bipods and two monopods.) This mounting 
configuration cleanly partitions optical and mechanical interfaces 
to simplify integration and test. During normal observations, 
the FGS provides the pointing error signal used by the telescope and 
spacecraft to maintain a stable and accurate optical line of sight.

\paragraph{Packaging interface:}

The Observatory architecture provides a 19.9 m$^{3}$ volume for 
the ISIM instruments, with 8 m$^{2}$ of radiator surface area for 
cooling the NIR detectors. The focal plane electronics boxes 
are mounted near the ISIM in a nominal room temperature environment 
to keep the cable line length below 6 meters to ensure cable 
noise is minimized. The ISIM computers are mounted within the warm spacecraft. 

\paragraph{Thermal interface:}

In order to achieve a 70\% observational efficiency, the science 
instruments that are not making science observations will 
make calibration exposures such as darks in parallel.
The supporting structure of the instrument payload has very 
tight optical alignment requirements which necessitate 
stable constant power dissipation. Therefore, the 
baseline operations concept calls for all of the instruments 
to be on all of the time (similar to HST). The ISIM thermal, 
power, and data handling systems are designed to support 
simultaneous operation of the four science instruments 
and the fine guidance sensor with appropriate margin. 

\paragraph{Wavefront sensing and control interface:}

The alignment and phasing of the telescope optics require data gathered 
by the NIRCam. For WFS\&C procedures, there will be six lenses, 
filters and grisms in the NIRCam filter 
wheels. The physical positioning of these optical elements is no more exacting 
than for science imaging through filters. During the WFS\&C process, the 
NIRCam will be operated in the same manner as that used for science 
(i.e., no special modes are necessary). In addition, NIRCam has a 
pupil imaging lens which can be used to image the primary mirror.

\subsubsection{Wavefront Sensing and Control Subsystem (WFS\&C)}

The ability to perform on-orbit alignment of the telescope is one of 
the enabling technologies that allow JWST to be built. 
The large telescope area is made up of 18 smaller segments 
that can be manufactured and tested more easily than larger ones. 
The WFS\&C subsystem aligns these segments so that their wavefronts 
match properly, creating a diffraction-limited 6.6-m telescope, rather 
than overlapping images from 18 individual 1.3 m telescopes. 

\begin{figure*}
\centering
\includegraphics[width=1.00\textwidth]{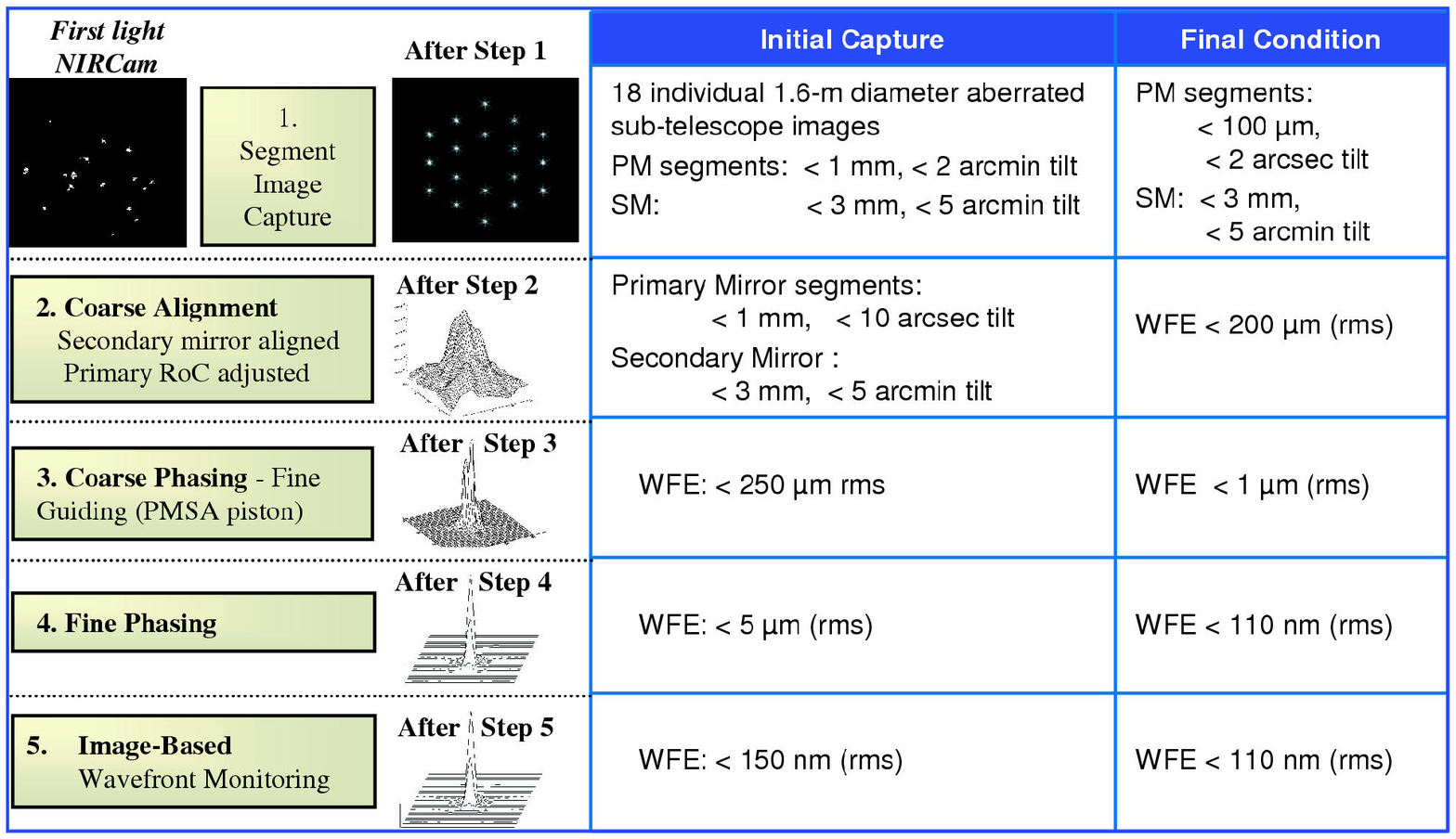}
\caption{
A summary of the WFS\&C commissioning and maintenance phases.}
\label{fig036}
\end{figure*}

The 1.3-m semi-rigid segments limit segment-level wavefront errors 
to those controllable through fabrication, radius-of-curvature 
adjustment, and rigid-body positioning. The WFS\&C does not require 
nor use any other deformation of optical-structural surfaces. Such 
deformation would cross-couple surface corrections and limit 
the testability of the optics in the ground 1-g environment. The 
semi-rigid mirror architecture supports the use of a set of deterministic 
WFS\&C algorithms that control the fundamental and critical 
initial telescope alignment operation (Acton et al. 2004). Another 
benefit is the use of a closed-form reconstructor algorithm to 
measure and correct segment-to-segment piston errors, a crucial 
initial operation in creating a phased primary mirror. Determination 
of the wavefront error and the necessary telescope mirror commands is 
done on the ground using the downlinked image data. 
The algorithm uses least-squares fits to images taken in and out of 
focus, at different wavelengths, and at multiple field points. The focus
is adjusted using weak lenses in the NIRCam filter wheel. This 
algorithm was used to diagnose and measure the spherical 
aberration of the HST primary mirror. Mirror adjustment 
commands are uplinked to the observatory. Fig.~\ref{fig036} summarizes 
the WFS\&C process. 

The initial telescope commissioning process occurs in four phases: (1) image capture 
and identification, (2) coarse alignment, (3) coarse phasing, 
and (4) fine phasing. Shi et al. (2004) analyzed the WFS\&C architecture 
and implemented hardware and software demonstrations of 
the more critical aspects of this architecture. Their analyses 
and simulations show that the residual errors from one step (final 
condition) are well within the capture range of the next step 
(initial condition.) The WFS\&C process is straightforward, has 
a singular deterministic solution, and will be fully tested on 
the ground.

To accomplish commissioning, the telescope is pointed at a celestial 
region with specific characteristics at the 2 \ensuremath{\mu}m NIRCam 
operation wavelength. A bright (magnitude 10 or brighter) isolated 
commissioning star is located on the NIRCam detector. A second star 
(guide star), with magnitude and position restrictions relative 
to the commissioning star, is used with the fine guiding sensor. There 
are many star fields that meet these criteria to ensure 
availability regardless of the JWST launch date (Green et al. 
2004).

\subsubsection{Sunshield}

The sunshield enables the passively-cooled thermal design. It 
reduces the $\sim$ 200 kW incident solar radiation that impinges 
on the sunshield to milliwatts incident on the telescope and ISIM. 
This solar attenuation is a result of the five-layer configuration 
of the sunshield. Its physical size and shape determine the celestial field of regard 
for the observatory. By reducing the solar radiation 
to the milliwatt level, the observatory has an intrinsically 
stable point spread function as it is pointed over its FOR. 
The orientation and angular separation of the 
individual layers can be seen in the artist's rendition of the 
JWST Observatory in Fig.~\ref{fig027}.

A sunshield with such a large surface area will result in angular momentum 
build up from solar radiation. The angle between the forward and 
aft sunshield was selected to minimize this 
momentum build up. The momentum will be dumped by firing the thrusters that
are used for orbit maintenance.

\subsubsection{Spacecraft Bus}

\begin{figure*}
\centering
\includegraphics[width=1.00\textwidth]{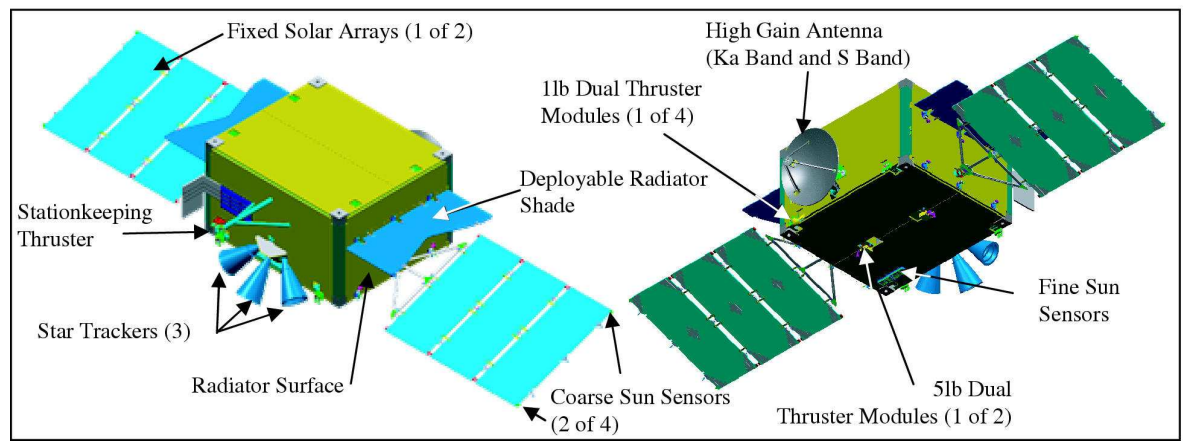}
\caption{
JWST spacecraft bus external configuration.}
\label{fig037}
\end{figure*}

The spacecraft (Fig.~\ref{fig037}) provides the housekeeping function 
of the Observatory. It has a 471 Gbit solid state recorder 
to hold the science and engineering data collected between 
and during the daily contacts with the ground station. The 
recorder can hold two days of collected data, providing redundancy 
against a missed contact. Three star trackers (including one 
for redundancy) are used to point the observatory toward the 
science target prior to guide star acquisition and to provide 
roll stability about the telescope line of sight. Six 
reaction wheels (including two for redundancy) are mounted on 
isolators near the center of gravity of the bus to reduce disturbances 
to the observatory. These reaction wheels offload the fine steering 
control (operating with a 16 Hz update rate from the FGS), maintain 
the fine steering mirror near a central position and limit blurring from
differential field distortion. Power is provided by two solar 
arrays that are canted toward the sun when the observatory is pointed 
in the middle of its FOR. The downlink operates 
at Ka band and has a selectable rate of 7, 14 or 28 Mbps. A pair 
of omni-directional antennas (at S band) provide nearly complete spherical 
coverage for emergency communications.

\subsection{Integration and Test}

Incremental verification will be used to integrate and test JWST 
in a way similar to that used for the Chandra X-ray Observatory. Performance parameters 
are verified incrementally as the observatory is integrated, stopping 
at the highest level of integration that adequately verifies 
the system performance. This will culminate in a final end-to-end 
cryogenic optical test of the observatory.

Once each mirror element, with its actuators installed, successfully
completes cryogenic tests, the mirror elements will be attached to
the backplane support structure and the telescope aligned (Wells
et al. 2004). The telescope and ISIM combination will be cryogenically
tested at Chamber A at NASA's Johnson Space Center and verified
prior to integration to the spacecraft and sunshield. This
telescope/ISIM cryogenic test will be the end-to-end optical
performance verification for the Observatory. The test chamber is
16.8 m diameter and 35.7 m high, giving adequate room to configure
cost-effective assembly, test, and verification equipment.

\begin{figure*}
\centering
\includegraphics[width=1.00\textwidth]{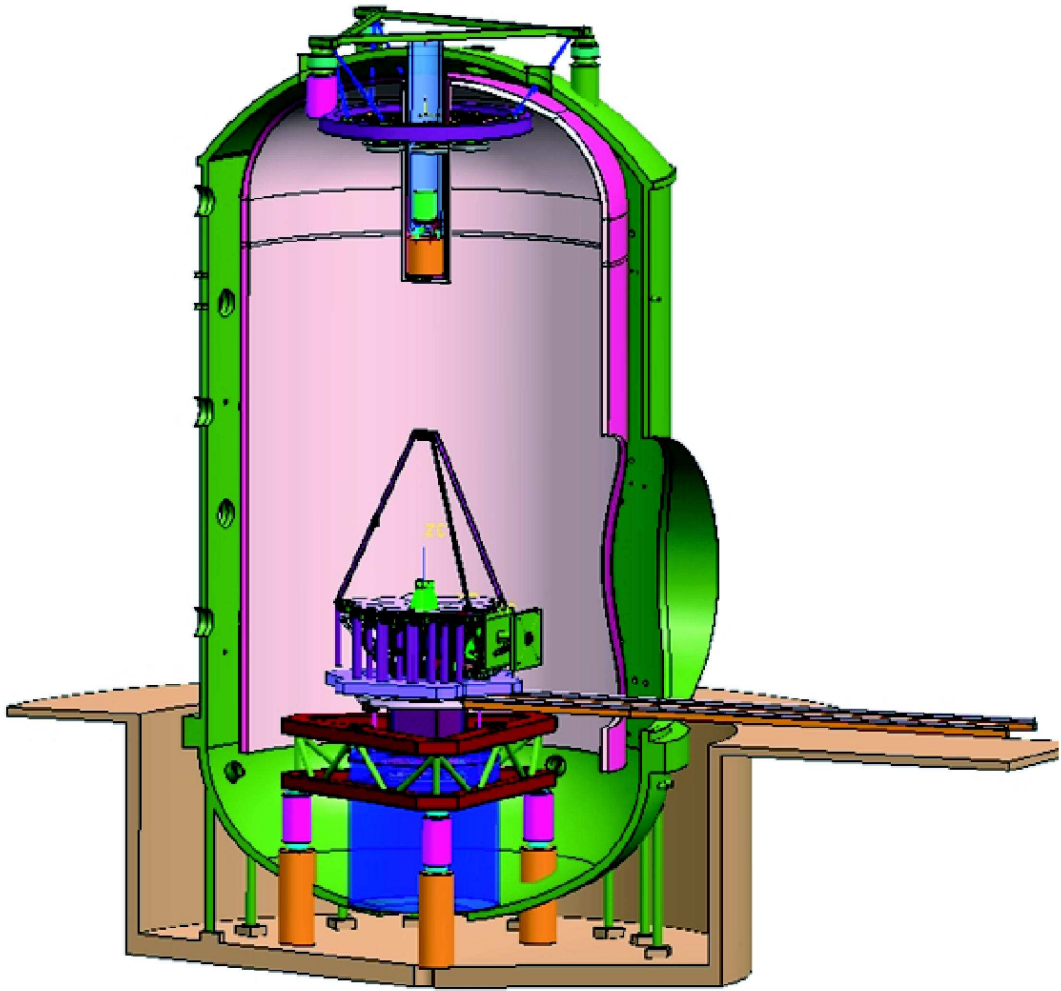}
\caption{
The JWST observatory pictured in the thermal vacuum test 
chamber at NASA's Johnson Space Center. The test chamber is 16.8 m diameter
and 35.7 m high. The full telescope will be tested at its center of curvature,
and a sampled full-aperture test which will test the wavefront from each segment.}
\label{fig038}
\end{figure*}

The telescope will be tested with the optical axis vertical (see Fig.~\ref{fig038}) 
to minimize gravity effects. Testing the optics in this orientation 
minimizes the moments induced in the mirrors at the actuator 
attachment interface. The completed primary mirror will be tested
interferometrically at its center of curvature. In addition, 
a sampled full-aperture test is performed 
where each segment is sampled in a single wavefront test. The 
semi-rigid hexagonal primary mirror architecture allows use of 
six 1.5-m flats rather than an expensive and high risk 7-m flat 
or collimator. Another benefit of the sparsely sampled 
approach is that the autocollimating flats need not be phased 
to each other. The test is operated using individual triads of 
three primary mirror segments, allowing the mirror rigid body 
actuation to correct for phasing of the flats. Only stability 
of the flats is required. 

The sunshield deployment will be tested at room temperature.
A scale version of the sunshield will be tested in the deployed 
configuration with a helium-cooled shroud. By performing a test 
of the sunshield with a helium-cooled shroud background, we will 
validate the thermal model of the sunshield at operational temperatures. 

\subsection{Instrumentation}

Table~\ref{tab008} lists the main characteristics of JWST's science
instruments. The sensitivities of the instruments are listed in
Table~\ref{tab009}. Although stated as the sensitivity achievable
for 10,000 s exposures, it is expected that cosmic ray hits will
limit the maximum exposure time for an individual integration to
about 1000 s, and that longer total exposure times will be achieved
through co-adding. Based on experience with Hubble data, for example 
in the Hubble Deep Field, we expect the errors to scale as the square root of the 
exposure time in co-adds as long as 10$^5$ or even 10$^6$ s. 
The absolute photometric accuracy is expected
to be 5\% for imaging and 10 to 15\% for coronagraphy and spectroscopy,
based on calibration observations of standard stars.

\begin{table}[t]
\caption{Science Instrument Characteristics\label{tab008}}
\begin{tabular}{lcccc}
\hline\noalign{\smallskip}
{Instrument} &
{Wavelength} &
{Detector} &
{Plate Scale} &
{Field of View}\\
&
{(\ensuremath{\mu}m)}&
\multicolumn{3}{c}
{(milliarcsec/pixel)}
\\[3pt]
\tableheadseprule\noalign{\smallskip}
NIRCam&
0.6 - 2.3&
Eight&
32&
2.2\ensuremath{\times}4.4 arcmin\\
\hspace{0.25in}short&
&
2048\ensuremath{\times}2048&
\\
&&&\\
\hspace{0.25in}long$^{\rm a}$&
2.4 - 5.0&
Two
&
65&
2.2\ensuremath{\times}4.4 arcmin\\
&
2048\ensuremath{\times}2048&
\\
&&&\\
NIRSpec&
0.6 - 5.0&
Two&
100&
\\
\hspace{0.25in}MSA$^{\rm b}$&
&
2048\ensuremath{\times}2048&
&
3.4\ensuremath{\times}3.1 arcmin\\
\hspace{0.25in}slits$^{\rm c}$&
&
&&
$\sim$0.2$\times$4 arcsec\\
\hspace{0.25in}IFU&
&
&&
3.0$\times$3.0 arcsec\\
&&&\\
MIRI&
5.0 - 29.0&
1024\ensuremath{\times}1024&
110&
\\
\hspace{0.25in}imaging&
&
&&
1.4\ensuremath{\times}1.9 arcmin\\
\hspace{0.25in}coronagraphy&
&
&&
26$\times$26 arcsec\\
\hspace{0.25in}spectra$^{\rm d}$&
5.0 - 10.0&
&&
0.2$\times$5 arcsec\\
\hspace{0.25in}IFU&
5.0 - 29.0&
Two&
200 to 470&
3.6\ensuremath{\times}3.6 
\\
&
&
1024\ensuremath{\times}1024&
&
to 7.5\ensuremath{\times}7.5 arcsec
\\
&&&\\
TFI&
1.6 - 4.9$^{\rm e}$&
2048\ensuremath{\times}2048&
65&
2.2\ensuremath{\times}2.2 arcmin\\
&
&
&&
\\
\noalign{\smallskip}\hline
\end{tabular}

NOTE --- (a) Use of a dichroic renders the NIRCam long-wavelength 
field of view co-spatial with the short wavelength channel, and 
the two channels acquire data simultaneously. (b) NIRSpec includes a 
micro-shutter assembly (MSA) with four 365 $\times$ 171 micro-shutter arrays. The
individual shutters are each 203 (spectral) $\times$ 463 (spatial) milliarcsec clear
aperture on a 267 $\times$ 528 milliarcsec pitch.
(c) NIRSpec also includes several fixed slits which provide redundancy and 
high contrast spectroscopy on individual targets, and an integral field unit (IFU).
(d) MIRI includes a fixed slit for low-resolution (R$\sim$100) spectroscopy over the 5 to
10 $\mu$m range, and an integral field unit for R$\sim$3000 spectroscopy over the full 5
to 29.0 $\mu$m range. The long wavelength cut-off for MIRI spectroscopy is set by the
detector performance, which drops longward of 28.0 $\mu$m. (e) The wavelength range for the 
TFI is 1.6 to 2.6 $\mu$m and 3.1 to 4.9 $\mu$m. There is no sensitivity between 2.6 and 
3.1 $\mu$m.

\end{table}

\begin{table}[t]
\caption{Instrument Sensitivities\label{tab009}}
\begin{tabular}{lllll}
\hline\noalign{\smallskip}
{Instrument/Mode} &
{$\lambda$ ($\mu$m)} &
{Bandwidth} &
{Sensitivity}
\\[3pt]
\tableheadseprule\noalign{\smallskip}
NIRCam&
2.0&
R=4&
11.4 nJy, AB=28.8
\\
TFI&
3.5&
R=100&
126 nJy, AB=26.1
\\
NIRSpec/Low Res&
3.0&
R=100&
132 nJy, AB=26.1
\\
NIRspec/Med Res&
2.0&
R=1000&
1.64 $\times$ 10$^{-18}$ erg s$^{-1}$ cm$^{-2}$
\\
MIRI/Broad-Band&
10.0&
R=5&
700 nJy, AB=24.3
\\
MIRI/Broad-Band&
21.0&
R=4.2&
8.7 \ensuremath{\mu}Jy, AB=21.6
\\
MIRI/Spect.&
9.2&
R=2400&
1.0 $\times$ 10$^{-}$$^{17}$ erg s$^{-1}$ cm$^{-2}$
\\
MIRI/Spect.&
22.5&
R=1200&28.8
5.6 $\times$ 10$^{-}$$^{17}$ erg s$^{-1}$ cm$^{-2}$
\\
\noalign{\smallskip}\hline
\end{tabular}

NOTE --- Sensitivity is defined to be the brightness of
a point source detected at 10 $\sigma$ in 10000 s. Longer or shorter
exposures are expected to scale approximately as the square root
of the exposure time. Targets at the North Ecliptic Pole are
assumed. The sensitivities in this table represent the best estimate
at the time of submission and are subject to change.

\end{table}

\subsubsection{Near-Infrared Camera}

\begin{figure*}
\centering
\includegraphics[width=1.00\textwidth]{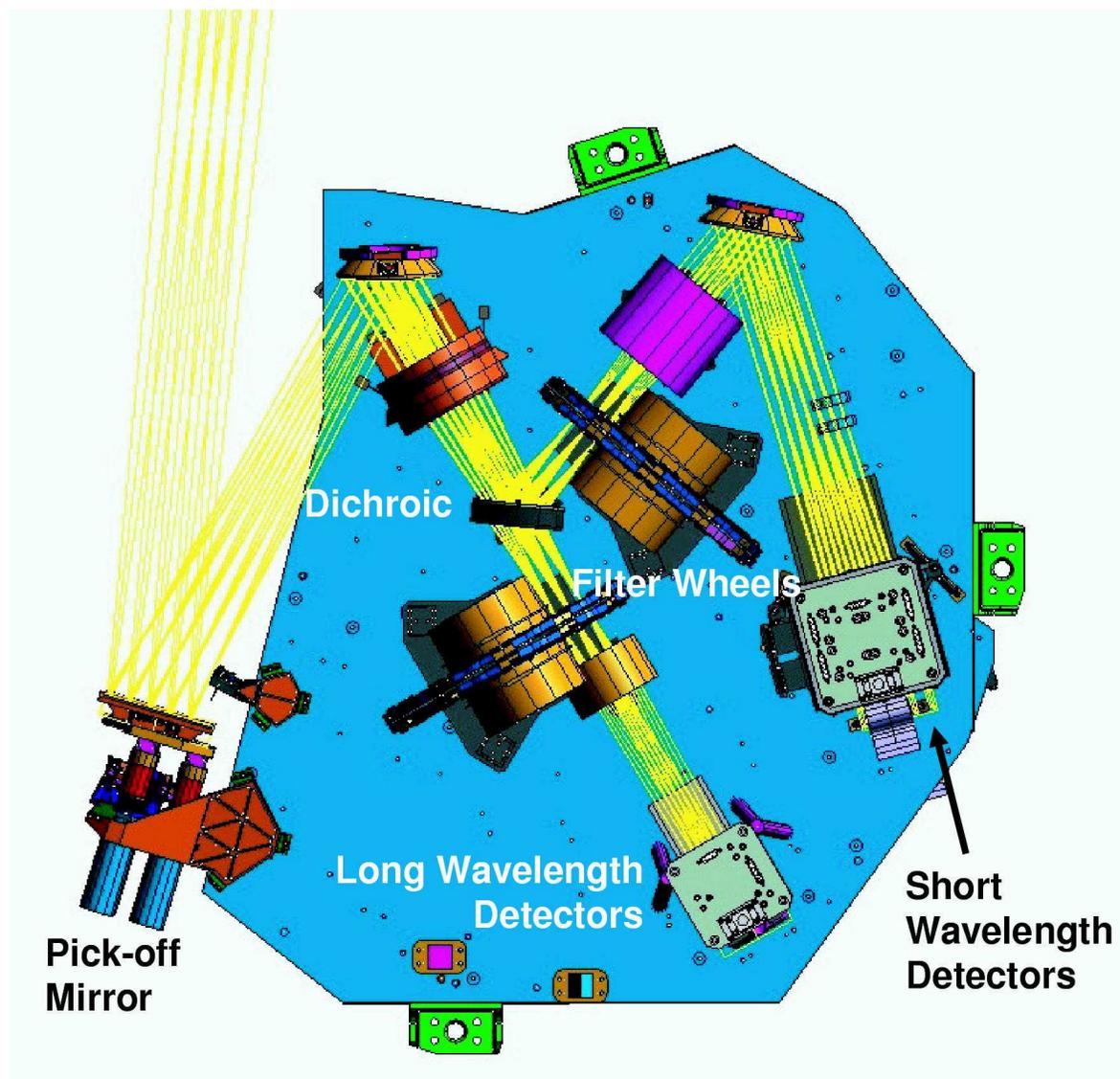}
\caption{
Optical layout of one of two NIRCam imaging modules. A dichroic directs the light
into short and long wavelength detectors. There are two identical imaging modules, providing
redundancy and a large field of view.}
\label{fig039}
\end{figure*}

NIRCam provides filter imaging in the 0.6 to 5.0 \ensuremath{\mu}m
range with wavelength multiplexing. It includes the ability to
sense the wavefront errors of the observatory. NIRCam consists of
an imaging assembly within an enclosure that is mounted in the
ISIM. The imaging assembly consists of two fully redundant, identical
optical trains mounted on two beryllium benches, one of which is
shown in Fig.~\ref{fig039}. The incoming light initially reflects
off the pick-off mirror. Subsequently it passes through the collimator
and the dichroic, which is used to split the light into the short
(0.6 to 2.3 \ensuremath{\mu}m) and long (2.4 to 5.0 \ensuremath{\mu}m.)
wavelength light paths. Each of these two beams then passes through
a pupil wheel and filter wheel combination, each beam having its
own pair of pupil and filter wheel. After this, the light passes
through the camera corrector optics and is imaged (after reflecting
off a fold flat in the short wavelength beam) onto the detectors.

\begin{table}[t]
\caption{Preliminary list of NIRCam filters and pupils\label{tab010}}
\begin{tabular}{cccc}
\hline\noalign{\smallskip}
\multicolumn{2}{c}{Short wavelength channel}\\
{Filter Wheel} & 
{Pupil Wheel}
\\[3pt]
\tableheadseprule\noalign{\smallskip}
Broad-band 0.7 $\mu$m&
Imaging pupil
\\
Broad-band 0.9 $\mu$m&
TBD
\\
Broad-band 1.15 $\mu$m&
Flat field pinholes
\\
Broad-band 1.5 $\mu$m&
10\% H$_2$O 1.62$\mu$m
\\
Broad-band 2.0 $\mu$m&
10\% H$_2$O 1.40$\mu$m
\\
10\% CH$_4$ 2.10$\mu$m&
WFS DHS 1
\\
1\% H$_2$ 2.12$\mu$m&
WFS weak lens 1
\\
WFS 1.15 to 2.3$\mu$m&
WFS weak lens 2
\\
WFS weak lens 3&
Outward pinholes
\\
1\% H$_2$ 2.25$\mu$m/1\% [FeII] 1.644$\mu$m$^{\rm a}$&
Coronagraphic pupil 1 with wedge
\\
10\% H$_2$O 1.82$\mu$m&
Coronagraphic pupil 2 with wedge
\\
1\% P$\alpha$ 1.875 $\mu$m&
WFS DHS 2
\\
\hline\noalign{\smallskip}
\multicolumn{2}{c}{Long wavelength channel}\\
{Filter Wheel} & 
{Pupil Wheel}
\\[3pt]
\tableheadseprule\noalign{\smallskip}
Broad-band 2.7 $\mu$m&
Imaging pupil
\\
Broad-band 3.6 $\mu$m&
Flat field pinholes
\\
Broad-band 4.4 $\mu$m&
1\% H$_2$ 3.23$\mu$m
\\
10\% H$_2$O 3.00$\mu$m&
1\% H$_2$ 4.18$\mu$m
\\
10\% PAH 3.35$\mu$m&
1\% CO 4.60$\mu$m
\\
10\% 3.60$\mu$m&
1\% H$_2$ 4.69$\mu$m
\\
10\% 3.90$\mu$m&
TBD
\\
10\% CO$_2$ 4.30$\mu$m&
Outward pinholes
\\
10\% CO 4.60$\mu$m&
WFS grism 1
\\
10\% 4.80$\mu$m&
WFS grism 2
\\
1\% Br$\alpha$ 4.05$\mu$m&
Coronagraphic pupil 1
\\
10\% CH$_4$ 2.50$\mu$m&
Coronagraphic pupil 2
\\
\noalign{\smallskip}\hline
\end{tabular}

{NOTE --- WFS is wavefront sensing optics, DHS is dispersed Hartmann sensing optics, PAH is polycyclic aromatic hydrocarbon,
TBD is to be determined. (a) This position will have two different filters in the two channels of NIRCam, resulting in half the field of view
in each filter.}

\end{table}

\begin{figure*}
\centering
\includegraphics[width=1.00\textwidth]{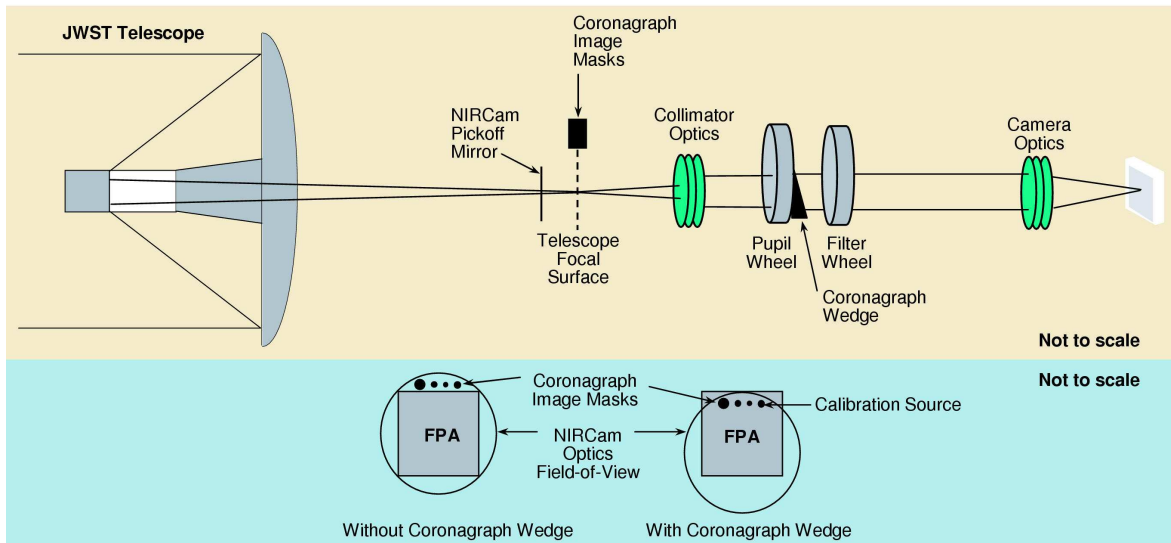}
\caption{
Schematic of NIRCam coronagraph design. An optical wedge in the pupil wheel brings the 
coronagraphic spots into the field of view. The spots are matched with Lyot stops.}
\label{fig040}
\end{figure*}

The instrument contains a total of ten 2k\ensuremath{\times}2k
detector chips, including those in the identical redundant optical
trains. The short wavelength arm in each optical train contains a
2\ensuremath{\times}2 mosaic of these detectors, optimized for the 0.6
- 2.3 \ensuremath{\mu}m wavelength range, with a small gap
(\ensuremath{\sim}3 mm = \ensuremath{\sim}5 arcsec) between adjacent
detectors. The detectors arrays are HgCdTe of HAWAII II heritage built
by Rockwell Science Center. The detectors will all have thinned
substrates to avoid cosmic ray scintillation issues, as well as to
extend their sensitivity below 0.85 \ensuremath{\mu}m.

\begin{figure*}
\centering
\includegraphics[width=1.00\textwidth]{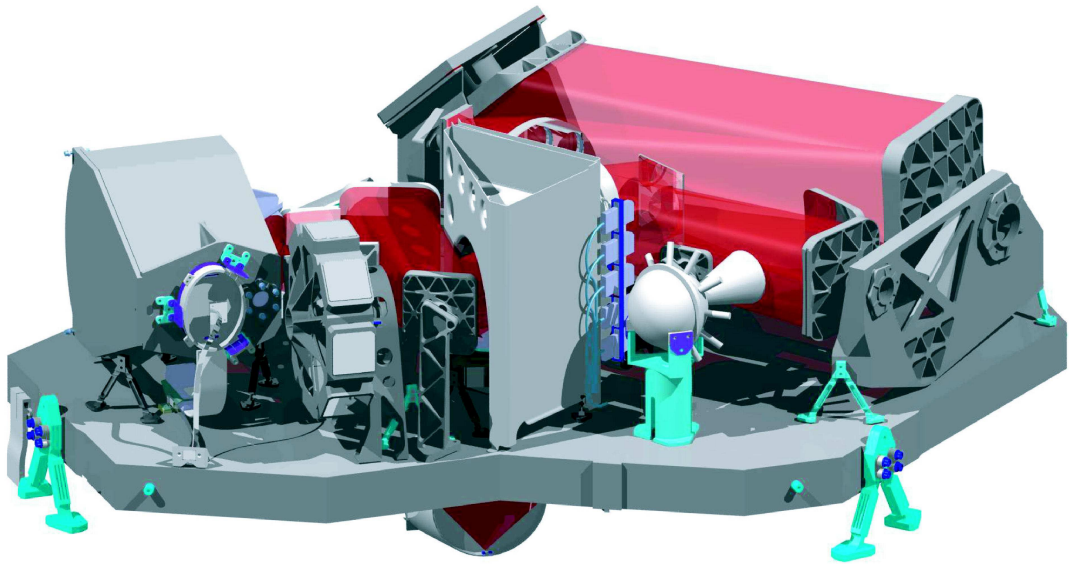}
\caption{
The NIRSpec instrument.}
\label{fig041}
\end{figure*}

Each optical train contains a dual filter and pupil wheel, containing 
a range of wide-, medium- and narrow-band filters and the WFS\&C 
optics. Each wheel has 
12 slots. The preliminary filter selection is given in Table~\ref{tab010}. 

\paragraph{Coronagraphy:}

To enable the coronagraphic imaging, each of the two identical 
optical trains in the instrument also contains a traditional 
focal plane coronagraphic mask plate held at a fixed distance 
from the detectors, so that the coronagraph spots are always in focus 
at the detector plane. Each coronagraphic plate is transmissive, 
and contains a series of spots of different sizes, including linear
and radial-sinc occulters, to block the 
light from a bright object. Each coronagraphic plate also includes 
a neutral density spot to enable centroiding on bright stars, 
as well as calibration sources at each end that can send light through 
the optical train of the imager to enable internal alignment 
checks. Normally these coronagraphic plates are not in the optical 
path for the instrument, but they are selected by rotating into 
the beam a mild optical wedge that is mounted in the pupil wheel 
(see Fig.~\ref{fig040}), which translates the image plane so that the 
coronagraphic masks are shifted onto the active detector area. 
Diffraction can also be reduced by apodization at the pupil mask, 
thus the pupil wheels will be equipped with both a classical 
and an apodized pupil with integral wedges in each case.
Current models predict a contrast of $\sim 10^4$ at 0.5 arcsec, 
at a wavelength of 4.6 $\mu$m.

\subsubsection{Near-Infrared Spectrograph}

NIRSpec is a near infrared multi-object dispersive spectrograph
capable of simultaneously observing more than 100 sources over a
field-of-view (FOV) larger than 3 \ensuremath{\times} 3 arcmin. In
addition to the multi-object capability, it includes fixed slits
and an integral field unit for imaging spectroscopy. Six gratings
will yield resolving powers of R \ensuremath{\sim} 1000 and
\ensuremath{\sim} 2700 in three spectral bands, spanning the range
1.0 to 5.0 \ensuremath{\mu}m. A single prism will yield R
\ensuremath{\sim} 100 over 0.6 to 5.0 \ensuremath{\mu}m.
Fig.~\ref{fig041} shows a layout of the instrument.

The region of sky to be observed is transferred from the JWST 
telescope to the spectrograph aperture focal plane by a pick-off 
mirror and a system of fore-optics that includes a filter 
wheel for selecting bandpasses and introducing internal calibration 
sources.

Targets in the FOV are normally selected by opening groups of
shutters in a micro-shutter assembly (MSA) to form multiple apertures.
The micro-shutter assembly itself consists of a mosaic of 4 subunits
producing a final array of 730 (spectral) by 342
(spatial) individually addressable shutters with 203 \ensuremath{\times}
463 milliarcsec openings and 267 \ensuremath{\times} 528 milliarcsec
pitch. Sweeping a magnet across the surface of the micro shutters
opens all of the shutters. Individual shutters may then be
addressed and released electronically, and the return path of the
magnet closes the released shutters. The minimum aperture size is
1 shutter (spectral) by 1 shutter (spatial) at all wavelengths.
Multiple pointings may be required to avoid placing targets near
the edge of a shutter and to observe targets with spectra that
would overlap if observed simultaneously at the requested roll
angle. The nominal slit length is 3 shutters in all wavebands. In
the open configuration, a shutter passes light from the fore-optics
to the collimator. A slitless mode can be configured by opening
all of the micro shutters. As the shutters are individually
addressable, long slits, diagonal slits, Hadamard-transform masks
(Riesenberg \& Dillner 1999), and other patterns can also be
configured with them.

In addition to the slits defined by the micro-shutter assembly,
NIRSpec also includes five fixed slits that can be used for high-contrast
spectroscopy. They are placed in a central strip of the aperture
focal plane between sub-units of the micro-shutter assembly and
also provide redundancy in case the micro-shutters fail. Three
fixed slits are 3.5 arcsec long and 200 milliarcsec wide. One fixed
slit is 4 arcsec long and 400 milliarcsec wide for increased
throughput at the expense of spectral resolution. One fixed slit
is 2 arcsec long and 100 milliarcsec wide for brighter targets.

The strip between micro-shutter sub-units also contains the 3 by
3 arcsec entrance aperture for an integral field unit (IFU). The
IFU has 30 slices, each 100 milliarcsec wide. The IFU relay optics
introduce a 2:1 anamorphic magnification of each slicer such that
the matching projected virtual slits are properly sampled on the
detector by two 50 milliarcsec pixels in the dispersion direction
and at the (nominal) 100 milliarcsec per pixel in the spatial
direction.

\begin{figure*}
\centering
\includegraphics[width=1.00\textwidth]{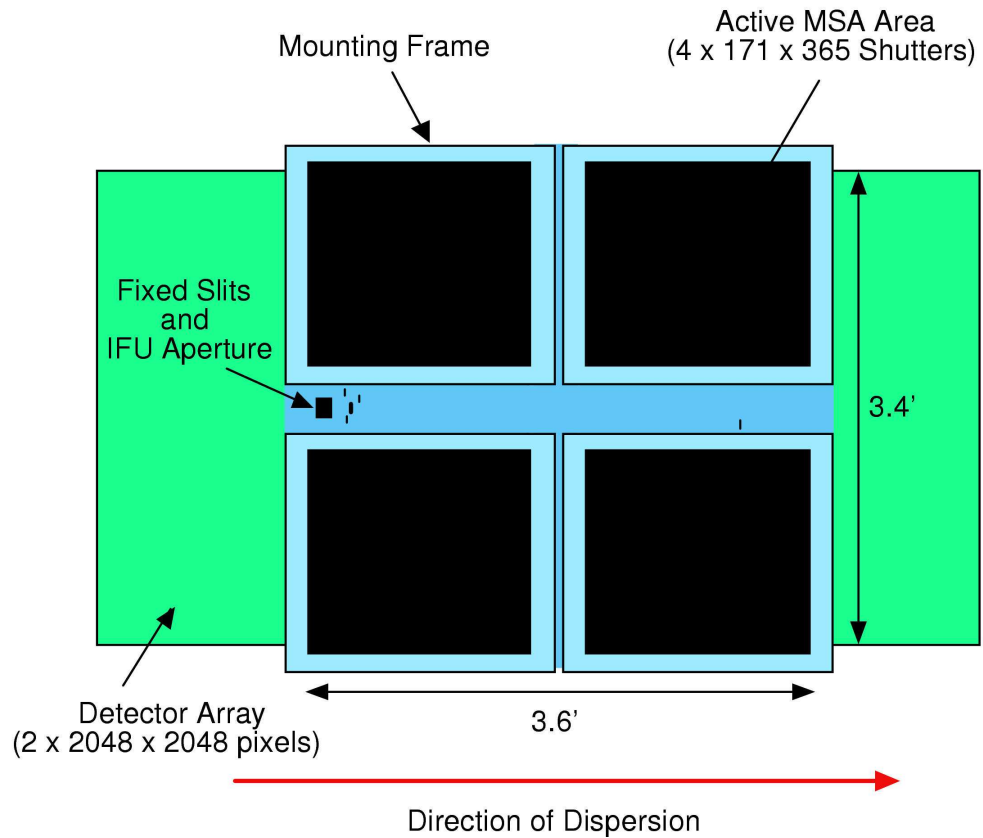}
\caption{
Schematic layout of the NIRSpec slit mask overlaid on the detector
array and projected to the same angular scale.}
\label{fig042}
\end{figure*}

The aperture focal plane is re-imaged onto a mosaic of two NIR
detectors by a collimator, a dispersing element (gratings or a
double-pass prism) or an imaging mirror, and a camera. Three gratings
are used for first-order coverage of the three NIRSpec wavebands
at R \ensuremath{\sim} 1000 (1.0 to 1.8 \ensuremath{\mu}m; 1.7 to
3.0 \ensuremath{\mu}m; 2.9 to 5.0 \ensuremath{\mu}m). The same
three wavebands are also covered by first-order R \ensuremath{\sim}
2700 gratings for objects in a fixed slit or in the IFU. The prism
gives R \ensuremath{\sim} 100 resolution over the entire NIRSpec
bandpass (0.6 to 5 \ensuremath{\mu}m) but can optionally be
blocked below 1 \ensuremath{\mu}m with one of the filters. Any of
the aperture selection devices (micro-shutter assembly, fixed slits or IFU)
can be used at any spectral resolution.

The focal plane array is a mosaic of two detectors (see Fig.~\ref{fig042}), 
each 2k \ensuremath{\times} 2k, 
forming an array of 2k \ensuremath{\times} 4k 100 milliarcsec pixels. The 
detectors will be thinned HgCdTe arrays built by Rockwell Science 
Center. NIRSpec contains a calibration unit with a number 
of continuum and line sources.

\subsubsection{Mid-Infrared Instrument}

\begin{figure*}
\centering
\includegraphics[width=1.00\textwidth]{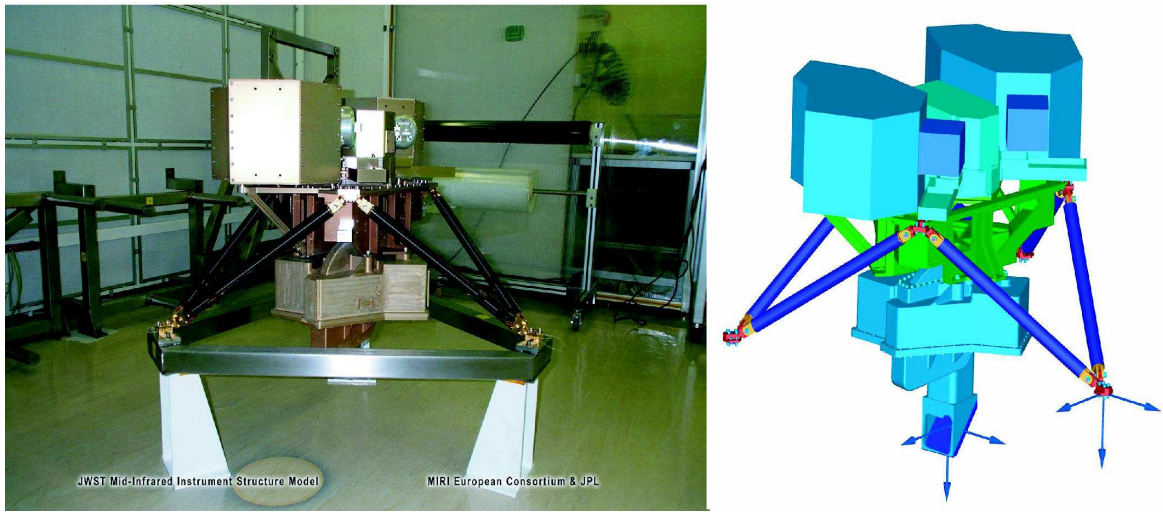}
\caption{
The MIRI structural and thermal model (left) compared to a
computer design of the instrument (right).}
\label{fig043}
\end{figure*}

The Mid-Infrared Instrument (MIRI) on JWST provides imaging and
spectroscopic measurements over the wavelength range 5 to 29
\ensuremath{\mu}m. MIRI consists of an optical bench assembly
(Fig.~\ref{fig043}) with associated instrument control electronics,
actively cooled detector modules with associated focal plane
electronics, and a cryo-cooler with associated control electronics.
The cryo-cooler electronics interface with the spacecraft command
and telemetry processor, while the instrument control electronics 
interface with the ISIM command and data handling. The 
optical bench assembly contains
two actively cooled subcomponents, an imager and an Integral Field
Unit (IFU) spectrograph, plus an on-board calibration unit.

\paragraph{Imaging:}

\begin{figure*}
\centering
\includegraphics[width=1.00\textwidth]{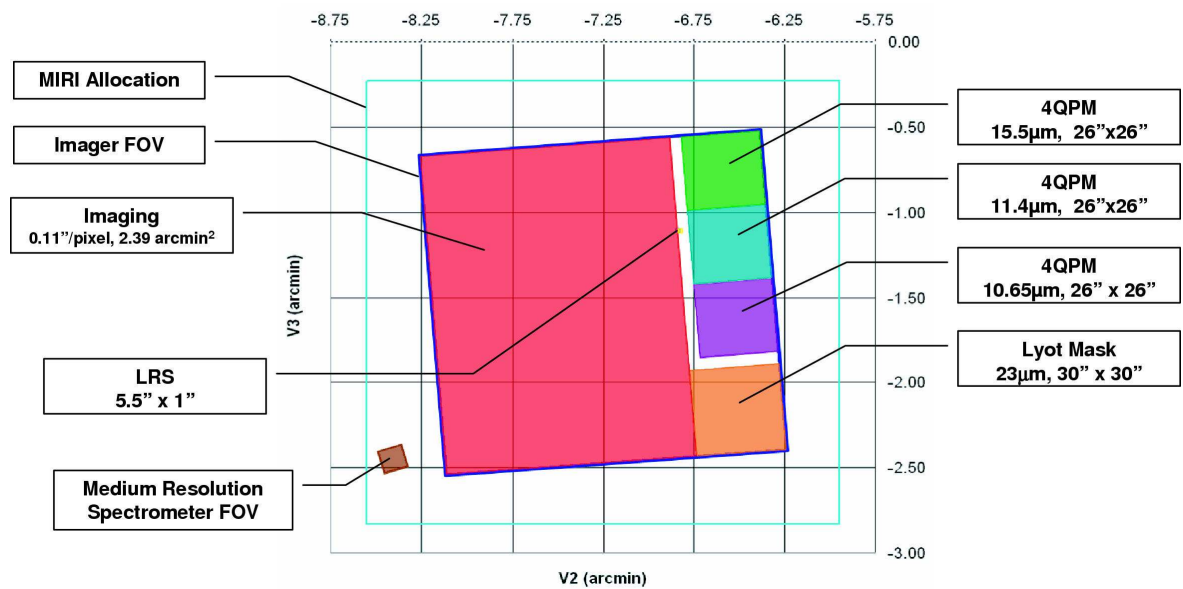}
\caption{
MIRI's imaging focal plane. One quarter of the field of view is devoted to
coronagraphy, there is a 5 arcsec wide low-resolution spectroscopy strip, and 
the remaining 1.4 $\times$ 1.9 arcmin$^2$ is available for broad-band imaging.}
\label{fig044}
\end{figure*}

The imager module provides broad-band imaging (see Table~\ref{tab011}
for preliminary filter selection), coronagraphy, and low-resolution
(R \ensuremath{\sim} 100, 5 to 10 \ensuremath{\mu}m) slit spectroscopy
using a single 1024\ensuremath{\times}1024 pixel Raytheon Si:As
detector with 25 \ensuremath{\mu}m pixels. Fig.~\ref{fig044} shows
the focal plane arrangement of the elements of the MIRI imager.
The region on the left is the clear aperture available for imaging.
The gray region on the right marks the mechanical frame that supports
the coronagraphic masks and the slit for the low-resolution
spectrometer. The coronagraphic masks include three phase masks
for a quadrant-phase coronagraph and one opaque spot for a Lyot
coronagraph. The coronagraphic masks each have a square field of
view of 26 \ensuremath{\times} 26 arcsec and are optimized for
particular wavelengths. The imager's only moving part is an
18-position filter wheel. Filter positions are allocated as follows:
12 filters for imaging, 4 filter and diaphragm combinations for
coronagraphy, 1 ZnS-Ge double prism for the low-resolution
spectroscopic mode and 1 dark position. The imager will have a
pixel scale of 0.11 arcsec/pixel and a total field of view of 113
\ensuremath{\times} 113 arcsec; however, the field of view of its
clear aperture is 84 \ensuremath{\times} 113 arcsec because the
coronagraph masks and the low-resolution spectrograph are fixed on
one side of the focal plane.

\begin{table}[t]
\caption{MIRI filters\label{tab011}}
\begin{tabular}{lccc}
\hline\noalign{\smallskip}
& 
{$\lambda_0$ ($\mu$m)} &
{$\Delta\lambda$ ($\mu$m)} &
{comment}
\\[3pt]
\tableheadseprule\noalign{\smallskip}
B1&5.6&1.2&Broad band\\
B2&7.7&2.2&PAH, broad band\\
B3&10&2&Silicate, broad band\\
I1&11.3&0.7&PAH, broad band\\
I2&12.8&2.4&Broad band\\
B4&15&3&Broad band\\
I3&18&3&Silicate, broad band\\
B5&21&5&Broad band\\
B6&25.5&\ensuremath{\sim}4&Broad band\\
B6'&25.5&\ensuremath{\sim}4&Redundant filter, risk reduction\\
ND\#&neutral density&&Coronagraphic acquisition\\
NIR&near-IR, TBD&&Testing \\
&blackened blank& N/A&Darks
\\
\noalign{\smallskip}\hline
\end{tabular}

\end{table}

\paragraph{Integral-Field Spectroscopy:}

\begin{figure*}
\centering
\includegraphics[width=1.00\textwidth]{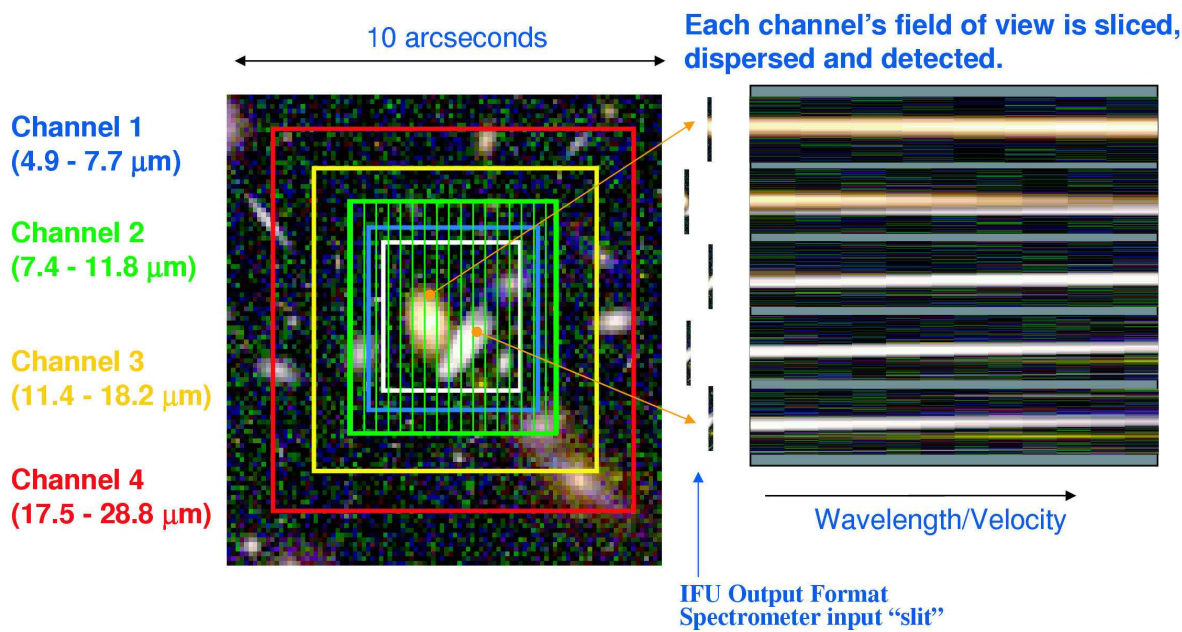}
\caption{
Schematic illustration of the MIRI IFU image slicer format (left)
and dispersed spectra on detector (right).}
\label{fig045}
\end{figure*}

The integral-field spectrograph obtains simultaneous spectral and
spatial data on a small region of sky. The spectrograph field of
view is next to that of the imager so that accurate positioning of
targets will be possible by locating the image with the imager
channel and off-setting to the spectrograph. The light is divided
into four spectral ranges by dichroics, and two of these ranges
are imaged onto each of two detector arrays. A full spectrum is
obtained by taking exposures at each of three settings of the
grating wheel. The spectrograph uses four image slicers to produce
dispersed images of the sky on two 1024 \ensuremath{\times} 1024
detectors, providing R \ensuremath{\sim} 3000 integral-field
spectroscopy over the full 5 to 29 \ensuremath{\mu}m wavelength range, 
although the sensitivity of the detectors drops 
longward of 28 \ensuremath{\mu}m. As shown in
Fig.~\ref{fig045} and Table~\ref{tab012}, the IFUs provide four
simultaneous and concentric fields of view. The slice widths set
the angular resolution perpendicular to the slice. The pixel sizes
set the angular resolution along the slice.

\begin{table}[t]
\caption{MIRI integral field spectroscopy parameters\label{tab012}}
\begin{tabular}{cccccc}
\hline\noalign{\smallskip}
{Wavelength} &
{Pixel Size} &
{Slice Width} &
{Slices} &
{FOV} &
{Resolving} \\
{$\mu$m} &
{arcsec} &
{arcsec} &
{\#} &
{arcsec$^2$} &
{Power}
\\[3pt]
\tableheadseprule\noalign{\smallskip}
5.0 to 7.7&0.196&0.176&22&3.00 $\times$ 3.87&2400 to 3700\\
7.7 to 11.9&0.196&0.277&16&3.50 $\times$ 4.42&2400 to 3600\\
11.9 to 18.3&0.245&0.387&16&5.20 $\times$  6.19&2400 to 3600\\
18.3 to 28.3&0.273&0.645&12&6.70 $\times$ 7.73&2000 to 2400
\\
\noalign{\smallskip}\hline
\end{tabular}

\end{table}

The spectral window of each IFU channel is covered using three
separate gratings (i.e., 12 gratings to cover the four channels).
Each grating is fixed in orientation and can be rotated into the
optical path using a wheel mechanism (there are two wheel mechanisms
which each hold 3 pairs of gratings). The optics system for the
four IFUs is split into two identical sections (in terms of optical
layout). One section is dedicated to the two short wavelength IFUs
and the other handles the two longer wavelength IFUs, with one
detector for each section. The two sections share the wheel mechanisms
(each mechanism incorporates three gratings for one of the channels
in the short wavelength section and three gratings for one of the
channels in the long wavelength section). As shown figuratively in
Fig.~\ref{fig045}, the image slicers in the MIRI IFU dissect the
input image plane. The dispersed spectra from two IFU inputs are
placed on one detector side-by-side. The spatial information from
the IFU is spread out into two adjacent rows with the information
from each slice separated by a small gap.

\paragraph{Coronagraphy:}

MIRI features a coronagraph designed for high contrast imaging in 
selected mid-infrared bandpasses. Three quadrant phase masks
(Boccaletti et al 2004; Gratadour et al. 2005) provide high contrast
imaging to an inner working angle of $\lambda/d$, with bandpass of
$\lambda/20$, centered at 10.65 $\mu$m, 11.4 $\mu$m, and 15.5 $\mu$m
respectively. A fourth, traditional Lyot mask of radius 0.9 arcsec,
will provide R $\sim$ 5 imaging at a central wavelength of 23
$\mu$m. Simulations predict that the quadrant phases masks will
achieve a contrast of $\sim 10^4$ at $3 \lambda/D$. The Lyot stop
is predicted to deliver a contrast of $2 \times 10^3$ at $3
\lambda/D$.

\subsubsection{Tunable Filter Imager}

The tunable filter imager (TFI) provides narrow-band near-infrared
imaging over a field of view of  2.2 \ensuremath{\times} 2.2
arcmin$^2$ with a spectral resolution R $\sim$ 100. The etalon
design allows observations at wavelengths of 1.6 $\mu$m to 2.6
$\mu$m and 3.1 $\mu$m to 4.9 $\mu$m, although this design is still
preliminary. The gap in wavelength coverage allows one channel to
reach more than one octave in wavelength. Characteristics of the
TFI are listed in Table~\ref{tab013}.

\begin{table}[t]
\caption{Tunable Filter Imager Characteristics\label{tab013}}
\begin{tabular}{lcl}
\hline\noalign{\smallskip}
{Parameter} & 
{Value} &
{Comments}
\\[3pt]
\tableheadseprule\noalign{\smallskip}
Wavelength range&1.6 to 4.9 \ensuremath{\mu}m&Gap in coverage at 2.6 to 3.1 $\mu$m\\
Field of view&2.2 \ensuremath{\times} 2.2 arcmin&Half the FOV of NIRCam\\
Pixel size&65 milliarcsec&Nyquist sampled at 4.2 \ensuremath{\mu}m\\
Detector array&2048 \ensuremath{\times} 2048&Same as NIRCam long-wavelength\\
&&\hspace{0.25in}detectors\\
Spectral resolution &R \texttt{>} 80&Etalon intrinsic resolution higher\\
Clear aperture&56 mm&Pupil size \ensuremath{\sim} 40 mm. Clear aperture\\
&&\hspace{0.25in}set by etalon location\\
Finesse&\ensuremath{\sim} 30&Minimizes \# of blocking filters\\
Surface figure (P-V)&\texttt{<} 30 nm&Coated etalon surface figure must\\
&&\hspace{0.25in}support reflectance finesse\\
Transmittance&\texttt{>} 75\%&Set by achieved surface figure\\
Contrast&\texttt{>} 100&Peak spectral transmittance divided\\
&&\hspace{0.25in}by minimum between peaks\\
Passband shift&\texttt{<} 1\%&Typical designs have \texttt{<} 0.5\%\\
Blocking filters&\texttt{<} 8&Minimize filter wheel size and\\
&&\hspace{0.25in}simplify operations\\

\noalign{\smallskip}\hline
\end{tabular}

{NOTE --- From Rowlands et al. 2004a, 2004b.}

\end{table}

The TFI uses dielectric-coated Fabry-Perot etalon plates with a
small air (vacuum) gap. The finesse is about 30 and the filters
are used in third order. The finesse was chosen to be a compromise
between the surface figure requirements and the need to minimize
the number of blocking filters, while providing a contrast ratio
of about 100. The filters are scanned using piezo-electric
actuators, consisting of lead-zirconia-titanite ceramic transducers.
The air (vacuum) gap ranges from 2.0 to 8.0 \ensuremath{\mu}m plate
spacings. The Fabry-Perot operates at a nominal temperature of
\ensuremath{\sim} 35 K.

\begin{figure*}
\centering
\includegraphics[width=1.00\textwidth]{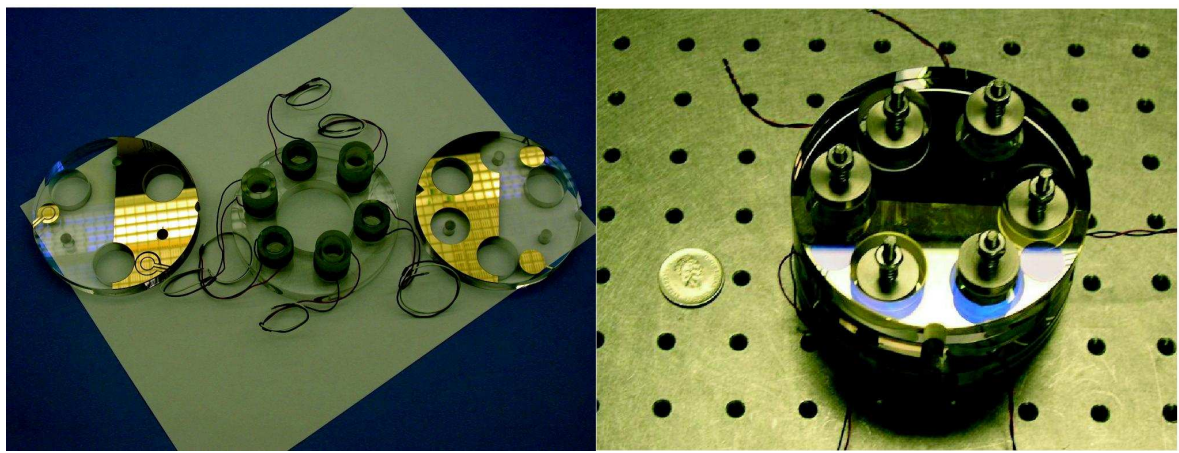}
\caption{
Prototype etalon structure for TFI. Disassembled (left) and assembled
(right). The disassembled view shows the gold coatings for diagnostic
measurements (half circles) and the gold coatings forming the capacitive
displacement pads (small circles) (From Rowlands et al. 2004b).}
\label{fig046}
\end{figure*}

To demonstrate the feasibility of a cryogenic Fabry-Perot etalon
that meets the requirements, Rowlands et al. (2004b) fabricated a
prototype. The etalon surface figure is the most critical requirement,
and is influenced by the coatings, as well as structural and
cryogenic issues (Fig.~\ref{fig046}).

The TFI incorporates four coronagraphic occulting spots permanently
to one side of the field of view, and occupying a region 20 by 80
arcsec. A set of selectable apodization masks is located at the
internal pupil images of each channel by the filter wheels. The
coronagraph will deliver a contrast ratio of $\sim$ 10$^{4}$
(10\ensuremath{\sigma}) at 1 arcsec separation. The sensitivity is
limited by speckle noise. Contrast ratios of 10$^{5}$ may be
achievable at sub-arcsec scales using roll or spectral deconvolution
techniques (Doyon et al. 2004; Sparks \& Ford 2002).

\subsubsection{Fine Guidance Sensor}

The fine guidance sensor (FGS) instrument uses two fields of 
view in the JWST focal plane to provide fine guidance for the 
telescope. The field of view locations are chosen to provide 
optimum lever arms to all instruments for roll about a single 
guide star. Roll is sensed by star trackers on the spacecraft bus.

The FGS consists of an optical assembly and a set of focal plane
and instrument control electronics. The optical assembly of the
FGS instrument consists of two channels, imaging separate regions
of the sky onto independent 2k $\times$ 2k detectors. The
detectors will be HgCdTe, similar to those in NIRCam. The plate
scale is 67 milliarcsec/pixel and the field of view is 2.3 $\times$ 2.3
arcmin$^2$.

The FGS will provide continuous pointing information to the
observatory that is used to stabilize the line of sight, allowing
JWST to obtain the required image quality. Each of the independent
FGS channels provides \texttt{>}90\% probability of obtaining a
useable guide star for any observatory pointing and roll angle.
With both channels, the probability is \texttt{>}95\%. The wavelength
region and pixel size have been optimized so that in fine-guidance
mode the FGS will provide pointing information to a precision of
\texttt{<}5 milliarcsec updated at 16 Hz. The guide stars will be chosen
from existing catalogs with AB \texttt{<} 19.5 mag in the J band
(see section 6.7.1). Fainter stars may be used for location
identification. In the event that a suitable guide star is not
available for a particular desired pointing and roll angle combination,
alternate choices of roll angles can be considered during scheduling.

JWST will be capable of relative pointing offsets with an accuracy 
of 5 milliarcsec rms, which will enable sub-pixel dithering and coronagraphic 
acquisition. Absolute astrometric accuracy will be limited to 1 arcsec 
rms by the accuracy of the guide star catalog.

JWST is expected to be capable of moving object tracking, at 
rates up to 30 milliarcsec s$^{-1}$ relative to the fixed guide stars, although 
some degradation of the image quality may occur. The JWST Project 
is currently doing a cost-benefit analysis of this capability. 
There is no expectation of the ability to follow curved trajectories, 
to track objects continuously as the guide stars cross sensor 
chip boundaries, or 
to observe moving targets in special orientations.

The FGS is designed to be completely redundant in terms of the 
guiding function. The loss of any single component would at most 
result in the loss of one FGS channel. This would reduce the 
probability of guide star acquisition to \ensuremath{\sim} 90\% if using 
current catalogs.

\subsection{Launch, Orbit and Commissioning}

\subsubsection{Launch and Orbit}

\begin{figure*}
\centering
\includegraphics[width=1.00\textwidth]{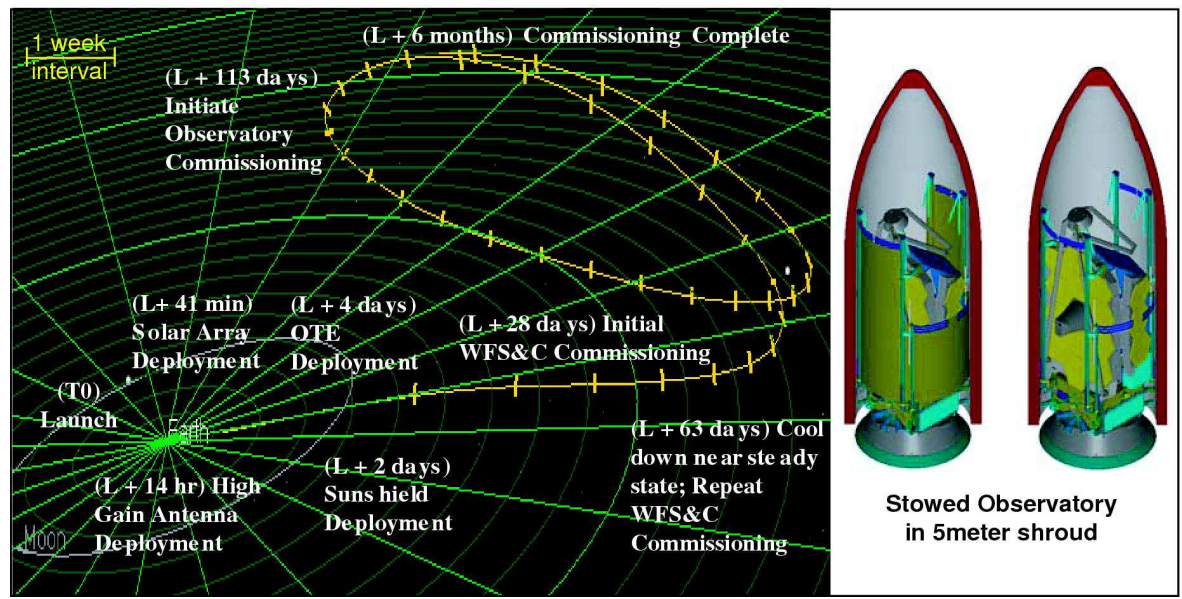}
\caption{
JWST orbit and trajectory to L2, and stowed view in 5 meter
shroud. The deployments will be complete 4 days after launch, the spacecraft will
arrive in the vicinity of L2 in about a month, and commissioning will be 
complete by six months after launch. JWST will carry propellant sized for a 10-year mission.}
\label{fig047}
\end{figure*}

JWST will be launched on an Arianespace Ariane 5 Enhanced Capability-A rocket into orbit about 
the second Lagrange (L2) point in the Earth-Sun system, approximately 1.5 x 10$^{6}$ km 
from the Earth. The orbit (shown in Fig.~\ref{fig047}) 
lies in a plane out of, but inclined slightly with respect to, 
the ecliptic plane. This orbit avoids Earth and Moon eclipses 
of the Sun, ensuring continuous electrical power. JWST will have a 6-month
orbit period about the L2 point in the rotating coordinate system moving with the 
Earth around the Sun. Station-keeping maneuvers 
are performed after the end of each orbit determination period 
of 22 days. 

The observatory fits into launch vehicles with a 5-m fairing
(Fig.~\ref{fig047}), such as the planned Ariane 5. In the figure,
the left view shows the sunshield stowed about the folded telescope.
The right view eliminates the sunshield to show the optical telescope
in its folded configuration. Because the secondary mirror faces
down during launch, it is shielded from particulate redistribution
during launch. This configuration with the stowed sunshield protecting
the folded telescope reduces contamination which would degrade
sensitivity and increase stray light. The total observatory mass
is 6500 kg, including station-keeping propellant sized for a 10-year lifetime.

\subsubsection{Deployment}

\begin{figure*}
\centering
\includegraphics[width=1.00\textwidth]{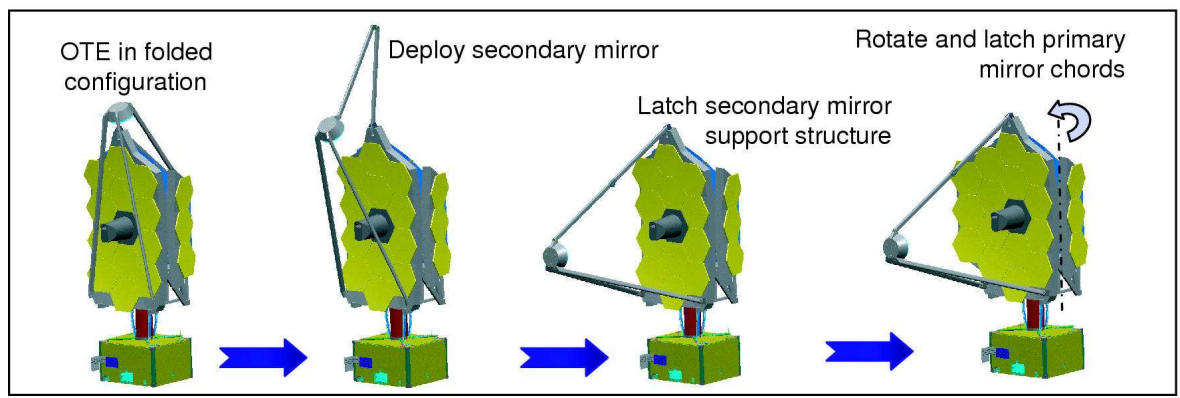}
\caption{
Telescope Deployment Sequence. In deployment steps 4 and 5, as described in the text,
the secondary and primary mirrors are deployed. These deployments will be followed
by wavefront sensing and control operations to align the primary mirror segments.}
\label{fig048}
\end{figure*}

The observatory has the following five deployments: (1) deploy spacecraft 
bus appendages, including solar arrays and the high-gain antenna,
(2) deploy sunshield, (3) extend telescope tower, (4) 
deploy secondary mirror support structure, and (5) deploy primary 
mirror wings. The deployment mechanism design includes heaters 
and other protections that eliminate the need for time-critical 
events and allow for unlatching and re-latching to relieve any 
residual long-term stress in the structure. The secondary-mirror 
deployment and primary-mirror deployment sequence are shown in Fig.~\ref{fig048}.

\subsubsection{Commissioning}

There will be pre-commissioning activities during the transfer 
to L2. About 28 days after launch (Fig.~\ref{fig047}), the observatory 
will cool down to a temperature that permits pre-commissioning 
activities to begin. Almost continuous ISIM availability for 
preliminary science observations is provided from this time until 
106 days after launch, when a final trajectory burn is required 
to achieve orbit about L2. The intrinsic passive-thermal stability 
of the semi-rigid mirror segment and the two-chord-fold primary 
allows early operation of science instruments, by providing a 
stable optical image to the ISIM. These pre-commissioning activities 
will develop an operational experience database that allows formal 
commissioning and science operations to be conducted efficiently. 

A final checkout of all systems is initiated after the L2 orbit 
is achieved. Commissioning (complete 6 months after launch) includes 
ISIM, telescope, sunshield, and spacecraft operations that were not 
feasible during the transfer to L2. It also repeats selected 
operations performed during the transfer, to ensure adequate 
knowledge of system performance in the final orbital-thermal 
conditions and allow comparisons with previous measurements. 
Examples of repeated operations include final optical distortion mappings 
for the observatory and the characterization of the wavefront 
sensing and control actuators' transfer function. Allotting 
76 days for ISIM pre-commissioning activities allows for operations 
rehearsal and increases familiarity with observatory operations 
without reducing operational availability.

\subsubsection{Mission Lifetime}

JWST will operate with all science instruments for at least five 
years after completion of commissioning. In order to exploit 
the full scientific potential of the mission, a lifetime of ten 
years or longer is desired. Although we will not require mission 
assurance to guarantee a lifetime greater than five years, JWST 
will maintain the possibility of a longer mission lifetime, and 
will carry propellant sized for at least 10 years of operation 
after launch. There are no other consumables which would limit 
lifetime.

\subsection{Operations}

JWST will be operated from a Science and Operations Center (S\&OC)
located at the Space Telescope Science Institute (STScI\footnote{STScI
is operated for NASA by the Association of Universities for Research
in Astronomy, Inc. (AURA).}), the organization that operates HST
for NASA. From experience with HST, Spitzer and other astronomical
missions, NASA has learned that operations can be expensive and
inefficient if the scientists and engineers who will operate an
observatory for NASA are not involved in the design and development
of the mission, and so STScI is supporting JWST during the design
and development phases.

Although the capabilities of JWST are being developed to address
science themes discussed in the previous sections, almost all of
the observing time on JWST will be competitively selected.
Approximately 10\% of the observing time for the first 5 years of
the mission has already been awarded to the science instrument
teams and to other members of the science working group. An additional
5\% will be director's discretionary time allocated by the director
of the S\&OC. The remaining $\sim$ 85\% of the observing time will
be awarded through a series of proposal solicitations, which will
be open to any astronomer in the world. The proposal solicitations
will begin a year or two before the anticipated launch of JWST.
This will take advantage of the scientific progress that has been
made in the intervening time both from ground-based and spaced-based
observatories. The scope of JWST's competitively-selected investigations
will range from large legacy-style projects, which last months and
address a range of science goals simultaneously, to small programs
that target important but very specific science objectives. To
maximize JWST's science productivity, the operations concept has
to make JWST easily accessible to those who will use it.

\subsubsection{Operations Concept}

JWST will be located at L2 in order to allow effective cooling 
of the telescope and ISIM. On HST and other low-Earth orbiting 
satellites about half of the heat load is reflected light and 
thermal radiation from the Earth. Other orbits for JWST were 
considered during the design phases of the mission, including 
a drift-away orbit similar to the one utilized by Spitzer, and 
highly elliptical Solar orbits that could have taken a smaller telescope 
out to the orbit of Jupiter. The advantage of L2 is that the 
distance to the observatory is approximately the same throughout 
the mission, so the data rate to and from the observatory can 
be maintained with the same large ground antennae and 1-m class 
downlink antennae.

The L2 orbit creates a number of simplifications for operations 
of a general-purpose astronomical observatory. Unlike HST's low-Earth 
orbit where target visibilities are interrupted every 95 minutes 
by earth occultation, astronomical targets from an L2 orbit are 
visible for long periods of time once or twice a year. Targets 
near the ecliptic plane are visible for two months twice a year; 
targets within 5 degrees of the ecliptic pole are visible continuously
throughout the year. (In low-Earth orbit, observatories such as HST and 
FUSE have continuous viewing zones but these zones actually move 
on the time scale of the orbital precession period of \ensuremath{\sim}57 
days.) The continuous viewing zone of JWST will enable long-term 
monitoring of supernovae and other time variable objects, and 
will allow us to establish relative calibration standards that 
can be observed anytime. 

L2 is a saddle point in the gravitational potential of the Earth-Sun 
system, and as a result propellant is required to maintain JWST 
at this position. In addition, solar pressure on the sunshield 
causes torques on the spacecraft that cause the momentum wheels 
that control the spacecraft pointing to spin up. In low-Earth orbit, the 
angular momentum that builds up (primarily from the residual atmospheric 
drag) is usually dumped using devices that interact with the 
Earth's geo-magnetic field. At L2, this momentum buildup requires 
thruster firings. Since propellant is limited, its usage must 
be carefully monitored to meet the design goal of a 10-year operational 
lifetime.

From an operations perspective, the instruments on JWST are comparable 
in complexity to those on HST, but more complex than those on Spitzer.
This complexity is needed to accomplish the science 
mission of JWST, but has the potential to make the Observatory 
costly to operate if it translates into large numbers of observing 
modes. Therefore the S\&OC, the instrument developers and NASA expect 
to provide a relatively small number of standard modes for each 
instrument for use by observers, with a limited selection of 
readout and dither patterns. Limiting the number of observing 
modes also allows better calibrations of the instruments within 
the limited resources of the JWST lifetime. Limiting the number 
of modes is also important for assuring the quality of the archive, 
and the ease with which archived observations can be used for 
science other than that intended by the original observer. From 
this perspective, it is fortunate that the basic readout architecture 
of all of the detectors on JWST is very similar, if not identical. 

The acquisition of targets with JWST will differ in detail from 
that of HST. The spacecraft will be able to position the observatory 
to an accuracy of about 5 to 7 arcsec using its star trackers, but targets 
must be positioned to accuracies of (in some cases) 5 milliarcsec 
relative to the instruments' apertures. The JWST optical system 
cannot support large-field bore-sight trackers, so the guide stars 
are relatively faint. The new version of the 
guide star catalog (GSC) originally developed for Hubble acquisitions, 
the so-called GSC II (e.g., Spagna et al. 2004), will be used 
for JWST. GSC II extends to AB $\sim$ 19 mag in the J band, at which magnitude 
the areal density of guide stars is high enough for the two 
2.3 $\times$ 2.3 arcmin$^2$ fields of the FGS to reliably contain at least one guide 
star with over 95\% confidence anywhere on the sky. To reduce 
the possibility of acquisition failures, several candidate guide 
stars will be included in command loads to the FGS. Once a suitable 
guide star has been located, acquisition of the science targets 
with the NIRCam, MIRI and TFI will be very similar to acquisition 
procedures used for HST's imaging, long-slit and coronagraphic 
applications. For fine-target positioning, this involves obtaining 
one or more acquisition images with the science instrument, finding 
the centroid of the desired target, and using the centroid position 
to offset the science target to a specific position, such as 
the center of a slit in the science instrument. NIRSpec acquisitions 
(and observation planning) will likely be the most complex since 
the programmable microshutter array is intended to allow an astronomer 
to obtain spectra of up to 100 discrete objects simultaneously. 
This means that one must accurately determine both the desired 
offset and the desired roll from acquisition images obtained 
through the micro-shutter array and multi-star centroiding algorithms 
on-board the spacecraft. In contrast, the imaging modes will 
not require an acquisition sequence, as the 1 arcsec positioning 
accuracy of the FGS will be sufficient.

The observatory is designed to require very little real-time 
commanding. This is an important factor in reducing mission costs 
because continuous communications with the spacecraft will not 
be required. Fewer people will be needed to staff the S\&OC, 
and after commissioning, S\&OC operations will be sufficiently 
automated so that staffing only is needed during a normal workweek. 

\subsubsection{Event-Driven Architecture}

Commanding HST and most other low-Earth-orbiting satellites is 
based upon absolute time. Command loads are uplinked and the 
spacecraft's computer executes each command at a specific absolute 
time. In some cases, a sequence of relative time commands is 
issued, but the underlying principle of this approach is that 
one can determine exactly when commands start and stop. This 
commanding concept was well suited to a time when on-board computers 
were extremely primitive by modern standards, and is still appropriate 
in situations where external factors, such as regular Earth occultation 
in low-Earth orbit of an inertially-pointed satellite, require 
that many commands be issued at specific times. 

JWST's observing program will have few absolute time requirements. 
Earth blockage is not a problem at L2. Data downlinks and command 
up-links will not interrupt science observations, as is the case 
for Spitzer, since JWST will have a gimbaled antenna. Real-time 
commanding will be used only for critical operations, such as 
station-keeping maneuvers or recovery from anomalies. JWST will 
use a different commanding approach that will be event driven 
rather than absolute-time driven. The commanding concept for 
JWST is like a command queue; commands in the queue are executed 
sequentially and the next command starts when an indication is 
received that the previous command is completed (successfully 
or unsuccessfully). This approach sacrifices certainty, but simplifies 
the software system that is needed for operations planning and 
lowers operations cost, both in the mission's development and 
operational phases. It avoids the need to accurately model execution 
times for every command and then to update that model as the 
flight software and hardware evolve. It also improves observational 
efficiency since a basic tenet of absolute-time commanding is 
that if a command does not complete normally, the spacecraft 
must either ignore the failure or wait until the next observation 
is scheduled. For JWST, that wait could be as long as 24 hours. 
By contrast, if an observation fails in an event-driven approach 
(for example, by failing to acquire a guide star), the observatory 
can move on to the next observation or the next target. Similarly, 
with an event-driven approach, if a single instrument goes into 
a failure or ``safe'' mode, observations with that instrument 
are skipped and observations with other instruments are brought 
forward in time. This allows JWST to continue to conduct high-quality 
science without loss of overall efficiency. 

\subsubsection{Visits and Observation Plan}

The event-driven architecture for JWST will be implemented through 
a construct of visits and an observation plan. A visit is simply 
a logically grouped series of activities, along with a set of 
conditions that can be identified at the beginning of the series 
and which must be met for the visit to be executed. Typically, 
a visit would comprise a slew to a new target, the steps required 
to finely point at the target, the setup of the instrument for 
the observation itself, and the acquisition of all of the science 
data at that pointing position. The constraints for executing 
the visit would include the availability of the prime science 
instrument for the observation and a time window during which 
the visit had to start. The observation plan constitutes the 
single queue that orders the visits. Software in the on-board computer 
executes the observation plan. Although the visit activities 
can be relatively complex, the role of the software is to initiate 
the next visit after the previous visit is complete, to wait 
until the start time for the next visit, or to skip to the next 
visit in the queue. If a visit ends early because of a failure 
to identify the appropriate guide stars, for example, the on-board 
software starts the next visit in the sequence. The only time 
the observatory waits is when the earliest start of the next 
visit is still in the future. 

The operational approach is science driven and intended to be 
simple and understandable to the user. Astronomers will not need 
to visit the S\&OC for their observations. The interaction of astronomers 
with JWST will be very similar to those of other space-based 
missions designed for use by large numbers of astronomers.

The S\&OC will solicit proposals for JWST observations annually, 
on behalf of NASA, ESA and CSA. Following the Spitzer model, it is expected 
that the first year's observations will be dominated by a small
number of large Legacy-style programs and the guaranteed-time
observations submitted by the guaranteed-time observers (the
instrument teams and other science working group members). In that
case, a transition to a larger number of investigations (probably
100 to 200 programs annually) will occur after the first year.
Those proposing observations will use an integrated planning tool,
consisting of a graphical user interface and associated widgets,
including exposure-time calculators and tools for importing sky
maps and accurately positioning the JWST apertures on targets. When
writing a proposal, the astronomer will complete those portions
required for scientific and technical assessment, including any
special requirements for timing and/or fixed orientation of the
observatory (Phase I). As with Spitzer, time will effectively be
allocated in ``wall-clock time'', including estimates for slew and
setup times. Once the proposal is selected, the scientist will fill
in the remaining details that would be required to execute the
approved observations on JWST (Phase II). This two-phase approach
was largely developed by STScI for HST to allow astronomers to
spend most of their time before selection creating
scientific justifications for their ideas. Only those whose 
proposals are selected must
develop the additional details needed to execute the observations
on the spacecraft.

Following program selection, the S\&OC will construct a long-range 
plan for observing with JWST. The long-range plan will consist 
of the approved science programs and the calibration and maintenance 
activities for the observatory. The same planning tools used 
by general observers for their science programs will be used 
for the calibration and maintenance programs. This will reduce 
cost and complexity within the ground system by limiting the 
total number of planning systems. It will also provide better 
service to users since calibration and operations scientists 
and operators will use the same tools. The long-range plan will 
specify possible scheduling windows for all the observations 
and calibration activities for the coming year. The scheduling 
windows are not simply the times during the year when a target 
is in the field of regard of the observatory, but are limited 
to assure that there is a high probability that most (90\%) of 
the approved observations will be carried out during the year. 
This involves assuring that there is a good mix of ``easy'' and 
``hard'' observations that are available for scheduling when short-term 
scheduling takes place. Examples of ``hard'' observations include 
sets of observations that are extremely long or that have tight 
timing or orientation constraints. The long-range plan will be updated 
as observations are completed and modified to reflect new proposals 
as they are approved and made ready for observation with JWST.

The long-range plan provides a pool of observations that must 
be turned into a sequence of observations to be executed on the 
observatory. Development of this sequence, which is known as 
the observational plan, will normally begin about one month before 
observations are executed on the observatory. Scheduling is carried 
out as close to the actual observing time as practical so that 
the latest information about the observatory and the potential 
observations can be incorporated into the short-term scheduling
process. A typical observation plan will likely last about 22 days, the
timescale on which orbit maintenance would be carried out. Shorter
observation plans will be used during commissioning, for contingency
operations, and for some time-critical activities. Automated tools
will help planners create the observation plan from the available
targets using a priority-based system that ranges from targets that
must be scheduled during the 22 days, to those that can be scheduled
in a number of scheduling intervals. Imaging observations at $\lambda
< 10 \mu$m are limited in sensitivity by the zodiacal light, and
so sensitivity is maximized when the observations are done at high
solar elongation.

Since JWST does not have to contend with Earth eclipses or South
Atlantic Anomaly passages as HST does, the optimal schedule is one
that maximizes the number of higher priority observations while
minimizing slew lengths, and managing the amount of momentum stored
in the momentum wheels and the peak data volume on the recorder.
JWST is intended to be efficient and science exposures are currently
expected to use 70\% of wall-clock time, although this depends on
the mix of short and long observations. To achieve this high
efficiency, JWST has been designed so that certain calibration data
can be obtained from one instrument while another instrument, the
prime science instrument, is used for science. During creation of
the observation plan, parallel calibration observations will be
integrated with the science-observing plan in a way that maximizes
their utility, but does not adversely affect the science observation.
No ``science parallels'' are currently planned, although there are
no restrictions in the observatory that would prevent them.

The observation plan, along with all of the associated information
describing each individual pointing, will be broken into segments
and sent to JWST at weekly intervals, during one of the daily
contacts with the observatory. If for some reason observations are
missed, it will be possible to uplink a new plan whenever JWST is
in ground contact. Usually the S\&OC will let JWST continue with
the same operations plan even if a few observations have been
missed. Communications with JWST will be through NASA's Deep Space
Network. Real-time communications and commanding of the
spacecraft will be conducted via S band, while high-speed data
downlink will use Ka band. The solid-state recorder and downlink
will be sized for a typical daily load of 232 Gbit (compressed)
for science and engineering data. This data volume is required
because of the 80 Mpix detector arrays used for science observations
with readouts every 20 to 200 s. These non-destructive rapid readouts
are required to compensate for the effects of cosmic rays on the
detectors at L2. Typically, daily ground contacts lasting about three
hours in total duration will occur to uplink new commands and
downlink up to 232 Gbit of data. These contacts will also provide
Doppler tracking and ranging data for orbit determination.

On rare occasions, it will be necessary to interrupt the JWST 
observing plan to make target of opportunity observations of 
time-critical events, such as supernovae or gamma-ray burst sources. 
By constructing and uploading a new observation plan, JWST will 
be able to make observations of targets of opportunity on timescales 
as short as 2 days from the decision to do so.

HST is protected from cosmic rays and high-energy particles from 
the Sun by the Van Allen belts. At L2, however, the quiescent 
flux is expected to be 5 to 10 particles cm$^{-}$$^{2}${\nobreakspace}s$^{-}$$^{1}$. 
The charge that is deposited as these particles pass through 
the detectors corrupts about 5 to 10 percent of the pixels in 
1000 s. To limit the effect of this on sensitivity, images must 
be collected for transmission to the ground over a shorter (20 
to 200 s) time period, although this uses non-destructive readouts 
and the detectors are not reset this often. The portions of each image
that are affected by cosmic rays can be removed as part of data processing on the ground. 
An alternative would have been to attempt to remove the bad pixels 
on the spacecraft; however, this is computationally expensive 
and risks the possibility that the detectors 
will not perform exactly as predicted on orbit. Consequently, 
the raw images will be sent to the ground after lossless compression. 

During the commissioning phase of JWST, the telescope's flight 
operations system will be staffed 24 hours a day, 7 days a week, 
and real-time contact will be maintained with the observatory 
for all critical operations. Engineers and scientists from Northrop 
Grumman, the instrument teams, and NASA, ESA and CSA will participate fully 
in a joint mission operations team in the commissioning of the 
observatory and its instruments. Much of the deployment and initial 
turn-on will require real-time commanding, but a gradual transition 
to semi-autonomous operations will commence as soon as feasible. 

The ground operations software will be constructed around the
largely commercial-off-the-shelf command and telemetry system that
will have been used in JWST's integration and testing. Although
real-time contact with the spacecraft will typically be
three hours a day in normal operations, much of the data will be
cached at ground stations and transmitted within 24 hours back to
the S\&OC. Data received by the flight operations data management
system will normally have been processed to create error-corrected,
compressed data packets but the S\&OC will have the capability to
complete the initial level of processing in the event of contingencies.
The operations system will automatically monitor the performance
and state of health of the observatory, integrate mission scheduling
with and oversee execution of the observation plan, and handle the
initial receipt of data. It will include capabilities for automatic
notification of staff in the case of anomalies, since the experts
required to deal with specific problems may not be physically
present at the S\&OC when an anomaly is discovered. The flight
operations team within the S\&OC will be responsible for uplink
and verification of flight software tables and loads required to
update and maintain the observatory. The uplinks will be carried
out in real time, but are not expected to interrupt normal operations.

The S\&OC is responsible for preserving the image quality of 
the telescope through a sequence of wavefront sensing and control 
visits. During the commissioning phase, most of the initial telescope 
adjustments will be carried out via real-time commanding, but 
in normal operations the wavefront-sensing visits, and other 
JWST calibration observations, will be executed in the same way 
as science observations, and carried out under the supervision 
of the flight software. The visits will usually consist of NIRCam 
observations of a field with one or more bright stars through 
a set of special-purpose filters and lenses. The wavefront-sensing 
visits will occur weekly, or on a time scale that is short compared 
with the expected changes in the mirror. Data from the wavefront-sensing
visits will be retrieved from the observatory and through the Deep
Space Network on a priority basis. The S\&OC will analyze the
images. When the quality of the images degrades to the degree that
warrants an update, mirror-actuator corrections will be generated
and sent to the spacecraft. At the next wavefront-sensing and
control visit, the corrections will be carried out and data before
and after the update recorded.

Maintaining the observatory's L2 orbit is the only regular activity 
that will require real-time commanding. Because orbital maintenance 
is so critical to the safety of the mission, it will be carried 
out during ground contact when operations engineers are able 
to monitor the performance of the observatory.

An archive of all of the data obtained from the observatory will
be maintained in the S\&OC. The JWST data management system will
use the capabilities of the Multi-Mission Archive at Space Telescope.
Originally developed for HST, this archive now houses data from a
variety of NASA space astronomy missions. The JWST archive will
contain the data stored there in its raw form, the calibration
files necessary to calibrate the data, and databases that describe
the data. As with HST, a safe-store archive will be maintained at
a separate site. Scientific data received from the spacecraft will
be processed within 48 hours to assess quality using the calibration
files available at the time of the observations and to make an
initial version of the science data available to the principal
investigator.

When the HST archive was initially built in the early 1990s, 
users received science data that were calibrated at the time of 
the observation. A user who retrieved data a year later did not 
benefit from ongoing improvements in the calibration pipeline 
or the results of calibrations that took place at the time of 
the observations, unless he or she recalibrated the data. Since 
then, however, the price of processing has dropped considerably. 
Now, HST archive data are normally reprocessed ``on-the-fly'' 
each time a request arrives. The same approach will be used for 
JWST. Users will request data from the archive using Web-based 
tools similar to those used for HST data. Although a few users 
may want their data on a physical medium, most are expected to 
retrieve data directly to their home institution via the Internet.

\subsection{Management}

JWST is a partnership of NASA, ESA and CSA. NASA's Goddard Space 
Flight Center (GSFC) provides overall project management, systems 
engineering, and scientific leadership for the project. The prime 
contractor is Northrop Grumman Space Technologies (NGST), with 
major sub-contracts to Ball Aerospace, ITT and Alliant Techsystems. 
The Space Telescope Science Institute (STScI) is the science and 
operations center. The responsibilities 
of the major JWST partners are given in Table~\ref{tab014} (Sabelhaus \& 
Decker 2004). 

\begin{table}[t]
\caption{Responsibilties of the major JWST Partners\label{tab014}}
\begin{tabular}{ll}
\hline\noalign{\smallskip}
{Organization} & 
{Responsibilities}
\\[3pt]
\tableheadseprule\noalign{\smallskip}
Goddard Space Flight&Overall project management\\
\hspace{0.25in}Center&Overall systems engineering\\
&Scientific leadership\\
&Integrated science instrument module (ISIM)\\
&ISIM assembly, integration and test (AI\&T)\\
&Micro shutter assembly and detectors for\\
&\hspace{0.25in}NIRSpec\\
Northrop Grumman&Observatory systems engineering 
and interfaces\\
\hspace{0.25in}Space Technology&Observatory AI\&T\\
&Telescope, spacecraft bus and sunshield design,\\
&\hspace{0.25in}manufacturing, 
AI\&T\\
&Launch site processing, observatory launch and\\
&\hspace{0.25in}commissioning\\
&Observatory performance and programmatics\\
&Spacecraft, sunshield, and deployables\\
&Support ground segment and operations\\
Ball Aerospace&Telescope optical design and optics\\
&Beryllium mirror segment development and\\
&\hspace{0.25in}cryogenic testing\\
&Wavefront sensing \& control (WFS\&C) design\\
&\hspace{0.25in}and algorithms\\
&Telescope and observatory AI\&T support\\
ITT&Telescope ground AI\&T\\
&Thermal vacuum test configuration and\\
&\hspace{0.25in}interfaces\\
Alliant Techsystems&Telescope backplane and secondary mirror\\
&\hspace{0.25in}support structure\\
Space Telescope&Ground systems development\\
\hspace{0.25in}Science Institute&Flight and science operations\\
&Optics and instrument support\\
&Science program peer review and selection\\
European Space Agency/&\\
\hspace{0.25in}European Consortium&MIRI optical bench assembly\\
\hspace{0.25in}Arianespace&Ariane 5 launch vehicle\\
\hspace{0.25in}EADS Astrium&NIRSPec Instrument\\
Canadian Space Agency/&Fine guidance sensor\\
\hspace{0.25in}COM DEV&Tunable filter imager\\
University of Arizona/&\\
\hspace{0.25in}Lockheed Martin ATC&NIRCam instrument\\
Jet Propulsion Laboratory&MIRI management\\
&MIRI detectors, cooler, software and end-to-end\\
&\hspace{0.25in}verification\\
&Wave front sensing and control technology\\
Rockwell Science Center&NIR detectors\\
Raytheon Vision Systems&MIR detectors\\
Marshall Space Flight&Primary mirror technology development and\\
&\hspace{0.25in}testing\\
\hspace{0.25in}Center&Environmental analysis\\
Ames Research Center&Detector technology development\\
Johnson Space Center&Thermal vacuum test facility
\\
\noalign{\smallskip}\hline
\end{tabular}
\end{table}

\section{Summary}

The JWST science requirements are divided into four themes. The
key objective of The End of the Dark Ages: First Light and Reionization
theme is to identify the first luminous sources to form and to
determine the ionization history of the early universe. The key
objective of The Assembly of Galaxies theme is to determine how
galaxies and the dark matter, gas, stars, metals, morphological
structures, and active nuclei within them evolved from the epoch
of reionization to the present day. The key objective of The Birth
of Stars and Protoplanetary Systems theme is to unravel the birth
and early evolution of stars, from infall on to dust-enshrouded
protostars to the genesis of planetary systems. The key objective
of the Planetary Systems and the Origins of Life theme is to
determine the physical and chemical properties of planetary systems
including our own, and investigate the potential for the origins
of life in those systems. Within these themes and objectives, we
have derived representative astronomical observations.

To enable these observations, JWST consists of a telescope, an
instrument package, a spacecraft and a sunshield. The telescope
consists of 18 beryllium segments, some of which are deployed after
launch. The segments will be brought into optical alignment on-orbit
through a process of periodic wavefront sensing and control. The
instrument package contains the four science instruments and a fine
guidance sensor. The spacecraft provides pointing, orbit maintenance
and communications. The sunshield provides passive thermal control.
The JWST operations plan is based on that used for previous space
observatories, and the majority of JWST observing time will be
allocated to the international astronomical community through annual
peer-reviewed proposal opportunities.

In this paper we have described the astronomical observations 
that JWST is designed to make and the implementation that will 
enable those observations. We have provided scientific justifications 
for these observations within 4 science themes. The themes are 
chosen to span the range of science that we expect JWST will 
do, and the observations are chosen to be representative of that 
science. We wish to emphasize, however, that these observations 
will not necessarily be done. The majority of observing time 
on JWST will be allocated to the worldwide astronomical community 
through competitive selection of peer-reviewed proposals. Opportunities 
for proposals will begin one or two years before launch, and 
will continue annually for the lifetime of the mission. JWST 
will be a highly capable general-purpose observatory able to 
address a very wide range of scientific investigations. It represents 
a major contribution to scientific progress by the governments 
of the United States of America, of the European nations and of Canada. 
Regular, competitive peer-reviewed proposal selection will ensure 
that this international resource will address the most relevant 
and strongly justified scientific questions, and will leave a 
legacy of knowledge and discovery for future generations.

\end{document}